\documentclass[referee,onecolumn]{aa}

\usepackage{graphicx}
\usepackage{txfonts}
\usepackage{lscape}
\usepackage{caption}
\usepackage{stackengine}
\usepackage{hyperref}
\hypersetup{colorlinks=true, linkcolor=blue, filecolor=blue, urlcolor=blue, citecolor=blue}
\usepackage{gensymb}
\usepackage{textgreek}
\usepackage{natbib}
\begin{document}
    \title{A New Multi-Wavelength Census of Blazars}
    \author{A. Paggi
        \inst{1}\fnmsep\inst{2}
        \and
        M. Bonato\inst{3}\fnmsep\inst{4}
        \and
        C. M. Raiteri\inst{1}
        \and
        M. Villata\inst{1}
        \and
        G. De Zotti\inst{4}
        \and
        M. I. Carnerero\inst{1}
    }
    \institute{
        INAF, Osservatorio Astrofisico di Torino, via Osservatorio 20, 10025 Pino Torinese, Italy\\
        \email{alessandro.paggi@inaf.it}
        \and
        INFN – Istituto Nazionale di Fisica Nucleare, Sezione di Torino, via Pietro Giuria 1, I-10125 Turin, Italy
        \and
        INAF, Istituto di Radioastronomia, via Piero Gobetti 101, I-40129 Bologna, Italy
        \and
        INAF, Osservatorio Astronomico di Padova, Vicolo dell’Osservaorio 5, I-35122 Padova, Italy
    }
    \date{Received May 16, 2020; accepted June 14 2020}
    \abstract
        {Blazars are the rarest and most powerful active galactic nuclei, playing a crucial and growing role in today multi-frequency and multi-messenger astrophysics. Current blazar catalogs, however, are incomplete and particularly depleted at low Galactic latitudes.}
		{We aim at augmenting the current blazar census starting from a sample of ALMA calibrators that provides more homogeneous sky coverage, especially at low Galactic latitudes, to build a catalog of blazar candidates that can provide candidate counterparts to unassociated \(\gamma\)-ray sources and to sources of high-energy neutrino emission or ultra-high energy cosmic rays.}
		{Starting from the ALMA Calibrator Catalog we built a catalog of 1580 blazar candidates (ALMA Blazar Candidates, ABC) for which we collect multi-wavelength information, including \textit{Gaia}, SDSS, LAMOST, WISE, X-ray (\textit{Swift}-XRT, \textit{Chandra}-ACIS and \textit{XMM-Newton}-EPIC), and \textit{Fermi}-LAT data. We also compared ABC sources with existing blazar catalogs, like 4FGL, 3HSP, WIBRaLS2 and the KDEBLLACS.}
		{The ABC catalogue fills the lack of low Galactic latitude sources in current blazar catalogues. ABC sources are significantly dimmer than known blazars in \textit{Gaia} g band, and they appear bluer in SDSS and WISE colors than known blazars. In addition, most ABC sources classified as QSO and BL Lac fall into the SDSS colour regions of low redshift quasars.
		Most ABC sources (\(\sim 90\%\)) have optical spectra that classify them as QSO, while the remaining sources resulted galactic objects.
		ABC sources are on average similar in X-rays to known blazar, while in \(\gamma\)-rays they are on average dimmer and softer than known blazars, indicating a significant contribution of FSRQ sources. Making use of WISE colours, we classified 715 ABC sources as candidate \(\gamma\)-ray blazar of different classes.}
		{We built a new catalogue of 1580 candidate blazars with a rich multi-wavelength data-set, filling the lack of low Galactic latitude sources in current blazar catalogues. This will be particularly important to identify the source population of high energy neutrinos or ultra-high energy cosmic rays, or to verify the \textit{Gaia} optical reference frame. In addition, ABC sources can be investigated both through optical spectroscopic observation campaigns or through repeated photometric observations for variability studies. In this context, the data collected by the upcoming LSST surveys will provide a key tool to investigate the possible blazar nature of these sources.}		
    \keywords{
    	Catalogs--
        Galaxies: active
    }
   \maketitle
    \section{Introduction}\label{sec:introduction}
    
        Blazars are the rarest and most powerful active galactic nuclei. Their emission is dominated by variable, non-thermal radiation that extends over the entire electromagnetic spectrum, high and variable polarization, apparent superluminal motion, and high luminosities characterized by intense and rapid variability \citep[e.g.,][]{1995PASP..107..803U, 2013MNRAS.431.1914G}. These observational properties are generally interpreted in terms of a relativistic jet aligned within a small angle to our line of sight \citep{1978bllo.conf..328B}. Traditionally blazars have been classified in two main subclasses, as BL Lac objects and FSRQs, with the former showing featureless optical spectra, while the latter are characterized by strong quasar emission lines, as well as higher radio polarization. More specifically, if the only spectral features observed are emission lines with rest-frame equivalent width \(EW<5\,\AA\), the object is classified as BL Lac \citep{1991ApJ...374..431S, 1997ApJ...489L..17S}, otherwise it is classified as FSRQ \citep{1999ApJ...525..127L}. The blazar spectral energy distributions (SEDs) typically show two peaks: one in the range of radio-soft X-rays due to synchrotron emission by highly relativistic electrons within the jet, and another one at hard X-ray or \(\gamma\)-ray energies. The latter is interpreted as inverse Compton upscattering by the electrons in the jet on the seed photons provided by the synchrotron emission \citep[synchrotron self-Compton, SSC, see for example][]{1996ApJ...463..555I} which dominates the high energy output in BL Lacs. The possible addition of seed photons from outside the jets yields contributions to the non-thermal radiations in the form of external inverse Compton scattering \citep[EC, see][]{1993ApJ...416..458D, 2009ApJ...692...32D} often dominating the \(\gamma\)-ray outputs in FSRQs \citep{2009A&A...502..749A, 2011ApJ...743..171A}.
    
        Blazars are playing a crucial and growing role in today multi-frequency and multi-messenger astrophysics: they dominate the high-energy (from MeV to TeV) extragalactic sky \citep{DiMauro2018, 2019AA...632A..77C, Chiaro2019} and recently have been associated \citep{IceCubeCollaboration2018, Garrappa2019} to high-energy astrophysical neutrinos. In turn, these neutrinos arise from interactions of ultra-high energy cosmic rays \citep[UHECR;][]{Sarazin2019} via charged pion decay. Thus blazars jets may also be among the long sought accelerators of the UHECR.

        The 5th edition of the \textit{Roma-BZCAT} Multifrequency Catalogue of Blazars \citep[BZCAT,][]{2015Ap&SS.357...75M} is the most comprehensive list of blazars confirmed by means of published spectra. This catalogue contains 3561 entries, and for each source it lists radio coordinates obtained from very-long-baseline interferometry measurements \citep[VLBI,][]{2009AA...506.1477T, 2011AA...529A..91T, 2011AJ....142...89P}, augmented with available multi-wavelength information like optical magnitude from USNO B1 or SDSS DR10, radio flux density from NVSS, \citep{1998AJ....115.1693C}, FIRST \citep{1997ApJ...475..479W}, SUMSS \citep{2003MNRAS.342.1117M}, GB6 \citep{{1996ApJS..103..427G}} or PMN \citep{1994ApJS...91..111W}), the microwave flux density from \textit{Planck} \citep{2014AA...571A..28P}, the soft X-ray flux from ROSAT archive or Swift-XRT catalogues, the hard X-ray flux from Palermo BAT Catalogue \citep{2010AA...524A..64C}, the \(\gamma\)-ray flux from the first \citep[1FGL,][]{2010ApJS..188..405A} and second \citep[2FGL,][]{2012ApJS..199...31N} \textit{Fermi} catalogues, and the redshift.
        
        Confirmed blazars listed in the \textit{BZCAT} are subdivided in:
        \begin{itemize}
            \item 1059 \textbf{BZBs}: BL Lac objects,
            \item 1909 \textbf{BZQs}: Flat Spectrum Radio Quasars.
        \end{itemize}
        In addition the 5th edition of \textit{BZCAT} lists 92 BL Lac candidates, that is sources classified as BL Lac in the literature, but without published optical spectra. In the following we will consider these sources as BZBs.
        Other kind of sources listed in \textit{BZCAT} are:
        \begin{itemize}
            \item 274 \textbf{BZGs}: sources classified as BL Lacs in the literature but with a host galactic emission dominating the nuclear one
            \item 227 \textbf{BZUs}: blazars of uncertain type.
        \end{itemize}

        Blazar catalogues as large and complete as possible are necessary to provide candidate counterparts to unassociated \(\gamma\)-ray sources detected by the \textit{Fermi} satellite and to sources of high-energy neutrino emission or UHECRs. In particular, a complete sky coverage of such catalogues is critically important in the case of rare events such as detections of high-energy neutrinos. The identification of the neutrino-emitting source population requires to utilize as fully as possible the all-sky neutrino sample by means of stacking analyses towards candidates. To this end we obviously need catalogues covering the whole sky.

\begin{figure*}
	\centering
	\includegraphics[scale=0.23]{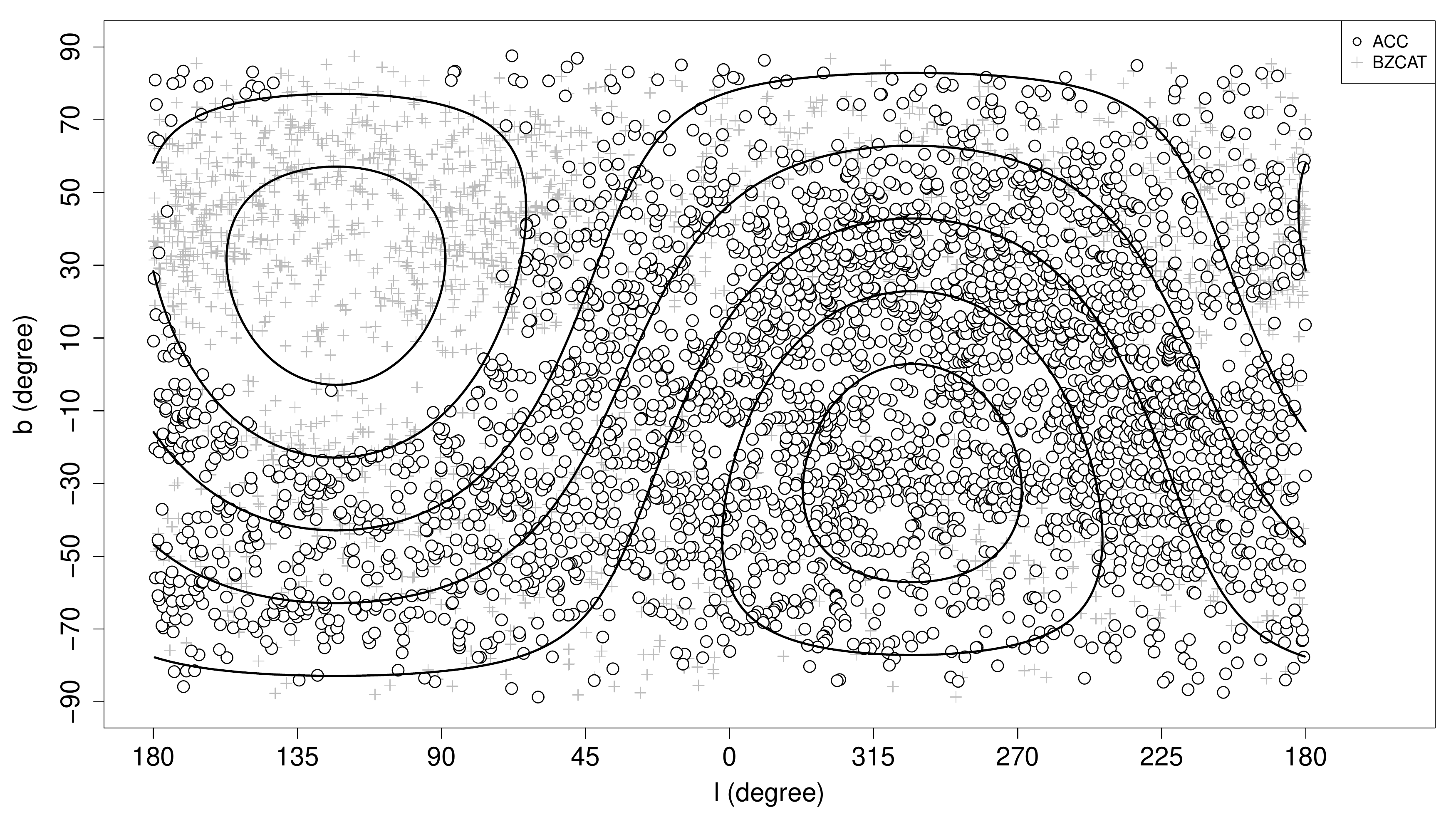}
	\includegraphics[scale=0.30]{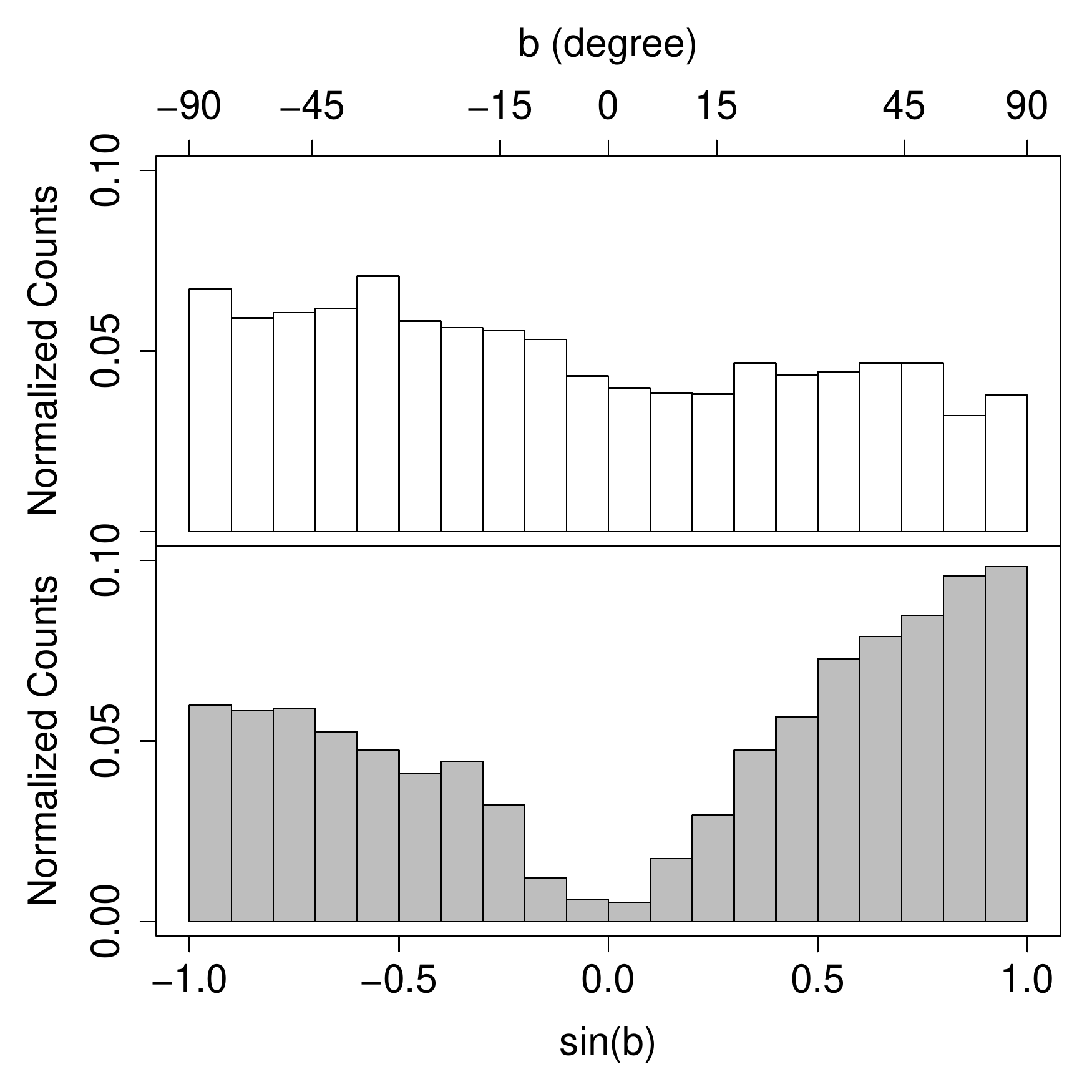}
	\caption{(Left Panel) The white circles indicate the Galactic coordinates of the ACC sources. The grey crosses in the background represent the Galactic coordinates of the \textit{BZCAT} sources. The black lines represent equatorial coordinates with declinations from \(-60\degree\) to \(60\degree\) with increment of \(20\degree\). (Right Panel) Normalized Galactic latitude distribution of ACC (top) and \textit{Bzcat} (bottom) source.}\label{fig:galactic_projection}
\end{figure*}

        The Atacama Large Millimeter Array (ALMA) calibrators are compact sources bright at mm and sub-mm wavelengths. They were primarily drawn from "seed" catalogues, such as those of the Very Long Array (VLA), of the SubMillimeter Array (SMA), of the Australia Telescope Compact Array (ATCA), and of the Combined Radio All-Sky Targeted Eight-GHz Survey (CRATES)\footnote{\href{https://almascience.eso.org/alma-data/calibrator-catalogue}{https://almascience.eso.org/alma-data/calibrator-catalogue}}. The ALMA Calibrator Catalogue \citep[ACC;][]{2019MNRAS.485.1188B} contains 3364 sources, whose distribution in Galactic coordinates is presented in the left panel of Fig. \ref{fig:galactic_projection}. Because of the ALMA location in the Southern Hemisphere, its calibration sources have declination lower than \(\sim +60\degree\); this explains the "hole" centered at about \(l=-120\degree\) and \(b=30\degree\) in the left panel of the figure.

        The distribution of the Galactic latitude of ACC sources is presented in the upper part of the right panel of the same figure. The Galactic north-south asymmetry in latitude distribution is evident, with \(\sim 50\%\) of the ACC sources having \(b<-10\degree\) and \(\sim 35\%\) having \(b>10\degree\). In particular we note that we have \(\sim 15\%\) of the sources at low Galactic latitudes, that is, \(\left|{b}\right|<{10}\degree\). This is at variance with the sources listed in the \textit{BZCAT}, whose Galactic latitude distribution is shown in the lower part of the right panel of Fig. \ref{fig:galactic_projection}. In this case we have \(\sim 40\%\) of the \textit{BZCAT} sources with \(b<-10\degree\) and \(\sim 57\%\) with \(b>10\degree\), while only \(\sim 3\%\) of the \textit{BZCAT} sources have \(\left|{b}\right|<{10}\degree\). This is manly due to the fact that the firm blazar classification adopted in \textit{BZCAT} catalogue is based on optical spectroscopy, which is problematic along the Galactic plane due to strong absorption and source crowding.

        The ALMA Calibrator Catalogue represents therefore an excellent starting point from which we can build a new catalogue of blazar candidates, with the goal of completing the census of blazars especially a low galactic latitudes where the current blazar catalogues are depleted.

        In this paper, after an analysis of the sample of ALMA calibrators to select bona fide blazar candidates (Sect. \ref{sec:sample_selection}), we compare the sample of selected sources, referred to as ALMA Blazar Candidates (ABC) catalogue with other blazar catalogues (Sect. \ref{sec:catalog_comparison}). Next (Sect. \ref{sec:multi_lambda}) we collect multi-wavelength data on ABC sources by cross-matching our catalogue with public infrared, optical, and \(\gamma\)-ray catalogues, and by performing an extensive X-ray analysis of available data. In Sect. \ref{sec:candidates}, following \citet{2019ApJS..242....4D}, we use Wide-Field Infrared Survey Explorer \citep[WISE,][]{2010AJ....140.1868W} data to select ABC sources whose mid-infrared colours are consistent with those of confirmed \(\gamma\)-ray emitting blazars and to assign them to blazar sub-classes. Finally, in Sect. \ref{sec:conclusion} we summarize our conclusions.

    \section{Sample selection}\label{sec:sample_selection}
        In this section we compare the radio properties of ACC sources with those of \textit{BZCAT} source, and select a sample of ACC sources not included in \textit{BZCAT} to build a catalog of ALMA blazar candidates.
        
\begin{figure*}
	\centering
	\includegraphics[scale=0.38]{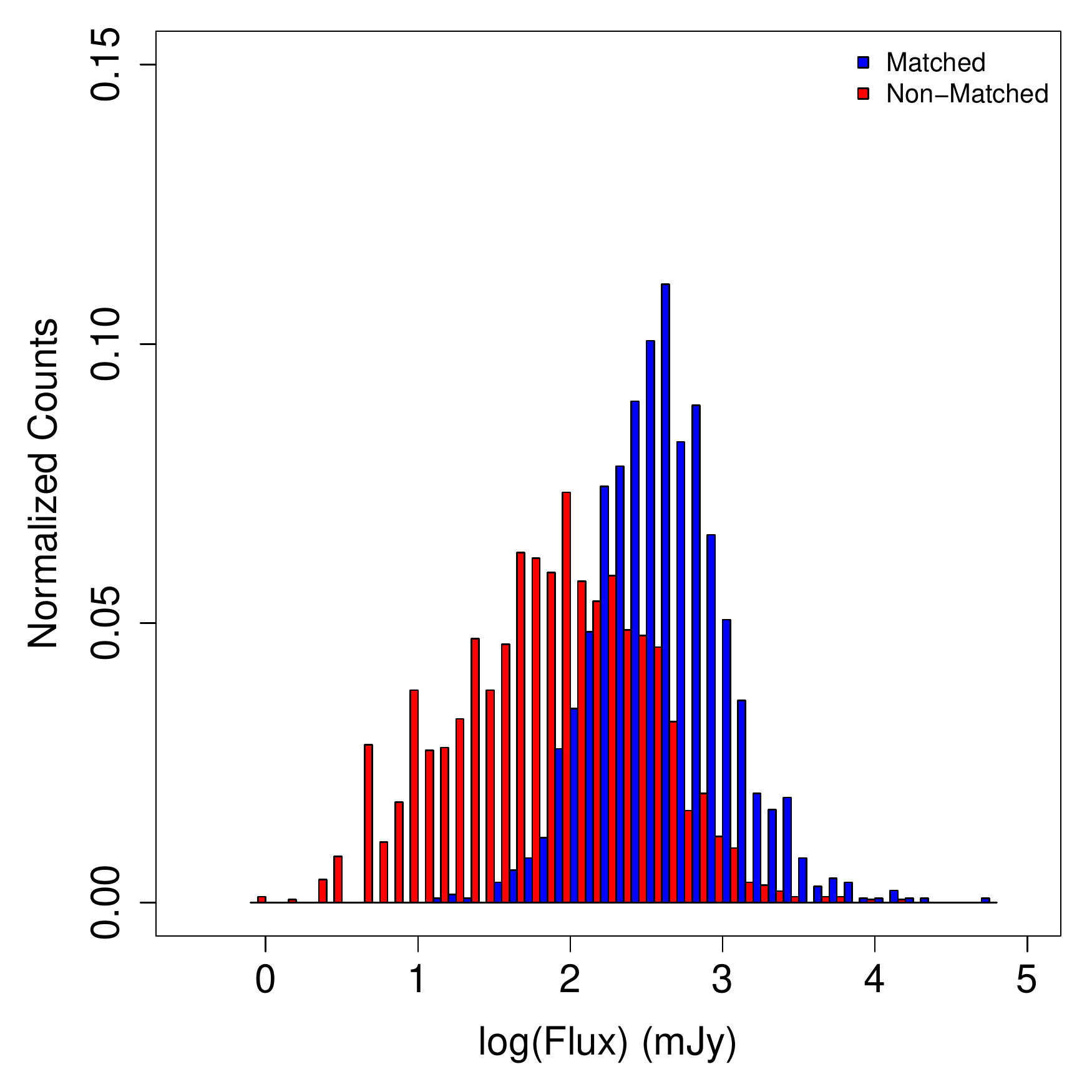}
	\includegraphics[scale=0.38]{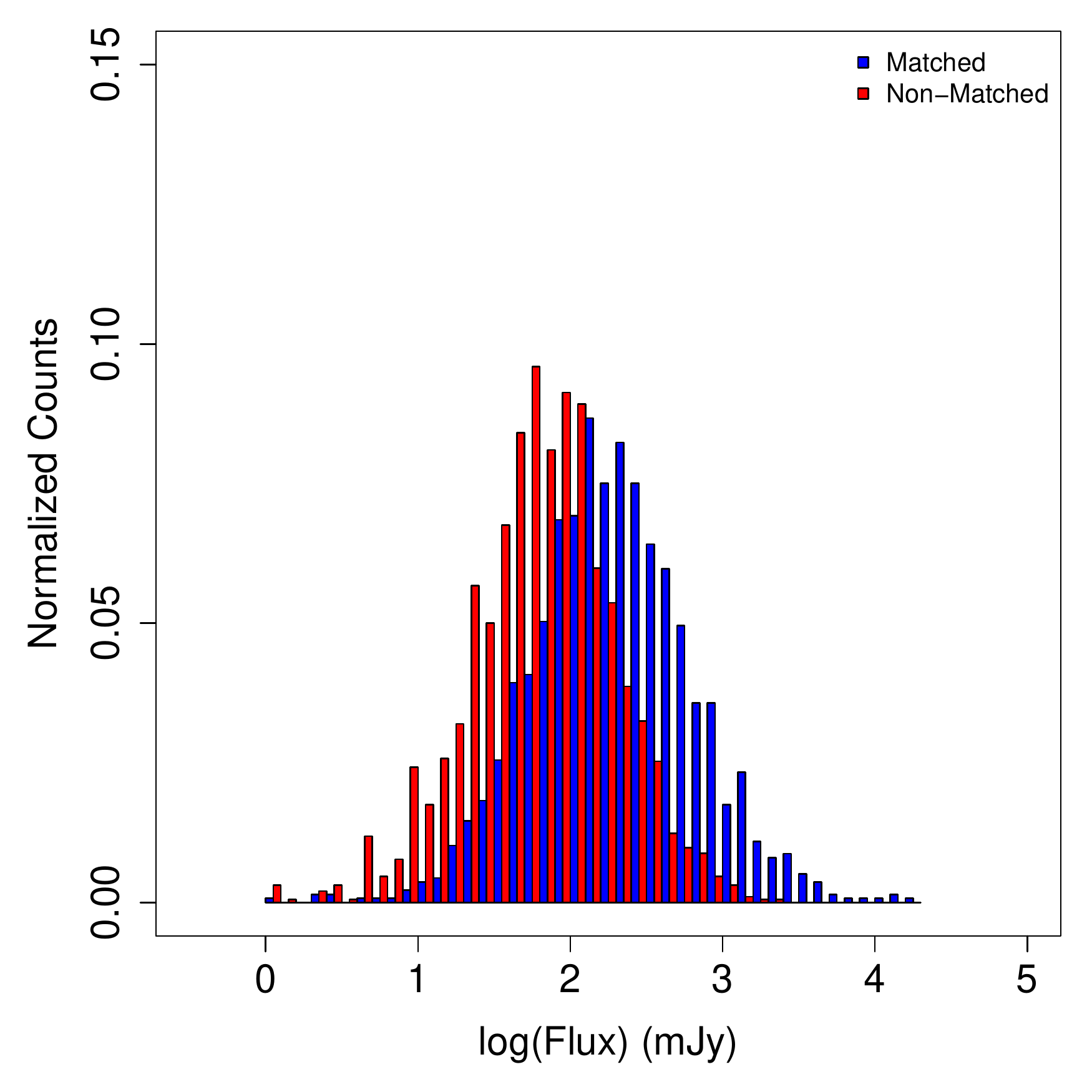}
	\caption{(Left panel) Comparison of the normalized distributions of the radio fluxes densities at \(1.4\) or \(0.843\textrm{ GHz}\) (as listed in the \textit{BZCAT}) for the \textit{BZCAT} sources with a match in the ACC catalogue (blue) and for the \textit{BZCAT} sources below \(<60\degree\) of declination without a match in the ACC catalogue (red). (Right panel) Comparison of the normalized distributions of the average ALMA band 3 radio fluxes densities for the ACC sources with a match in the \textit{BZCAT} (blue) and for the ACC sources without a match in the \textit{BZCAT} (red).}\label{fig:acc_bzcat_radio_flux}
\end{figure*}

        As mentioned before, the ACC contains 3364 sources, with declination \(\delta <60\degree\)). The sources in \textit{BZCAT} below such declination are 3340. Cross-matching them with the ACC using a search radius of \(10\arcsec\) we find 1391 matches. The comparison between the radio flux densities at \(1.4\) or \(0.843\textrm{ GHz}\) (as listed in the \textit{BZCAT}) for these 1391 matches and the remaining 1949 \textit{BZCAT} sources with declination \(<60\degree\) without a match in the ACC catalogue is presented in the left panel of Fig. \ref{fig:acc_bzcat_radio_flux}, from which it is evident that the ACC sources represent the brightest radio end of the blazar population. A similar result is found when comparing the average ALMA band 3 (\(65-90\textrm{ GHz}\)) flux density of 1391 ACC sources that have a match with \textit{BZCAT}, with the remaining 1973 ACC sources without a \textit{BZCAT} match, as shown in the right panel of Fig. \ref{fig:acc_bzcat_radio_flux}.

        We then excluded from the sample of 3364 ACC sources the 1391 sources with a match in the \textit{BZCAT}, narrowing our sample to 1973 sources. For the sources in ACC \citet{2019MNRAS.485.1188B} the authors evaluated the low frequency spectral index \(\alpha_{low}\) between \(1\) and \(5\textrm{ GHz}\) using the 1.4 GHz flux densities from NVSS, SUMSS, GB6 and PMN survey catalogues. We then considered only the sources classified in the ACC as possible blazars, that is, sources with \(\alpha_{low}<-0.5\) and/or with evidences of variability or \(\gamma\)-ray emission. This narrows our sample to 1646 sources. To get a preliminary characterization of these sources we searched in the SIMBAD\footnote{\href{http://simbad.u-strasbg.fr/simbad/}{http://simbad.u-strasbg.fr/simbad/}} database \citep{2000AAS..143....9W} for literature information, and collected them in Table \ref{tab:catalog}. In this table, in addition to source name (column 1) and coordinates (columns 2 and 3), we list the source alternate name (column 4), the redshift (column 5), the source class given in literature according to SIMBAD object classification (column 6, see Table \ref{tab:catalog} note for more details) and the relative reference (column 7). We note that some sources are listed in the SIMBAD as \textit{candidates} (AGN candidates, 2 sources; BL Lac candidates, 1 source; blazar candidates, 169 sources; quasar candidates, 3 sources). For the sake of simplicity, in the following we will consider candidates together with sources that have a secure classification.
        
\begin{figure*}
	\centering
	\includegraphics[scale=0.4]{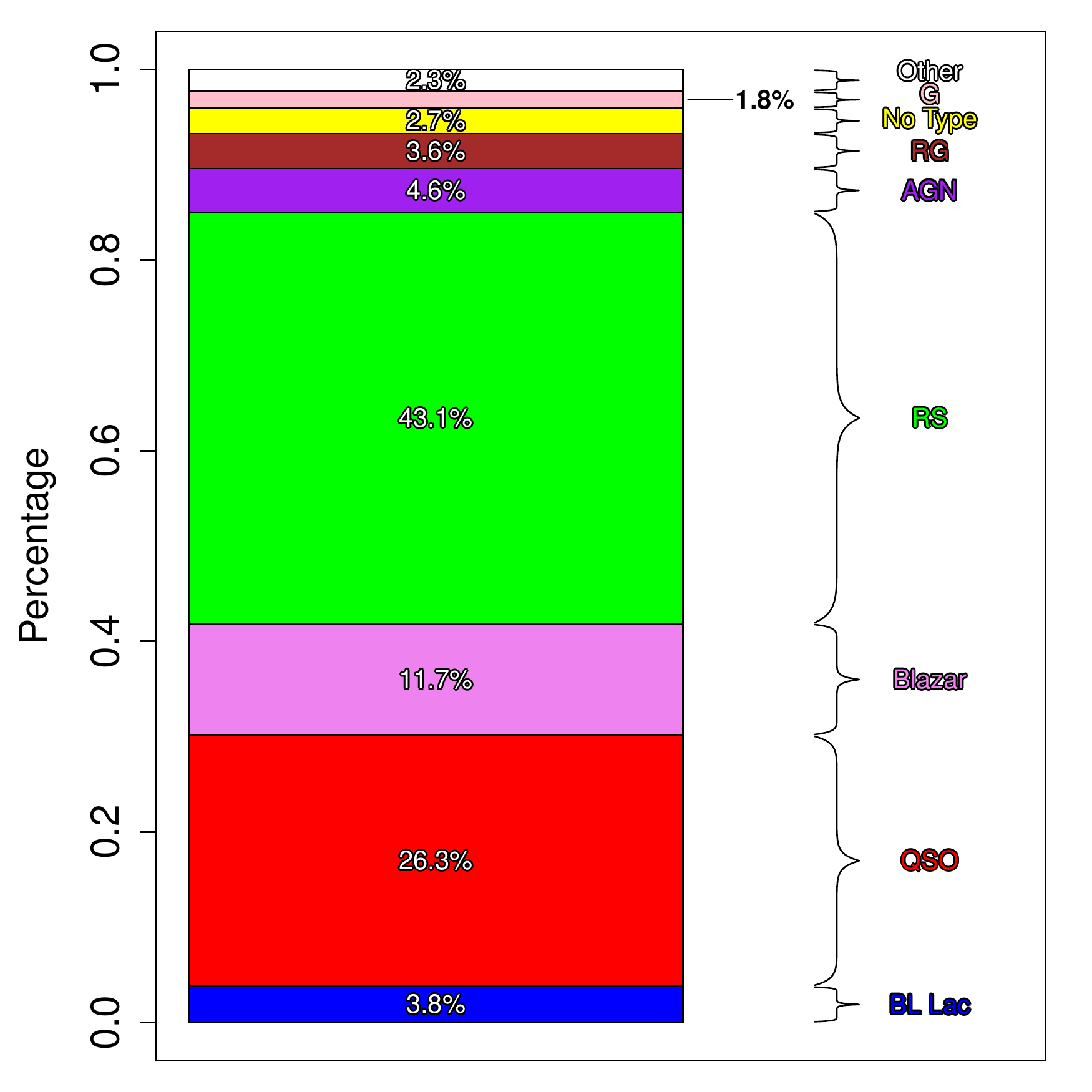}
	\caption{Composition of the sample of 1646 ALMA calibrators with flat radio spectrum and without a \textit{BZCAT} counterpart in terms of their SIMBAD type. In blue we represent BL Lacs, in red quasars (QSO), in violet blazars, in green radio sources (RS), in purple AGN, in brown radio galaxies (RG), in yellow sources without classification (No Type), in pink galaxies (G), and in white other type of sources (including galaxies in clusters, planetary nebulae, Seyfert galaxies, stars, etc.).}\label{fig:source_types}
\end{figure*}

        The composition of this sample of 1646 sources in terms of source class is presented in Fig. \ref{fig:source_types}. It is evident that the two main contributors to the sample are generic radio sources (RS, 43.1\%), quasars (QSO, 26.3\%), and blazars (11.7\%).
        
        Since we are interested in selecting a sample of blazar candidates, in the following we will exclude from our analysis the sources classified as galaxies (labeled as G in Fig. \ref{fig:source_types}), galaxies in clusters, planetary nebulae, Seyfert galaxies, stars, etc. (labeled as Others in Fig. \ref{fig:source_types}). In addition, we will also include the only source with a \(\gamma\)-ray source classification (\textsc{gam} in SIMBAD), namely J0055-1217, since lying at high Galactic latitude \(\sim -75\degree\) it is likely to be a blazar. Being only one source, in the following we will include it in the RS group. This narrows our sample to 1580 source, which represent about \(\sim 96\%\) of the ACC sample, referred to \textit{ALMA Blazar Candidates} (ABC).

\begin{figure*}
	\centering
	\includegraphics[scale=0.4]{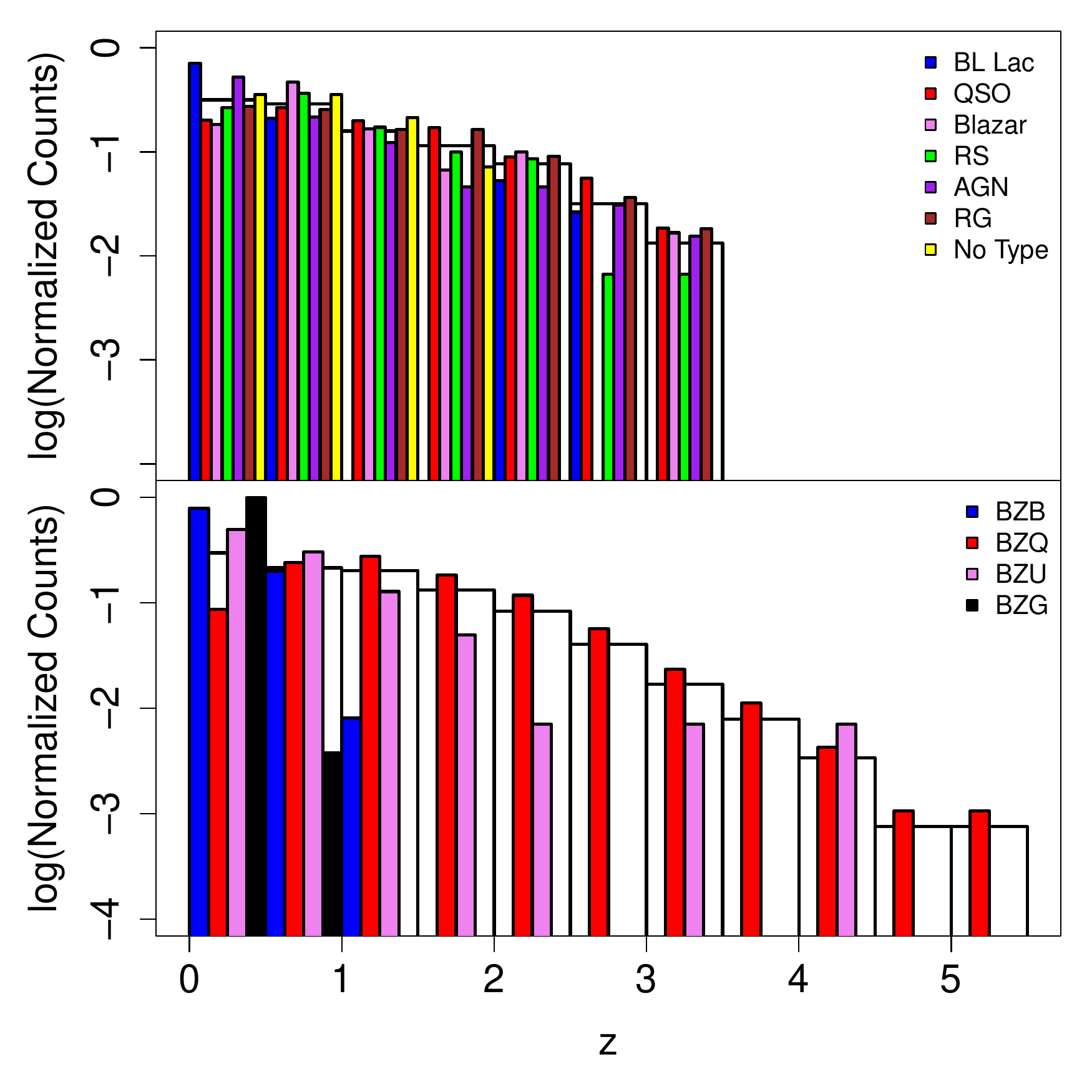}
	\caption{(Top panel) Redshift distribution of \textit{BZCAT} sources. The colored bars represent the different source classes indicated in the legend, while the white bars in the background show the redshift distribution of the whole \textit{BZCAT}. (Lower panel) Same as the top panel, but for the ABC sources.}\label{fig:redshift_hist}
\end{figure*}

        In Fig. \ref{fig:redshift_hist} we compare the redshift distribution of sources in \textit{BZCAT} (top panel) and ABC (lower panel). We note that redshift estimates are available for 706 ABC sources (\(\sim 46\%\)) and for 2566 \textit{BZCAT} sources (\(\sim 72\%\))\footnote{We excluded from this analysis the redshift estimate \(z=6.802\) of \textit{BZCAT} source 5BZQJ1556+3517, since its \href{https://dr16.sdss.org/optical/spectrum/view?plateid=4965&mjd=55721&fiberid=548&run2d=v5_13_0}{SDSS spectrum} shows that this estimate based on tentative detections of Ly-\(\alpha\) and NV\(\lambda\)1240 lines is doubtful.}. We see that \textit{BZQ} sources extend up to \(z\sim 5.5\), with BZG and BZB showing smaller redshifts \(z\leq 0.6\) and \(z\leq 1.3\), respectively. On the other hand ABC sources only reach \(z\sim 3.5\), with BL Lacs extending up to \(z\leq 2.6\) and sources without classification reaching \(z\leq 1.8\), while the other source types extend up to \(z>3.5\).

    \section{Comparison with other Blazar candidate Catalogs}\label{sec:catalog_comparison}
        In this section we compare the catalogue of ABC with other catalogues of blazar sources. The full results of this comparison are presented in Table \ref{tab:comparison}.

        We start from the \textit{Fermi} Large Area Telescope Fourth Source Catalog \citep[4FGL,][]{2020ApJS..247...33A}, the most recent release of the \textit{Fermi} mission \(\gamma\)-ray source catalog. Blazars represent the largest population of identified \(\gamma\)-ray sources. The 4FGL contains 5065 sources in the \(50\textrm{ MeV}-1\textrm{ TeV}\), including 3915 sources associated with lower-energy counterparts with the Bayesian source association method \citep{2010ApJS..188..405A} and the Likelihood Ratio method \citep{2011ApJ...743..171A, 2015ApJ...810...14A}, both based on the spatial coincidence between the \(\gamma\)-ray sources and their potential counterparts, and 358 sources identified basing on periodic variability, correlated variability at other wavelengths, or spatial morphology.

        Sources in 4FGL are classified as BLL (BL Lac objects), FSRQs, BCU (blazar of uncertain type), RDG (radio galaxies), AGNs, etc., according to the properties of their counterpart at other wavelengths. In particular, 3137 sources in the 4FGL are classified as blazars, which therefore represent \(\sim 62\%\) of the whole 4FGL and \(\sim 80\%\) of the identified sources.

        We therefore cross-matched the catalogue of ABC with the associated/identified sources in the 4FGL, adopting a search radius of \(3\farcs 0\) around the coordinates of the associated counterparts, finding 259 matches. We then looked for ABC falling in the \(90\%\) uncertainty ellipse of unassociated/unidentified sources in the 4FGL, finding no matches.

\begin{figure*}
	\centering
	\includegraphics[scale=0.38]{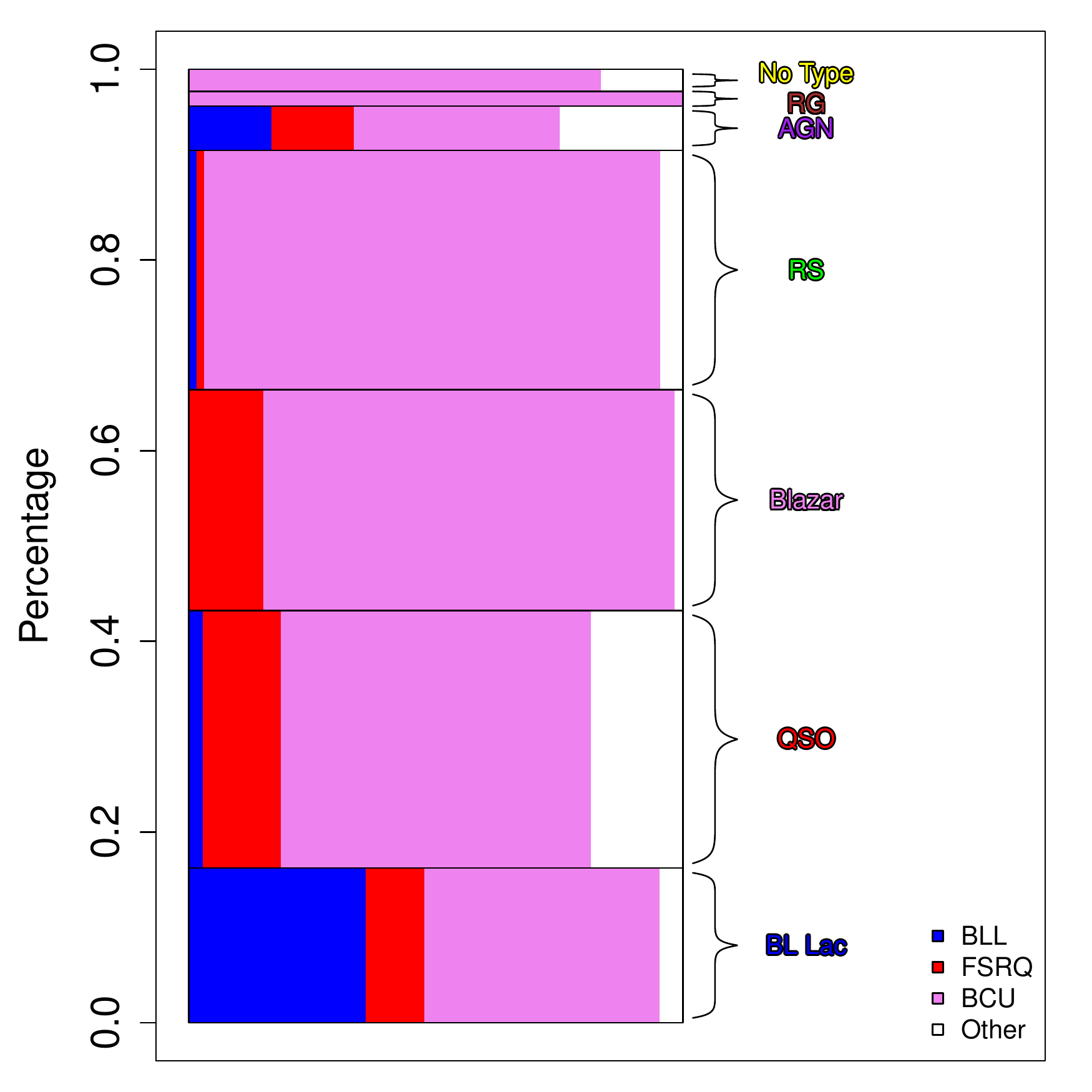}
	\includegraphics[scale=0.38]{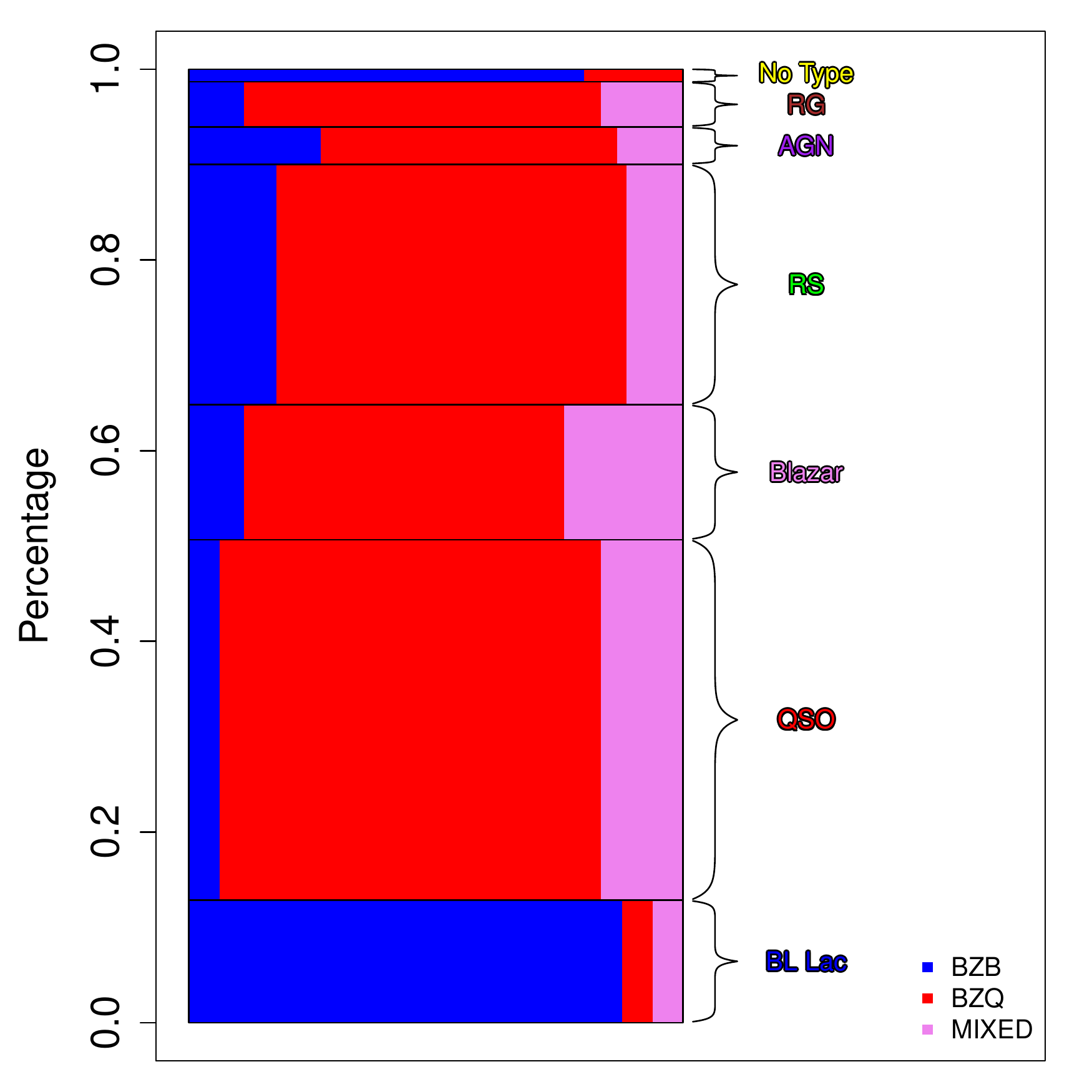}
	\caption{(Left panel) Distribution of the 4FGL \citep{2020ApJS..247...33A} source types in the ABC source classes. Blue rectangles indicate BLLs, red rectangles indicate FSRQs, violet rectangles indicate BCUs, and white rectangles indicate other kind of sources (including compact steep spectrum radio sources, radio galaxies, starburst galaxies, generic active galactic nuclei, or unknown sources). (Right panel) Distribution of the WIBRaLS2 \citep{2019ApJS..242....4D} source types in the ABC source classes. Blue rectangles indicate BZBs, red rectangles indicate BZQs and violet rectangles indicate MIXED sources.}\label{fig:source_type_cand_catalog}
\end{figure*}

        The distribution of these matches in the different source classes of ABC is shown in the left panel of Fig. \ref{fig:source_type_cand_catalog}. In this figure each rectangle represents an ABC class, and the coloured bars inside it represent the distribution of sources in this class among the 4FGL types.

        We see that all ABC types are dominated by \textit{Fermi} BCUs, which represent most (\(\sim 73\%\)) of the 4FGL sources matching ABC. On the other hand, the majority of sources classified as BLL in the 4FGL fall in the BL Lac class, while FSRQs mainly fall in QSO and Blazar classes. In addition, we note that 6 AGN and 4 RDG sources fall in the QSO class, while the ABC AGN type contains 2 BLLs, 2 FSRQs, 5 BCUs, and 3 RDGs.
        
        We then compare the catalogue of ABC with the third catalogue of extreme and high-synchrotron peaked blazars \citep[3HSP,][]{2019AA...632A..77C}. This is a catalogue containing 2013 high-synchrotron peaked blazars (HSPs), i.e. BL Lacs with the synchrotron component of their SED peaking at frequencies larger than \({10}^{15} \textrm{ Hz}\). They were selected through multi-wavelength analysis, and 657 of these sources are in common with the 5th edition of \textit{BZCAT}

        We note that the Galactic latitude distribution of sources in 3HSP is similar to that of \textit{BZCAT} (see Fig. \ref{fig:galactic_projection}), that is, peaking outside of the Galactic plane (\(\left|{b}\right|>{10}\degree\)). By cross-matching the ABC and the 3HSP catalogues adopting a search radius of \(3\farcs 0\) we find 9 matches. In particular, one object has no classification in the ABC catalog, one object is classified ad Radio Galaxy, and 7 are classified as BL Lacs.
        
        We then consider two catalogues of blazar candidates presented in \citet{2019ApJS..242....4D}, namely the second WISE Blazar-like Radio-Loud Sources (WIBRaLS2) catalogue and the KDEBLLACS.
        
        The WIBRaLS2 catalogue includes 9541 blazar candidates selected on the basis of their radio-loudness and of their WISE colours. Candidates were required to be detected in all four WISE bands. Their radio-loudness was defined on the basis of the ratio between the mid-infrared \(22\,\textrm{{\textmu}m}\) flux density \(S_{22\textrm{{\textmu}m}}\) and the radio flux density \(S_{\textrm{Radio}}\), defined as \(q_{22} = \log{\left({S_{22\textrm{{\textmu}m}}/S_{\textrm{Radio}}}\right)}\). Radio flux densities were from NVSS \citep{1998AJ....115.1693C}, FIRST \citep{1997ApJ...475..479W} and SUMSS \citep{2003MNRAS.342.1117M}. In fact, \citet{2012ApJ...748...68D} discovered that blazar listed in the second edition of \textit{BZCAT} \citep{2009AA...495..691M} and associated with \(\gamma\)-ray sources listed in the \textit{Fermi} Large Area Telescope Second Source Catalog \citep[2FGL,][]{2012ApJS..199...31N} occupy a specific region of the two-dimensional WISE colour-colour planes.
        
        \citet{2019ApJS..242....4D} defined the region of the three-dimensional principal component space generated by the three independent WISE colours occupied by blazar listed in the 5th edition of \textit{BZCAT} and associated with \(\gamma\)-ray sources listed in the \textit{Fermi} Large Area Telescope Third Source Catalog \citep[3FGL,][]{2015ApJS..218...23A}. The WIBRaLS2 catalogue contains radio-loud sources located in this region. Such sources are classified as candidates BZB and BZQ depending if they are compatible with the region occupied by BL Lacs or FSRQs. In case they are compatible with both, they are classified as candidate MIXED\footnote{The thresholds on radio loudness are selected as \(q_{22}\leq -0.61\), \(q_{22}\leq -0.97\) and \(q_{22}\leq -0.79\) for BZB, BZQ and MIXED candidates, respectively.}. Finally, blazar candidates in WIBRaLS2 are assigned a class going from D to A depending on how compatible a source is with the respective blazar region.
        
        By cross-matching the ABC catalogue with WIBRaLS2 adopting a search radius of \(3\farcs 0\) we find 381 matches. The distribution on WIBRaLS2 source types among the various ABC classes is presented in the right panel of Fig. \ref{fig:source_type_cand_catalog}. Again, each rectangle in this figure represents an ABC class, and the coloured bars inside it represent the amount of sources in this class falling the WIBRaLS2 class types, according to the colour code indicated in the legend. BZB sources are found mostly among BL Lacs, while other classes are dominated by BZQs which represent \(\sim 64\%\) of the WIBRaLS2 sources associated with ABC catalog.

        KDEBLLACS is a catalogue of BL Lac candidates selected among radio-loud sources using criteria analogous to those of the WIBRaLS2 catalog, except for requiring detection in only the first three WISE bands and, consequently, for redefining the radio-loudness on the basis of their \(S_{12\textrm{{\textmu}m}}\) \(12\,\textrm{{\textmu}m}\) flux density, that is, on the ratio \(q_{12} = \log{\left({S_{12\textrm{{\textmu}m}}/S_{\textrm{Radio}}}\right)}\), where \(S_{\textrm{Radio}}\) is the radio flux density\footnote{In particular, BL Lac candidates are selected with \(-1.85<q_{12}<-1\), \(-1.64<q_{12}<-1.01\) and \(-1.64<q_{12}<-1\) for sources with NVSS, FIRST and SUMSS counterparts, respectively.}. BL Lac candidates are located in the region of the two dimensional WISE colour-colour plane occupied by BL Lacs listed in the 5th edition of \textit{BZCAT} and associated with \(\gamma\)-ray sources listed in the 3FGL. This region is bounded by the isodensity contour, obtained with kernel density estimation (KDE), containing \(90\%\)of \textit{BZCAT} \(\gamma\)-ray BL Lacs. Finally, the KDEBLLACS catalogue is restricted to the sources outside the Galactic plane (\(\left|{b}\right|>{10}\degree\)). By cross-matching the ABC catalogue with KDEBLLACS adopting a search radius of \(3\farcs 0\), we find only one match, namely the source J0835-5953, which is classified in the ABC catalogue as Blazar.

    \section{Multi-wavelength Analysis}\label{sec:multi_lambda}
        In this Section we aim at better characterizing our ABC sources by looking for their multi-wavelength counterparts in the available catalogues obtained from major surveys.

        \subsection{Optical Data}\label{sec:optical}
            Optical observations of blazars both in photometry \citep[e.g.][]{2016AA...591A..21M, 2019MNRAS.489.1837R, 2020ApJ...890...97A} and spectroscopy \citep[e.g.][]{2015AJ....149..160R, 2019Ap&SS.364...85P, 2020Ap&SS.365...12D} provide an excellent tool to characterized their broad band emission and a to pinpoint their classification through the observation (or lack of it) of strong emission lines. In this Section we collect photometric and spectroscopic data available in public catalogs with the goal of obtaining a better characterization of the ABC sources.

            \subsubsection{Galactic extinction}\label{sec:absorption}
                As we saw in Sect.\ \ref{sec:sample_selection}, nearly one fourth of the ABC sources are located close to the Galactic plane, where extinction is a major issue. We used the Galactic Dust Reddening and Extinction tool of the NASA/IPAC Infrared Science Archive\footnote{\href{https://irsa.ipac.caltech.edu/applications/DUST/}{https://irsa.ipac.caltech.edu/applications/DUST/}} to obtain the value of Galactic absorption along the line of sight of our ABC sources according to the analysis by \citet{2011ApJ...737..103S}.

\begin{figure*}
	\centering
	\includegraphics[scale=0.38]{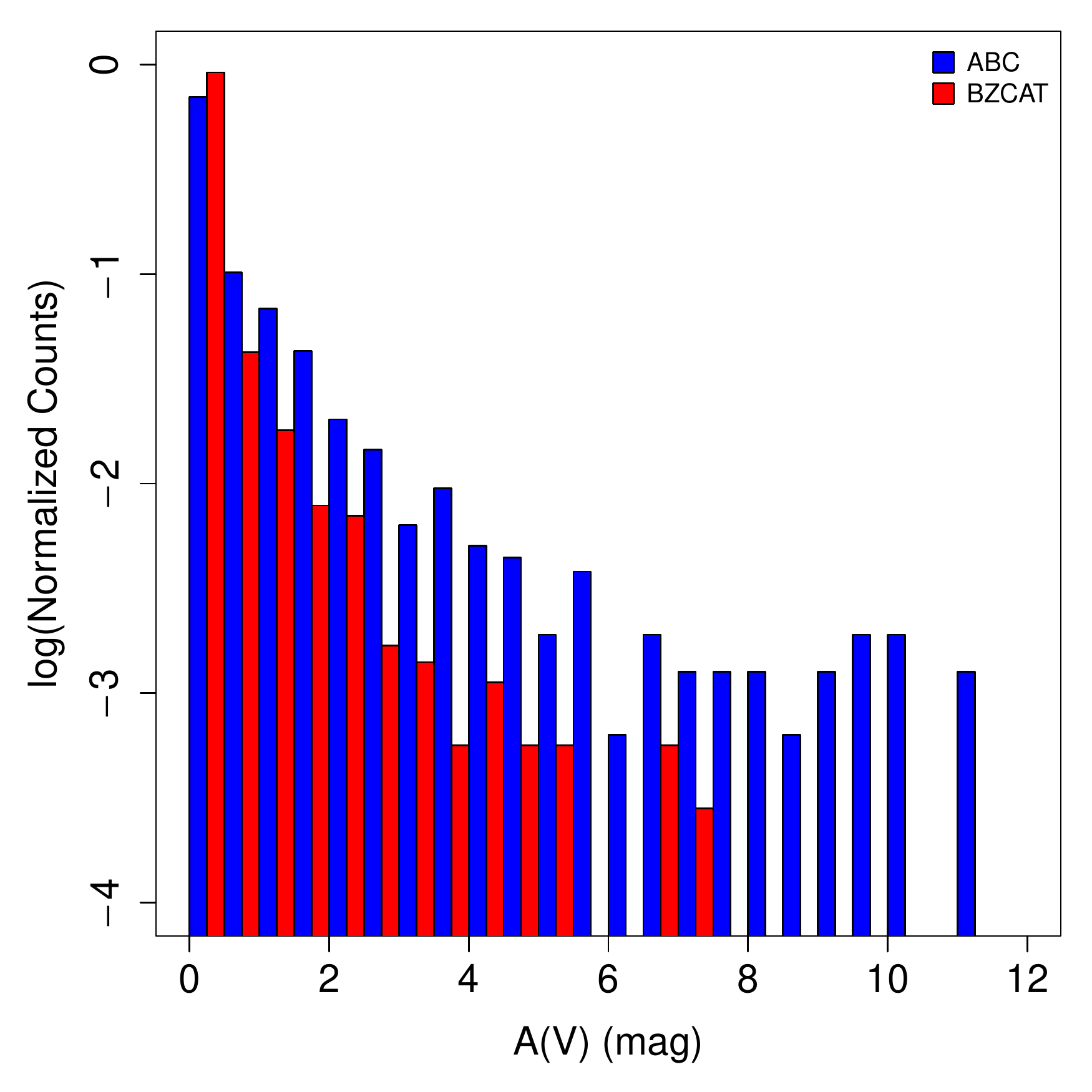}
	\includegraphics[scale=0.38]{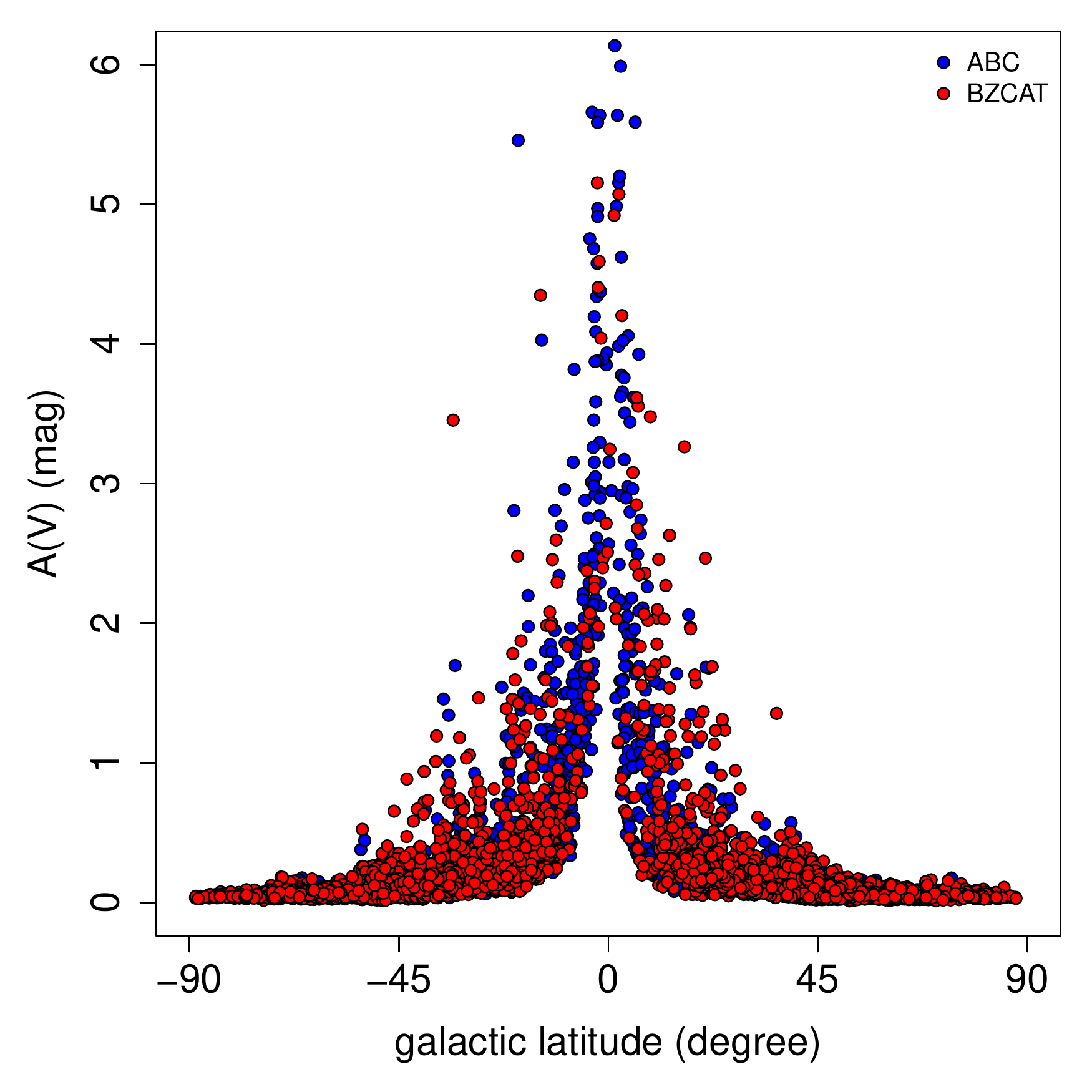}
	\caption{(Left panel) Distribution of the Galactic extinction in the \(V\) band \(A(V)\) at the coordinates of ABC sources (blue bars) and at the coordinates of \textit{BZCAT} sources (red bars). (Right panel) Galactic extinction in the \(V\) band \(A(V)\) versus Galactic latitude for ABC sources (blue circles) and for \textit{BZCAT} sources (red circles).}\label{fig:absorption}
\end{figure*}

                In Fig. \ref{fig:absorption} we show the distribution of the Galactic extinction in the \(V\) band, \(A(V)\) (left panel), and the distribution of \(A(V)\) as a function of the Galactic latitude (right panel) for both the ABC and \textit{BZCAT} sources. We see that \textit{BZCAT} objects have, on average, lower \(A(V)\) values than ABC ones, confirming that an important fraction of ABC sources are probable blazars so far unrecognized because they lie close to the Galactic plane, where strong absorption makes optical observations difficult.

            \subsubsection{GAIA}\label{sec:gaia}
                The \textit{Gaia} satellite \citep{2016AA...595A...1G} was launched in 2013 with the aim to perform the largest, most precise 3D map of our Galaxy. A first data release (DR1) was issued in 2016, covering the first 14 months of observations. A second data release (DR2) followed in 2018, including 22 months of observations \citep{2018AA...616A...1G}. \textit{Gaia} DR2 contains high-precision parallaxes and proper motions for over 1 billion sources together with precise multiband photometry.
            
                As stressed by \citet{2019MNRAS.490.5615B}, \textit{Gaia} selects point-like sources, so most galaxies are missed. Moreover, parallaxes and proper motions of galaxies are liable to incorrect fitting by the astrometric model. Therefore, most BL Lac sources with dominant host galaxies (i. e., BZGs) may have no \textit{Gaia} counterparts or may suffer for overestimates of parallaxes and proper motions.
            
\begin{figure*}
	\centering
	\includegraphics[scale=0.4]{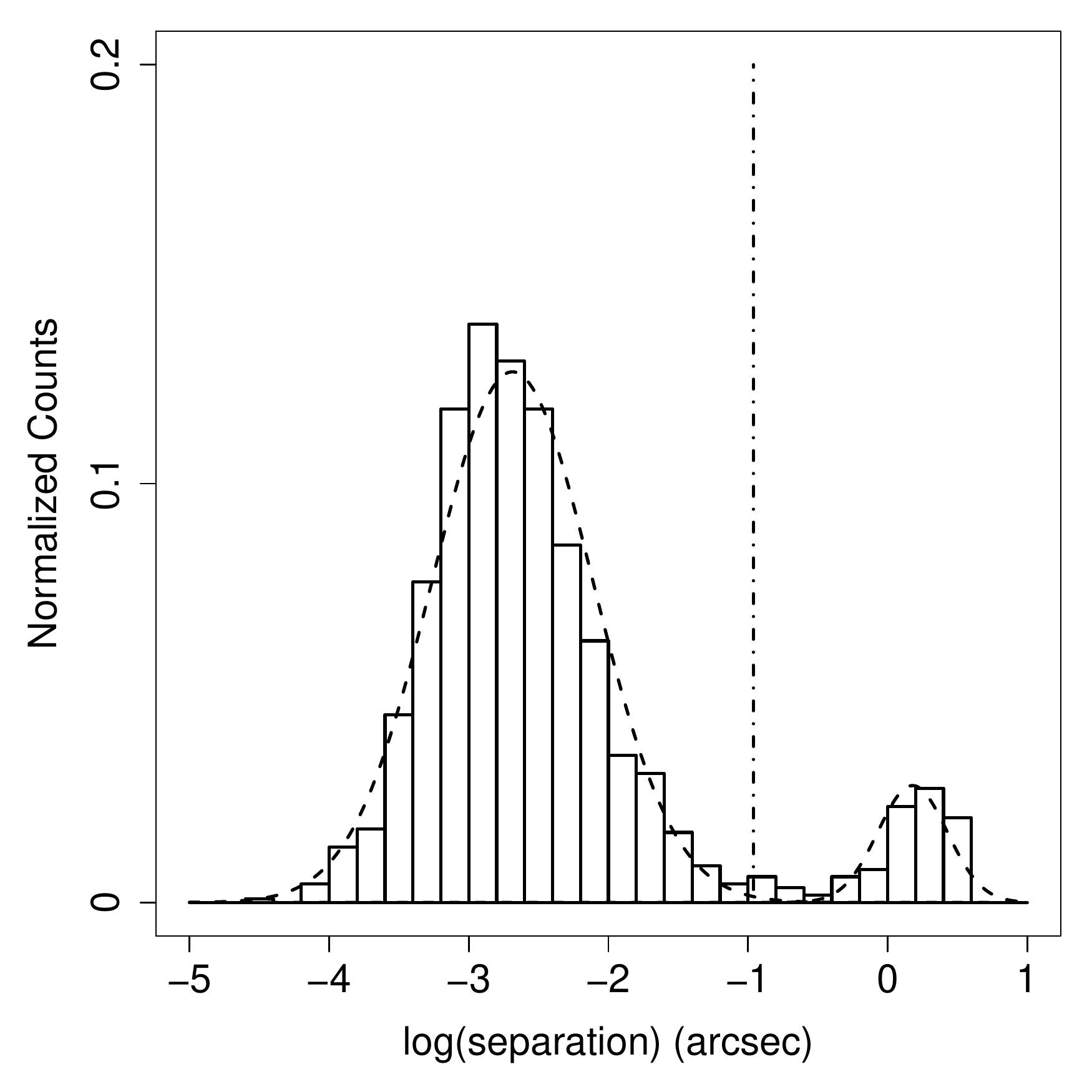}
	\caption{Distribution of the separations between the ABC source positions and its closest \textit{Gaia} match. The dashed lines represent the two gaussians fitting the separation distribution. The vertical dot-dashed line represents the 3 sigma deviation from the mean of the largest gaussian, chosen as the boundary for reliable matches (see Sect. \ref{sec:gaia}).}\label{fig:gaia_sep_hist}
\end{figure*}

                We looked for \textit{Gaia} counterparts of the ABC sources in DR2 using the standard \(3\farcs 0\) search radius, and finding 1137 matches. However, we expect that the \textit{Gaia} counterparts can be shifted by no more than a fraction of arcsec with respect to the ALMA position. Therefore, in Fig. \ref{fig:gaia_sep_hist} we plot the distribution of the separations between the ABC source positions and its closest \textit{Gaia} match. The separation distribution was fitted with Expectation-Maximization algorithm for mixtures of univariate normals (\textsc{normalmixEM}\footnote{\href{https://www.rdocumentation.org/packages/mixtools/versions/1.0.4}{https://www.rdocumentation.org/packages/mixtools/versions/1.0.4}} R package). We can clearly see that this distribution is adequately represented by two gaussians, the smallest one with mean \(\mu_1=\sim 1\farcs 5\) and the largest one with mean \(\mu_2=\sim 2 \textrm{ mas}\), indicating that the majority of these matches are accurate to the sub-arcsecond scale. We therefore select as reliable only the matches that show a separation smaller than \(\sim 0\farcs 1\), that corresponds to a deviation of \(3\) sigmas from \(\mu_2\), narrowing down the number of associations to 1030.
            
\begin{figure*}
	\centering
	\includegraphics[scale=0.38]{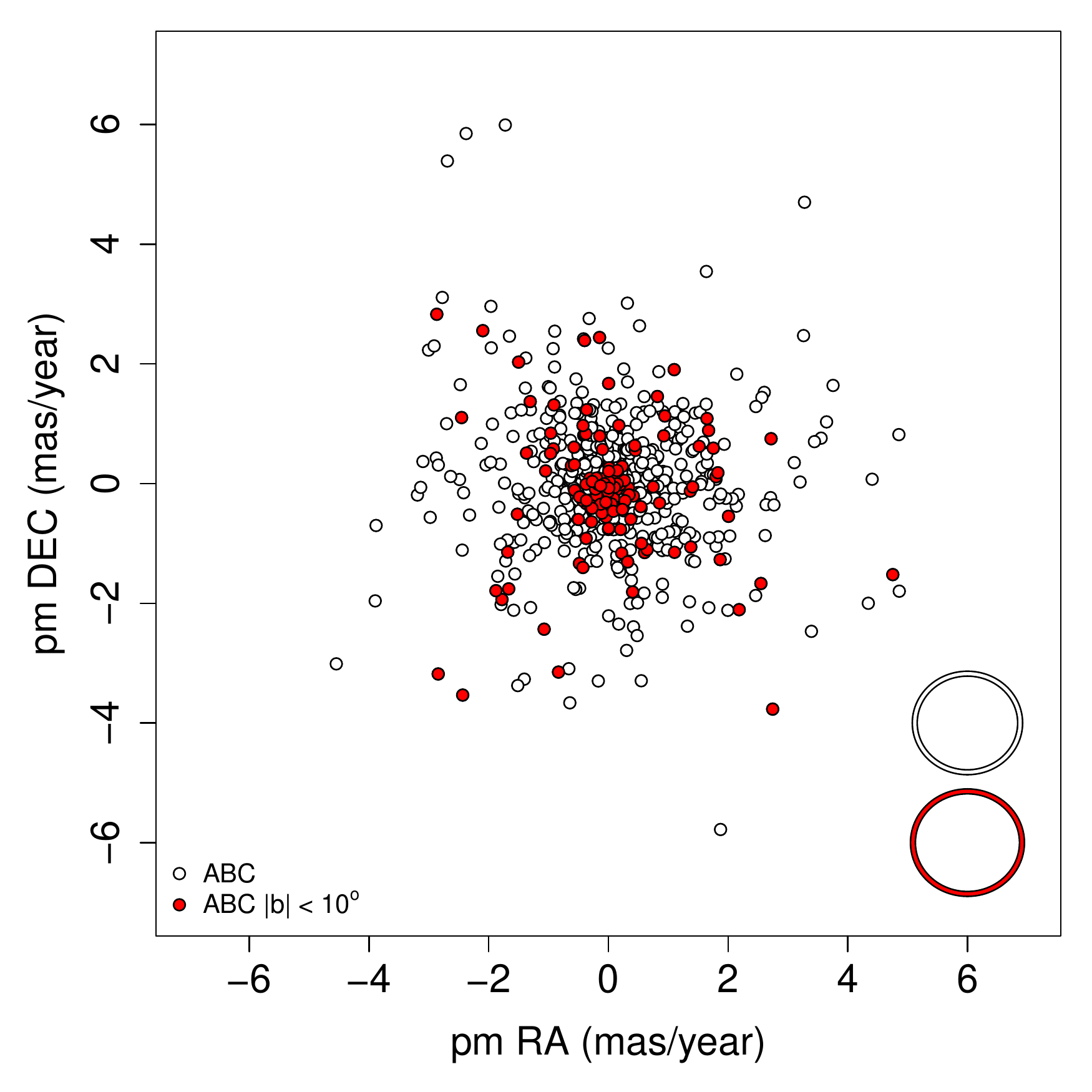}
	\includegraphics[scale=0.38]{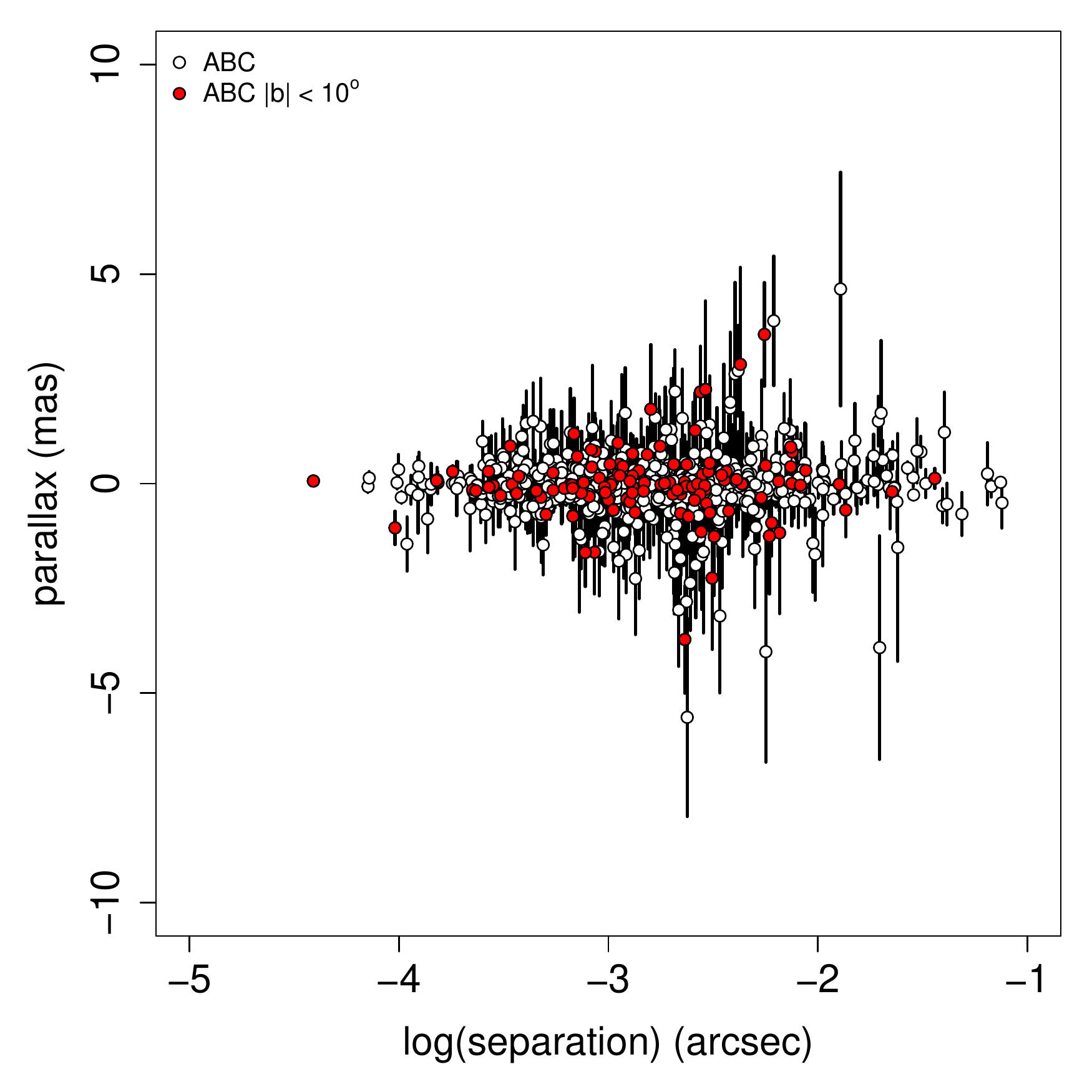}
	\caption{(Left panel) Distribution of the 805 possible \textit{Gaia} identifications in the motion plane (see Sect. \ref{sec:gaia}). The white circles represent all possible \textit{Gaia} identifications, and the red circles represent the sources close to the Galactic plane (\(\left|{b}\right|<{10}\degree\)). The ellipses in the lower-right corner represent the average uncertainties for all possible \textit{Gaia} identifications (white ellipse) and for the sources close to the Galactic plane (red ellipse). (Right panel) \textit{Gaia} parallaxes of the possible associations versus separation between the ABC sources and their \textit{Gaia} matches. The vertical bars represent the 1-\(\sigma\) errors on parallax, the white circles represent all possible \textit{Gaia} identifications, and the red circles represent the sources close to the Galactic plane.}\label{fig:gaia}
\end{figure*}

                In addition, extragalactic objects should ideally have null parallaxes and proper motions (even though not all sources with null parallax and proper motion are necessarily extragalactic objects). We then consider as possible identifications all associations where the proper motion and parallaxes are compatible with zero at 3 sigma level, narrowing down the number of associations to 805 possible identifications. These are shown in the left panel of Fig. \ref{fig:gaia} in the RA DEC proper motion plane, where we see that there is no significant difference between the proper motions of all possible identifications and sources close to the Galactic plane (\(\left|{b}\right|<{10}\degree\)).

                In the right panel of the same figure we show \textit{Gaia} parallaxes of the possible associations versus separation between the ABC sources and their \textit{Gaia} matches. We note that negative parallaxes are a consequence of how \textit{Gaia} data are treated and can be safely used \citep{2018AA...616A...1G}. Again, we do not see a significant difference between the parallaxes of all possible identifications and sources close to the Galactic plane.

                In Fig. \ref{fig:colour_mag} we compare the distribution of ABC sources (blue circles) and \textit{BZCAT} sources (red circles) in the \textit{Gaia} g vs b-r colour-magnitude plot. The great majority (\(\sim 98\%\)) of the \textit{Gaia} counterparts selected for ABC sources are detected in g, b and r bands. \textit{Gaia} counterparts for \textit{BZCAT} sources have been selected adopting the same matching criteria used for ABC sources. All magnitudes have been corrected for Galactic absorption using reddening estimates from \citet{2011ApJ...737..103S} and the extinction model from \citet{2007ApJ...663..320F}. In the same plot we show with gray points in the background \(\sim 5\times{10}^4\) random \textit{Gaia} sources together with KDE isodensity curves containing \(60\%\), \(70\%\), \(80\%\) and \(90\%\) of these \textit{Gaia} random sources. On top and on the right of the main panel of this figure we show the normalized distributions of the b-r colour and g magnitude, respectively, for the ABC, \textit{BZCAT} and random sources. It is evident that neither ABC nor \textit{BZCAT} sources are clearly separated on the colour-magnitude diagram from the random \textit{Gaia} sources. ABC sources are, on average, slightly bluer and dimmer than the \textit{BZCAT} ones, although spanning a similar range of magnitudes. A Kolmogorov–Smirnov \citep[KS;][]{1933kol,1939smir} test shows that the distributions of the \textit{Gaia} g magnitudes of ABC sources and \textit{BZCAT} sources have a p-chance \(p=3.0\times{10}^{-11}\) of having been randomly sampled from a common parent distribution, while for the b-r colour distributions this value is \(p=1.1\times{10}^{-5}\).

                We can then conclude that ABC sources are significantly dimmer than \textit{BZCAT} sources in \textit{Gaia} g band, while the difference in the \textit{Gaia} b-r colour between the two populations is less pronounced, while still significant.

            \subsubsection{SDSS DR12 and LAMOST DR5}\label{sec:sdss}
                The 12th data release of the Sloan Digital Sky Survey \citep[SDSS DR12,][]{2015ApJS..219...12A} covers over \(14000\textrm{ deg}^2\) of the sky, providing optical multi-band photometric information for more than 450 millions unique objects, and optical spectra for more than 5 millions sources.

                To determine the optimal cross-match radius of ABC sources with the SDSS DR12 catalogue we considered values from \(0\farcs 1\) to \(3\farcs 0\) in steps of \(0\farcs 1\). We then repeated the same procedure around random positions in the sky.

\begin{figure*}
	\centering
	\includegraphics[scale=0.38]{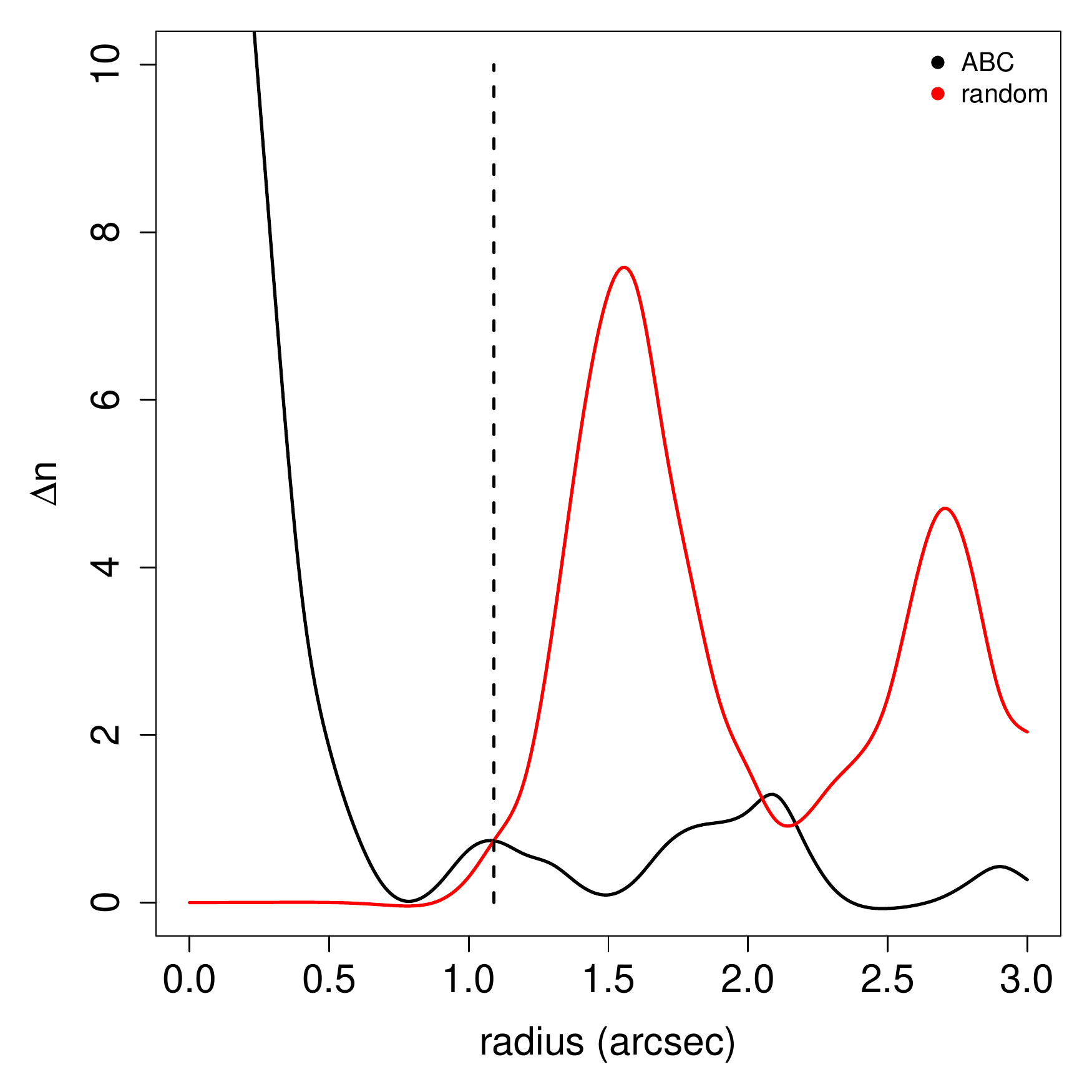}
	\includegraphics[scale=0.38]{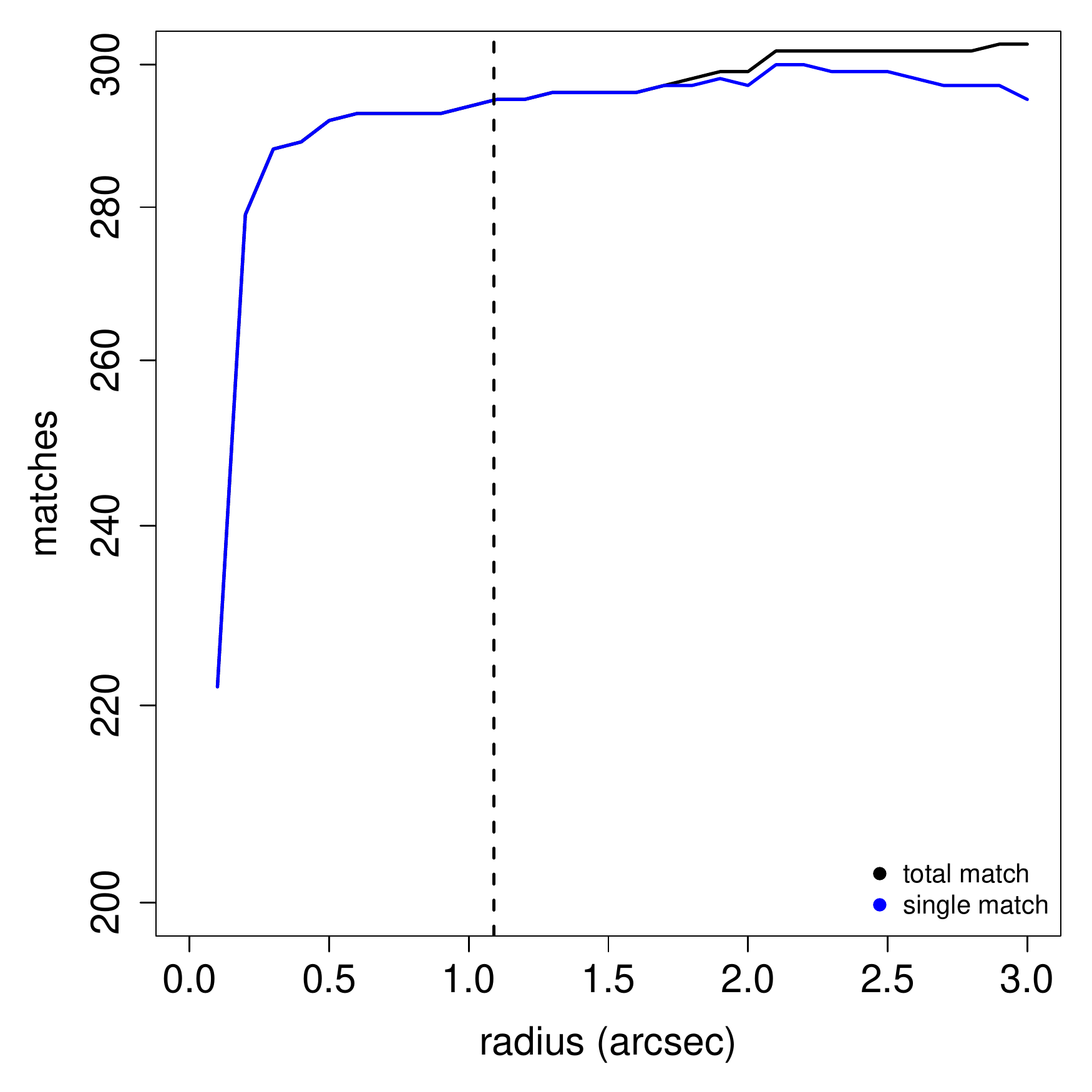}
	\caption{(Left panel) Increase of sources with at least one SDSS counterpart (\(\Delta n\)) plotted versus the search radius for ABC sources (black line) and for random positions in the sky (red line). Lines representing (\(\Delta n\)) were smoothed for better visualization. The optimal search radius is represented with a vertical dashed line. (Right panel) Number of matches between ABC sources and SDSS DR12 sources plotted vs the search radius. The number of ABC sources with at least one match in SDSS DR12 is indicated with a black line, while the number of ABC the sources with a single match in SDSS DR12 is indicated with a blue line. As in the right panel, the optimal search radius is represented with a vertical dashed line.}\label{fig:sdss_radius}
\end{figure*}

                In the left panel of Fig. \ref{fig:sdss_radius} we compare the increase of sources with at least one SDSS counterpart (\(\Delta n\)) with increasing search radius for the ABC sources (black line) and for the random positions (red line). Since these variations are rather noisy, for better visualization we smoothed (\(\Delta n\)) using the \textsc{smooth.spline} R tool with a smoothing parameter \(0.4\). For ABC sources the majority of matches is already found with a search radius of \(0\farcs 6\), and further increases in the search radius only add few matches, while for the random positions (\(\Delta n\)) is initially \(0\), and then it increases after \(1\farcs 0\). We therefore choose as an optimal search radius for SDSS DR12 counterpart the radius of \(1\farcs 1\) at which (\(\Delta n\)) the increase of matches at random positions in the sky becomes larger than that at the positions of ABC sources. As shown in the right panel of Fig. \ref{fig:sdss_radius}, at this radius we have a total of 295 SDSS DR12 matches for the ABC sources, without multiple matches. All these 295 sources are detected in the five SDSS bands.

                The Large sky Area Multi-Object fiber Spectroscopic Telescope \citep[LAMOST][]{2012RAA....12.1197C} survey is a large-scale spectroscopic survey which follows completely different target selection algorithms than SDSS, and is therefore complementary to the latter. The 5th data release of LAMOST survey (DR5) provides optical spectra for more than 9 millions sources.

                We adopted the same approach to determine the optimal radius to cross-match ABC sources with the LAMOST DR5 catalogue we looked for matches in the latter with search radii, increasing from \(0\farcs 1\) to \(3\farcs 0\), with increases of \(0\farcs 1\), around both the coordinates of ABC sources and random positions in the sky. We find that for ABC sources the majority of matches is already found with a search radius of \(0\farcs 3\). Due to the difference in surface density of sources between SDSS DR12 (\(\sim 470\) million) and LAMOST DR5 (\(\sim 7\) million), we start finding some matches around random sources only around \(4\arcsec\). We then choose as an optimal search radius for LAMOST DR5 counterpart the radius of \(0\farcs 5\) at which (\(\Delta n\)) reaches \(0\), and above which fluctuations in (\(\Delta n\)) for ABC sources and random positions are similar. At this radius we have a total of 31 LAMOST DR5 matches for the ABC sources, without multiple matches. Of these, 29 have also SDSS DR12 counterparts.

\begin{figure*}
	\centering
	\includegraphics[scale=0.25]{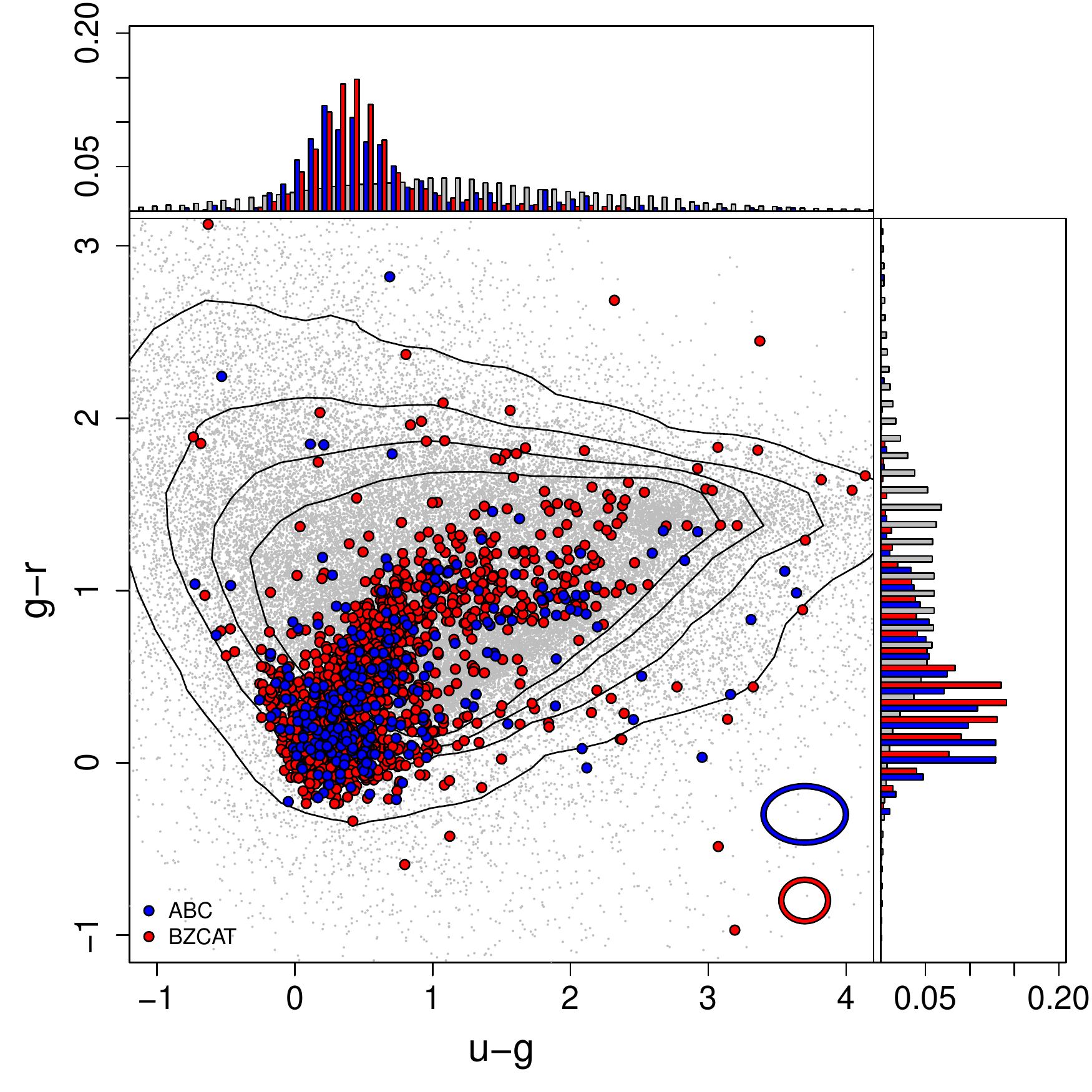}
	\includegraphics[scale=0.25]{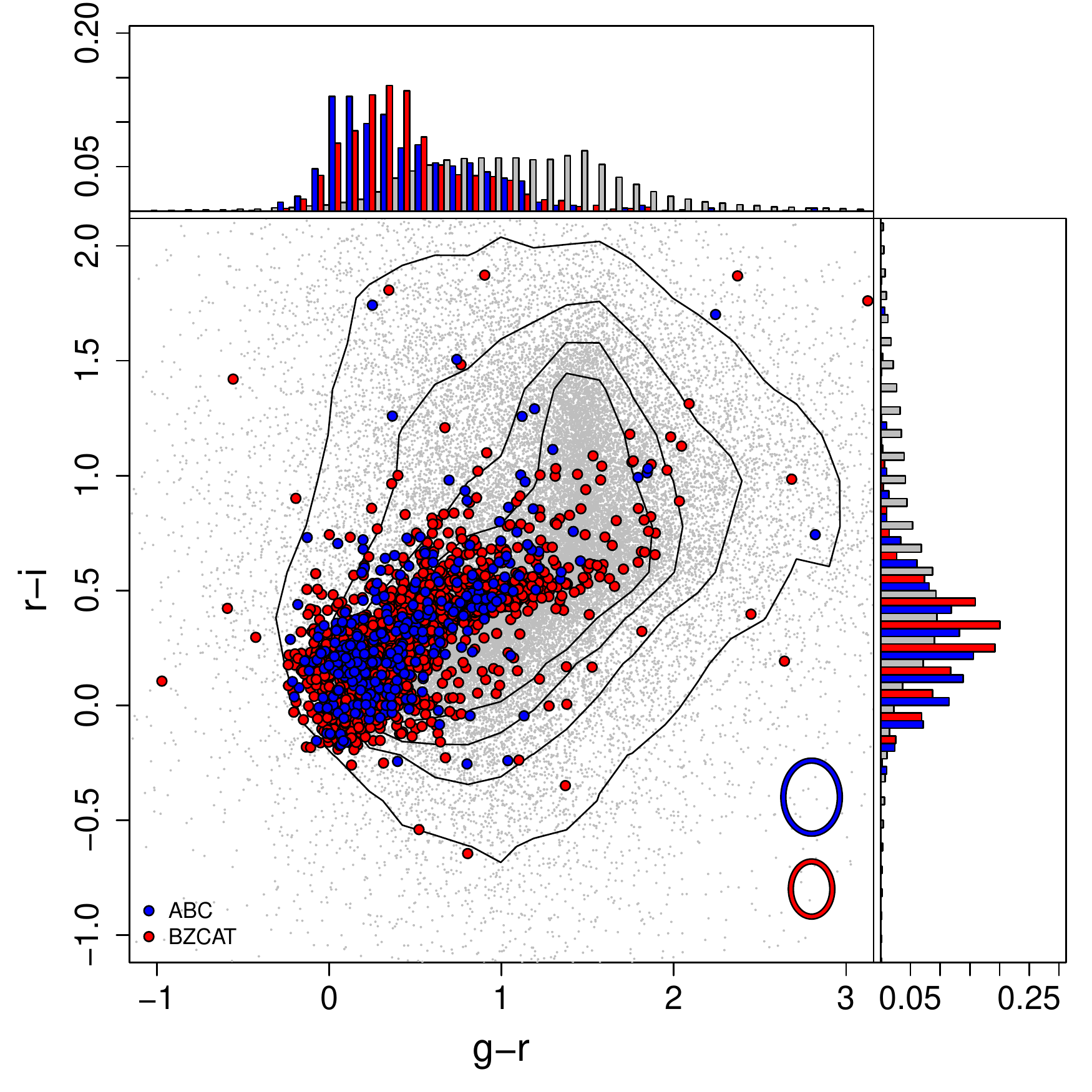}
	\includegraphics[scale=0.25]{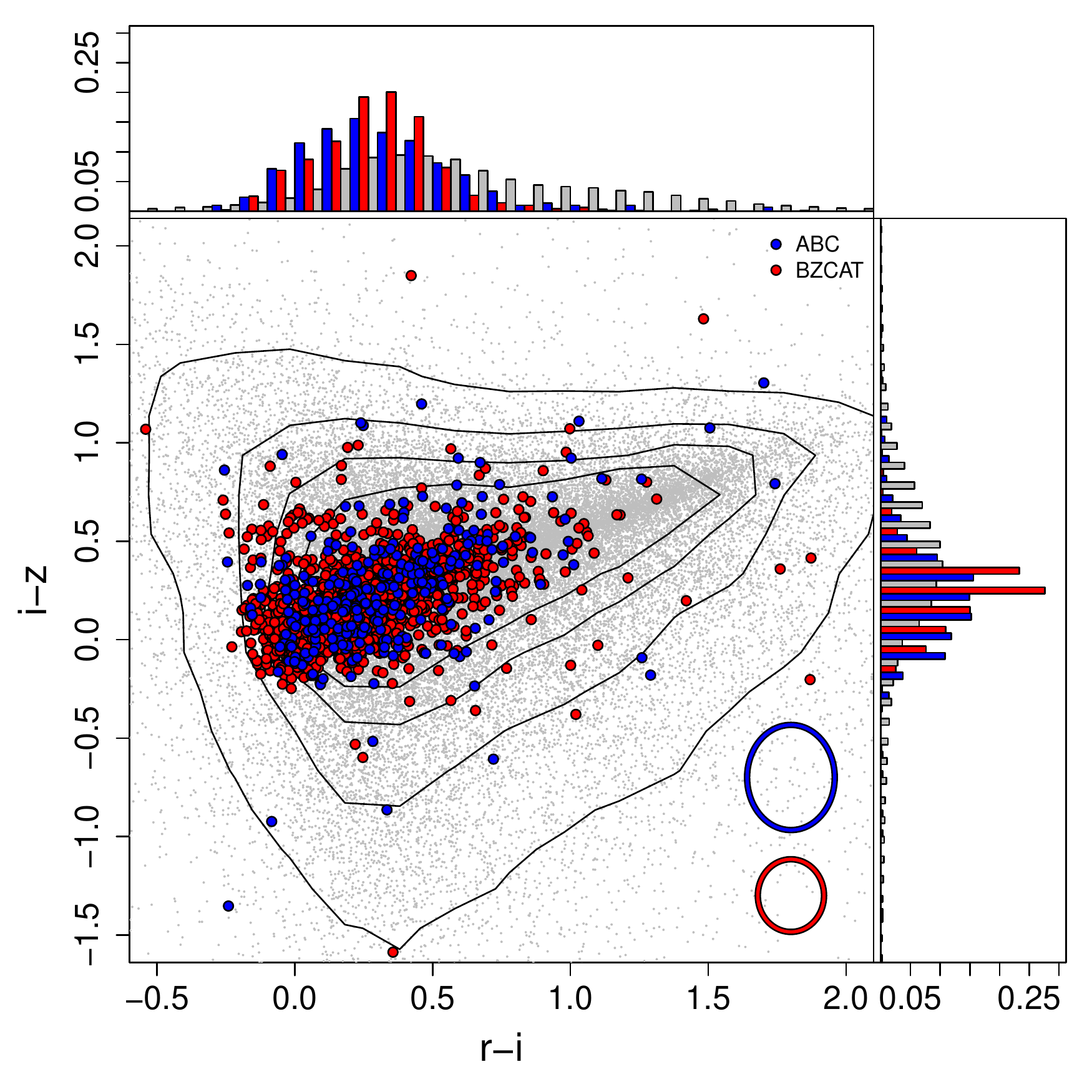}
	\caption{Comparison of ABC sources (blue circles) and \textit{BZCAT} sources (red circles) in the SDSS g-r vs u-g (left panel), r-i vs g-r (central panel) and i-z vs r-i (right panel) colour-colour diagrams. Gray points in the background represent random SDSS sources, and the black continuous lines indicate KDE isodensity curves containing \(60\%\), \(70\%\), \(80\%\) and \(90\%\) of the random sources. On the top and on the right of each main panel we show the normalized distributions of the SDSS colours, for the ABC, \textit{BZCAT} and random sources. The ellipses in the panels indicate the average uncertainties on the SDSS colours.}\label{fig:sdss_colours}
\end{figure*}

                In Fig. \ref{fig:sdss_colours} we compare ABC sources (blue circles) and \textit{BZCAT} sources (red circles) in the SDSS colour-colour plots, namely g-r vs u-g (left panel), r-i vs g-r (central panel), and i-z vs r-i (right panel). SDSS counterparts for \textit{BZCAT} sources have been selected adopting the same matching radius used for ABC sources. Again, all magnitudes have been corrected for Galactic absorption.

\begin{figure*}
	\centering
	\includegraphics[scale=0.4]{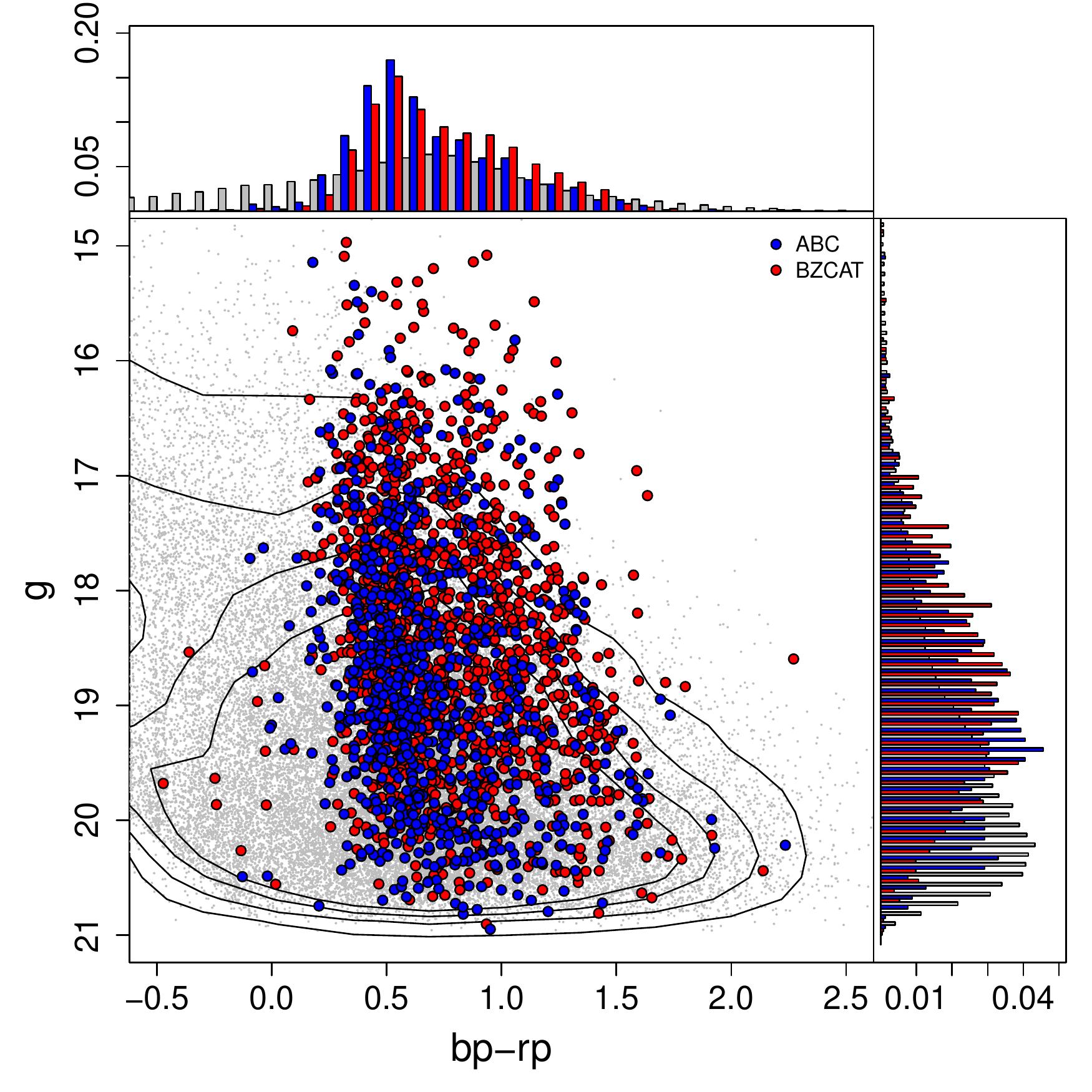}
	\caption{Comparison of ABC sources (blue circles) and \textit{BZCAT} sources (red circles) in the \textit{Gaia} g vs b-r colour-magnitude plot. Gray points in the background represent random \textit{Gaia} sources, and the black continuous lines indicate KDE isodensity curves containing \(60\%\), \(70\%\), \(80\%\) and \(90\%\) of the \textit{Gaia} random sources. On top and on the right of the main panel we present the normalized distributions of the b-r colour and g magnitude, respectively, for the ABC, \textit{BZCAT} and random sources.}\label{fig:colour_mag}
\end{figure*}

                Like for Fig. \ref{fig:colour_mag}, in the panels of Fig. \ref{fig:sdss_colours} we show with gray points in the background \(\sim 5\times{10}^4\) random SDSS sources together with KDE isodensity curves containing \(60\%\), \(70\%\), \(80\%\) and \(90\%\) of these random sources. Again, on the top and on the right of the main panels of this figure we show the normalized distributions of the SDSS colour for the ABC, \textit{BZCAT} and random sources. The ellipses in the panels indicate the average uncertainties on the SDSS colours.

                We see that neither ABC nor \textit{BZCAT} sources can be easily separated from random SDSS sources, with the possible exception of g-r colour, in which ABC and \textit{BZCAT} sources appear bluer than the majority of random sources, although with significant contamination. In general ABC sources appear bluer than \textit{BZCAT} sources. However, the KS test for the distributions of the SDSS colours of ABC sources and \textit{BZCAT} sources have p-chances of \(p=5.7\times{10}^{-3}\), \(p=5.5\times{10}^{-3}\), \(p=4.4\times{10}^{-3}\) and \(p=9.2\times{10}^{-5}\) for the u-g, g-r, r-i and i-z colours, respectively. Therefore we can conclude that although ABC sources are bluer than \textit{BZCAT} sources these differences are not statistically significant.

\begin{figure*}
	\centering
	\includegraphics[scale=0.38]{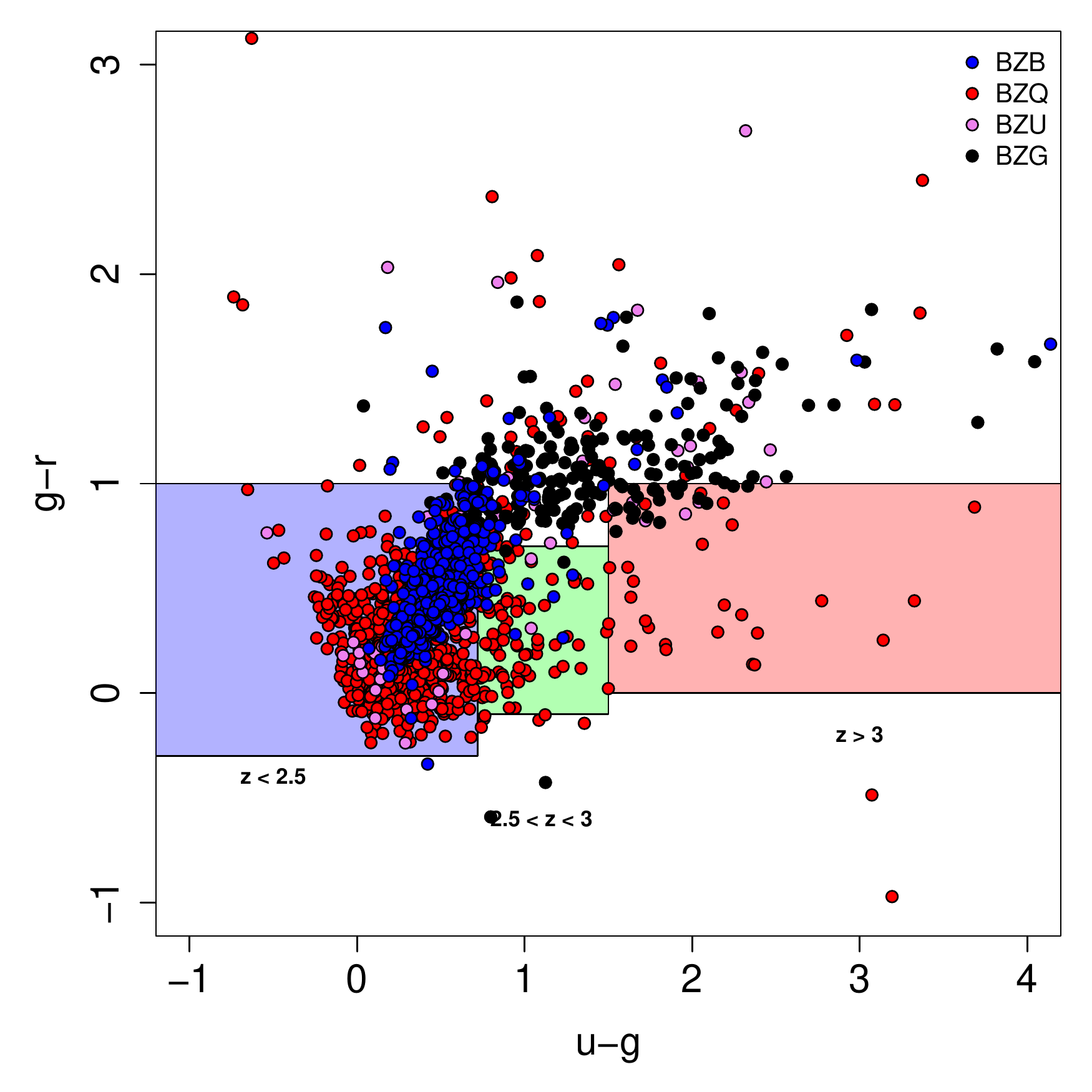}
	\includegraphics[scale=0.38]{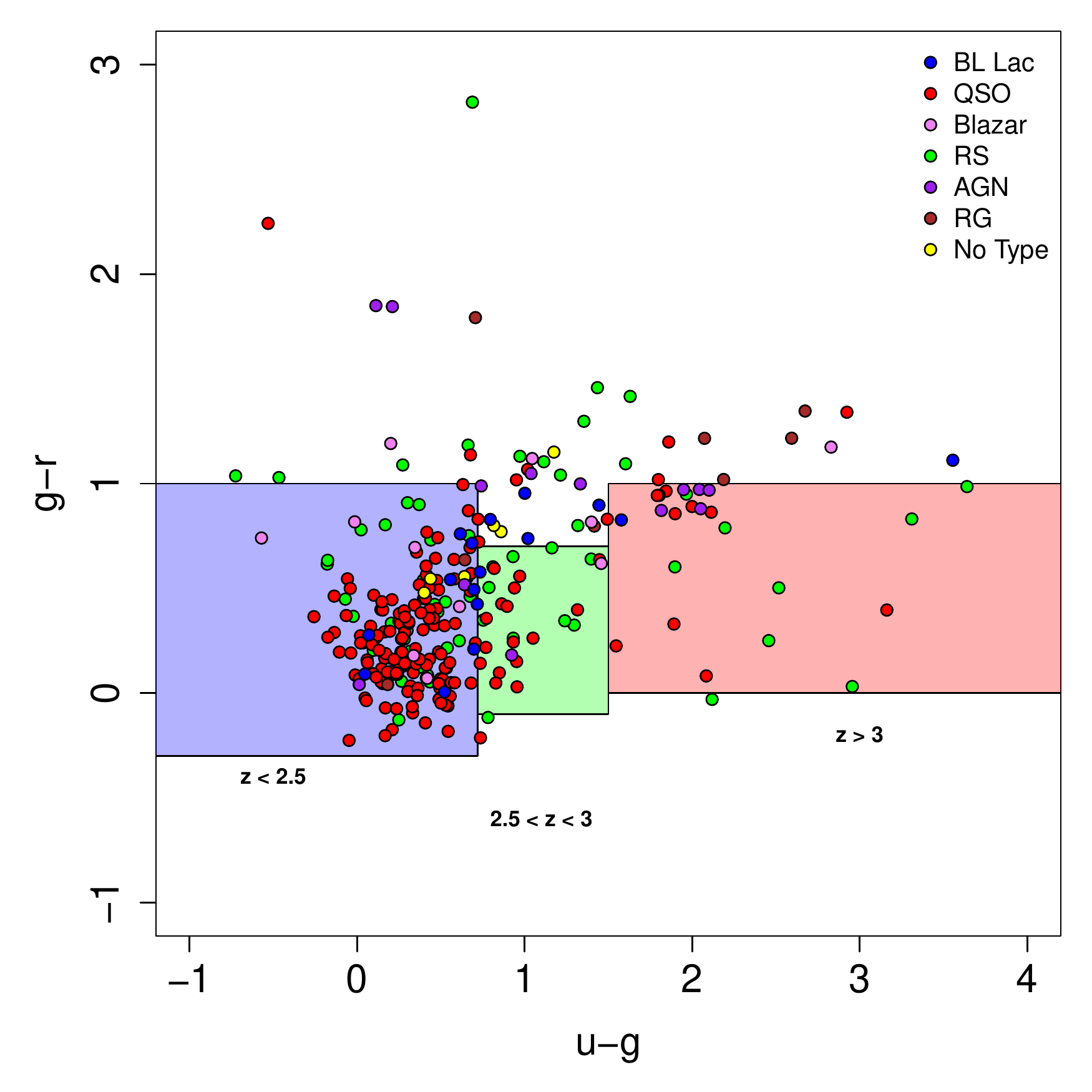}
	\caption{Distribution of \textit{BZCAT} (left panel) and ABC (right panel) sources in the SDSS g-r vs u-g colour-colour plane. The three coloured rectangles represent the regions indicated by \citet{2011AJ....141...93B} to select low (\(z<2.5\)), intermediate (\(2.5<z<3\)), and high (\(z>3\)) redshift quasars. \textit{BZCAT} and ABC subclasses are represented with circles of different colours, as indicated in the legend.}\label{fig:sdss_BB}
\end{figure*}

                In Fig. \ref{fig:sdss_BB} we show the \textit{BZCAT} (left panel) and ABC (right panel) sources on the SDSS g-r vs u-g colour-colour plane, with superimposed the three regions pinpointed by \citet{2011AJ....141...93B} to select low (\(z<2.5\)), intermediate (\(2.5<z<3\)), and high (\(z>3\)) redshift quasars. For \textit{BZCAT} sources, we see that the majority of BZBs and BZQs fall into the region of low redshift quasars, with some BZQs falling into the regions of higher redshift quasars.

                The BZGs, on the other hand, represent the majority of the \textit{BZCAT} sources falling outside the quasar regions from \citeauthor{2011AJ....141...93B}, with some of the u-g redder BZGs turning into the region of \(z>3\) quasars. For ABC sources we have a similar situation, with most QSOs and BL Lacs falling into the region of low redshift quasars, and some QSOs entering the regions of higher redshift quasars. Sources outside the quasar regions from \citeauthor{2011AJ....141...93B} are a mixed bag without a clear predominant class.

\begin{figure*}
	\centering
	\includegraphics[scale=0.4]{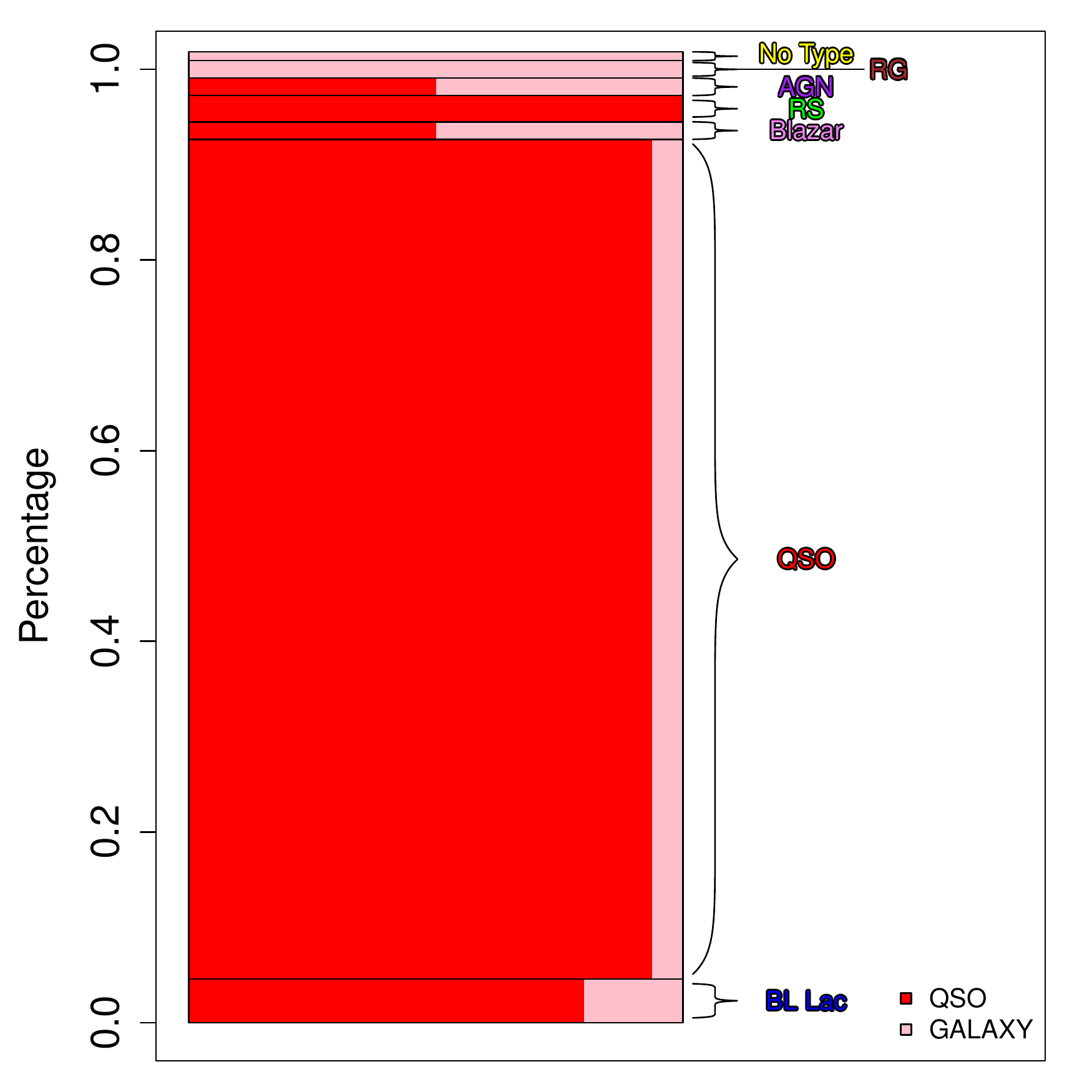}
	\caption{Distribution of the spectral classes (QSO or GALAXY) of the 110 sources with an optical spectra in SDSS DR12 or LAMOST DR5 in the ABC sources types.}\label{fig:source_type_optical}
\end{figure*}

                We then looked for available SDSS DR12 and LAMOST DR5 spectra for the previously identified counterparts, selecting only spectra without analysis flags. In SDSS DR12 we found 98 spectra, and in LAMOST DR5 we found 28 spectra (16 of which already found in SDSS), for a total of 110 sources with available optical spectra, and they are presented in Fig. \ref{fig:sdss_spectra} and Fig. \ref{fig:lamost_spectra}. These spectra are classified either as QSO (98) or GALAXY (12) spectra, and the distribution of these spectral classes in ABC sources types is presented in Fig. \ref{fig:source_type_optical}. The majority of spectral classes in all ABC sources types is of course represented by QSOs, because these represent more than \(90\%\) of the sources with optical spectral classification. We can see, however, that one of the two ABC sources classified as Blazars and as AGN have a GALAXY optical classification, like the two ABC sources classified as RG and the one without literature classification.
            
                The optical properties of ABC sources are summarized in Table \ref{tab:classification}, including the LAMOST and SDSS counterpart, spectral classification and redshift.

            \subsection{Infrared Data: WISE}\label{sec:wise}
                WISE was launched in 2009 and completed its double survey of the whole sky in 2011. Data were acquired with a \(40 \textrm{ cm}\) telescope in four filters, \(w1\), \(w2\), \(w3\), and \(w4\), with central wavelength at \(3.4\), \(4.6\), \(12\), and \(22\textrm{ \textmu m}\), respectively. As mentioned before, WISE data have been extensively used to select blazar candidates and study their properties \citep[see also][]{2012ApJ...745L..27P, 2015AA...580A..73C, 2016AA...587A...8R}. Released in 2013, the AllWISE catalogue \citep{2013wise.rept....1C} is a compilation of the data from the cryogenic and post-cryogenic survey phases of the WISE mission, and it contains more than seven hundred million celestial objects.

\begin{figure*}
	\centering
	\includegraphics[scale=0.38]{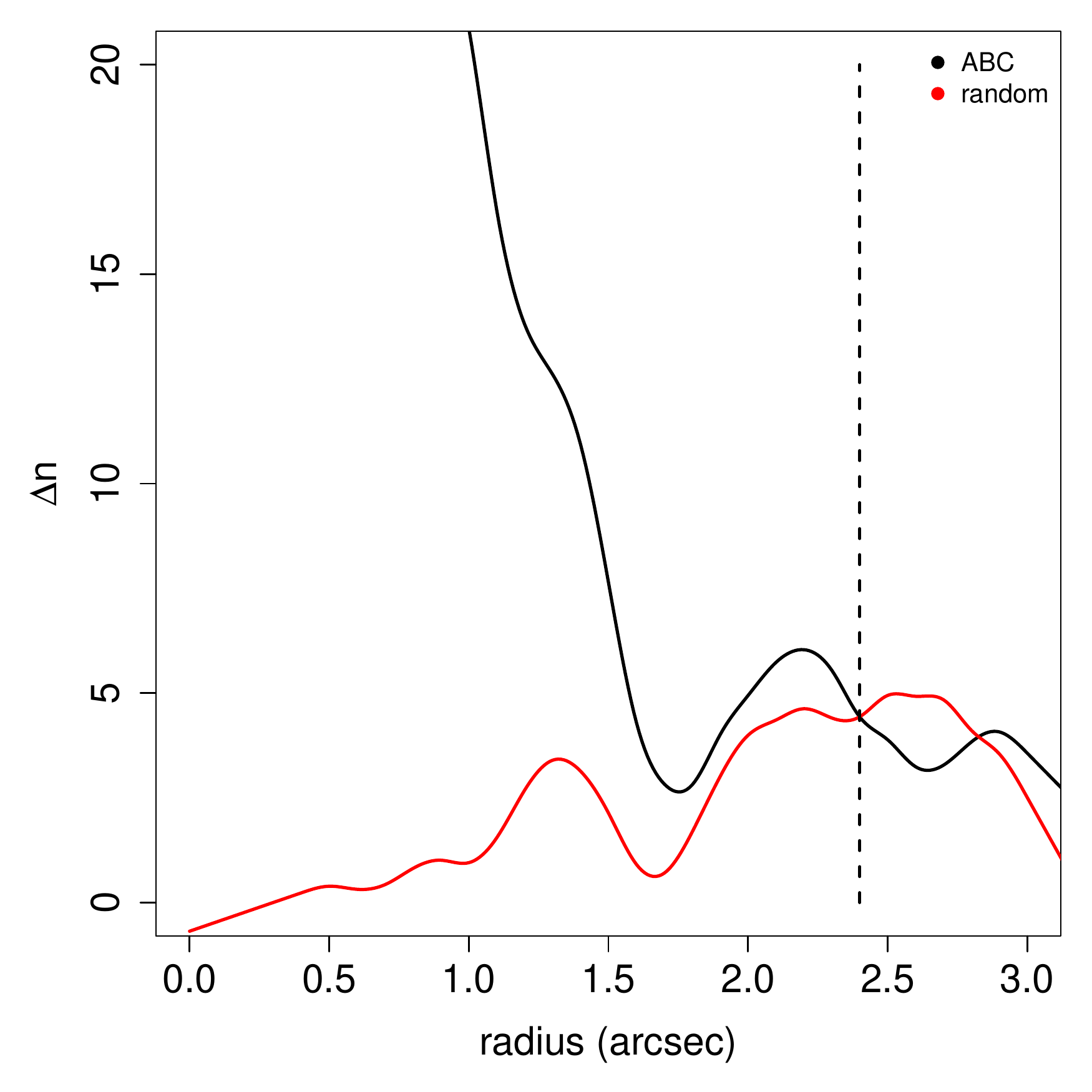}
	\includegraphics[scale=0.38]{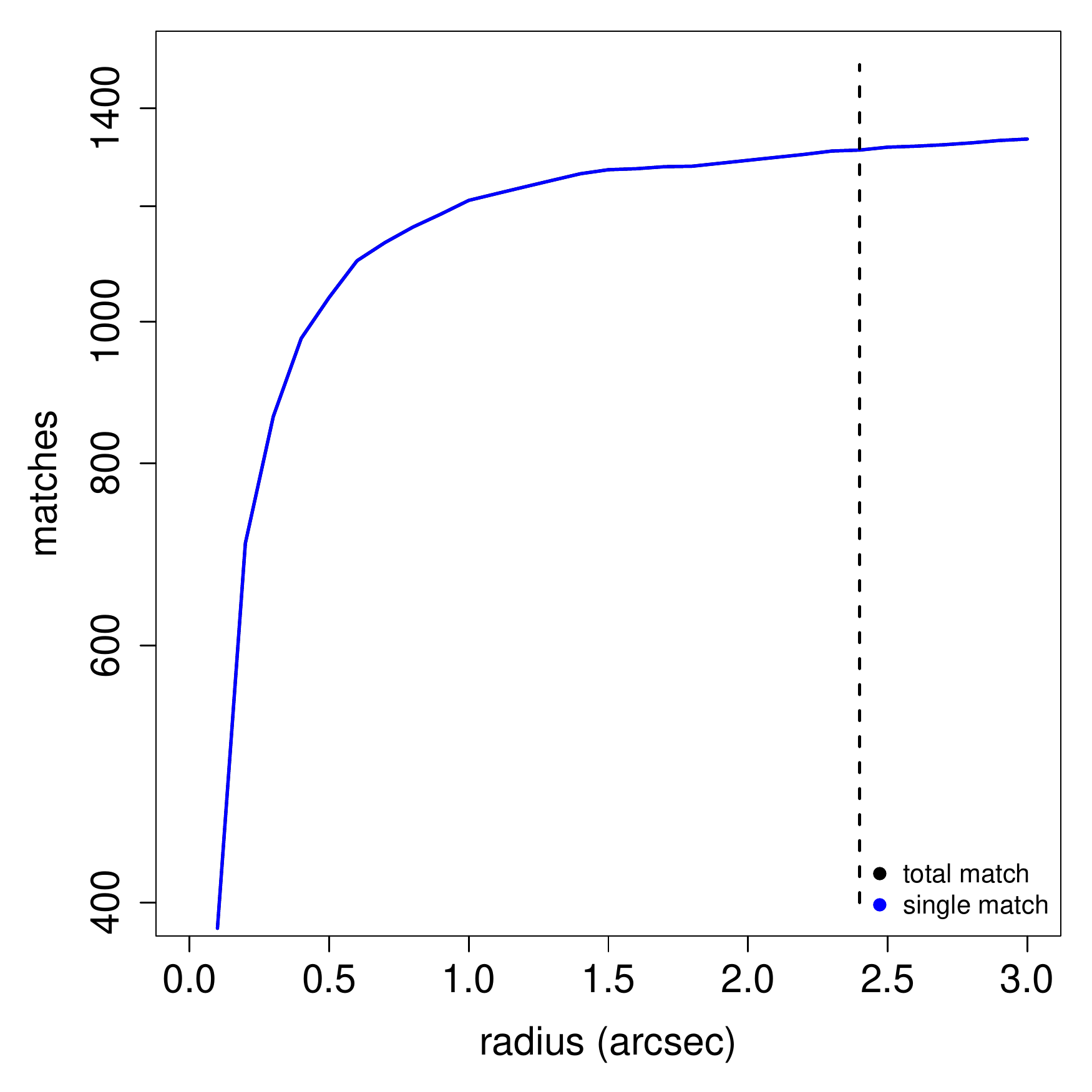}
	\caption{Same as Fig. \ref{fig:sdss_radius}, but for matches between ABC sources and AllWISE source.}\label{fig:wise_radius}
\end{figure*}

                The optimal search radius was determined using the same procedure as for SDSS. In the left panel of Fig. \ref{fig:wise_radius} we compare the increase of sources with at least one AllWISE counterpart (\(\Delta n\)) with increasing search radius for the ABC sources (black line) and for the random positions (red line). Again, (\(\Delta n\)) was smoothed for better visualization. For ABC sources the majority of matches is already found with a search radius of \(1\farcs 5\), while for the random positions (\(\Delta n\)) is initially \(0\), and then it increases after \(1\farcs 0\). We therefore choose as an optimal search radius for WISE counterpart the radius of \(2\farcs 4\) at which (\(\Delta n\)) the increase of matches at random positions in the sky becomes larger than that at the positions of ABC sources. As shown in the right panel of Fig. \ref{fig:wise_radius}, at this radius we have a total of 1311 AllWISE matches for the ABC sources, without multiple matches.

\begin{figure*}
	\centering
	\includegraphics[scale=0.38]{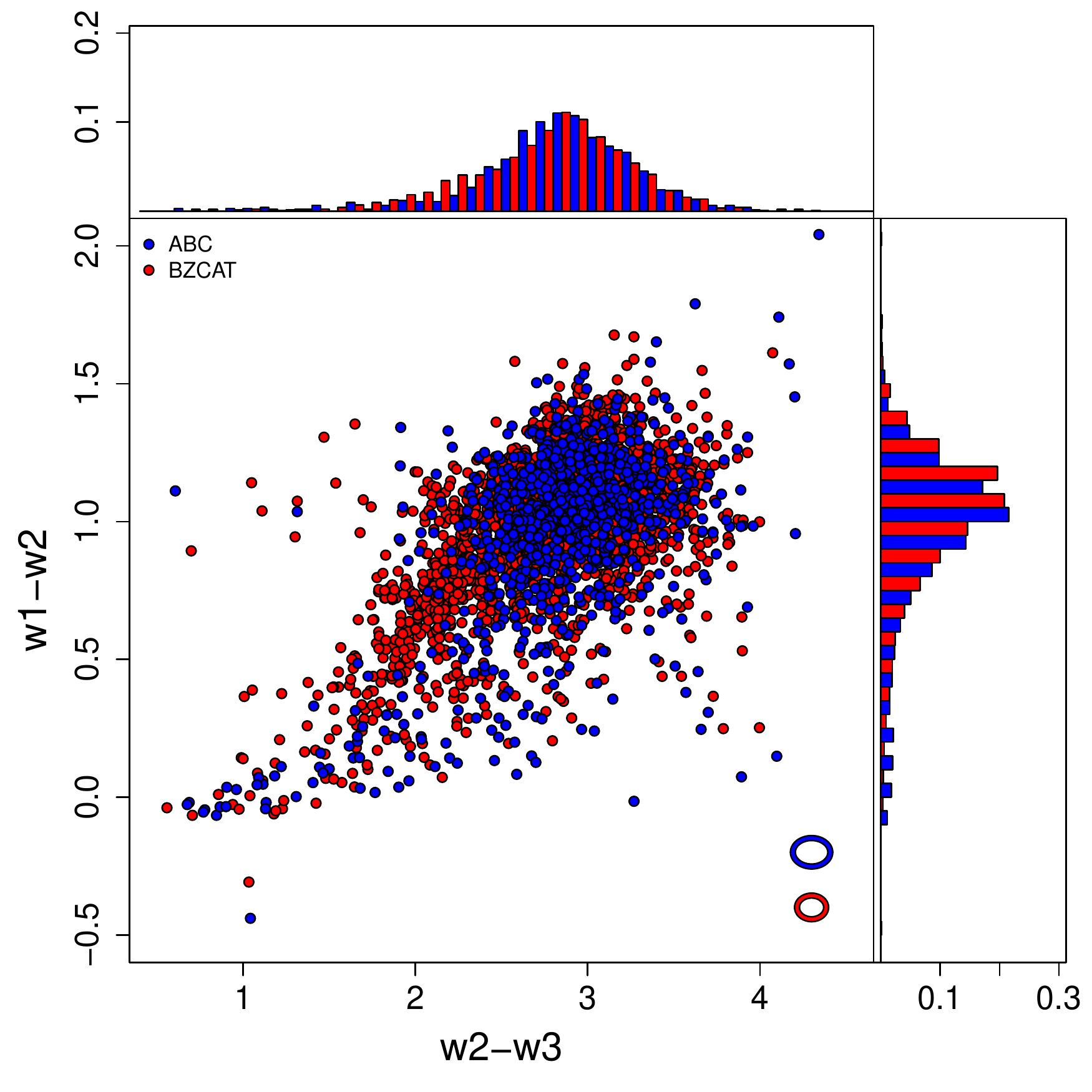}
	\includegraphics[scale=0.38]{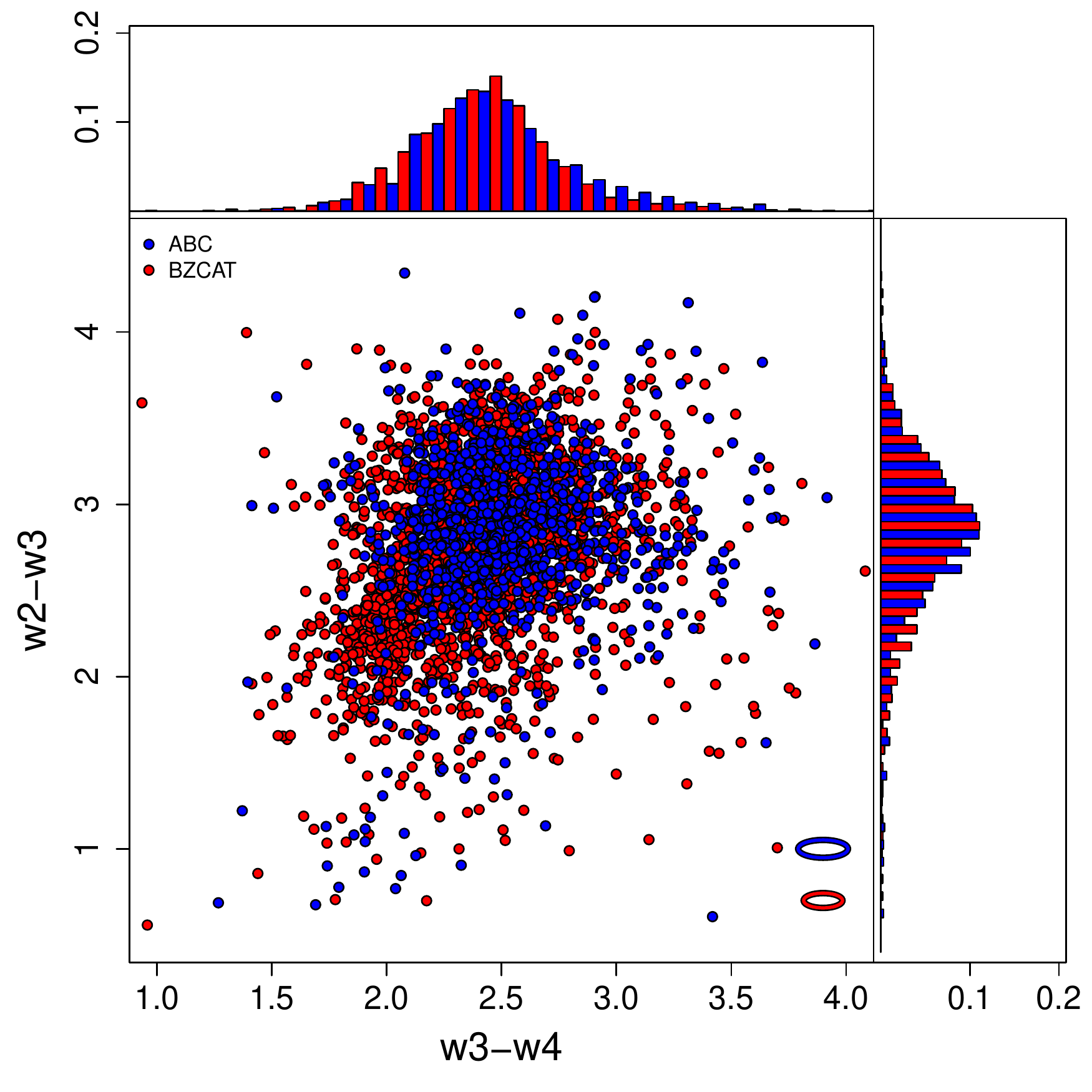}
	\caption{Distribution of ABC sources (blue circles) and \textit{BZCAT} sources (red circles) in the WISE w1-w2 vs w2-w3 (left panel) and w2-w3 vs w3-w4 (right panel) colour-colour diagrams. On the top and on the right of each main panel we show the normalized distributions of the WISE colours for the ABC and \textit{BZCAT} sources. The ellipses in the panels indicate the average uncertainties on the WISE colours.}\label{fig:wise_bzcat}
\end{figure*}

                In particular we have 906 source with a detection in all four WISE bands, 739 of which (\(\sim 82\%\)) have a detection with a signal to noise ratio of at least \(3\) in all four WISE bands. The distribution of the 906 ABC sources with a detection in all four WISE bands in the WISE colour-colour diagrams is presented with blue circles in Fig. \ref{fig:wise_bzcat}, where the distribution of the \textit{BZCAT} sources (red circles) is also shown for comparison. The w1-w2 vs w2-w3 and w2-w3 vs w3-w4 colour-colour diagrams are presented in the left and right panel of the figure, respectively. WISE counterparts for \textit{BZCAT} sources have been selected adopting the same matching radius used for ABC sources. As usual, all magnitudes have been corrected for Galactic absorption. We note however that the average Galactic absorption is smaller than the average photometric uncertainties for \(\left|{b}\right|>{35}\degree\) and \(\left|{b}\right|>{20}\degree\) for w1 and w2 bands, respectively, and in these two bands it becomes lager than \(0.1\textrm{ mag}\) for \(\left|{b}\right|<{20}\degree\). On the other hand Galactic absorption becomes relevant in w3 and w4 bands only very close to the Galactic center (\(\left|{b}\right|<{4}\degree\)). Again, on the top and on the right of the main panels we show the normalized distributions of the WISE colour for the ABC and \textit{BZCAT} sources. The ellipses in the panels indicate the average uncertainties on the WISE colours. Sources from ABC have similar w1-w2 colours to \textit{BZCAT} sources, while they appear bluer in the w2-w3 and w3-w4 colours. This is confirmed by the KS test for the distributions of the WISE colours of ABC sources and \textit{BZCAT} sources, that yield p-chances of \(p=0.04\), \(p=4.0\times{10}^{-4}\) and \(p=8.9\times{10}^{-11}\) for the w1-w2, w2-w3 and w3-w4 colours, respectively. Therefore we can conclude that the difference is significant only in the w2-w3 and w3-w4 colours.

        \subsection{X-ray Data}\label{sec:xray}
            Blazars are known X-ray sources since \textit{ROSAT} DXRBS \citep{1998AJ....115.1253P, 2001MNRAS.323..757L} and \textit{Einstein} IPC \citep{1992ApJS...80..257E, 1999AAS...195.1601P} surveys \citep[see also][]{2000AIPC..515...53P}. Since then, the X-ray properties of blazars have been deeply investigated by many authors \citep[see for example][]{1994MNRAS.268L..51G, 1995ApJ...444..567P, 1997ApJ...480..534C, 2000astro.ph..5479R, 2001AA...375..739D, 2011ApJ...739...73M, 2011ApJ...742L..32M}.

            The nature of X-ray emission in blazar is essentially non-thermal, and the physical processes responsible for this emission vary in the different blazar subclasses. For HSPs the synchrotron peak frequency \(\nu_p\) is larger than \({10}^{15}\textrm{ Hz}\), and therefore the X-ray emission in these sources is dominated by the synchrotron radiation from the relativistic electrons in the jets. As the synchrotron peak frequency moves to lower energies the SSC component enters the X-ray band, and for LSPs with \(\nu_p < {10}^{14}\textrm{ Hz}\) this is the dominant contribution to the observed X-ray emission. This is even more true in FSRQs, where the synchrotron peak frequency is \(\nu_p < {10}^{13}\textrm{ Hz}\) and the X-ray emission is dominated by the inverse Compton components (SSC and EC).

            To study the X-ray emission from the ABC sources we made use of the data available for this sample in the archives of \textit{Swift}, \textit{Chandra} and \textit{XMM-Newton} missions.

            \subsubsection{\textit{Swift}-XRT}\label{sec:swift}
                \textit{Swift} has proven to be an excellent multi-frequency observatory for blazar research, so far observing hundreds of sources \citep[e.g.,][]{2007SPIE.6688E..0GM, 2012AA...548A..87M, 2012AAS...21941506D}, providing an extremely rich and unique database of multi-frequency (optical, UV, X-ray), simultaneous blazar observations. Several papers on samples selected with different criteria have already been published, including: blazars detected at TeV energies \citep[e.g.,][]{2008AA...478..395M, 2011ApJ...739...73M, 2011ApJ...742L..32M}, simultaneous optical-to-X-ray observations of flaring TeV sources \citep[e.g.,][]{2007AA...462..889P, 2007AA...467..501T} as well as the investigation of low and high frequency peaked BL Lacs \citep[e.g.,][]{2010AA...512A..74M, 2012AA...541A.160G}. \textit{Swift} has also been used for UV-optical and X-ray follow-up observations of TeV flaring blazars \citep[e.g.,][]{2011ApJ...742..127A, 2012AA...544A.142A, 2013AA...552A.118H} and in the framework of broad-band multiwavelength campaigns \citep[e.g.][]{2011A&A...534A..87R, 2013MNRAS.436.1530R, 2017MNRAS.472.3789C}. It has also been useful in obtaining photometric redshift constraints for many \textit{Fermi}-detected BL Lacs \citep{2012AA...538A..26R}.

                The XRT data were downloaded from HEASARC\footnote{\href{https://heasarc.gsfc.nasa.gov/}{https://heasarc.gsfc.nasa.gov/}} data archive, and processed using the \textsc{XRTDAS} software \citep{capalbi2005} developed at the ASI Science Data Center and included in the HEAsoft package (v. 6.26.1) distributed by HEASARC, using a procedure similar to that illustrated in \citet{2013ApJS..209....9P}. \textit{Swift}-XRT photon counting (PC) data were available for 325 ABC sources. For each observation calibrated and cleaned PC mode event files were produced with the \textsc{xrtpipeline} task (ver. 0.13.5), producing exposure maps for each observation. In addition to the screening criteria used by the standard pipeline processing, we applied a further filter to screen background spikes that can occur when the angle between the pointing direction of the satellite and the bright Earth limb is low. In order to eliminate this so called bright Earth effect, due to the scattered optical light that usually occurs towards the beginning or the end of each orbit, we used the procedure proposed by \citet{2011AA...528A.122P} and \citet{2013AA...551A.142D}. We monitored the count-rate on the CCD border and, through the \textsc{xselect} package, we excluded time intervals when the count-rate in this region exceeded \(40 \textrm{ counts/s}\). In addition we selected only time intervals with CCD temperatures less than \(-50\degree\textrm{ C}\) (instead of the standard limit of \(-47\degree\textrm{ C}\)) since contamination by dark current and hot pixels, which increase the low energy background, is strongly temperature dependent \citep{2013AA...551A.142D}.

                To detect X-ray sources in the XRT images, we made use of the \textsc{ximage} detection algorithm \textsc{detect}, which locates the point sources using a sliding-cell method. The average background intensity is estimated in several small square boxes uniformly located within the image. The position and intensity of each detected source are calculated in a box whose size maximizes the signal-to-noise ratio. The algorithm was set to work in bright mode, which is recommended for crowded fields and fields containing bright sources, since it can reconstruct the centroids of very nearby sources. We then evaluated the net count-rates for the detected sources with the \textsc{sosta} algorithm that, besides the net count-rates and the respective uncertainties, yields the statistical significance of each source. We note that \textsc{sosta} requires the positions of the sources detected by \textsc{detect}, and the uncertainties in the count-rates returned by \textsc{sosta} are purely statistical - i.e. do not include systematic errors - and are in general smaller than those given by \textsc{detect}. We used count-rates produced by \textsc{sosta} because these are in most cases more accurate, because \textsc{detect} uses a global background for the entire image, whereas \textsc{sosta} uses a local background. Finally, we refined the source position and relative positional errors by the task \textsc{xrtcentroid} of the \textsc{XRTDAS} package, considering only the sources detected at position compatible with the ABC sources. In this way we detected X-Ray counterparts for 101 sources.

                In general XRT-PC source spectra - with the corresponding arf and rmf files - are obtained form events extracted with \textsc{xrtproducts} task using a 30 pixel radius circle centered on the detected source coordinates, while background spectra were estimated from a nearby source-free circular region of 60 pixel radius. When the source count-rate is above \(0.5 \textrm{ counts}/\textrm{s}^{-1}\), the data are significantly affected by pileup in the inner part of the PSF \citep{2005SPIE.5898..360M}. To remove the pile-up contamination, we extract only events contained in an annular region centered on the source \citep[][]{2007AA...462..889P}. The inner radius of the region was determined by comparing the observed profiles with the analytical model derived by \citet{2005SPIE.5898..360M} and typically has a 4 or 5 pixels radius, while the outer radius is 20 pixels for each observation. In this way we were able to obtain X-ray for spectra 43 ABC sources. %(39 PL + 4 peculiar)

            \subsubsection{\textit{Chandra}-ACIS}\label{sec:chandra}
                \textit{Chandra} X-ray telescope has been used in the past years to provide important information about the high-energy emission of blazars \citep[e.g.][]{2019MNRAS.489.2732I}. In addition, thanks to its unmatched angular resolution, Chandra has been used to resolve and study the X-ray jets of several blazar jets \citep[e.g.][]{2004ApJ...614..615J, 2007ApJ...662..900T, 2011ApJ...729...26M, 2011ApJ...730...92H}.

                \textit{Chandra}-ACIS data were available for fields containing 62 ABC sources, and they were retrieved from the \textit{Chandra} Data Archive\footnote{\href{http://cda.harvard.edu/chaser}{http://cda.harvard.edu/chaser}}, we run the ACIS level 2 processing with \textsc{chandra\_repro} to apply up-to-date calibrations (CTI correction, ACIS gain, bad pixels), and then excluded time intervals of background flares exceeding 3\(\sigma\) with the \textsc{deflare} task. We produced full-band exposure maps, psf maps to evaluate the psf size across the ACIS detector, and pileup maps with the \textsc{pileup\_map} task. We then run the \textsc{wavdetect} task to identify point sources in each observation with a \(\sqrt{2}\) sequence of wavelet scales (i.e., 1 1.41 2 2.83 4 5.66 8 11.31 16 pixels) and a false-positive probability threshold of \({10}^{-6}\). We then considered only the sources detected at position compatible with the ABC sources, and extracted count-rates making use of the \textsc{srcflux} task. In this way we detected X-ray counterparts for 56 ABC sources.

                ACIS source spectra and the corresponding arf and rmf files were extracted with the \textsc{specextract} tool from the source regions generated from \textsc{wavdetect}, excluding the inner pixels with pileup larger than \(5\%\) as estimated from the pileup maps, while background spectra were extracted from source-free circular regions with typical radii of \(80\arcsec\). We were able to extract spectra for 46 ABC sources. %(32 PL + 14 peculiar)

            \subsubsection{\textit{XMM-Newton}-EPIC}\label{sec:xmm}
                textit{XMM-Newton} space observatory, thanks to its large collecting area the ability to make long uninterrupted observations, provided important information for the multi-wavelength study of blazars \citep[e.g.][]{2006A&A...452..845R, 2009AstL...35..579F, 2015MNRAS.451.1356K, 2016NewA...44...21B}.

                \textit{XMM-Newton}-EPIC data, available for fields containing 64 ABC sources, were retrieved from the \textit{XMM-Newton} Science Archive\footnote{\href{http://nxsa.esac.esa.int/nxsa-web}{http://nxsa.esac.esa.int/nxsa-web}} and reduced with the \textsc{SAS}\footnote{\href{http://www.cosmos.esa.int/web/xmm-newton/sas}{http://www.cosmos.esa.int/web/xmm-newton/sas}} 18.0.0 software.
                
                Following \citet{2005ApJ...629..172N} we filtered EPIC data for hard-band flares by excluding the time intervals where the \(9.5-12 \textrm{ keV}\) (for MOS1 and MOS2) or \(10-12 \textrm{ keV}\) (for PN) count-rate evaluated on the whole detector FOV was more than 3\(\sigma\) away from its average value. To achieve a tighter filtering of background flares, we iteratively repeated this process two more times, re-evaluating the average hard-band count-rate and excluding time intervals away more than \(3\sigma\) from this value. The same procedure was applied to soft \(1-5 \textrm{ keV}\) band restricting the analysis to an annulus with inner and outer radii of \(12\arcmin\) and \(14\arcmin\) excluding sources in the field, where the detected emission is expected to be dominated by the background.
                
                When possible, we merged data from MOS1, MOS2 and PN detectors from all observations using the \textsc{merge} task, in order to detect the fainter sources that wouldn't be detected otherwise. Sources were detected on these merged images following the standard SAS sliding box task \textsc{edetect\_chain} that mainly consist of three steps: 1) source detection with local background, with a minimum detection likelihood of 8; 2) remove sources in step 1 and create a smooth background maps by fitting a 2-D spline to the residual image; 3) source detection with the background map produced in step 2 with a minimum detection likelihood of 10. The task \textsc{emldetect} was then used to determine the parameters for each input source - including the count-rate - by means of a maximum likelihood fit to the input images, selecting sources with a minimum detection likelihood of 15 and a flux in the \(0.3-10 \textrm{ keV}\) band larger than \({10}^{-14}\textrm{ erg}\textrm{ cm}^{-2}\textrm{ s}^{-1}\) (assuming an energy conversion factor of \(1.2\times {10}^{-11}\textrm{ cts}\textrm{ cm}^{2}\textrm{ erg}^{-1}\)). An analytical model of the point spread function (PSF) was evaluated at the source position and normalized to the source brightness. The source extent was then evaluated as the radius at which the PSF level equals half of local background. We then considered only the sources detected at position compatible with the ABC sources. In this way we detected X-ray counterparts for 55 ABC sources.
                
                The source spectra were extracted with the \textsc{evselect} task from the regions obtained with \textsc{emldetect}. The inner regions of high pileup were estimated using the \textit{epatplot} through the distortion of pattern distribution, following the procedure explained in the SAS Data Analysis Threads\footnote{\href{https://www.cosmos.esa.int/web/xmm-newton/sas-thread-epatplot}{https://www.cosmos.esa.int/web/xmm-newton/sas-thread-epatplot}}. The corresponding arf and rmf files were generated with the \textsc{rmfgen} and \textsc{arfgen} tasks to take into account time and position-dependent EPIC responses, and background spectra were extracted from source free regions of the sky. We were able to extract spectra for 51 ABC source. %(38 PL + 13 Peculiar)

            \subsubsection{X-Ray Spectral Fitting}\label{sec:xray-spectra}
                In total we detected X-ray counterparts for 173 ABC sources, and we were able to extract a total of 140 spectra for 92 ABC sources. Spectral fitting was performed with the Sherpa\footnote{\href{http://cxc.harvard.edu/sherpa}{http://cxc.harvard.edu/sherpa}} modeling and fitting application \citep{2001SPIE.4477...76F} in the \(0.3-7\textrm{ keV}\) energy range, adopting Gehrels weighting \citep{1986ApJ...303..336G}. Source spectra were binned to a minimum of 20 counts/bin to ensure the validity of \(\chi^2\) statistics. For the EPIC spectra we excluded from the spectral fitting the \(1.45-1.55\textrm{ keV}\) band due to variable Al K lines, and fitted simultaneously the MOS1, MOS2 and PN spectra.
            
                For the spectral fitting we used a model comprising an absorption component fixed to the Galactic value \citep{2005A&A...440..775K} and a power law, as expected in blazars. This model proved to adequately fit the majority (109) of the extracted spectra. The results of the power-law model fitting are presented in Table \ref{tab:xray_properties} where errors correspond to the \(1\)-\(\sigma\) confidence level for one interesting parameter (\(\Delta\chi^2 = 1\)). X-ray spectra fitted with power-law models are presented in Figs. \ref{fig:xrt_spectra}, \ref{fig:acis_spectra} and \ref{fig:epic_spectra}. The fluxes listed in Table \ref{tab:xray_properties} are estimated from the spectral fitting when a spectra was available, otherwise they have been estimated from the measured count-rates assuming a power-law model with a spectral index of \(2\).

                We note that 31 spectra for 24 ABC sources were not adequately fitted by a simple power-law model, but instead required more complex models comprising intrinsic absorption, thermal components, reflections components, and/or emission lines, and are presented in Figs. \ref{fig:xrt_spectra_peculiar}, \ref{fig:acis_spectra_peculiar}, and \ref{fig:epic_spectra_peculiar}. The results of the fit procedure on these spectra are summarized in Table \ref{tab:properties_peculiar}. Interestingly, only one of these 24 sources showing complex X-ray spectra (namely J1215-1731) has a Blazar classification in SIMBAD, while the others are AGNs (11 sources), QSOs (9), RSs (1), RGs (1) and objects without SIMBAD classification (1).

\begin{figure*}
	\centering
	\includegraphics[scale=0.38]{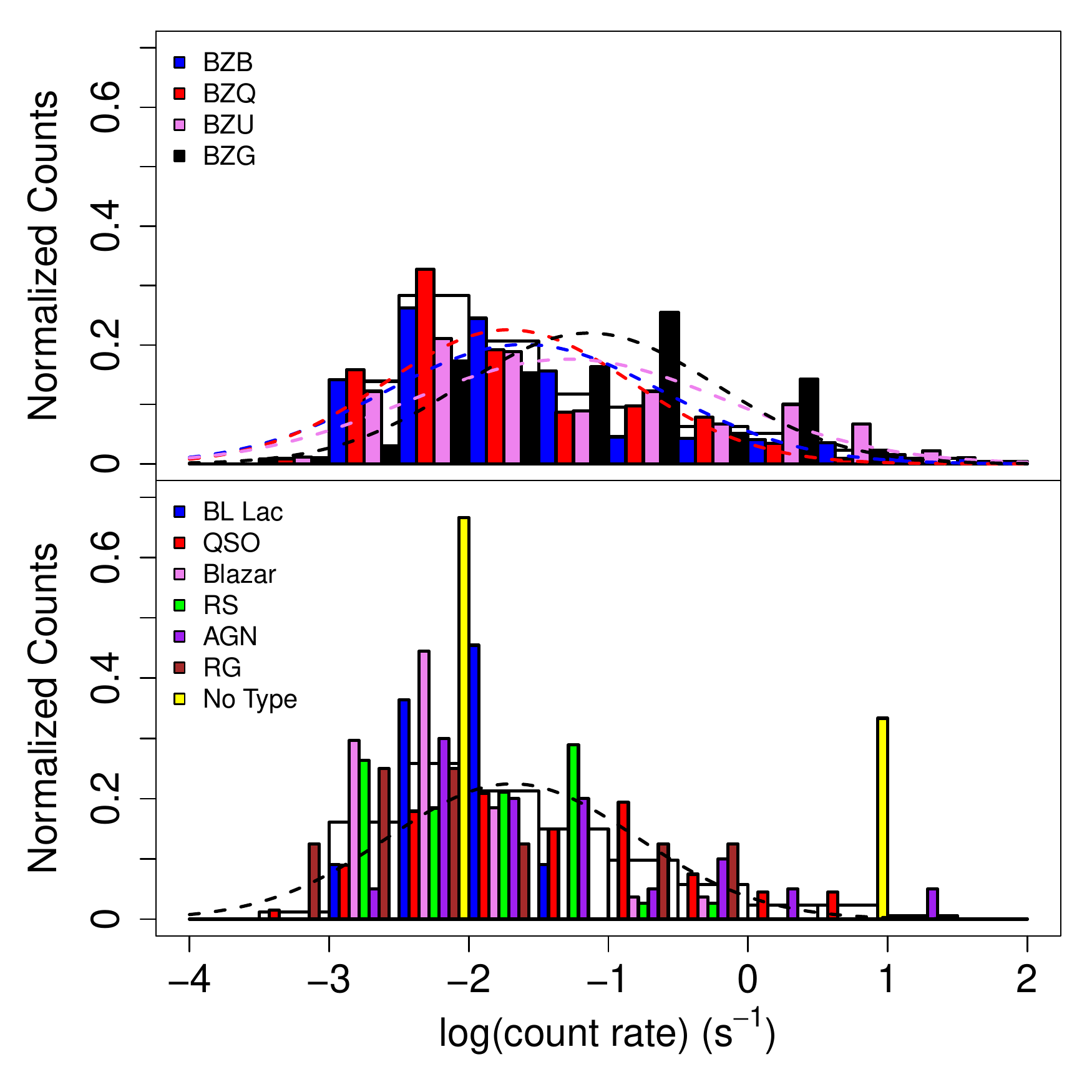}
	\includegraphics[scale=0.38]{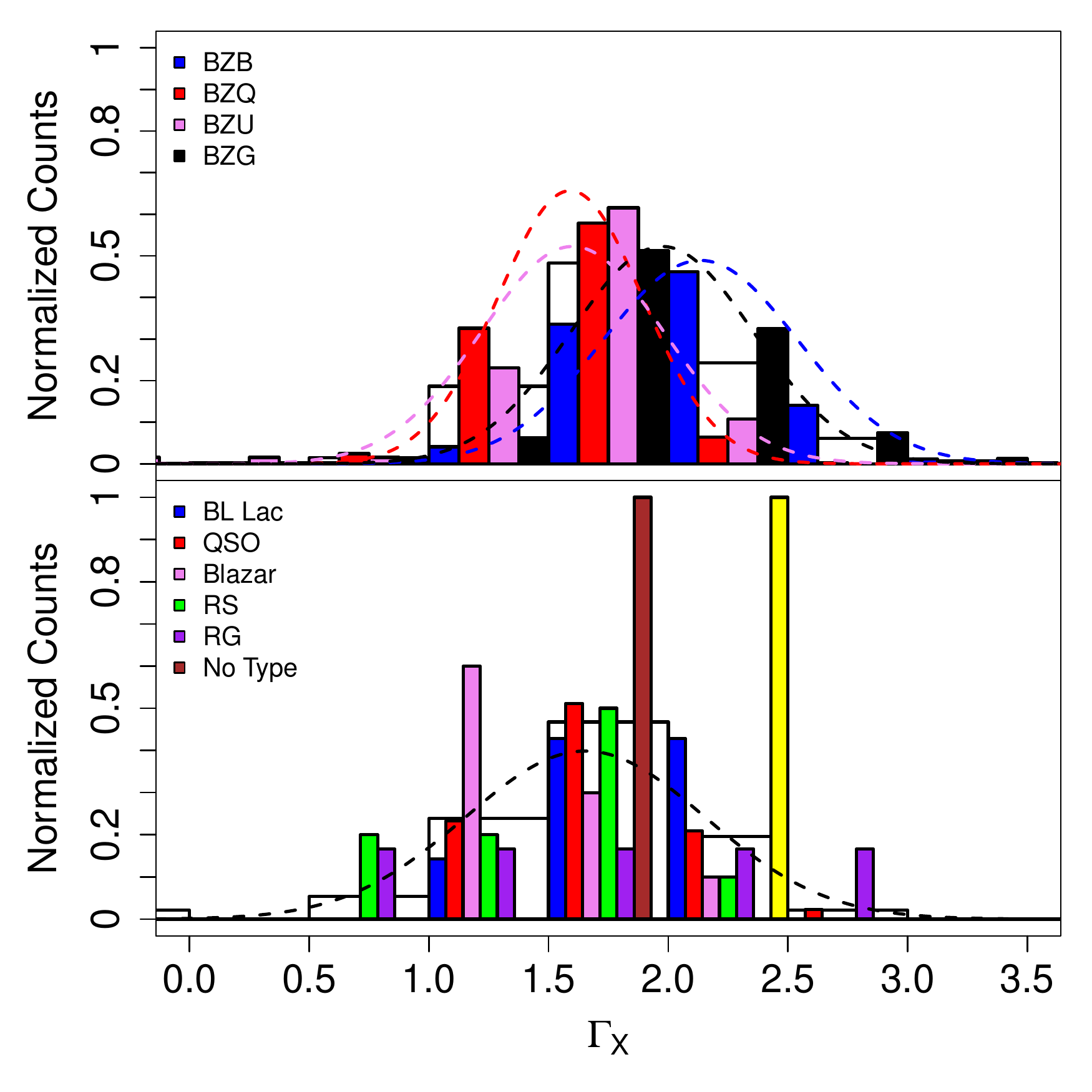}
	\caption{(Left panel) Normalized distributions of the \(0.3-7 \textrm{ figures/keV}\) count-rate for \textit{BZCAT} (top) and ABC (bottom) sources. The various subclasses of \textit{BZCAT} sources and types of ABC sources are presented with the colours indicated in the legend, and the white histograms on the background represent the distribution of the whole samples. For \textit{BZCAT} sources gaussian fits to the subclass distributions are indicated with dashed lines of the respective colours, while for ABC sources a gaussian fit to the distribution of the whole sample is indicated with a black dashed line. (Right panel) Same as the left panel, but for the distributions of the X-ray slope \(\Gamma_X\).}\label{fig:xray_hist}
\end{figure*}

                It is instructive to compare the X-ray properties of ABC and \textit{BZCAT} sources. To this end, we obtained and reduced \textit{Swift}-XRT, \textit{Chandra}-ACIS and \textit{XMM-Newton} data for \textit{BZCAT} sources in the same way we did for ABC sources. In the left panel of Fig. \ref{fig:xray_hist} we compare the normalized distributions of the X-ray count-rates for \textit{BZCAT} (top) and ABC (bottom) sources. When count-rates for a source were available for more than one instrument, we picked the count-rate corresponding to the detection with the higher signal to noise ratio. The \textit{BZCAT} of the different subclasses (BZB, BZQ, BZU and BZG) are presented with different colours, and gaussian fits to the count-rate distributions of each subclass are indicated with dashed lines of the respective colour. The peak logarithmic fluxes (in cgs units) of BZBs and BZQs are at \(-1.6\) and \(-1.7\), respectively, while BZUs and BZGs are slightly brighter on average, peaking at \(-1.3\) and \(-1.1\), respectively. For ABC sources the different source types are indicated with different colours. The dashed black line is a gaussian fit to the distribution of the whole sample is presented with a black dashed line. ABC sources show similar count-rates in X-rays compared to \textit{BZCAT} sources, at variance with what is observed at radio wavelengths (see Fig. \ref{fig:acc_bzcat_radio_flux}). In particular, the distribution of logarithmic count-rates for \textit{BZCAT} sources peaks at \(-1.6\), while that of ABC sources peaks at \(-1.7\). In addition the KS test shows that these two distributions have a p-chance \(p=0.44\) of having been randomly sampled from a common parent distribution, and are therefore statistically indistinguishable.

                The right panel of Fig. \ref{fig:xray_hist} shows the comparison of the power-law slopes \(\Gamma_X\) of the X-ray spectra in terms of count rates. Again, when \(\Gamma_X\) slopes for a source were available for more than one instrument, we picked the slope corresponding to the detection with the higher signal to noise ratio. BZBs and BZGs are the softer subclasses in X-rays, with their \(\Gamma_X\) distributions peaking at \(2.1\) and \(2.0\), respectively, while BZQs and BZUs are harder, both peaking at \(1.6\). The ABC sources have a \(\Gamma_X\) distribution similar to that of the \textit{BZCAT} as a whole. The ABC distribution peaks at \(\Gamma_X = 1.7\), while that of \textit{BZCAT} sources as whole peaks at \(1.8\). Also, the KS test shows that these two distributions have a p-chance \(p=0.10\) of having been randomly sampled from a common parent distribution, being therefore similar.

\begin{figure*}
	\centering
	\includegraphics[scale=0.4]{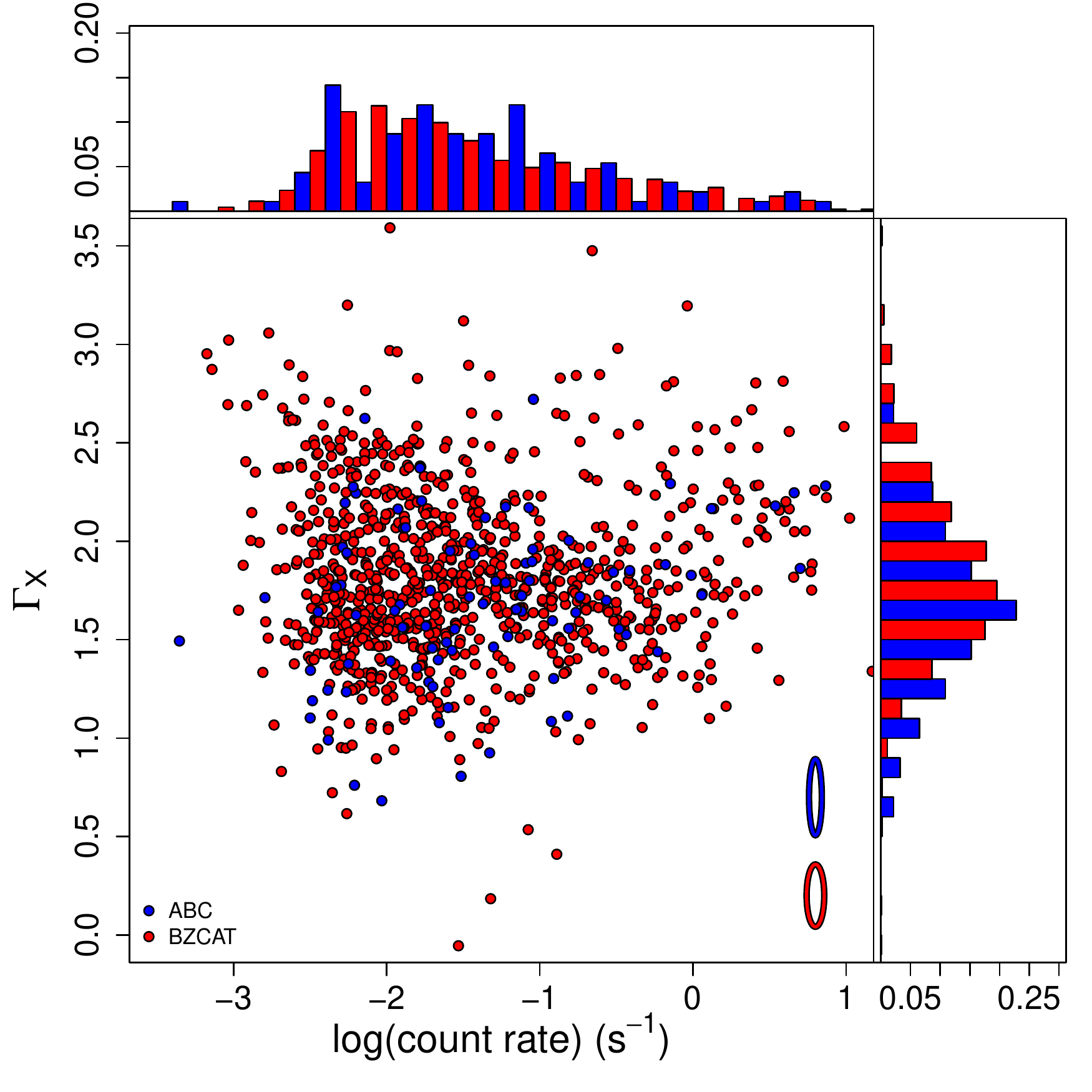}
	\caption{ABC (blue circles) and \textit{BZCAT} (red circles) sources represented on the X-ray slope vs. X-ray count-rate plot. The ellipses of the relative colour indicate the average uncertainties. On top and on the right of the main panels we show the normalized distributions of the X-ray count-rate and slope for ABC and \textit{BZCAT} sources, with the same colours used in the main panel.}\label{fig:xray_slope_vs_countrate}
\end{figure*}

                In Fig. \ref{fig:xray_slope_vs_countrate} we present on a X-ray slope vs. X-ray count-rate plot the ABC and \textit{BZCAT} sources that have an estimate for both count-rate and X-ray slope. On the top and right of this figure we present the normalized distributions of the X-ray count-rate and slope \(\Gamma_X\) for ABC and \textit{BZCAT} sources. Both count-rate and \(\Gamma_X\) distributions are similar between ABC and \textit{BZCAT} sources, and this is confirmed by the KS test, that yields a p-chance of \(0.11\) and \(0.10\) for the count-rate and \(\Gamma_X\) distributions, respectively.

        \subsection{\(\gamma\)-ray Data}\label{sec:gammaray}
            In this section we compare the \(\gamma\)-ray properties of ABC and \textit{BZCAT} sources, as reported in 4FGL catalog. In Sec. \ref{sec:catalog_comparison} we explained that we found 259 4FGL counterparts (mainly BCUs) for ABC sources. In addition we find 1506 \textit{BZCAT} sources with a counterpart in the 4FGL catalog. For each of these sources, we collected from the 4FGL the photon index obtained fitting the \textit{Fermi}-LAT spectra with a power-law (\textsc{PL\_Index Photon}) and the energy flux in the range \(100 \textrm{ MeV}-100 \textrm{ GeV}\) obtained by spectral fitting (\textsc{Energy\_Flux100}), mostly (\(\sim 70\%\)) with a power-law model. Although these two quantities are not independent, it is instructive to compare their distributions in \textit{BZCAT} and ABC sources.

\begin{figure*}
	\centering
	\includegraphics[scale=0.38]{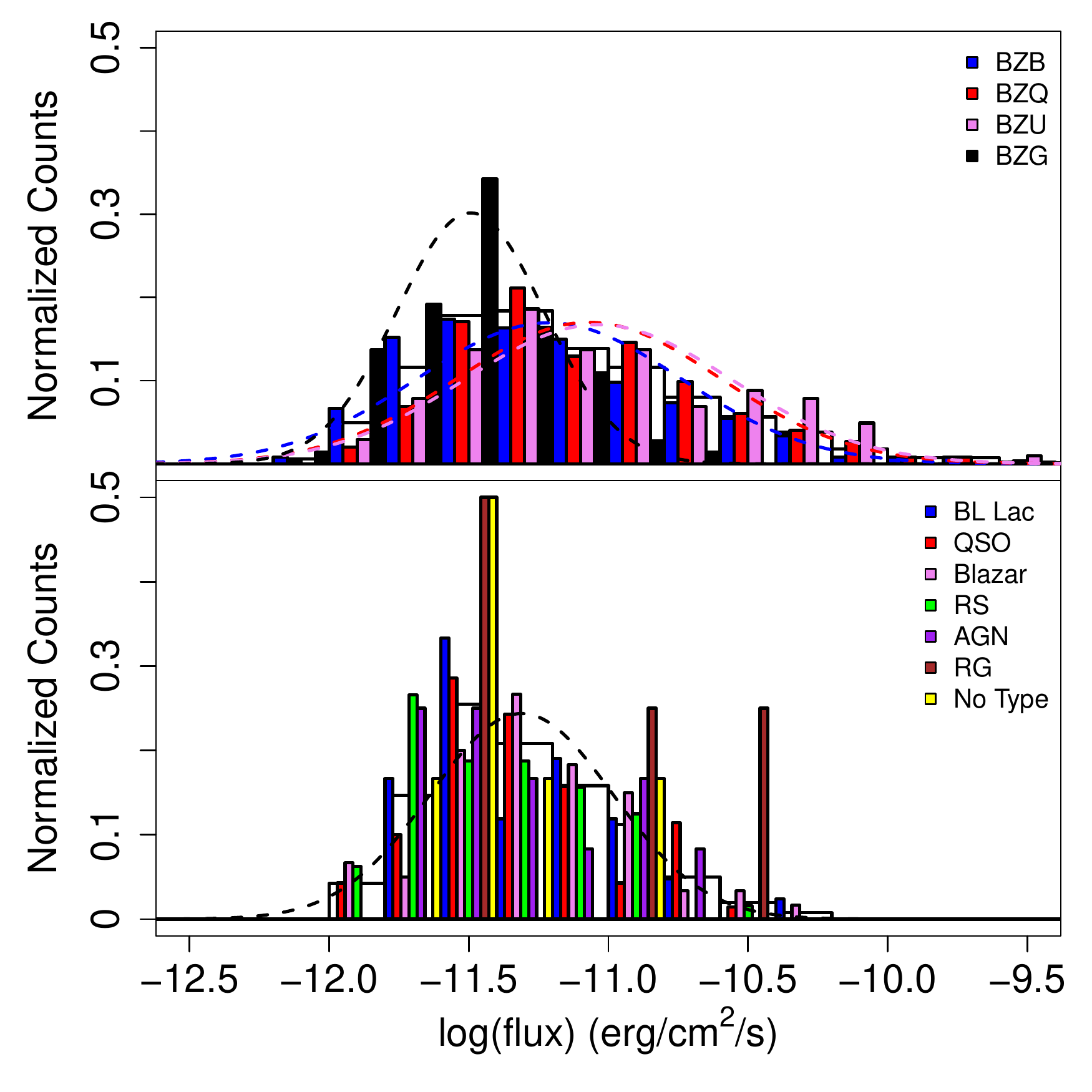}
	\includegraphics[scale=0.38]{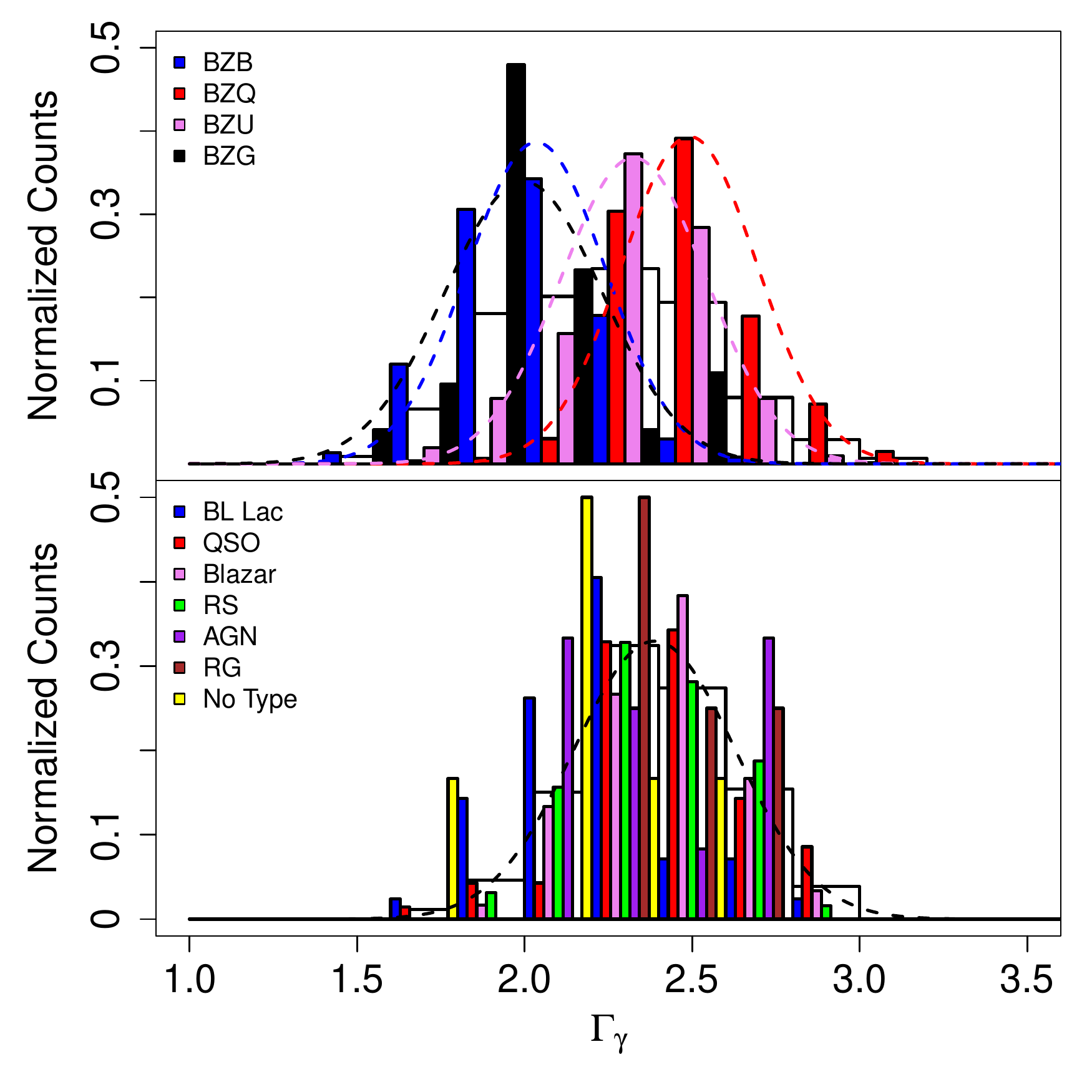}
	\caption{(Left panel) Normalized distributions of the \(100 \textrm{ MeV}-100 \textrm{ GeV}\) flux for \textit{BZCAT} (top) and ABC (bottom) sources. The various subclasses of \textit{BZCAT} sources and types of ABC sources are presented with the colours indicated in the legend, and the white histograms on the background represent the distribution of the whole samples. For \textit{BZCAT} sources gaussian fits to the subclass distributions are indicated with dashed lines of the respective colours, while for ABC source a gaussian fit to the distribution of the whole sample is indicated with a black dashed line. (Right panel) Same as the left panel, but for the distributions of the \(\gamma\)-ray slope \(\Gamma_\gamma\).}\label{fig:gammaray_hist}
\end{figure*}

            As in Fig. \ref{fig:xray_hist}, in the left panel of Fig. \ref{fig:gammaray_hist} we compare the normalized distributions of the \(\gamma\)-ray flux for \textit{BZCAT} (top) and ABC (bottom) sources. The \textit{BZCAT} of the different subclasses (BZB, BZQ, BZU and BZG) are presented with different colours, and gaussian fits to the flux distributions of each subclass are indicated with dashed lines of the respective colour. The peak logarithmic fluxes (in cgs units) of these fits are similar for BZQs and BZUs, being, \(-11.1\) and \(-11.0\), respectively, while BZGs are on average dimmer, peaking at \(-11.5\). BZBs sit somewhat in between, peaking at \(-11.2\). For ABC sources the different source types are indicated with different colours, but due to lower statistics with respect to the \textit{BZCAT} we overplot only a gaussian fit to the distribution of the whole sample. We see that \textit{BZCAT} sources extend to larger fluxes with respect to ABC sources, at variance with what is observed at radio wavelengths (see Fig. \ref{fig:acc_bzcat_radio_flux}). We note that both distributions of \(\gamma\)-ray logarithmic flux peak at similar values, with the \textit{BZCAT} sources peaking at \(-11.3\), and the ABC sources peaking at \(-11.3\), between BZG and BZB peaks. However a KS test shows that the two distributions have a p-chance \(p=1.3\times{10}^{-5}\) of having been randomly sampled from a common parent distribution, being therefore significantly different from the statistical point of view.

            The right panel of Fig. \ref{fig:gammaray_hist} shows a similar comparison, but for the power-law slopes \(\Gamma_\gamma\) of the \textit{Fermi}-LAT spectra. \textit{BZCAT} subclasses are clearly separated on the basis of their spectral shape, with BZQs being softer in \(\gamma\)-rays (their \(\Gamma_\gamma\) distribution peaks at \(2.5\)) than BZUs (peaking at \(2.4\)), while BZB and BZG spectra appear harder (both peaking at \(2.0\)). The ABC sources are on average softer than average \textit{BZCAT} sources, with their \(\Gamma_\gamma\) distribution peaking at \(2.4\) and \(2.2\), respectively. The KS test confirms however that the two distributions are completely different, with a p-chance \(p=4.0\times{10}^{-13}\) of having been randomly sampled from a common parent distribution. This p-chance is instead \(p=3.7\times{10}^{-2}\) when comparing ABC sources and BZUs from \textit{BZCAT}, indicating that ABC sources are probably a mixture of different sub-populations possibly dominated by softer FSRQs.

\begin{figure*}
	\centering
	\includegraphics[scale=0.4]{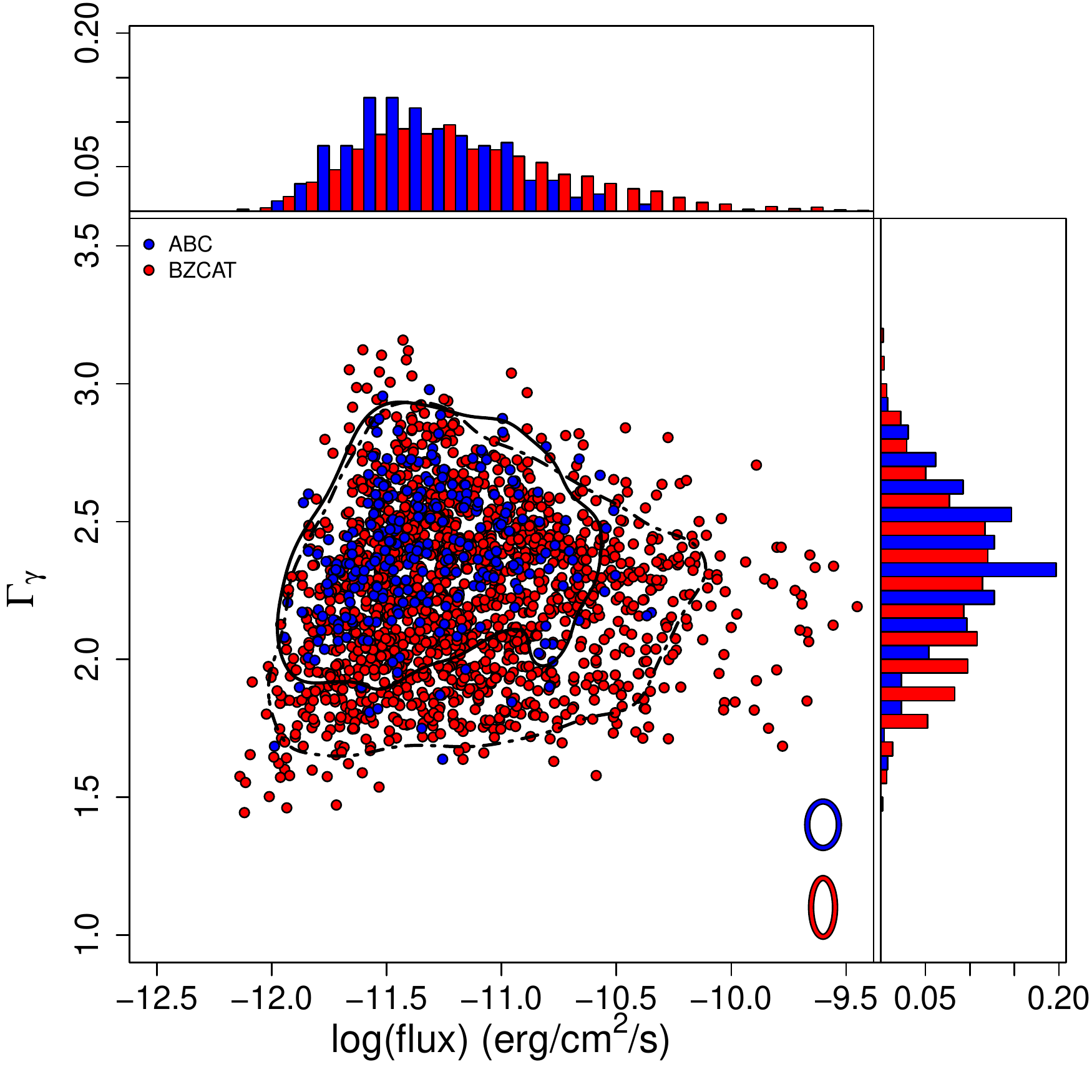}
	\caption{ABC (blue circles) and \textit{BZCAT} (red circles) sources represented on the \(\gamma\)-ray slope vs. \(\gamma\)-ray flux plot. The ellipses of the relative colour indicate the average uncertainties. The black full and dot-dashed lines represent the \(90\%\) KDE isodensity contours for ABC sources and \textit{BZCAT} sources, respectively. On the top and right panels are presented the normalized distributions of the \(\gamma\)-ray flux and slope for ABC and \textit{BZCAT} sources, with the same colours used in the main panel.}\label{fig:gammaray_gamma_flux}
\end{figure*}

            In Fig. \ref{fig:gammaray_gamma_flux} we plot ABC and \textit{BZCAT} sources with counterparts in 4FGL catalogue on a \(\gamma\)-ray slope vs. \(\gamma\)-ray flux plot. Again, we remind that these two quantities are not independent (since the flux is evaluated from a spectral fit), however this representation is useful to visualize the general properties of the sources. On the top and right of this figure we present the normalized distributions of the \(\gamma\)-ray flux and slope \(\Gamma_\gamma\) for ABC and \textit{BZCAT} sources. As noted before, ABC sources are on average softer and dimmer in \(\gamma\)-rays with respect to the blazar in \textit{BZCAT}. This is highlighted by the \(90\%\) KDE isodensity contours for ABC sources and \textit{BZCAT} sources represented with a black full and dot-dashed lines, respectively, that suggest that ABC sources occupy the same region of the \(\gamma\)-ray slope vs. \(\gamma\)-ray flux space, although clustering in the softer-dimmer region.

    \section{\(\gamma\)-ray Blazar Candidates Selection}\label{sec:candidates}
        In this section we make use of the WISE data collected in Sect. \ref{sec:wise} to select candidate \(\gamma\)-ray blazars in the ABC sample. As discussed by \citet{2019ApJS..242....4D}, WISE data provide an effective way to select \(\gamma\)-ray blazars, by comparing their colours with those of known \(\gamma\)-ray blazars.

\begin{figure*}
	\centering
	\includegraphics[scale=0.38]{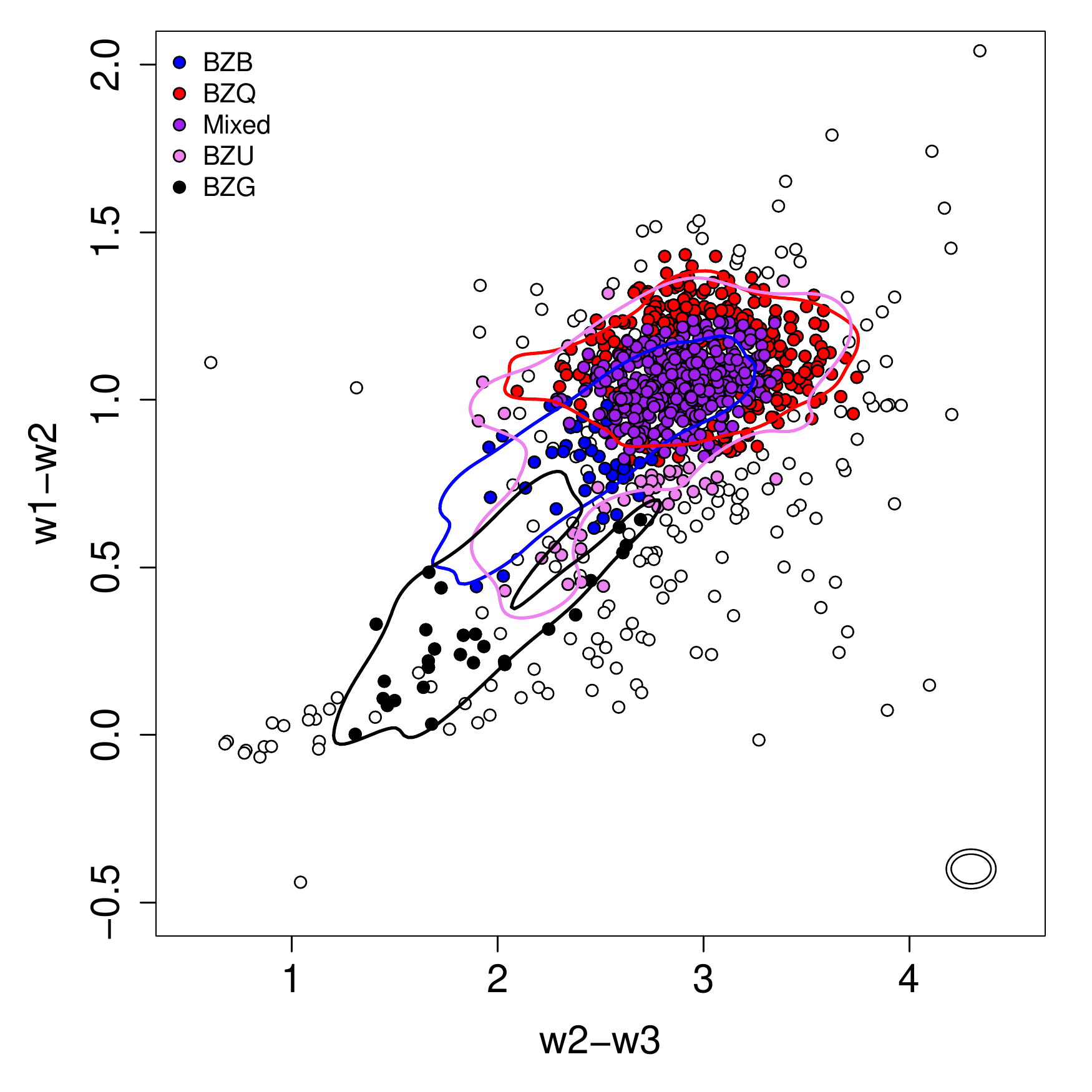}
	\includegraphics[scale=0.38]{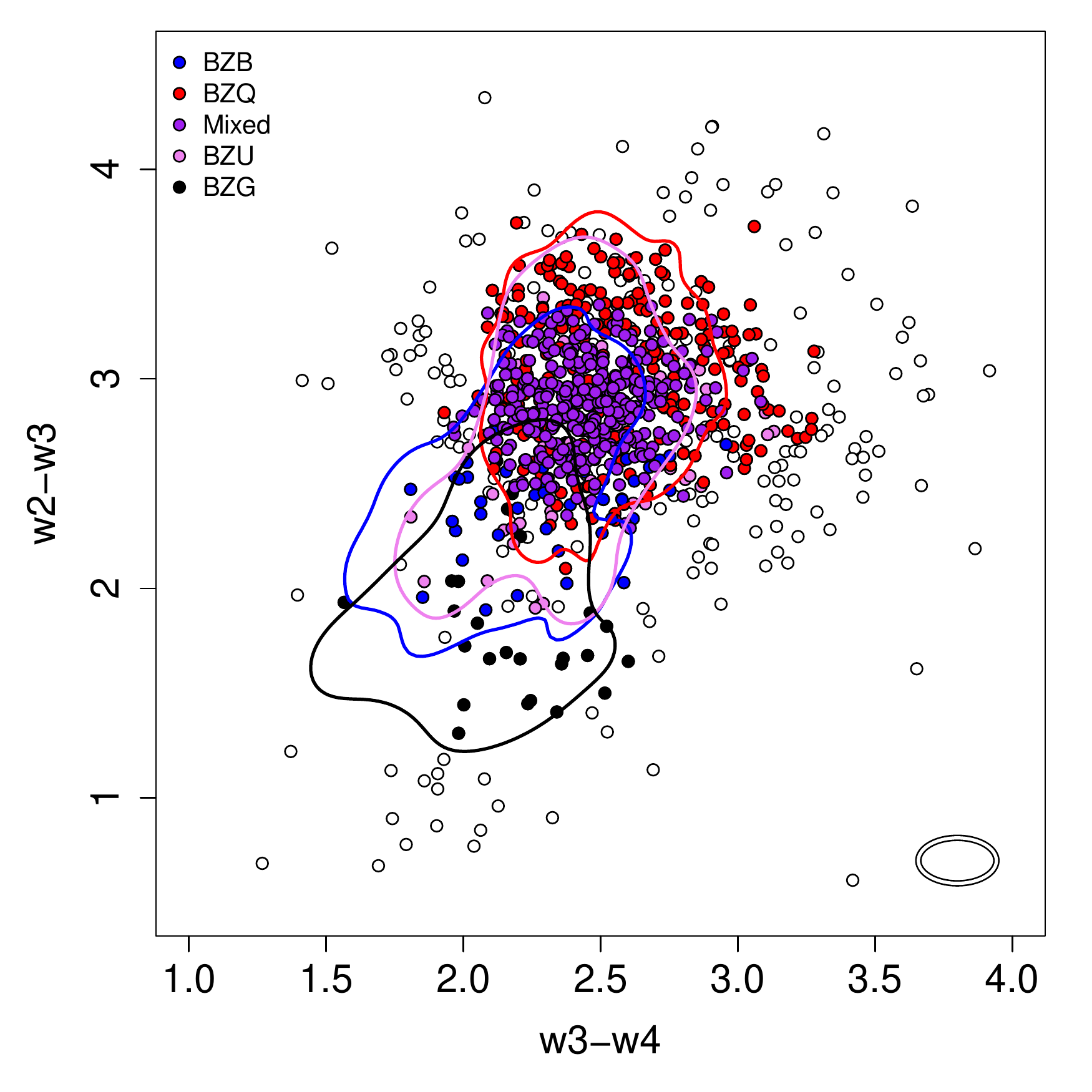}
	\caption{(Left panel) ABC sources with a WISE counterpart detected in all four WISE bands (white circles) represented in the w1-w2 vs. w2-w3 colour-colour plane. KDE isodensity contours containing \(90\%\) of the \textit{BZCAT} sources detected in \(\gamma\)-rays (as reported in 4FGL catalog) are presented with lines of different colours for the different \textit{BZCAT} subclasses, as indicated in the legend. Filled circles of the respective colours mark the \(\gamma\)-ray blazar candidates of the different subclasses (see main text). The ellipse indicates the average colour uncertainties. (Right panel) Same as the left panel, but for the w2-w3 vs. w3-w4 projection.}\label{fig:wise_cand}
\end{figure*}

        Here we will take an approach that is somewhat halfway between those adopted for the compilation of WIBRaLS2 and KDEBLLACS catalogues. First we select all \textit{BZCAT} sources detected in \(\gamma\)-rays, that is, with a counterpart in 4FGL catalogue (1506 sources) and with a WISE counterpart detected in all four WISE bands (selected as explained in Sect. \ref{sec:wise}), for a total of 1237 sources. Then, for each source subclass (BZB, BZQ, BZU and BZG), we evaluate the KDE isodensity contours containing \(90\%\) of the sources in both w1-w2 vs. w2-w3 and w2-w3 vs. w3-m4 colour-colour planes. We then compare the position of the 906 ABC sources with a WISE counterpart detected in all four bands in the two colour-colour planes, as shown in Fig. \ref{fig:wise_cand}. Here the KDE isodensity contours for different \textit{BZCAT} subclasses are indicated with lines of the relative colour indicated in the legend. To select \(\gamma\)-ray blazar candidates among these ABC sources we proceed as follows:
        \begin{itemize}
            \item since the regions occupied by BZBs and BZQs have a
                significant overlap, if a source is compatible with the
                \(90\%\) KDE isodensity contours of BZBs and BZQs on both
                colour-colour planes, it is classified as \(\gamma\)-ray
                blazar candidate of MIXED class,
            \item if a source is compatible with the \(90\%\) KDE isodensity
                contours of BZBs on both colour-colour planes but not with
                the \(90\%\) KDE isodensity contours of BZQs on both
                colour-colour planes, it is classified as \(\gamma\)-ray
                blazar candidate of BZB class,
            \item if a source is compatible with the \(90\%\) KDE isodensity
                contours of BZQs on both colour-colour planes but not with
                the \(90\%\) KDE isodensity contours of BZBs on both
                colour-colour planes, it is classified as \(\gamma\)-ray
                blazar candidate of BZQ class,
            \item if a source is compatible with the \(90\%\) KDE isodensity
                contours of BZUs on both colour-colour planes but neither
                with the \(90\%\) KDE isodensity contours of BZBs nor BZQs on
                both colour-colour planes, it is classified as \(\gamma\)-ray
                blazar candidate of BZU class,
            \item if a source is compatible with the \(90\%\) KDE isodensity
                contours of BZGs on both colour-colour planes but neither
                with the \(90\%\) KDE isodensity contours of BZBs, BZQs, nor
                BZUs on both colour-colour planes, it is classified as
                \(\gamma\)-ray blazar candidate of BZG class,
            \item if a source is not compatible with the \(90\%\) KDE isodensity contours of BZBs, BZQs, BZUs, nor BZGs, it is not classified as \(\gamma\)-ray blazar candidate.
        \end{itemize}

        We stress that to consider a source position on the colour-colour plane compatible with a KDE isodensity contour we take into account the WISE colour uncertainties, that is, the colour-colour uncertainty ellipse of the source must have an overlap with the isodensity contours. In this way we select 715 \(\gamma\)-ray blazar candidates, subdivided in 42 candidates BZBs, 247 candidates BZQs, 334 candidates MIXED, 46 candidates BZUs, and 28 candidates BZGs, indicated in Fig. \ref{fig:wise_cand} with circles of the respective colour. The properties of \(\gamma\)-ray blazar candidates are summarized in Table \ref{tab:classification}, including the WISE counterpart and its blazar candidate classification.

        As mentioned before,  this selection criterion is intermediate between those presented by \citet{2019ApJS..242....4D} for WIBRaLS2 catalog, that select sources detected in all four WISE bands based on their position in the three-dimensional principal component space generated by the three independent WISE colours, and for KDEBLLACS, that select sources detected in the first three WISE bands based on their compatibility with the \(90\%\) KDE isodensity contours of BZBs in the two-dimensional w1-w2 vs. w2-m3 colour-colour plane. In addition, the WIBRaLS2 method only selects BZB, BZQ and MIXED candidates, while with our method we select also BZU and BZG candidates. In addition, we note that \citet{2019ApJS..242....4D} based their selection methods on the third release of the \textit{Fermi}-LAT catalogue 3FGL, the latest that was available at the time of the publication, while for our method we use the results of the updated 4FGL. For these reasons the two methods are not equivalent, and we expect differences in the classification of the ABC sources.

\begin{figure*}
	\centering
	\includegraphics[scale=0.38]{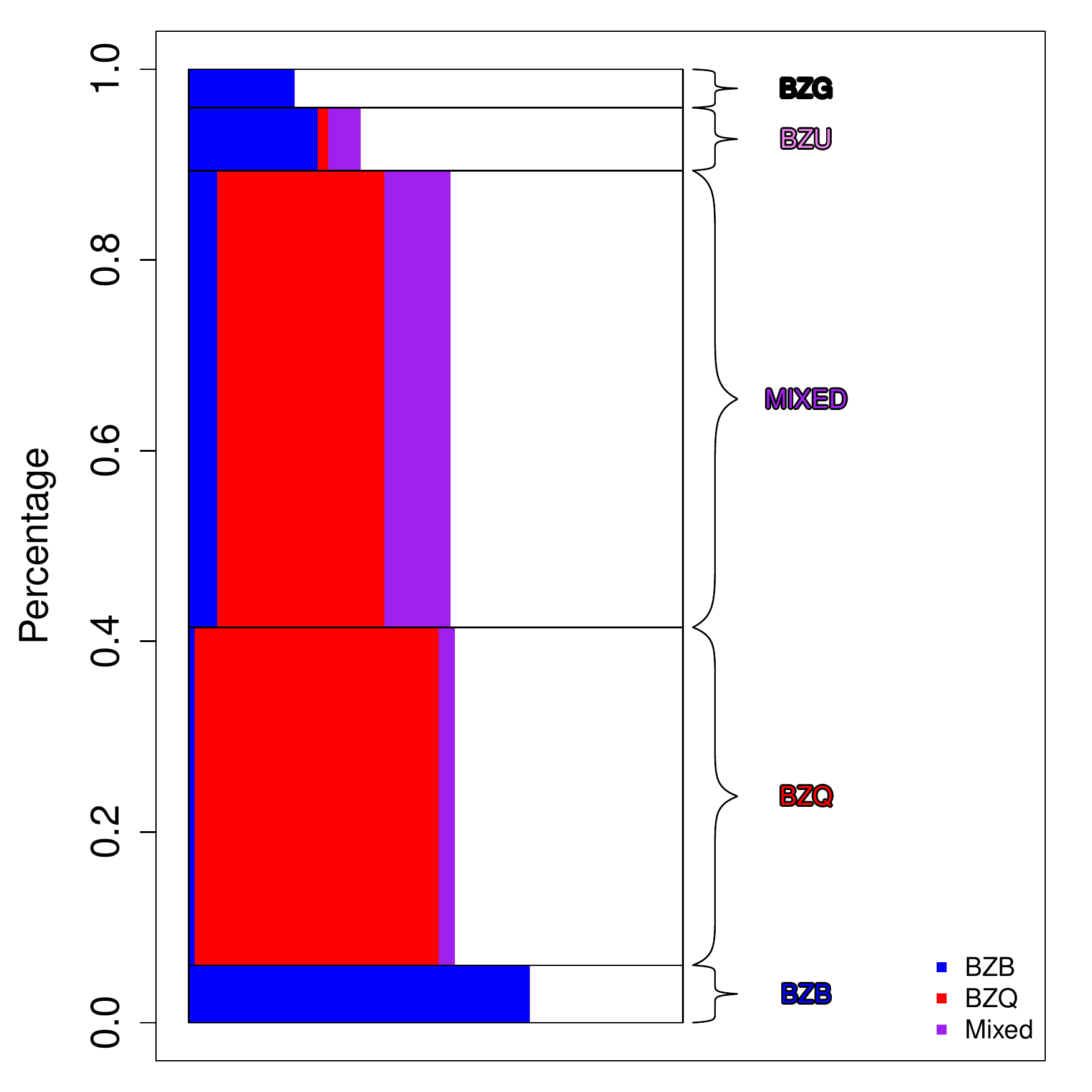}
	\includegraphics[scale=0.38]{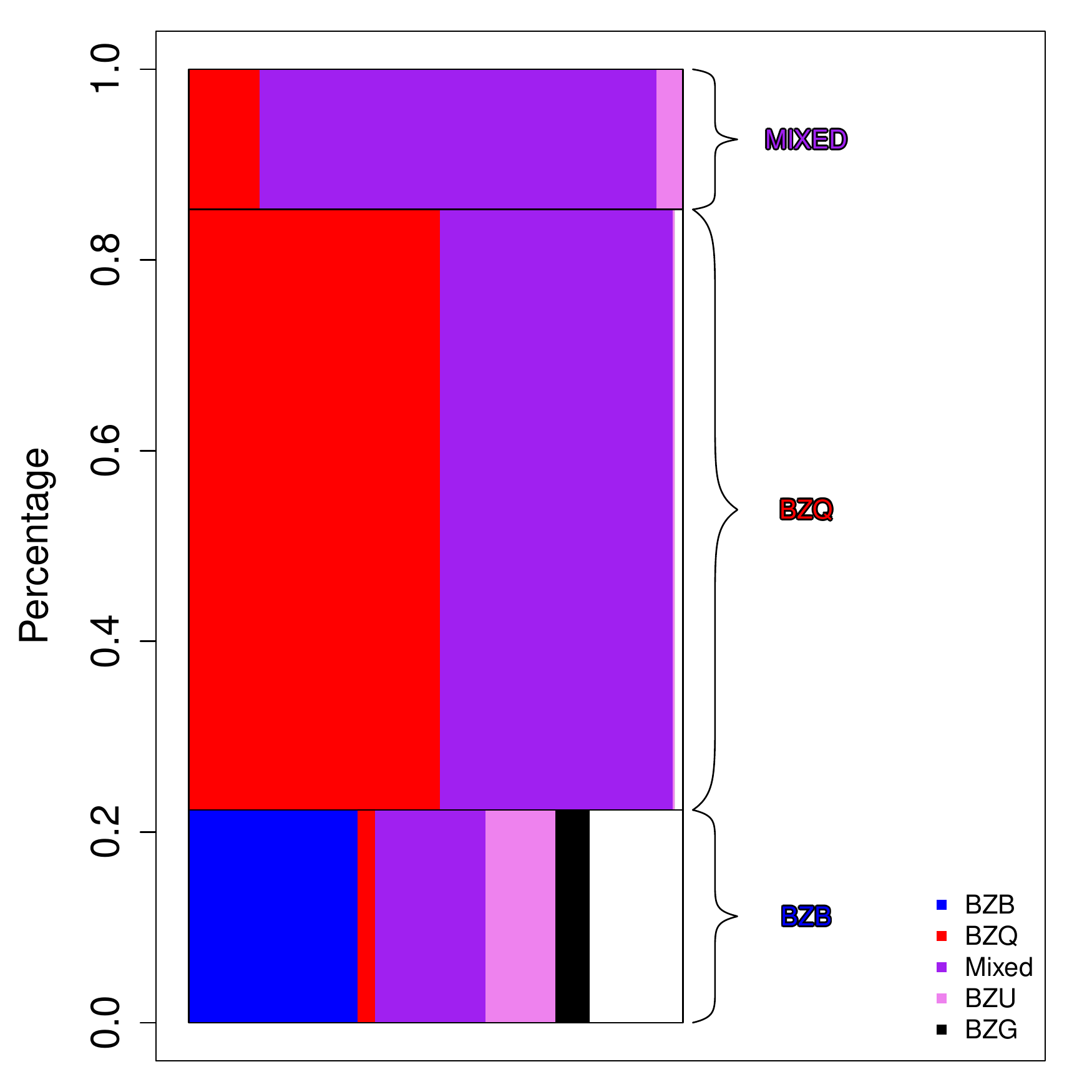}
	\caption{(Left panel) Comparison of out classification method and that proposed by \citet{2019ApJS..242....4D} for WIBRaLS2 catalog. The black rectangles represent the \(\gamma\)-ray blazar candidate classes selected by our method, while the coloured bars in each rectangle represent the percentage of sources also classified in WIBRaLS2 catalog. (Right panel) Same as the left panel, but with black rectangles representing the \(\gamma\)-ray blazar candidate classes of WIBRaLS2 catalog, and with the coloured bars representing the percentage of sources also classified in the present work.}\label{fig:abc_vs_wibrals}
\end{figure*}

        These differences are summarized in Fig. \ref{fig:abc_vs_wibrals}. In the left panel of this figure we present the 715 ABC \(\gamma\)-ray blazar candidates we selected with our method, 361 of which are also selected in the WIBRaLS2 catalog. For each subclass our classification method (BZB, BZQ, MIXED, BZU and BZG) we indicate with coloured rectangles the percentage of sources that have a WIBRaLS2 classification ((BZB, BZQ and MIXED). We see that \(\sim 70\%\) of our candidates BZBs are also selected as candidate BZBs in WIBRaLS2, while the remaining \(\sim 30\%\) of this subclass is not classified in WIBRaLS2 catalog. About \(\sim 50\%\) of our BZQ candidates is also classified as BZQ candidate in WIBRaLS2, while 3 are classified as BZBs and 8 as MIXED in WIBRaLS2. Only \(\sim 13\%\) of the sources we classify as MIXED candidates have the same classification in WIBRaLS2, while \(\sim 34\%\) and \(\sim 6\%\) of these sources are classified as BZQ and BZB candidate in WIBRaLS2, respectively. Finally, \(\sim 26\%\) and \(\sim 21\%\) of the sources that we classify as BZU and BZG, respectively (two classes non present inWIBRaLS2), are classified as BZB candidates in WIBRaLS2 catalog.

        In the right panel of the same figure we present the reverse comparison, that is, we study how the 381 ABC sources listed in the WIBRaLS2 catalogue are classified according to our selection method. The sources selected as BZB candidates in WIBRaLS2 catalogue are a mixed bag of different classifications for our method, while \(\sim 51\%\) and \(\sim 47\%\) of the sources classified as BZQ candidates in WIBRaLS2 are classified as candidate BZQs and MIXED according to our method, respectively. Finally, \(\sim 14\%\) and \(\sim 80\%\) of sources classified as candidate MIXED in WIBRaLS2 are classified as BZQs and MIXED candidates in our method.

\begin{figure*}
	\centering
	\includegraphics[scale=0.4]{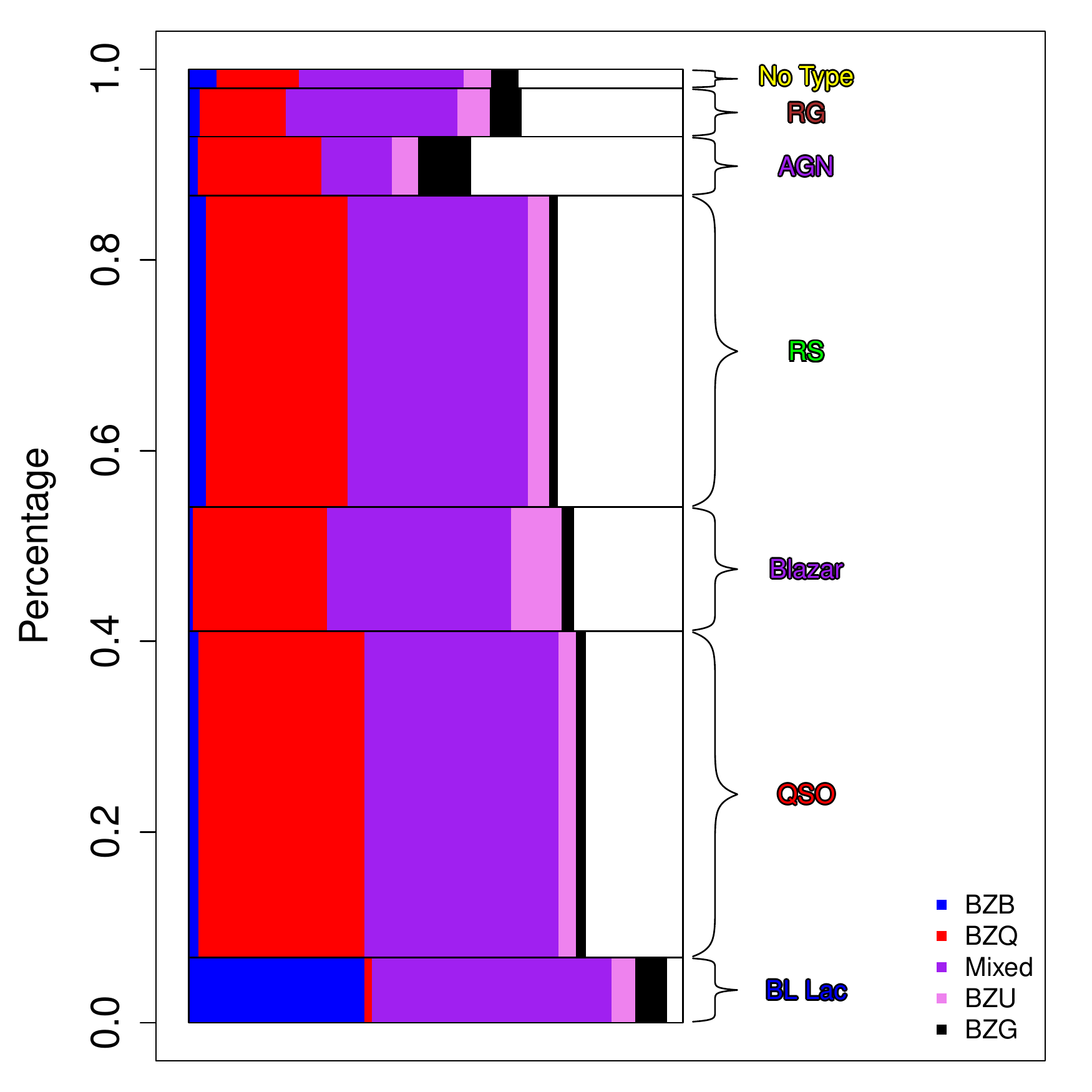}
	\caption{Summary of the 715 \(\gamma\)-ray blazar candidates selected according to the method presented in Sect. \ref{sec:candidates}. The black rectangles represent the different ABC source types, and the coloured bars in each rectangle represent the percentage of each source type classified as \(\gamma\)-ray blazar candidates of the different classes, with the colours indicated in
		the legend.}\label{fig:source_type_cand_wise}
\end{figure*}

        In Fig. \ref{fig:source_type_cand_wise} we summarize the results of our \(\gamma\)-ray blazar candidates selection in terms of ABC source types. We see that most BZB candidates belong to the BL Lac source type, while the other source types are mainly composed of candidate BZQs and MIXED candidates, that represent the majority (\(\sim 35\%\) and \(\sim 47\%\) respectively) of our \(\gamma\)-ray blazar candidates.

\begin{figure*}
	\centering
	\includegraphics[scale=0.38]{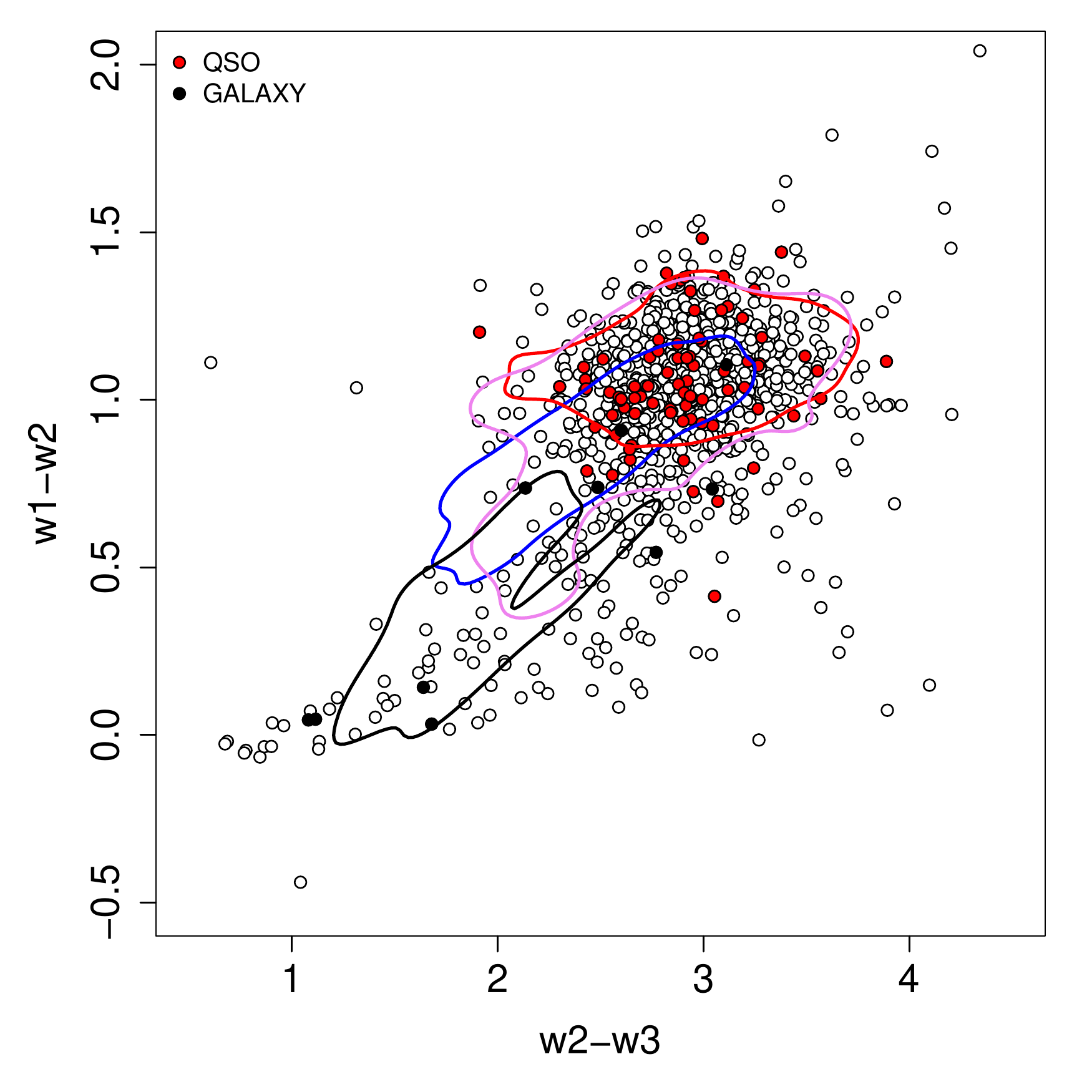}
	\includegraphics[scale=0.38]{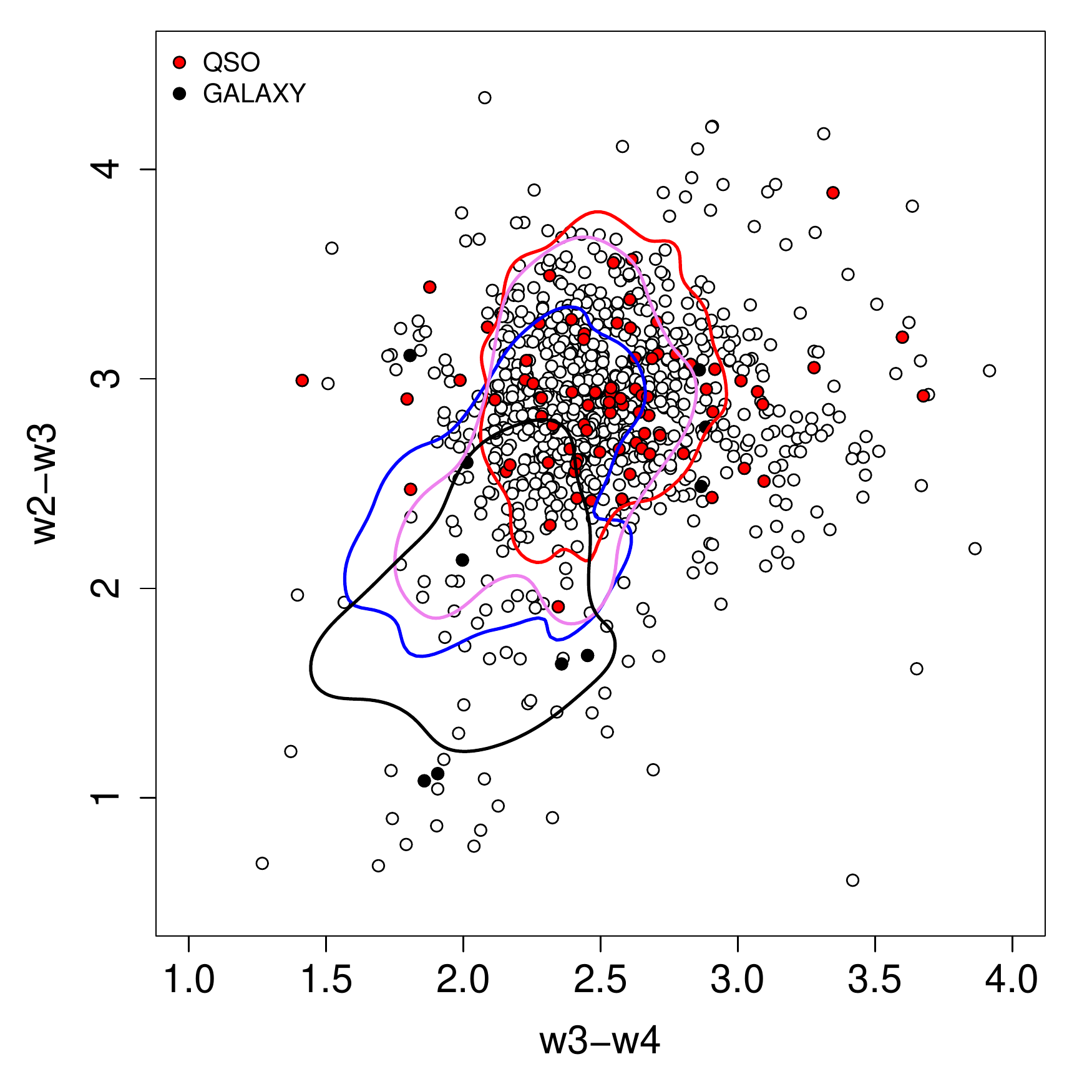}
	\caption{(Left panel) ABC sources with a WISE counterpart detected in all four WISE bands (white circles) represented in the w1-w2 vs. w2-w3 colour-colour plane. The same KDE isodensity contours of Fig. \ref{fig:wise_cand} are presented with lines of different colours for the different \textit{BZCAT} subclasses, as indicated in the legend. Filled red and black circles mark sources with an optical spectroscopic classification of QSO and GALAXY, respectively. (Right panel) Same as the left panel, but for the w2-w3 vs. w3-w4 projection.}\label{fig:wise_cand_optical}
\end{figure*}

        To conclude this section, in Fig. \ref{fig:wise_cand_optical} we see how the sources for which we have an optical spectroscopic classification (see Sect.\ref{sec:sdss}) compare with the regions occupied in the WISE colour-colour diagrams by the different classes of \textit{BZCAT} \(\gamma\)-ray blazars, represented as in Fig. \ref{fig:wise_cand}. Of the 110 sources for which we have an optical spectroscopic classification, 89 have a WISE counterpart detected in all four WISE bands, and they are presented in this figure, with red circles for sources with an optical spectroscopic classification of QSO, and with black circles for sources with an optical spectroscopic classification of GALAXY. We see that most QSO objects, as expected, lie in the region occupied by BZQs. GALAXY object, on the other hand, mostly lie in the region of BZBs - especially in the w1-w2 vs. w2-w3 projection - and in the region of BZGs (whose emission is dominated by the galactic one), extending toward the colour-colour region occupied by old elliptical galaxies \citep[see e.g.][]{2014MNRAS.442..629R}, suggesting that these objects are BL Lac objects whose host galaxy dominates the optical emission.

    \section{Conclusions}\label{sec:conclusion}
        We have built a new catalogue of candidate blazars, dubbed ABC, derived from the ALMA Calibrator Catalogue by \citet{2019MNRAS.485.1188B}. The ABC catalogue fills, at least partly, the lack of \(|b|<10\degree\) blazars in \textit{BZCAT}, providing low Galactic latitude candidate counterparts to unassociated high-energy sources. This is particularly important in the case of rare events like detections of high energy neutrinos for which a full exploitation of all-sky data is essential to identify the source population. Some of the ABC sources at low Galactic latitudes may also be useful to verify the \textit{Gaia} optical reference frame \citep{2016AA...595A...5M}.
        
        The ABC catalogue contains 1580 sources not included in the \textit{BZCAT}. It was cross-matched with \textit{Gaia} DR2, SDSS DR12, LAMOST DR5, AllWISE and 4FGL catalogues, finding 805, 295, 31, 1311 and 259 matches, respectively. These data were used for the classification of our sources and a comparison with the population of known blazars in \textit{BZCAT}.
        
        ABC sources are significantly dimmer than \textit{BZCAT} sources in \textit{Gaia} g band, while the difference in the \textit{Gaia} b-r colour between the two populations is less pronounced. Also, ABC sources appear bluer in SDSS than \textit{BZCAT} sources, although with low statistical significance. When comparing with the results of \citet{2011AJ....141...93B}, we see that most ABC sources classified as QSO and BL Lac fall into the region of low redshift quasars, with some QSOs entering the regions of higher redshift quasars. Regarding WISE colours, we find that ABC sources are significantly bluer than \textit{BZCAT} sources in the w2-w3 and w3-w4 colours. In addition, we collected 110 optical spectra in SDSS DR12 and LAMOST DR5, that mostly classify the corresponding  sources as QSO (98), while 12 sources resulted galactic objects.
        
        A fraction of ABC sources are located in fields covered by \textit{Swift}/XRT, \textit{Chandra}/ACIS and \textit{XMM-Newton}/EPIC observations. We have retrieved the archive data and made our own source extraction, achieving the detection of 101, 56 and 55 ABC sources, respectively. For 43, 46 and 51, respectively, of them we obtained the X-ray spectra. Our sources are, on average, similar in X-rays to \textit{BZCAT} blazars, implying that our sample is covering the same region of the blazar parameter space in this band.
        
        A comparison of \(\gamma\)-ray properties of ABC source with \textit{BZCAT} blazars has shown that ABC sources are, on average, dimmer, and their \(\gamma\)-ray spectra are, on average, softer, consistent with the ABC containing a significant fraction of FSRQs.

        About \(57\%\) (906 out of 1580) of ABC sources are detected in all four WISE bands. This has allowed us to re-examine the selection of \(\gamma\)-ray blazars by means of mid-IR colours, discussed by \citet{2019ApJS..242....4D}. Making use of the KDE contours in the WISE colour-colour diagrams containing \(90\%\) of WISE-detected 4FGL blazar subclasses, we were able to classify about \(80\%\) (715 out of 906) of the ABC WISE-detected sources as candidate \(\gamma\)-ray blazar. A comparison with the classification by \citet{2019ApJS..242....4D} for common sources has shown that these two methods yield different results, and can therefore be used in a complementary way.
        
        The main properties of the 879 ABC sources for which we were able to collect additional information are summarized in Table \ref{tab:alma_candidates_short}, including the presence in other blazar catalogues (see Sect. \ref{sec:catalog_comparison}), the optical spectroscopic classification (see Sect. \ref{sec:sdss}), the X-ray properties (see Sect. \ref{sec:xray-spectra}), and the blazar candidate classification (see Sect. \ref{sec:candidates}).
        
\begin{figure*}
	\centering
	\includegraphics[scale=0.4]{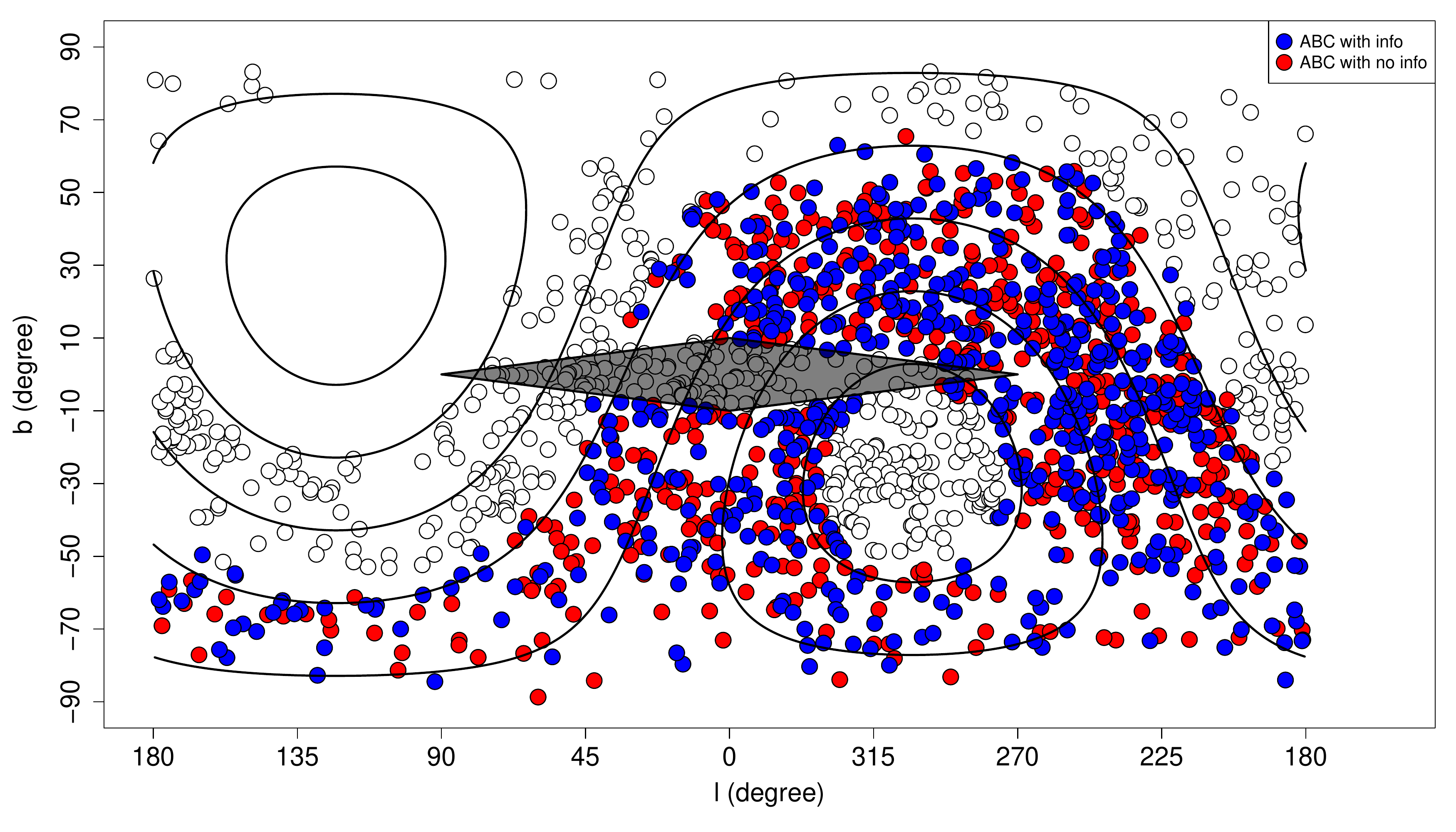}
	\caption{The white circles indicate the Galactic coordinates of the ABC sources. The colored circles represent the ABC sources falling in the area covered by the planned WFD survey, with the exception of the Galactic center indicated with a dark lozenge. In particular, blue circles represent ABC sources for which we were able to collect additional information (see Table \ref{tab:alma_candidates_short}), while the red circles represent ABC sources for which we could not collect additional information. The black lines represent equatorial coordinates with declinations from \(-60\degree\) to \(60\degree\) with increment of \(20\degree\).}\label{fig:galactic_projection_leftovers}
\end{figure*}

        The ABC provides a large sample of candidate blazars that can be investigated both through dedicated optical spectroscopic observation campaigns or through repeated photometric observations for variability studies. An ideal tool to perform the latter investigation will be the 10-year Legacy Survey of Space and Time (LSST) that, starting from 2022, will be performed at the Vera C. Rubin Observatory \citep{2019ApJ...873..111I}. LSST will repeatedly scan the whole southern sky, providing multi-epoch observations in six optical photometric bands (\textit{ugrizy}) for more than 35 billions objects. The main LSST survey will be the Wide, Fast, Deep (WFD) survey, observing the sky between \(-65\degree\) and \(+5\degree\) of declination, with an observation cadence of \(\sim 3\) days. In Fig. \ref{fig:galactic_projection_leftovers} we show the ABC sources in Galactic coordinates with white circles. The colored circles represent the ABC sources falling in the area covered by the planned WFD survey, with the exception of the Galactic center indicated with a dark lozenge. In particular, in the WFD survey area we have 556 ABC sources with additional information contained in Table \ref{tab:alma_candidates_short} (blue circles), and 529 ABC sources without additional information (red circles). Among the 125 ABC sources in the WFD survey area with an SDSS DR12 counterpart, 114 (more than \(90\%\)) have a \(r\) magnitude between the median single-visit \(5\sigma\) point sources depth of \(r=24.16\) for the planned WFD survey and the nominal LSST saturation limit of \(r\sim 16\) for \(15 \textrm{ s}\) exposures\footnote{See the \href{https://arxiv.org/abs/1812.00514}{white paper} on the LSST Observing Strategy.}.
        
        The data collected by the WFD survey will therefore provide a key tool to investigate the possible blazar nature of these sources.

    \begin{acknowledgements}
        This project has received funding from the European Union’S Horizon 2020 research and innovation programme under the Marie Sk\l odowska-Curie grant agreement NO 664931. MB acknowledges support from INAF under PRIN SKA/CTA FORECaST and from the Ministero degli Affari Esteri della Cooperazione Internazionale - Direzione Generale per la Promozione del Sistema Paese Progetto di Grande Rilevanza ZA18GR02. This work has made use of data from the European Space Agency (ESA) mission \textit{Gaia} (\href{https://www.cosmos.esa.int/gaia}{https://www.cosmos.esa.int/gaia}), processed by the \textit{Gaia} Data Processing and Analysis Consortium (DPAC, \href{https://www.cosmos.esa.int/web/gaia/dpac/consortium}{https://www.cosmos.esa.int/web/gaia/dpac/consortium}). Funding for the DPAC has been provided by national institutions, in particular the institutions participating in the \textit{Gaia} Multilateral Agreement. Funding for the Sloan Digital Sky Survey IV has been provided by the Alfred P. Sloan Foundation, the U.S. Department of Energy Office of Science, and the Participating Institutions. SDSS-IV acknowledges support and resources from the Center for High-Performance Computing at the University of Utah. The SDSS web site is \href{www.sdss.org}{www.sdss.org}. SDSS-IV is managed by the Astrophysical Research Consortium for the  Participating Institutions of the SDSS Collaboration including the  Brazilian Participation Group, the Carnegie Institution for Science,  Carnegie Mellon University, the Chilean Participation Group, the French Participation Group, Harvard-Smithsonian Center for Astrophysics,  Instituto de Astrof\'isica de Canarias, The Johns Hopkins University, Kavli Institute for the Physics and Mathematics of the Universe (IPMU) /  University of Tokyo, the Korean Participation Group, Lawrence Berkeley National Laboratory, Leibniz Institut f\"ur Astrophysik Potsdam (AIP),   Max-Planck-Institut f\"ur Astronomie (MPIA Heidelberg),  Max-Planck-Institut f\"ur Astrophysik (MPA Garching),  Max-Planck-Institut f\"ur Extraterrestrische Physik (MPE),  National Astronomical Observatories of China, New Mexico State University, New York University, University of Notre Dame,  Observat\'ario Nacional / MCTI, The Ohio State University, Pennsylvania State University, Shanghai Astronomical Observatory, United Kingdom Participation Group, Universidad Nacional Aut\'onoma de M\'exico, University of Arizona,  University of Colorado Boulder, University of Oxford, University of Portsmouth, University of Utah, University of Virginia, University of Washington, University of Wisconsin, Vanderbilt University, and Yale University. This research has made use of the SIMBAD database, operated at CDS, Strasbourg, France. Guoshoujing Telescope (the Large Sky Area Multi-Object Fiber Spectroscopic Telescope LAMOST) is a National Major Scientific Project built by the Chinese Academy of Sciences. Funding for the project has been provided by the National Development and Reform Commission. LAMOST is operated and managed by the National Astronomical Observatories, Chinese Academy of Sciences. This publication makes use of data products from the Wide-field Infrared Survey Explorer, which is a joint project of the University of California, Los Angeles, and the Jet Propulsion Laboratory/California Institute of Technology, funded by the National Aeronautics and Space Administration. We acknowledge the use of public data from the \textit{Swift} data archive. This research has made use of data obtained from the \textit{Chandra} Data Archive. This research has made use of observations obtained with \textit{XMM-Newton}, an ESA science mission with instruments and contributions directly funded by ESA Member States and NASA. This research has made use of data and/or software provided by the High Energy Astrophysics Science Archive Research Center (HEASARC), which is a service of the Astrophysics Science Division at NASA/GSFC. This research has made use of software provided by the \textit{Chandra} X-ray Center (CXC) in the application packages CIAO, ChIPS, and Sherpa. This research has made use of the TOPCAT software \citep{2005ASPC..347...29T}.
    \end{acknowledgements}

\newpage

\begin{landscape}
\begin{table}[ht]
\caption{List of ALMA sources with flat radio spectrum without counterparts in the \textit{BZCAT} (see Sect. \ref{sec:sample_selection}). For each source we present the source name (column 1), coordinates (columns 2 and 3), the source alternate name (column 4), the redshift (column 5), the source type given in literature (column 6, note for more details) and the reference (column 7).}\label{tab:catalog}
\centering
\begin{tabular}{l|l|l|l|c|c|l}
\hline\hline
Name ALMA & RA & DEC & Name NED & Redshift & Object Type & Reference \\
\hline
J0002-2153 & 00:02:11.98 & -21:53:09.87 & WISEA J000211.98-215310.0 &  & BLL & \citet{2014ApJS..215...14D} \\
J0003-1941 & 00:03:15.95 & -19:41:50.40 & WISEA J000315.94-194150.1 &  & QSO & \citet{2016AA...595A...5M} \\
J0004-4345 & 00:04:07.26 & -43:45:10.15 & PKS 0001-440 &  & rG & \citet{2017AJ....153..157T} \\
J0006-2955 & 00:06:01.12 & -29:55:50.10 & PKS 0003-302 &  & QSO & \citet{2016AA...595A...5M} \\
J0008+1144 & 00:08:00.84 & +11:44:00.77 & WISEA J000800.80+114400.9 &  & Rad & \\
J0008-1329 & 00:08:28.02 & -13:29:30.57 & PMN J0008-1329 &  & Rad & \\
J0008-3945 & 00:08:09.20 & -39:45:22.96 & WISE J000809.17-394522.8 &  & Rad & \\
J0010-0433 & 00:10:00.39 & -04:33:48.51 & WISEA J001000.38-043348.4 &  & Rad & \\
J0011+0823 & 00:11:35.27 & +08:23:55.59 & WISEA J001135.27+082355.5 &  & QSO & \citet{2016AA...595A...5M} \\
J0011-1434 & 00:11:40.46 & -14:34:04.63 & LQAC 002-014 001 &  & QSO & \citet{2016AA...595A...5M} \\
J0011-4105 & 00:11:52.40 & -41:05:45.17 & WISEA J001152.39-410545.1 &  & Rad & \\
J0011-8443 & 00:11:45.91 & -84:43:20.01 & WISE J001145.90-844320.0 &  & Rad & \\
J0011-8706 & 00:11:51.48 & -87:06:25.45 & PKS 0010-873 &  & rG & \citet{2017AJ....153..157T} \\
J0013+4051 & 00:13:31.13 & +40:51:37.14 & 4C +40.01 & 0.255 & Sy1 & \citet{2014ApJS..215...14D}
\end{tabular}
\tablefoot{SIMBAD object classification. AGN: Active Galaxy Nucleus, AG?: Possible Active Galaxy Nucleus, BCG: Brightest galaxy in a Cluster, BLL: BL Lac object, BL?: BL Lac candidate, Bla: Blazar, Bz?: Blazar candidate, G: Galaxy, gam: \(\gamma\)-ray source, GiC: Galaxy in cluster of galaxies, IR: infrared source, PaG: Pair of galaxies, PN: Planetary nebula, pr*: Pre-main sequence star, QSO: Quasar, Q?: Quasar candidate, Rad: Radio source, rG: Radio Galaxy, Sy1: Seyfert 1 galaxy, Sy2: Seyfert 2 galaxy, SyG: Seyfert galaxy, X:X-ray source. Full table available online.}
\end{table}
\end{landscape}

\begin{table}[ht]
\caption{X-ray properties of ABC sources. For each source we list the name (Name ALMA), the X-ray detector (Detector) and the observation exposure (EXP). If the source is detected we also list the count-rate (Countrate) and the signal to noise ratio (SNR). We then list the unabsorbed source flux (\(\textrm{Flux}_{0.1-2.4}\)) in the \(0.1-2.4\textrm{ keV}\) band, obtained from the count-rate assuming a power-law model with a spectral index \(1.8\) or, if a spectrum is available, from the spectral fit. For the sources adequately fitted with a power-law model, we also list the power-law spectral index \(\Gamma_X\) and normalization \(K\), together with the fit \(\chi^2\) and degrees of freedom (\(\chi^2\), D.O.F.).}\label{tab:xray_properties}
\centering
\resizebox{\textwidth}{!}{
\begin{tabular}{l|c|c|c|c|c|c|c|c}
\hline\hline
Name ALMA & Detector & EXP & Countrate & SNR & \(\textrm{Flux}_{0.1-2.4}\) & \(\Gamma_X\) & \(K\) & \(\chi^2\) (D.O.F.) \\
\hline
 & & ks & \({10}^{-3} \textrm{ s}^{-1}\) & & \({10}^{-13} \textrm{ erg} \textrm{ cm}^2 \textrm{ s}^{-1}\) & & \({10}^{-4}\textrm{ keV} \textrm{ cm}^{-2} \textrm{ s}^{-1}\)& \\
\hline
J0010-0433 & XRT & \(1.3\) &  &  &  &  &  & \\
J0011-8706 & XRT & \(4.9\) & \(2.55\pm 0.84\) & \(3.0\) & \(0.87\pm 0.29\) &  &  & \\
J0019-5641 & XRT & \(2.9\) &  &  &  &  &  & \\
J0024-6820 & XRT & \(0.7\) & \(13.25\pm 5.00\) & \(2.6\)  & \(3.82\pm 1.44\) & & & \\
J0025+3919 & ACIS & \(19.7\) & \(15.70\pm 1.00\) & \(15.7\) & \(1.39\pm 0.24\) & \({1.36}_{-0.22}^{+0.22}\) & \({0.53}_{-0.10}^{+0.11}\) & \(0.8(24)\)\\
J0028+2000 & XRT & \(3.0\) &  &  &  &  &  & \\
J0029+3456 & EPIC & \(28.2\) & \(66.77\pm 3.08\) & \(21.7\) & \(2.15\pm 0.10\) & & & \\
J0034-0054 & ACIS & \(2.0\) & \(14.60\pm 2.90\) & \(5.0\) & \(2.54\pm 0.50\) &  &  & \\
J0039-2220 & XRT & \(0.8\) & \(12.42\pm 4.30\) & \(2.9\) & \(3.51\pm 1.21\) &  &  & \\
J0045-3705 & XRT & \(2.2\) &  &  &  &  &  & \\
J0046-2631 & XRT & \(1.3\) &  &  &  &  &  & \\
J0054-1953 & XRT & \(1.1\) &  &  &  &  &  & \\
J0057-0123 & ACIS & \(7.9\)   & \(5.35\pm 0.88\)  & \(6.1\) & \(0.64\pm 0.10\) &  &  & \\
J0057+3021 & XRT & \(8.1\) & \(5.15\pm 0.91\) & \(5.7\) & \(6.29\pm 0.75\) & \({1.97}_{-0.19}^{0.20}\) & \({2.47}_{-0.29}^{0.29}\) & \(1.12(10)\)\\
 & ACIS & \(54.5\)  & \(89.10\pm 1.30\) & \(68.5\) & \(10.88\pm 0.16\) &  &  & \\
 & EPIC & \(40.6\) & \(24.02\pm 7.12\) & \(73.6\) & \(16.65\pm 0.22\) & & & \\
J0101-6233 & XRT & \(0.7\) &  &  &  &  &  & \\
J0102-5637 & XRT & \(2.7\) &  &  &  &  &  & \\
J0109-6049 & XRT & \(10.8\) &  &  &  &  &  & \\
J0110-0219 & XRT & \(1.0\) &  &  &  &  &  & \\
J0111-7302 & ACIS & \(49.4\)  & \(7.20\pm 0.44\)  & \(16.4\) & \(0.38\pm 0.05\)  & \({2.62}_{-0.36}^{+0.41}\) & \({0.49}_{-0.09}^{+0.10}\) & \(0.79(49)\) \\
 & EPIC & \(55.5\) & \(35.67\pm 2.40\) & \(14.9\) & \(1.24\pm 0.08\) & & & \\
J0119+3210 & XRT & \(2.3\) &  &  &  &  &  & \\
 & ACIS & \(4.7\) & \(0.62\pm 0.45\) & \(1.4\) & \(0.08\pm 0.05\) &  &  & \\
J0122-0018 & EPIC & \(9.0\) & \(45.42\pm 9.44\) & \(4.8\) & \(1.38\pm 0.29\) &  &  & \\
J0123-0923 & ACIS & \(10.0\) &  &  &  &  &  & \\
J0132-0804 & XRT & \(2.2\) & \(3.66\pm 1.70\) & \(2.2\) & \(19.00\pm 2.40\) & \({1.80}_{-0.22}^{+0.22}\) & \({6.42}_{-0.77}^{+0.76}\) & \(0.7(6)\)\\
 & ACIS & \(6.0\) & \(16.70\pm 1.80\) & \(9.3\) & \(4.70\pm 0.76\) & \({2.38}_{-0.21}^{+0.22}\) & \({1.57}_{-0.19}^{+0.20}\) & \(0.4(9)\)\\
J0134-0931 & ACIS & \(1.1\) & \(119.00\pm 11.00\) & \(10.8\) & \(3.88\pm 0.67\) & \({1.08}_{-0.26}^{+0.27}\) & \({1.30}_{-0.26}^{+0.28}\) & \(0.6(3)\)\\
J0137+3309 & XRT & \(2.1\) & \(15.23\pm 3.10\) & \(4.9\) & \(20.12\pm 2.22\) & \({1.71}_{-0.17}^{+0.17}\) & \({7.33}_{-0.83}^{+0.83}\) & \(0.5(7)\)\\
 & ACIS & \(9.2\) & \(713.00\pm 9.00\) & \(79.2\) & \(2.57\pm 0.15\) & \({2.29}_{-0.12}^{+0.12}\) & \({0.94}_{-0.05}^{+0.05}\) & \(0.6(21)\)\\
J0139+1753 & XRT & \(2.1\) & \(2.89\pm 1.40\) & \(2.0\) & \(0.90\pm 0.43\) &  &  & \\
J0156+3914 & XRT & \(5.4\) & \(3.16\pm 0.99\) & \(3.2\) & \(4.59\pm 0.68\) & \({1.10}_{-0.45}^{+0.44}\) & \({1.53}_{-0.31}^{+0.3}\) & \(1.3(3)\)
\end{tabular}
}
\tablefoot{Full table available online.}
\end{table}

\begin{table}[ht]
\caption{Result of the X-ray spectral fitting for ABC sources with a complex (non power-law) model. For each source we list the ALMA name (Name ALMA), the common name (Other Name), the object ABC type (Object Type), and the X-ray detector (Detector). The column Model shows the model components used to fit the spectrum, namely an intrinsic photoelectric absorption component \textsc{zwabs} (A), a power-law component (PL) a thermal component \textsc{apec} (T) a gaussian line \textsc{zgauss} (L) and a power law spectrum reflected from neutral material \textsc{pexrav} (R). We then list the hydrogen column density (\(N_H\), the power-law spectral index (\(\Gamma_X\) and normalization \(K\), the reflection component spectral index (\(\Gamma_R\) and normalization \(K_R\), the thermal component temperature (\(kT\)) and normalization \(K_T\), and the line(s) rest-frame energy (\(E_1\), \(E_2\)) and normalizations (\(K_1\), \(K_2\)). Finally, we list the fit \(\chi^2\) and degrees of freedom (\(\chi^2\), D.O.F.).}\label{tab:properties_peculiar}
\centering
\resizebox{\textwidth}{!}{
\begin{tabular}{c|c|c|c|c|c|c|c|c|c|c|c|c|c|c|c|c}
\hline\hline
Name ALMA & Other Name & Object Type & Detector & Model & \(N_H\) & \(\Gamma_X\) & \(K\) & \(\Gamma_R\) & \(K_R\) & \(kT\) & \(K_T\) & \(E_1\) & \(K_1\) & \(E_2\) & \(K_2\) & \(\chi^2\) (D.O.F.) \\
\hline
J0029+3456 & B2 0026+34 & AGN & EPIC & A * PL & \({0.88}_{-0.30}^{+0.37}\) & \({1.38}_{-0.19}^{+0.21}\) & \({0.40}_{-0.08}^{+0.12}\) & & & & & & & & & \(0.47(28)\) \\
\hline
J0057+3021 & NGC 0315 & AGN & ACIS & PL+T & & \({1.04}_{-0.06}^{+0.06}\) & \({0.47}_{-0.03}^{+0.03}\) & & & \({0.68}_{-0.02}^{+0.02}\) & \({0.69}_{-0.03}^{+0.03}\) & & & & & \(0.83(115)\) \\
 & & & EPIC & PL+T & & \({1.30}_{-0.04}^{+0.04}\) & \({1.08}_{-0.05}^{+0.05}\) & & & \({0.72}_{-0.01}^{+0.01}\) & \({1.12}_{-0.04}^{+0.04}\) & & & & & \(0.79(356)\) \\
\hline
J0111-7302 & PMN J0111-7302 & AGN & EPIC & PL+T & & \({1.72}_{-0.44}^{+0.43}\) & \({0.19}_{-0.07}^{+0.10}\) & & & \({0.56}_{-0.25}^{+0.18}\) & \({0.33}_{-0.11}^{+0.31}\) & & & & & \(0.81(14)\) \\
\hline
J0242-2132 & PKS 0240-217 & QSO & ACIS & A * PL & \({0.27}_{-0.08}^{+0.08}\) & \({2.36}_{-0.14}^{+0.15}\) & \({2.46}_{-0.28}^{+0.32}\) & & & & & & & & & \(0.88(37)\) \\
 & & & EPIC & A * (PL+L) & \({0.09}_{-0.01}^{+0.01}\) & \({2.03}_{-0.02}^{+0.02}\) & \({9.52}_{-0.15}^{+0.15}\) & & & & & \({6.71}_{-0.02}^{+0.02}\) & \({0.15}_{-0.02}^{+0.02}\) & & & \(0.80(969)\) \\
\hline
J0702-2841 & NGC 2325 & AGN & EPIC & PL+T & & \({1.54}_{-0.04}^{+0.04}\) & \({0.28}_{-0.01}^{+0.01}\) & & & \({0.76}_{-0.03}^{+0.03}\) & \({0.12}_{-0.01}^{+0.01}\) & & & & & \(0.64(323)\) \\
\hline
J0758+3747 & NGC 2484 & QSO & EPIC & PL+T & & \({2.16}_{-0.03}^{+0.03}\) & \({0.66}_{-0.02}^{+0.02}\) & & & \({0.83}_{-0.03}^{+0.03}\) & \({0.21}_{-0.01}^{+0.01}\) & & & & & \(0.57(360)\) \\
\hline
J0928-0409 & PKS J0928-0408 & QSO & EPIC & PL+L & & \({1.53}_{-0.22}^{+0.22}\) & \({0.29}_{-0.06}^{+0.06}\) & & & & & \({1.05}_{-0.06}^{+0.06}\) & \({0.32}_{-0.13}^{+0.13}\) & & & \(1.18(8)\) \\
\hline
J1044+0655 & PKS 1042+071 & QSO & ACIS & PL+L & & \({2.40}_{-0.25}^{+0.28}\) & \({0.33}_{-0.04}^{+0.04}\) & & & & & \({0.67}_{-0.05}^{+0.05}\) & \({12.89}_{-6.24}^{+13.19}\) & & & \(0.94(115)\) \\
\hline
J1109-3732 & NGC 3557 & AGN & ACIS & PL+T & & \({1.99}_{-0.34}^{+0.52}\) & \({0.38}_{-0.11}^{+0.19}\) & & & \({0.31}_{-0.21}^{+0.35}\) & \({0.67}_{-0.38}^{+48.42}\) & & & & & \(0.60(9)\) \\
\hline
J1215-1731 & PKS 1213-17 & Blazar & XRT & A * (PL+R) & \({2.48}_{-0.50}^{+0.23}\) & \(>7.79\) & \({19.34}_{-6.24}^{+5.53}\) & \(<-2.63\) & \(<0.01\) & & & & & & & \(0.86(9)\) \\
\hline
J1230+1223 & M87 & AGN & XRT & PL+T & & \({1.67}_{-0.09}^{+0.07}\) & \({18.98}_{-4.71}^{+4.71}\) & & & \({1.85}_{-0.08}^{+0.07}\) & & & & & & \(1.05(212)\) \\
 &  &  & ACIS & PL+T & & \({1.86}_{-0.05}^{+0.05}\) & \({0.73}_{-0.07}^{+0.07}\) & & & \({1.75}_{-0.12}^{+0.19}\) & \({1.08}_{-0.21}^{+0.22}\) & & & & & \(0.87(141)\) \\
 &  &  & EPIC & A*(PL+T) & \({0.05}_{-0.01}^{+0.01}\) & \({2.51}_{-0.07}^{+0.08}\) & \({280.33}_{-18.54}^{+19.12}\) & & & \({1.66}_{-0.01}^{+0.01}\) & \({1914.94}_{-32.90}^{+33.41}\) & & & & & \(1.14(1153)\) \\
\hline
J1305-4928 & NGC 4945 & RS & ACIS & A * (T+R+L+L) & \({2.07}_{-0.17}^{+0.16}\) & & & \({2.77}_{-0.15}^{+0.07}\) & \({305.94}_{-60.59}^{+31.85}\) & \({0.86}_{-0.04}^{+0.06}\) & \({2.93}_{-0.60}^{+0.78}\) & \({6.42}_{-0.01}^{+0.01}\) & \({0.34}_{-0.02}^{+0.02}\) & \({6.77}_{-0.02}^{+0.03}\) & \({0.10}_{-0.02}^{+0.02}\) & \(1.08(183)\) \\
 & & & EPIC & A * (T+R+L+L) & \({0.06}_{-0.05}^{+0.05}\) & & & \({2.80}_{-0.11}^{+0.05}\) & \({856.37}_{-118.34}^{+48.79}\) & \({0.79}_{-0.05}^{+0.06}\) & \({0.52}_{0.10}^{+0.15}\) & \({6.42}_{-0.01}^{+0.01}\) & \({0.88}_{-0.07}^{+0.06}\) & \({6.82}_{-0.06}^{+0.08}\) & \({0.21}_{-0.05}^{+0.04}\) & \(0.90(227)\) \\
 \hline
 J1319-1239 & NGC 5077 & AGN & ACIS & PL+T & & \({1.63}_{-0.36}^{+0.33}\) & \({0.08}_{-0.02}^{+0.02}\) & & & \({0.71}_{-0.13}^{+0.13}\) & \({0.03}_{-0.01}^{+0.01}\) & & & & & \(0.27(6)\) \\
 \hline
 J1321-4342 & NGC 5090 & RG & EPIC & A * (T+R) & \({0.18}_{-0.02}^{+0.02}\) & & & \({4.03}_{-0.08}^{+0.09}\) & \({1204.88}_{-110.86}^{+120.24}\) & \({0.71}_{-0.11}^{+0.03}\) & \({1.28}_{-0.10}^{+0.11}\) & & & & & \(0.66(189)\) \\
 \hline
 J1336-3357 & IC4296 & AGN & ACIS & PL+T & & \({0.82}_{-0.09}^{+0.08}\) & \({0.39}_{-0.04}^{+0.04}\) & & & \({0.76}_{-0.01}^{+0.01}\) & \({1.09}_{-0.04}^{+0.04}\) & & & & & \(0.78(89)\) \\
 &  &  & EPIC & PL+T & & \({1.01}_{-0.05}^{+0.05}\) & \({0.85}_{-0.05}^{+0.05}\) & & & \({0.82}_{-0.01}^{+0.01}\) & \({1.40}_{-0.04}^{+0.04}\) & & & & & \(0.69(392)\) \\
 \hline
 J1347+1217 & PKS 1345+12 & QSO & ACIS & A * (PL+L) & \({2.24}_{-0.40}^{+0.44}\) & \({0.94}_{-0.20}^{+0.21}\) & \({1.10}_{-0.28}^{+0.40}\) & & & & & \({0.76}_{-0.04}^{+0.04}\) & \({183.11}_{-141.83}^{+837.27}\) & & & \(0.82(53)\) \\
 \hline
 J1553+2348 & PKS 1551+23 & QSO & XRT & A * PL & \({2.13}_{-0.62}^{+0.71}\) & \({2.84}_{-0.49}^{+0.56}\) & \({36.64}_{-17.27}^{+36.20}\) & & & & & & & & & \(0.84(7)\) \\
 \hline
 J1558-1409 & PKS 1555-140 & AGN & ACIS & T & & & & & & \({4.27}_{-0.29}^{+0.57}\) & \({2.81}_{-0.08}^{+0.08}\) & & & & & \(0.85(75)\) \\
 &  &  & EPIC & PL+T & & \({1.62}_{-0.03}^{+0.03}\) & \({60.21}_{-5.70}^{+5.70}\) & & & \({3.74}_{-0.14}^{+0.13}\) & \({498.50}_{-22.92}^{+22.92}\) & & & & & \(0.90(1434)\) \\
 \hline
 J1627-2426 & PMN J1626-2426 & AGN & ACIS & R+L+L & & & & \(-1.72^*\) & \(<0.01\) & & & \(3.11^*\) & \(<0.01\) & \({4.35}_{-0.27}^{0.27}\)& & \(0.90(11)\) \\
 \hline
 J1711-3744 & PMN J1711-3744 & No Type & ACIS & A * PL & \({36.08}_{-20.11}^{+13.35}\) & \({8.02}_{-4.30}^{+7.43}\) & \(<28.93e4\) & & & & & & & & & \(1.10(22)\) \\
 \hline
 J1723-6500 & NGC 6328 & QSO & EPIC & A * PL & \({0.11}_{-0.02}^{+0.02}\) & \({1.90}_{-0.05}^{+0.05}\) & \({1.08}_{-0.05}^{+0.05}\) & & & & & & & & & \(0.65(309)\) \\
 \hline
 J2033+2146 & PKS 2031+21 & QSO & XRT & A * PL & \({0.40}_{-0.10}^{+0.12}\) & \({1.55}_{-0.13}^{+0.14}\) & \({32.32}_{-4.70}^{+5.70}\) & & & & & & & & & \(0.98(55)\) \\
 \hline
 J2157-6941 & PKS 2153-69 & QSO & ACIS & PL+T & & \({1.67}_{-0.02}^{+0.02}\) & \({2.70}_{-0.05}^{+0.05}\) & \({0.20}_{-0.01}^{+0.02}\) & \({0.64}_{-0.12}^{+0.12}\) & & & & & & & \(0.88(272)\) \\
 \hline
 J2212-2518 & CRATES J221220.71-251828.4 & AGN & ACIS & P+T & & \({1.94}_{-0.20}^{+0.18}\) & \({0.25}_{-0.03}^{+0.03}\) & & & \({0.74}_{-0.11}^{+0.10}\) & \({0.08}_{-0.02}^{+0.02}\) & & & & & \(0.30(16)\) \\
 \hline
\end{tabular}
}
\tablefoot{\(N_H\) is in \({10}^{22} \textrm{ cm}^{-2}\). \(K\) is \({10}^{-4}\textrm{ keV} \textrm{ cm}^{-2} \textrm{ s}^{-1}\). \(K_R\) is \({10}^{-2}\textrm{ keV} \textrm{ cm}^{-2} \textrm{ s}^{-1}\). \(kT\), \(E_1\) and \(E_2\) are in \(keV\). \(K_T\) is \({10}^{-4}\textrm{ cm}^{-5}\). \(K_1\) and \(K_2\) are in \({10}^{-4}\textrm{ cm}^{-2} \textrm{ s}^{-1}\).}
\end{table}

\begin{landscape}
\begin{table}[ht]
\caption{List of the main properties of ABC sources. For each source we list the name (Name ALMA), the coordinates (RA, Dec), the ABC source type (Object Type), the redshift (Redshift), the blazar candidate class obtained from WISE colors (Type WISE), the source type obtained from optical spectra (Type Optical), the source class listed in the 4FGL catalog (4FGL), if the source is present in the 3HSP catalog (3HSP), the blazar candidate class as reported in the WIBRaLS2 catalog (WIBRaLS2), if the source is present in the KDEBLLACS catalog (KDEBLLACS), and the X-ray information (X-ray), namely a source detection (det), a power-law spectrum (PL) or a more complex spectrum (COMPLEX).}\label{tab:alma_candidates_short}
\centering
\begin{tabular}{l|c|c|c|c|c|c|c|c|c|c|c}
\hline\hline
Name ALMA & RA & Dec & Object Type & Redshift & Type WISE & Type Optical & 4FGL & 3HSP & WIBRaLS2 & KDEBLLACS & X-ray \\
\hline
J0002-2153 & 00:02:11.98 & -21:53:09.87 & BL Lac & & MIXED & & & & & & \\
J0006-2955 & 00:06:01.12 & -29:55:50.10 & QSO & 0.683 & BZU & & & & & & \\
J0008-3945 & 00:08:09.20 & -39:45:22.96 & RS & 1.9 & MIXED & & BCU & & BZQ & & \\
J0011+0823 & 00:11:35.27 & +08:23:55.59 & QSO & 1.35 & MIXED & QSO & & & BZQ & & \\
J0011-4105 & 00:11:52.40 & -41:05:45.17 & RS & & MIXED & & & & & & \\
J0011-8443 & 00:11:45.91 & -84:43:20.01 & RS & 0.6 & MIXED & & & & BZQ & & \\
J0011-8706 & 00:11:51.48 & -87:06:25.45 & RG & 3.292       & & & &     & & det \\
J0019-5641 & 00:19:26.61 & -56:41:42.46 & RS & & & & BCU & & & & \\
J0024-0811 & 00:24:00.67 & -08:11:10.05 & QSO & 2.067 & BZQ & QSO & & & BZQ & & \\
J0024-4202 & 00:24:42.99 & -42:02:03.95 & QSO & 0.937 & BZQ & & & & BZQ & & \\
J0024-6820 & 00:24:06.72 & -68:20:54.59 & AGN & 0.354 & BZQ & & BCU & & BZQ & & det \\
J0025+3919 & 00:25:26.16 & +39:19:35.44 & QSO & 1.946 & MIXED & & & & & & PL \\
J0025-4803 & 00:25:45.82 & -48:03:55.10 & RS & & MIXED & & BCU & & BZQ & & \\
J0026-3512 & 00:26:16.39 & -35:12:48.79 & Blazar & 1.299 & MIXED & & & & & & \\
J0027+0929 & 00:27:05.79 & +09:29:57.76 & QSO & 1.4 & MIXED & & & & & & \\
J0028+2000 & 00:28:29.82 & +20:00:26.74 & QSO & 1.552 & MIXED & QSO & FSRQ & & BZQ & & 
\end{tabular}
\tablefoot{Full table available online.}
\end{table}
\end{landscape}

\begin{appendix}

\section{X-ray Spectra}
\begin{figure*}
   \centering
\includegraphics[scale=0.19]{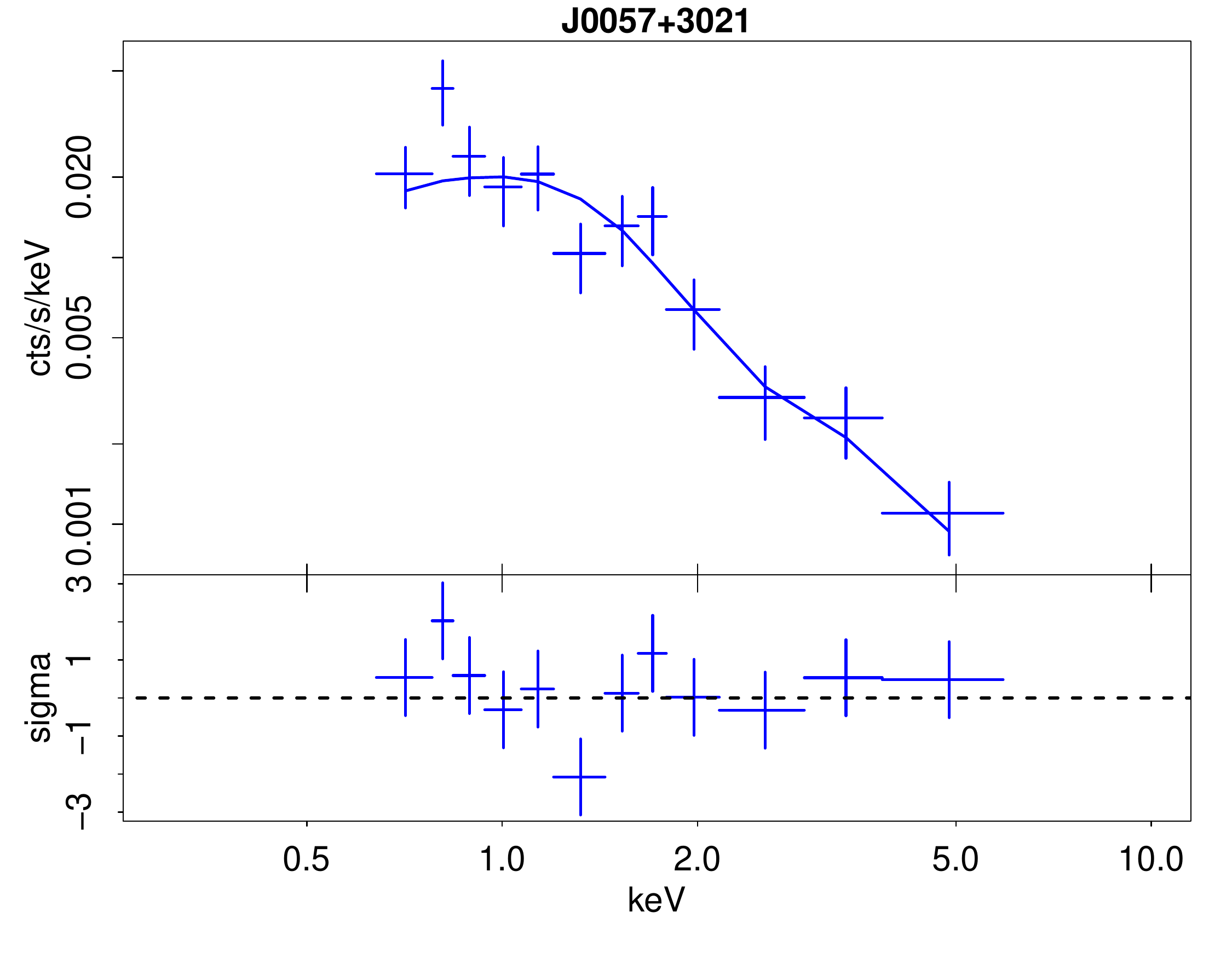}
\includegraphics[scale=0.19]{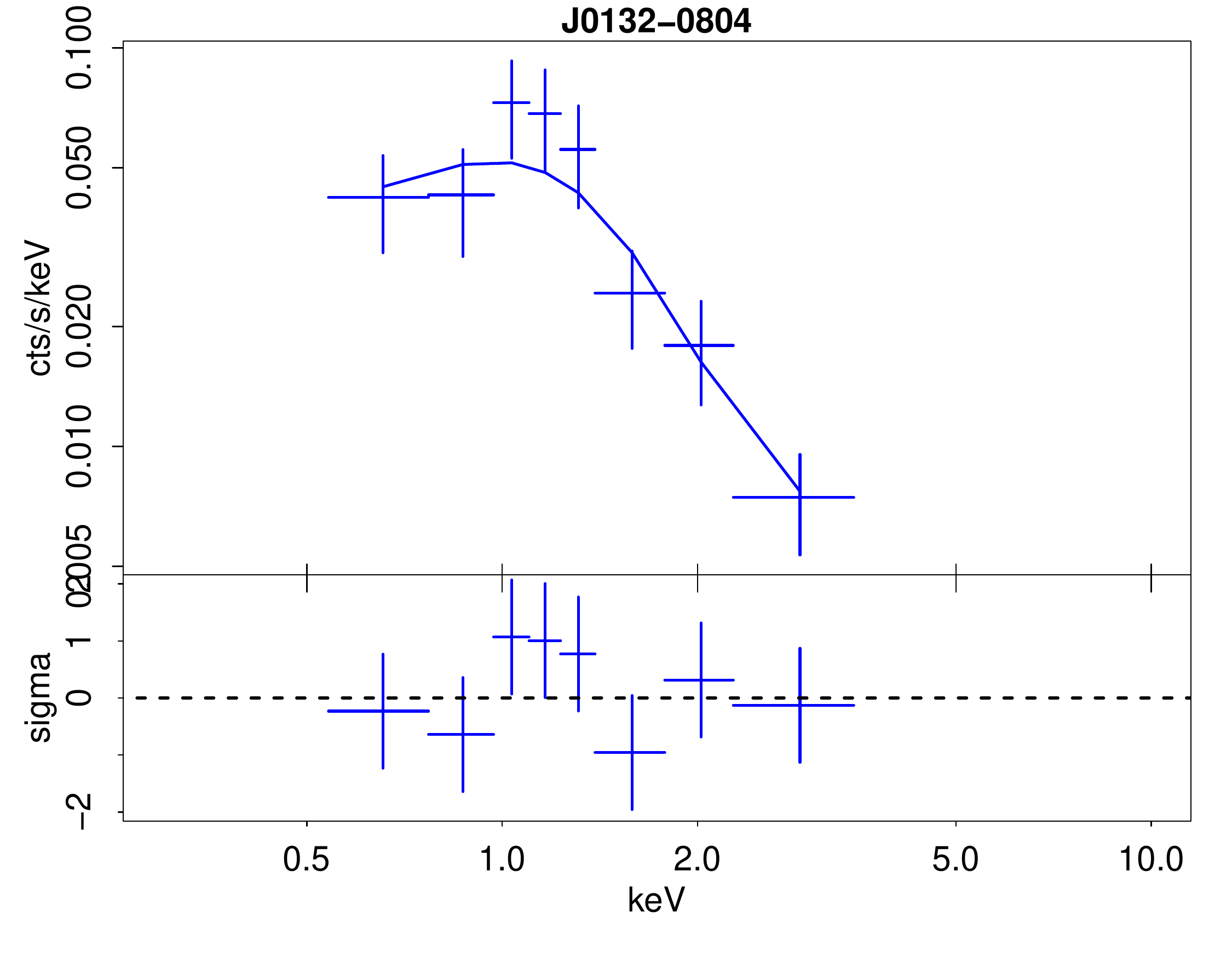}
\includegraphics[scale=0.19]{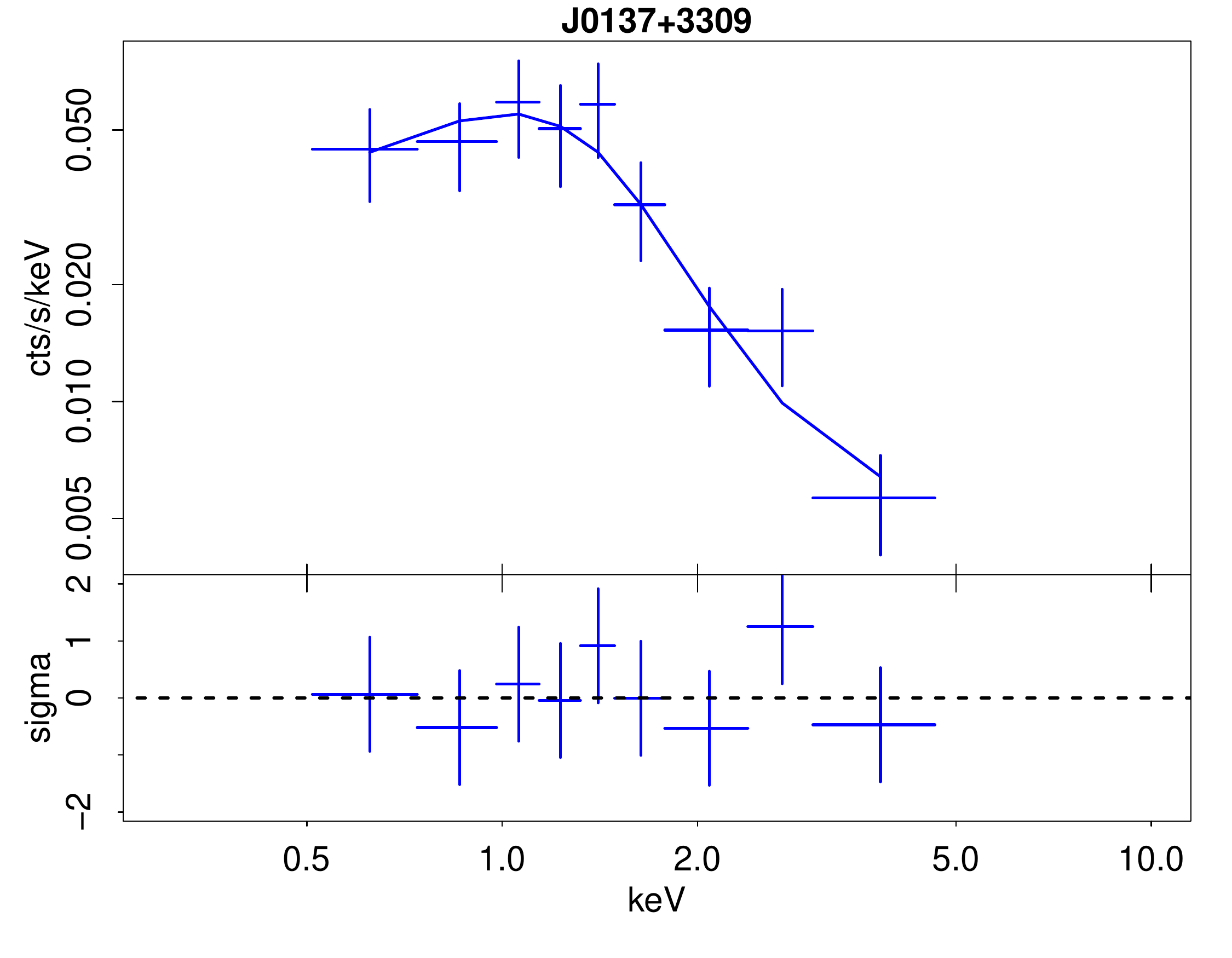}
\includegraphics[scale=0.19]{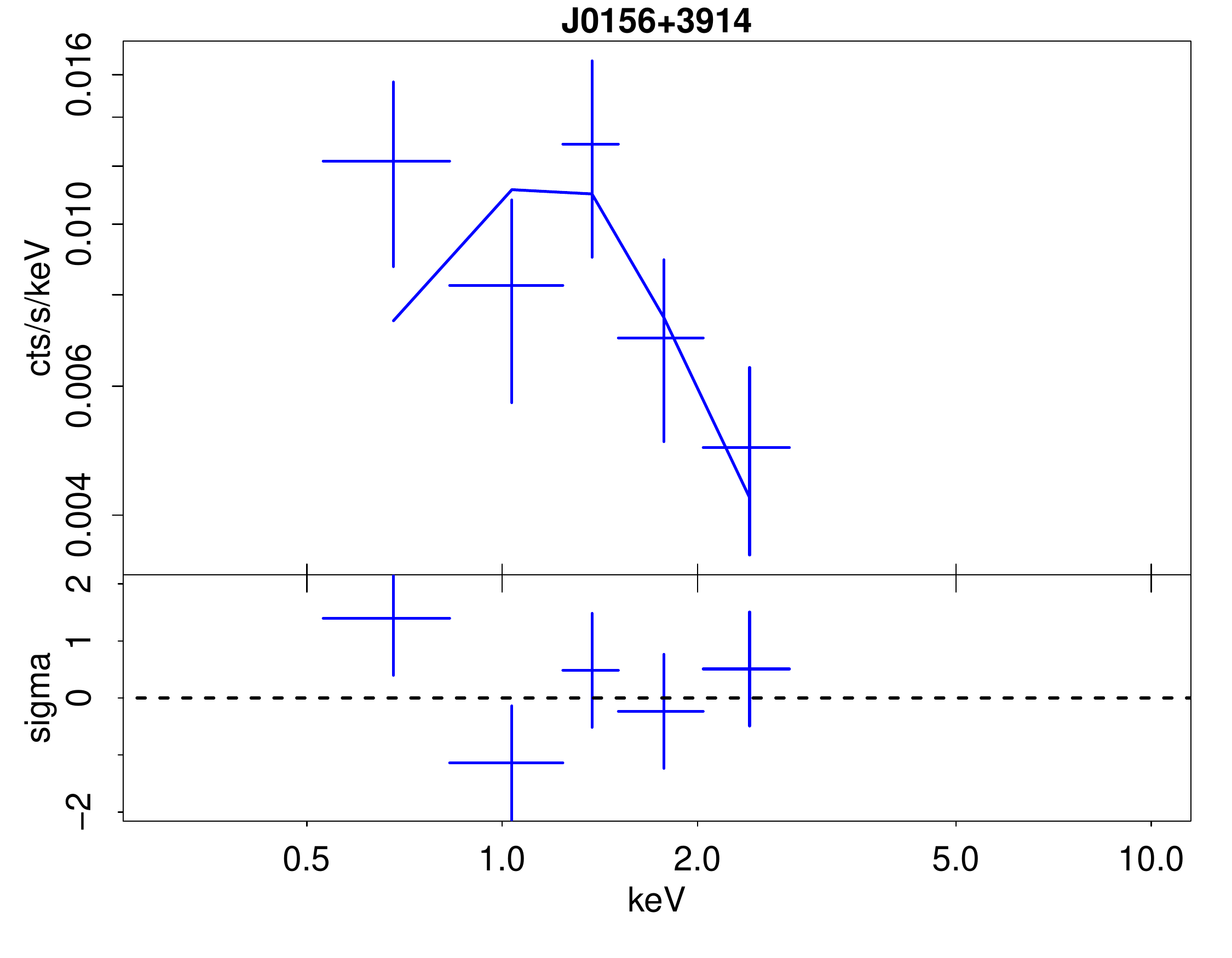}
\includegraphics[scale=0.19]{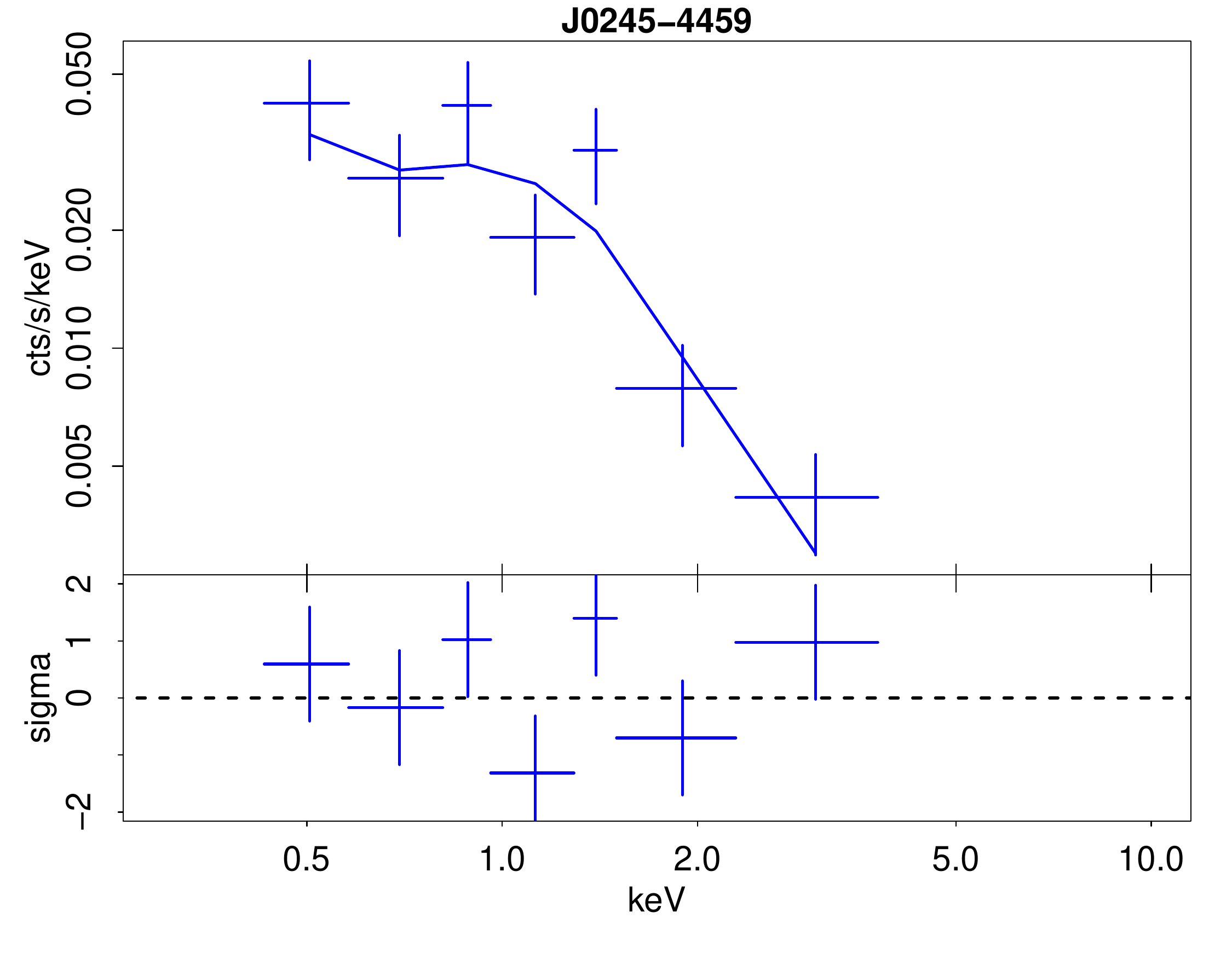}
\includegraphics[scale=0.19]{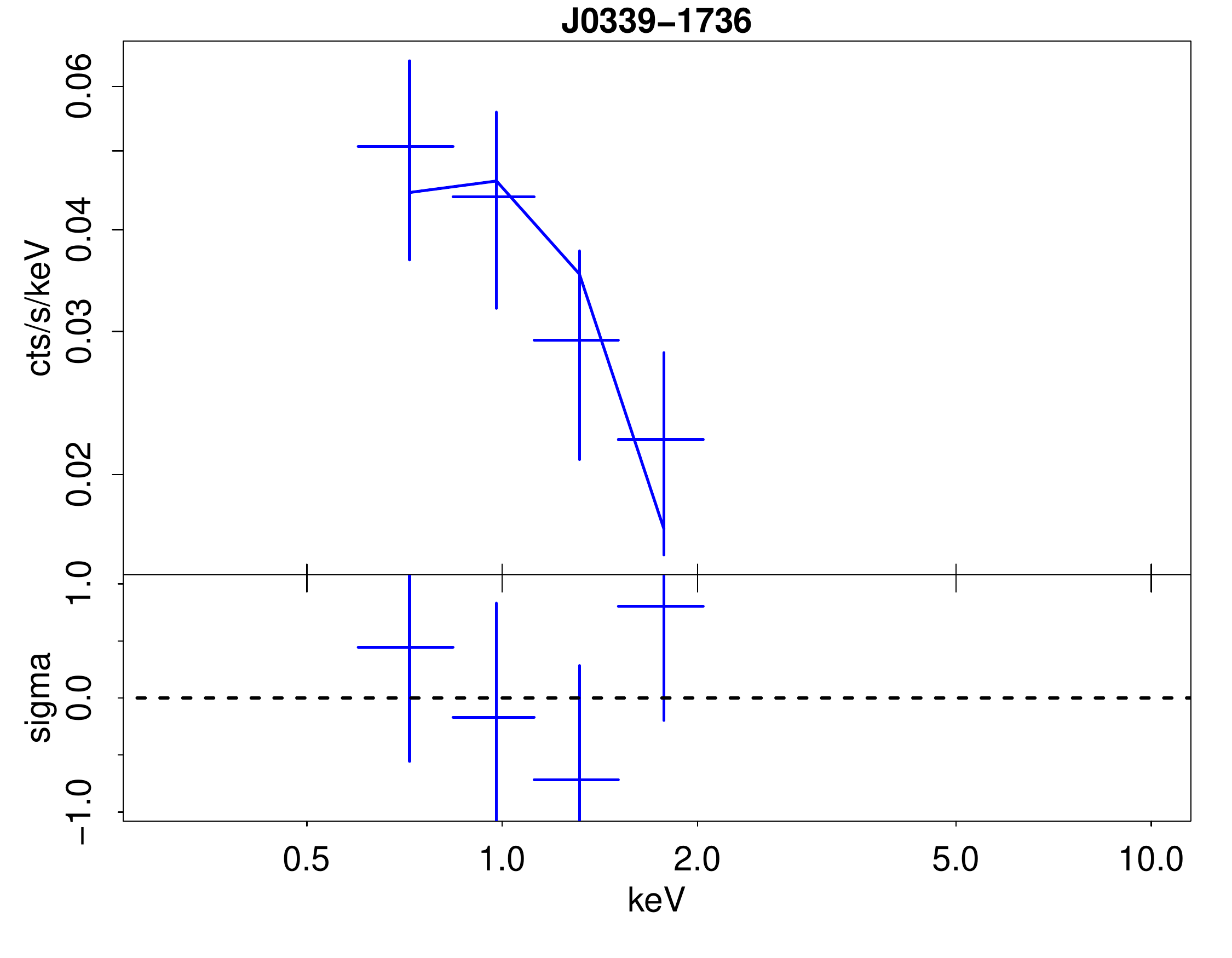}
\includegraphics[scale=0.19]{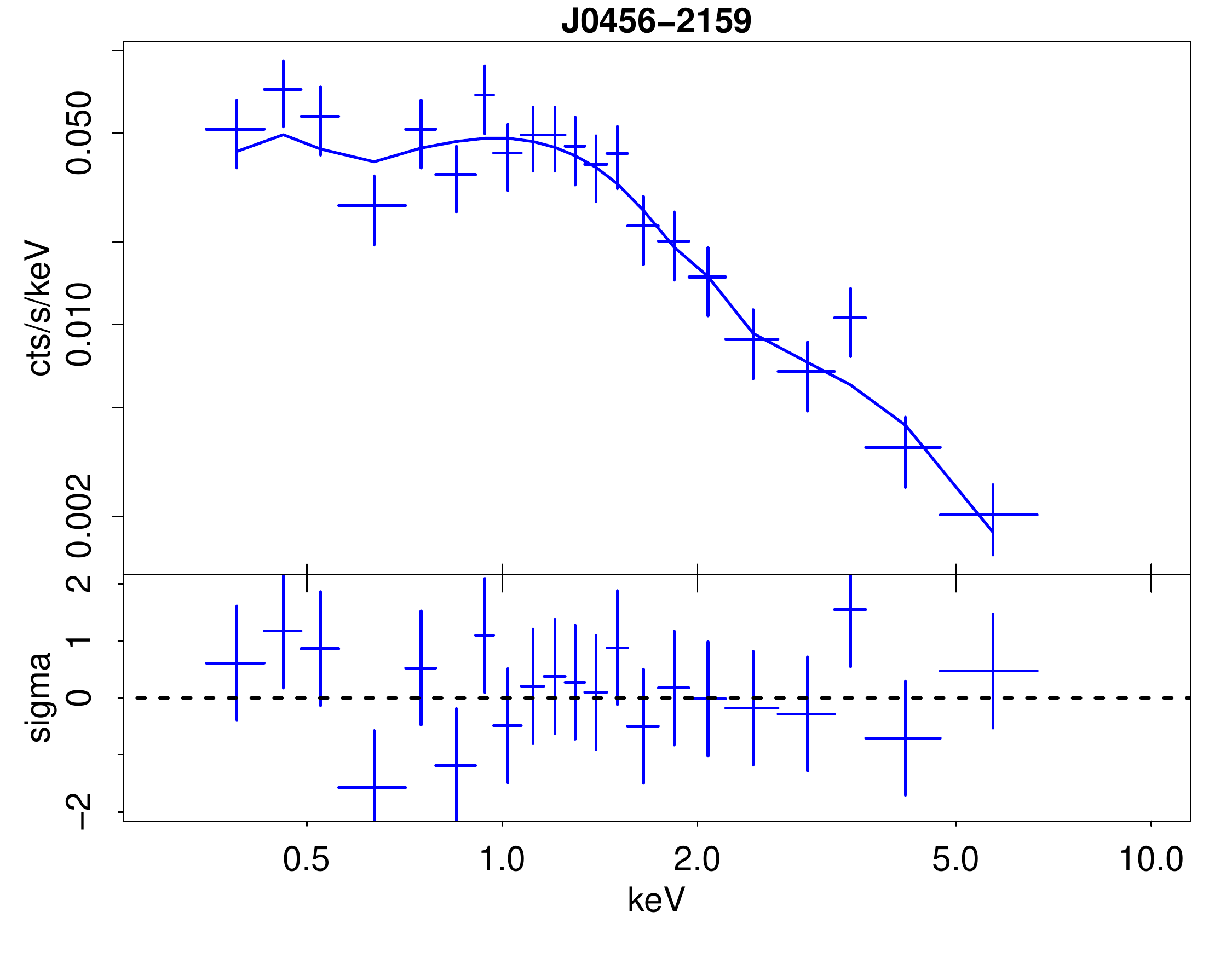}
\includegraphics[scale=0.19]{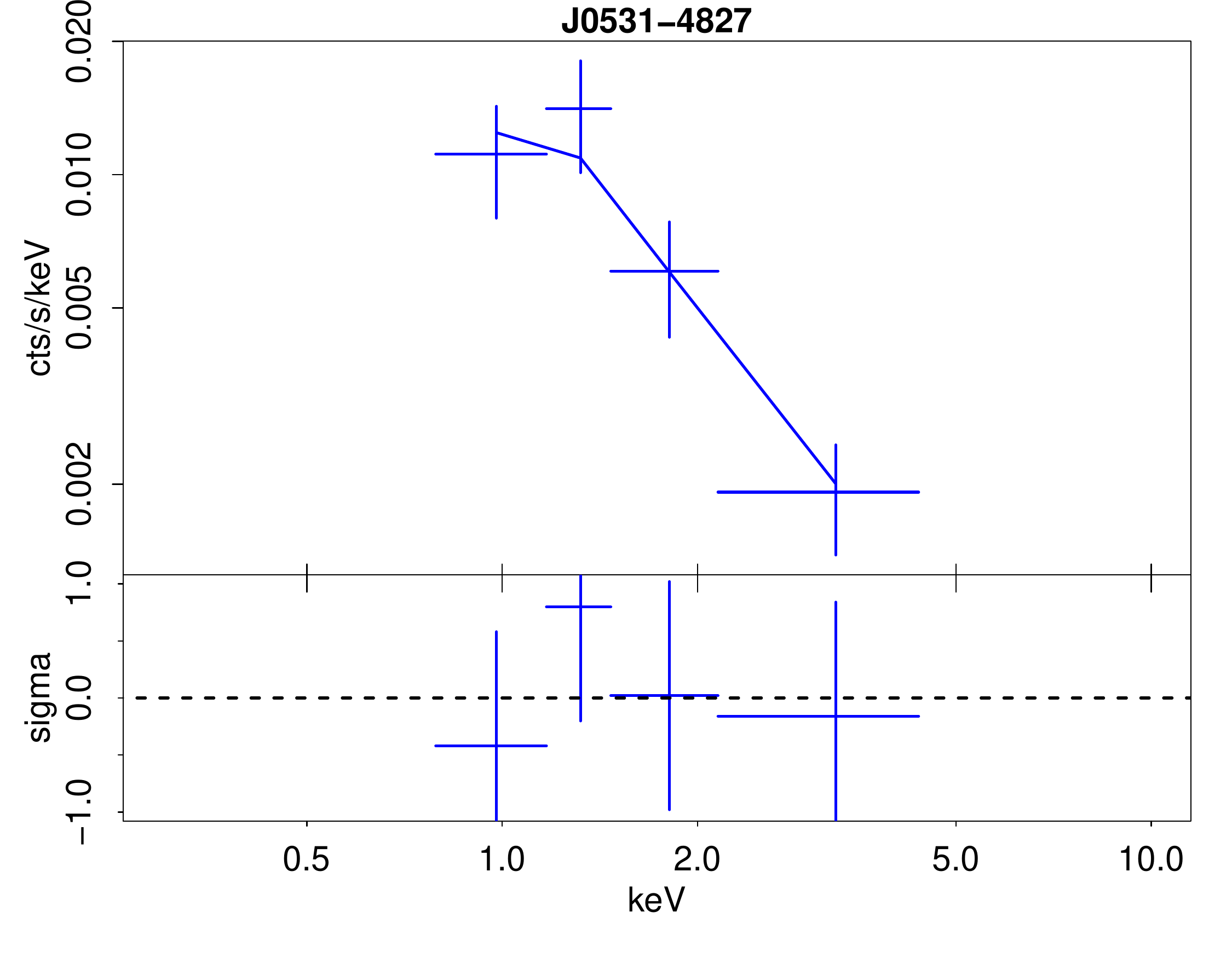}
\includegraphics[scale=0.19]{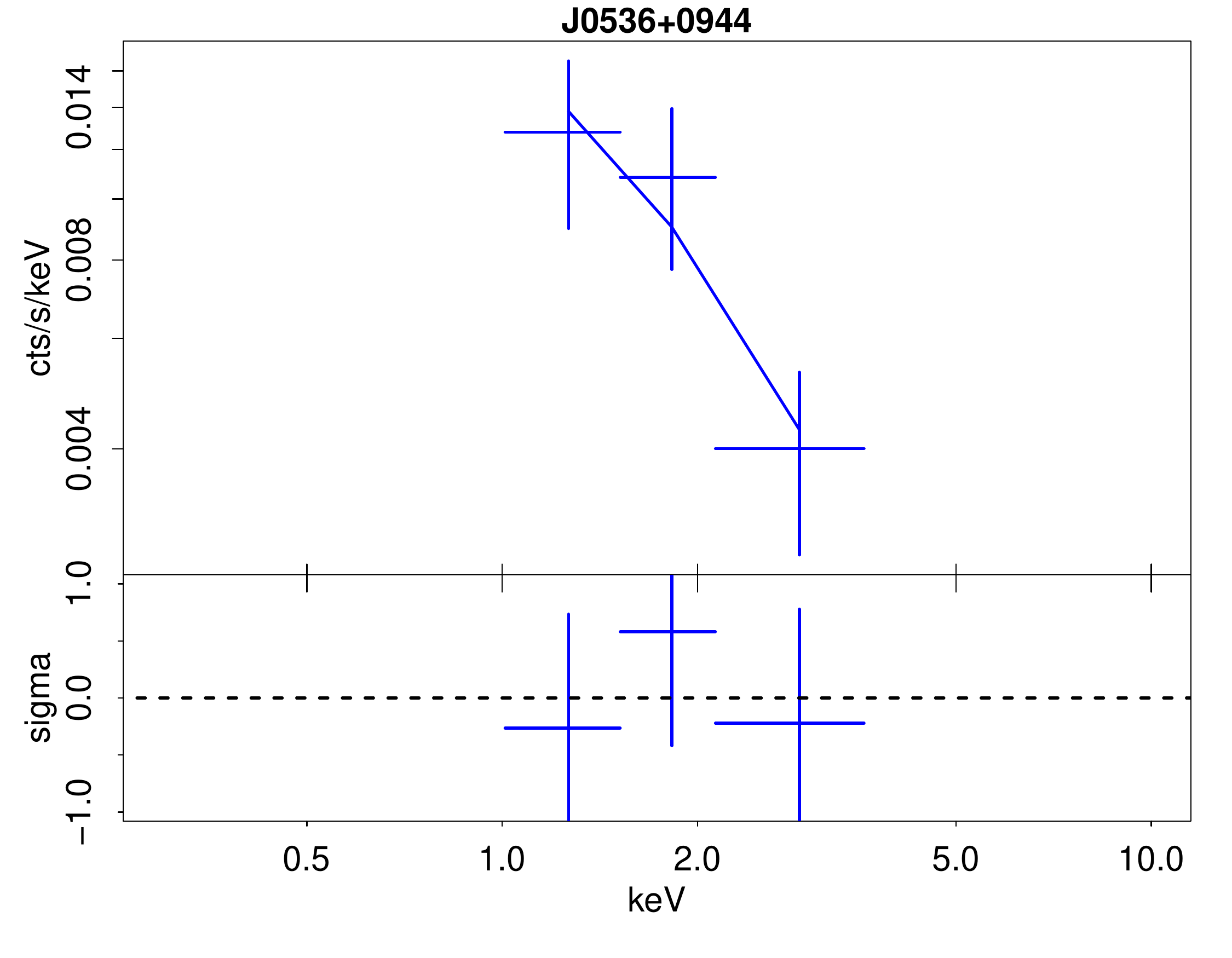}
\includegraphics[scale=0.19]{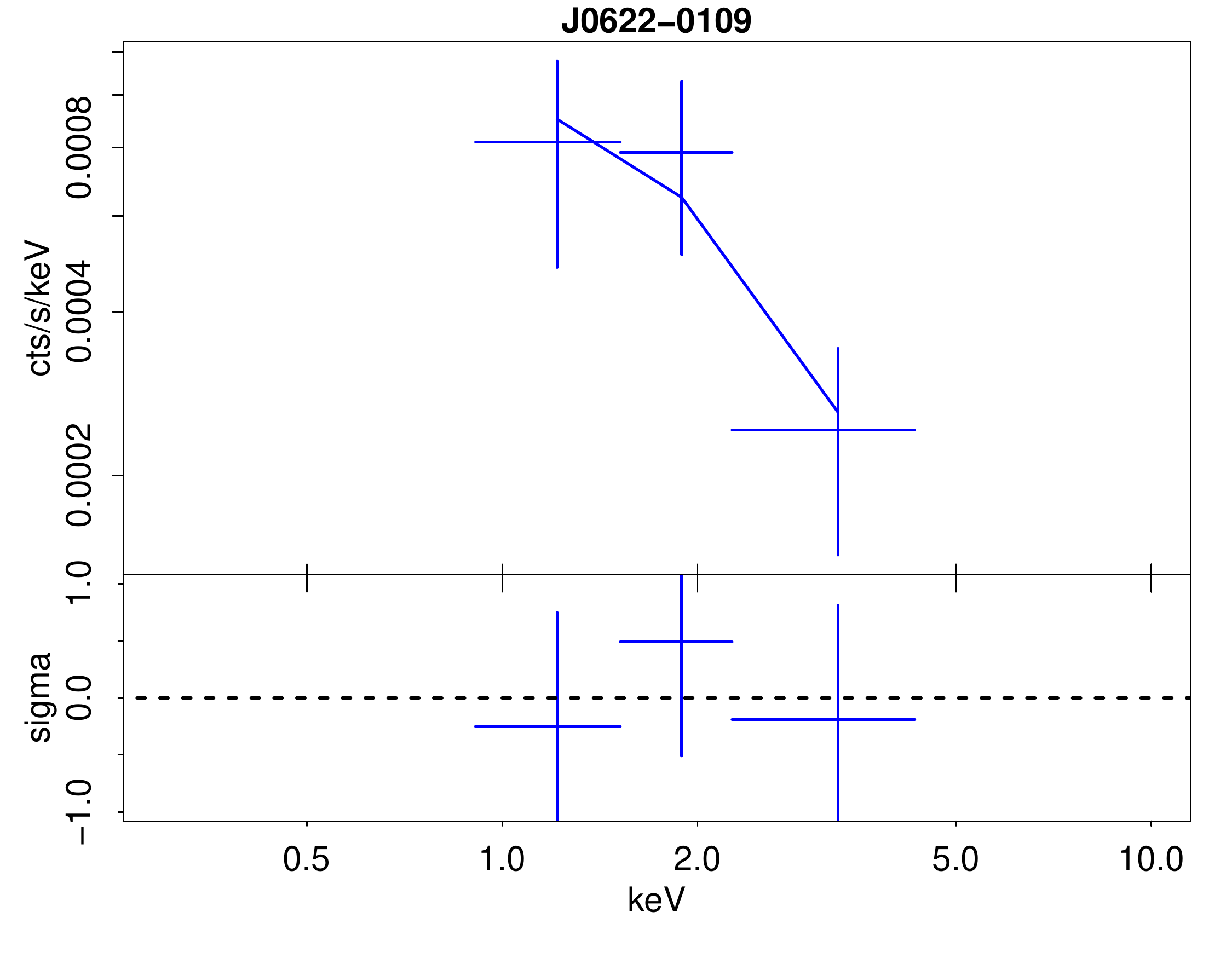}
\includegraphics[scale=0.19]{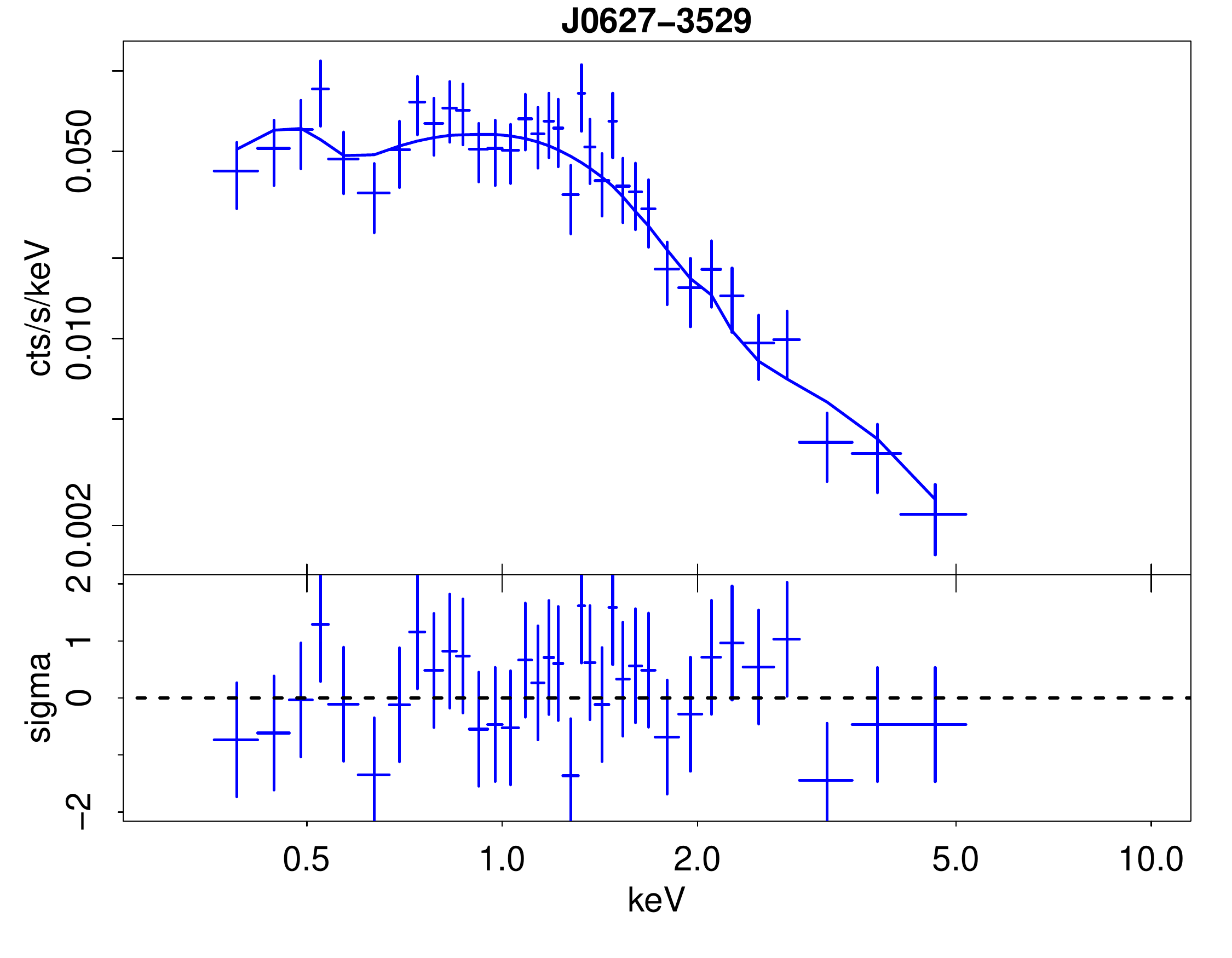}
\includegraphics[scale=0.19]{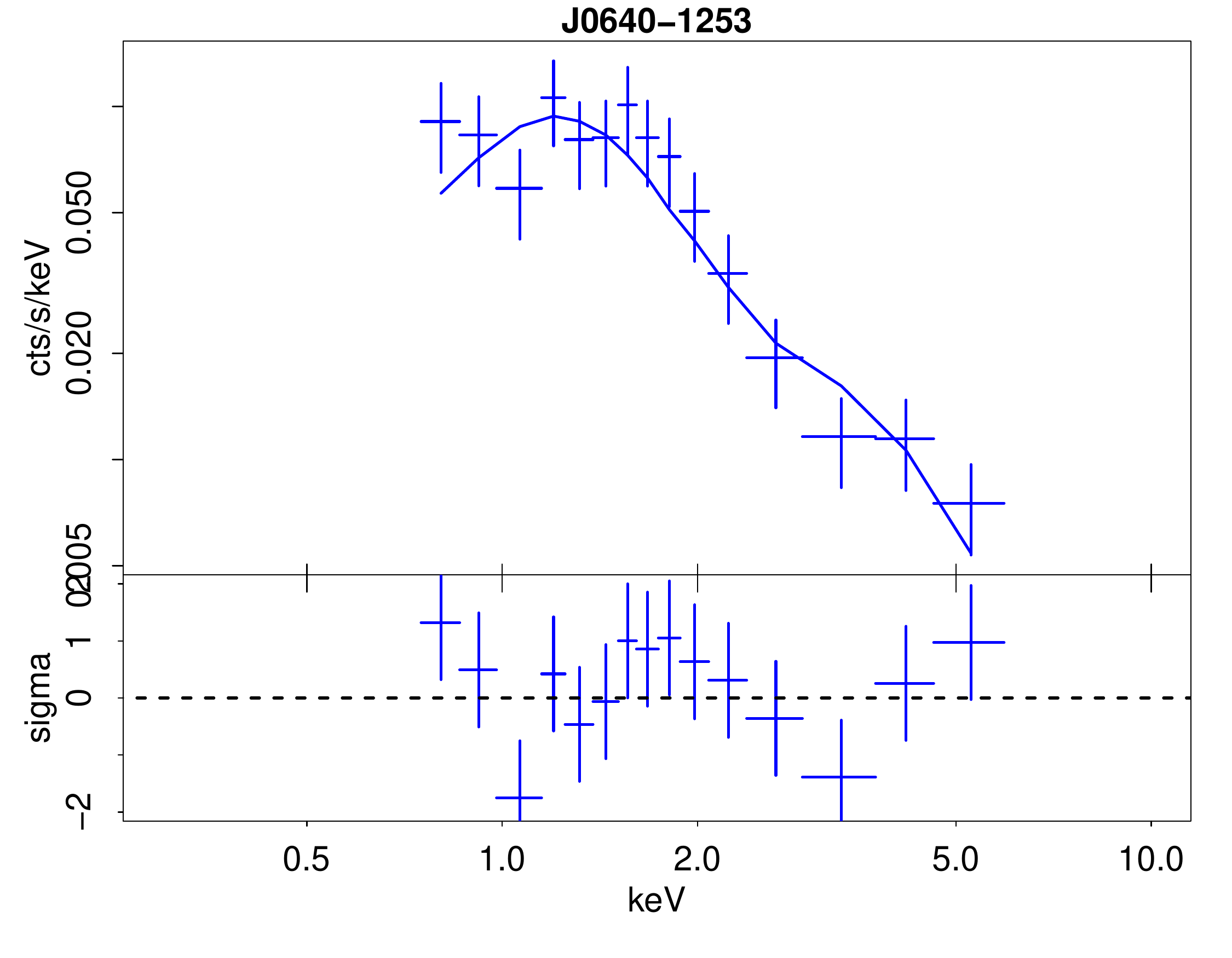}
\includegraphics[scale=0.19]{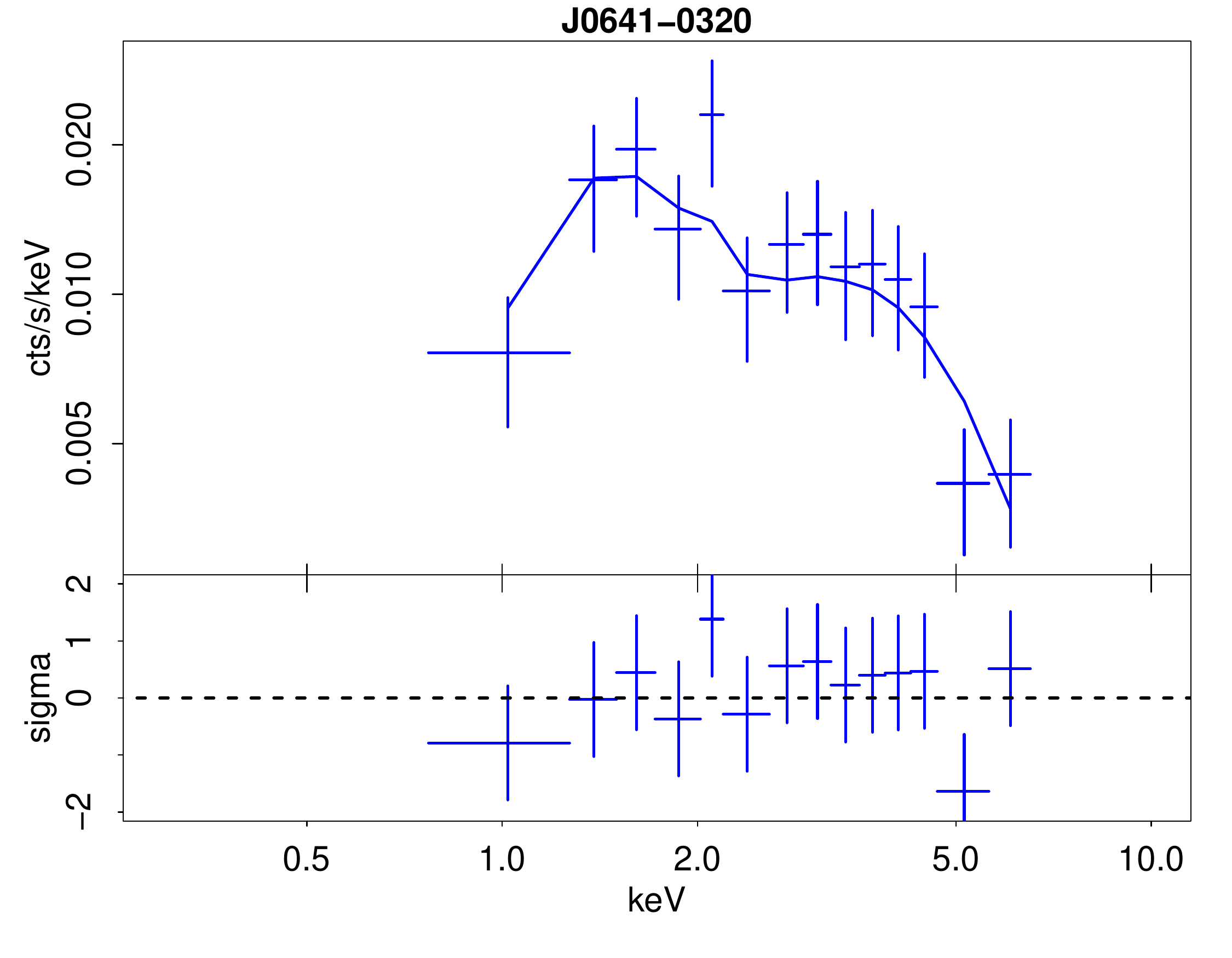}
\includegraphics[scale=0.19]{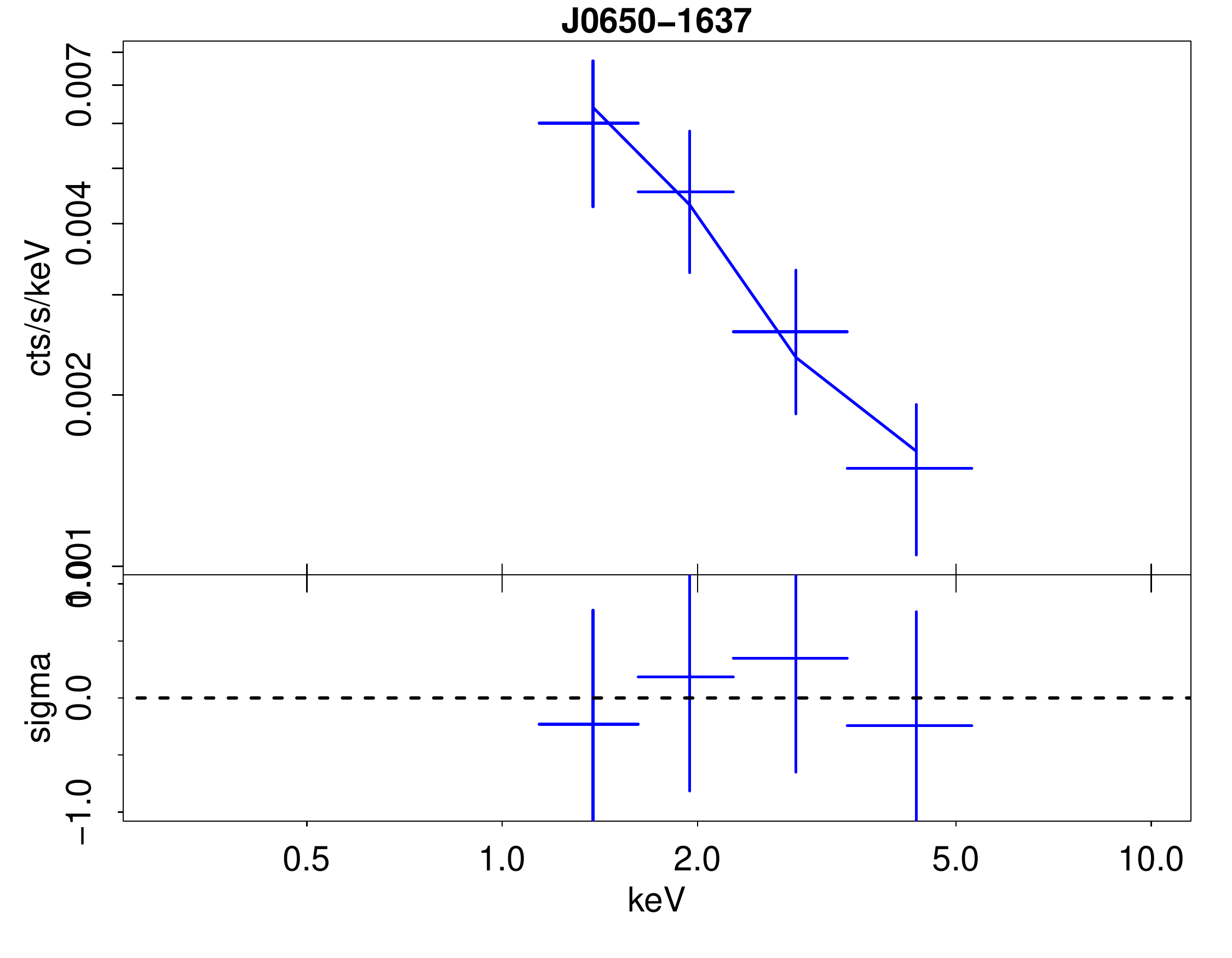}
\includegraphics[scale=0.19]{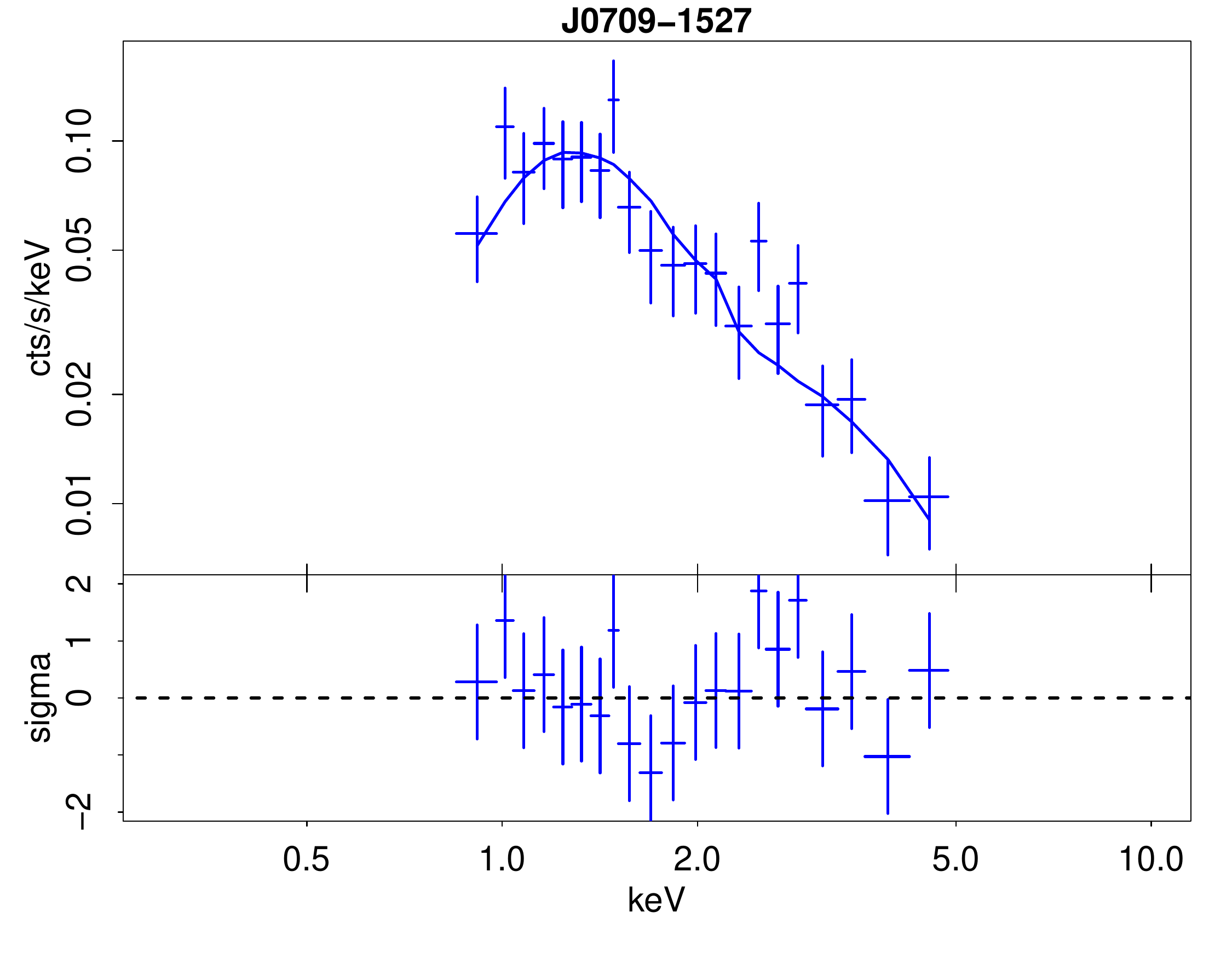}
\includegraphics[scale=0.19]{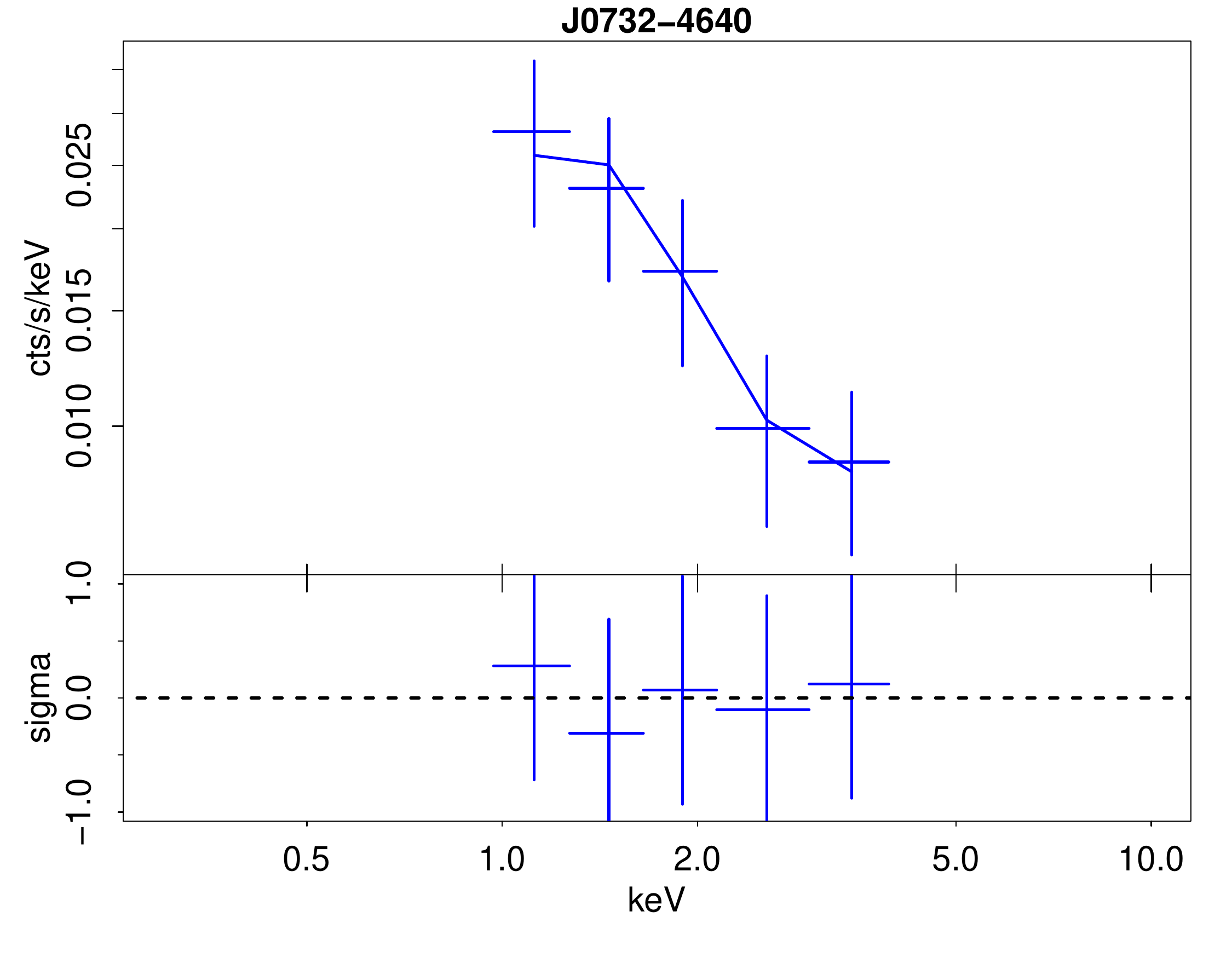}
\includegraphics[scale=0.19]{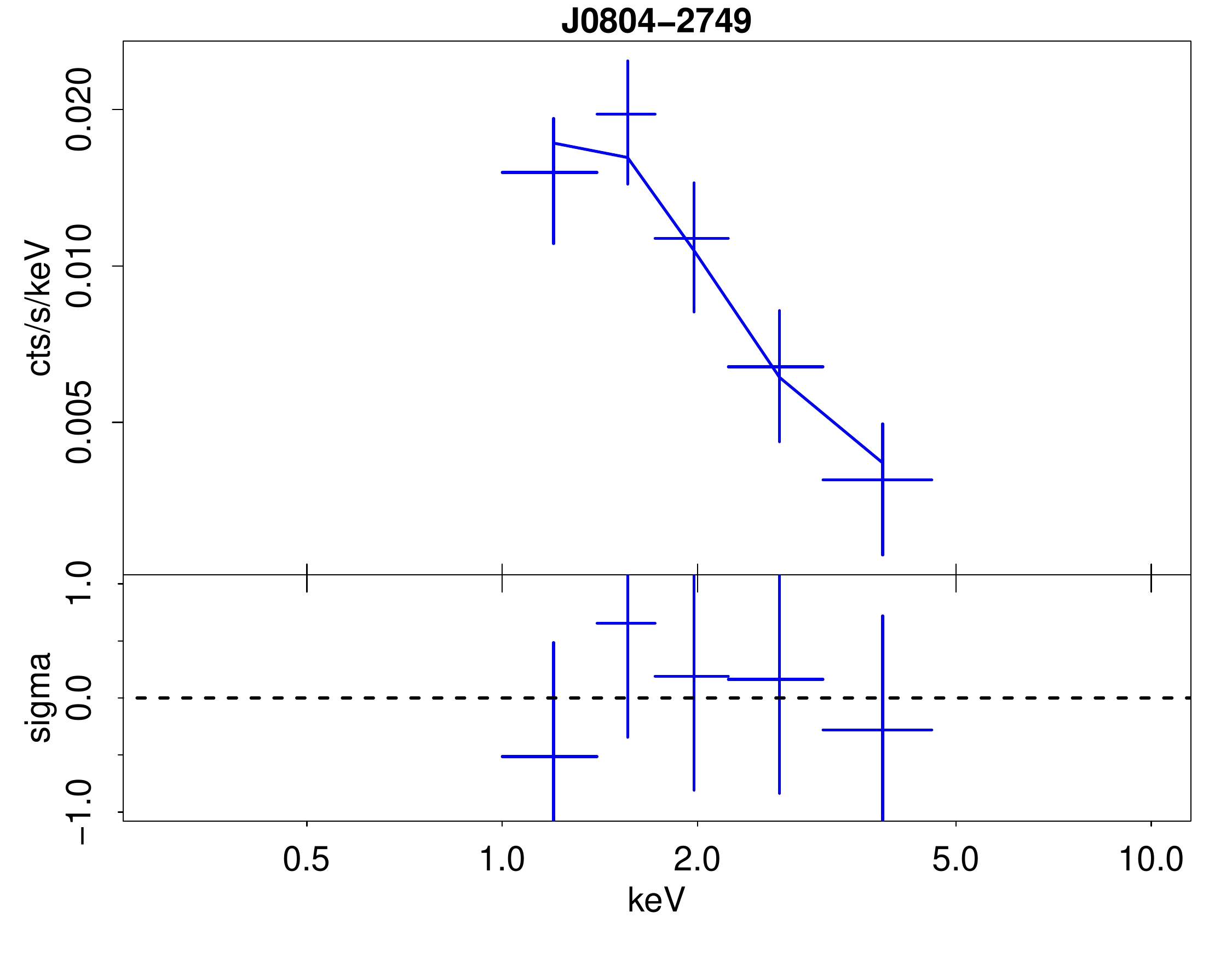}
\includegraphics[scale=0.19]{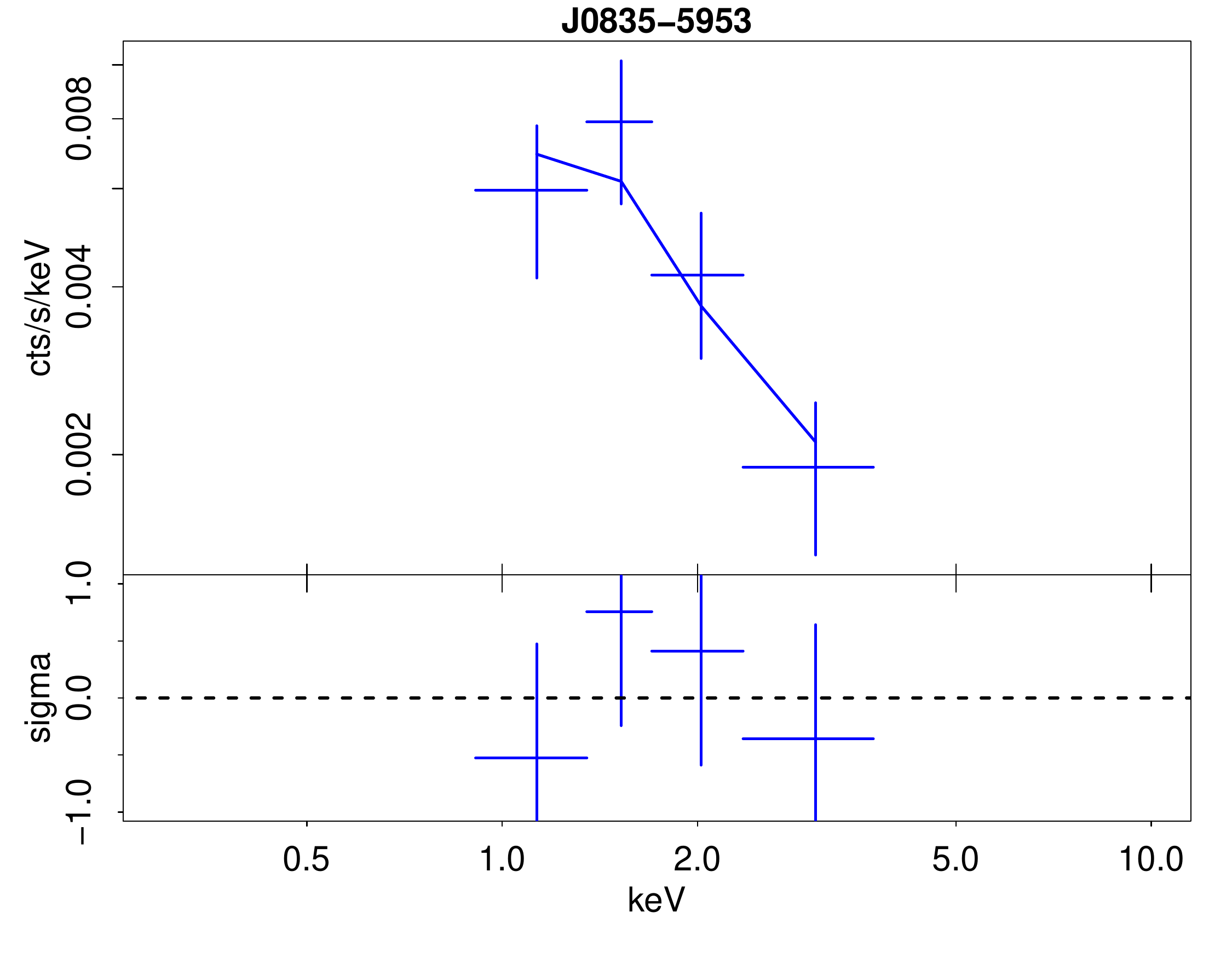}
\includegraphics[scale=0.19]{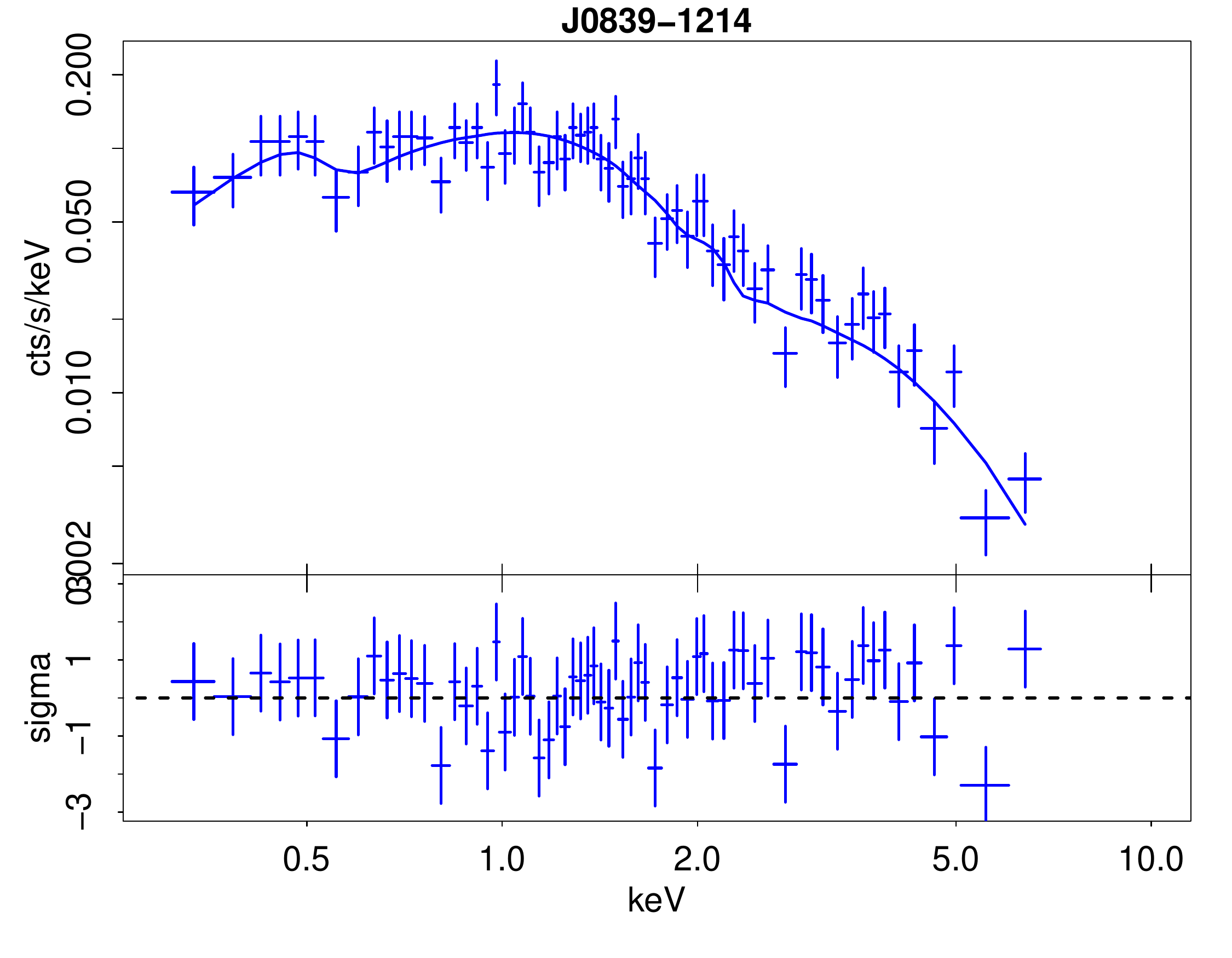}
\includegraphics[scale=0.19]{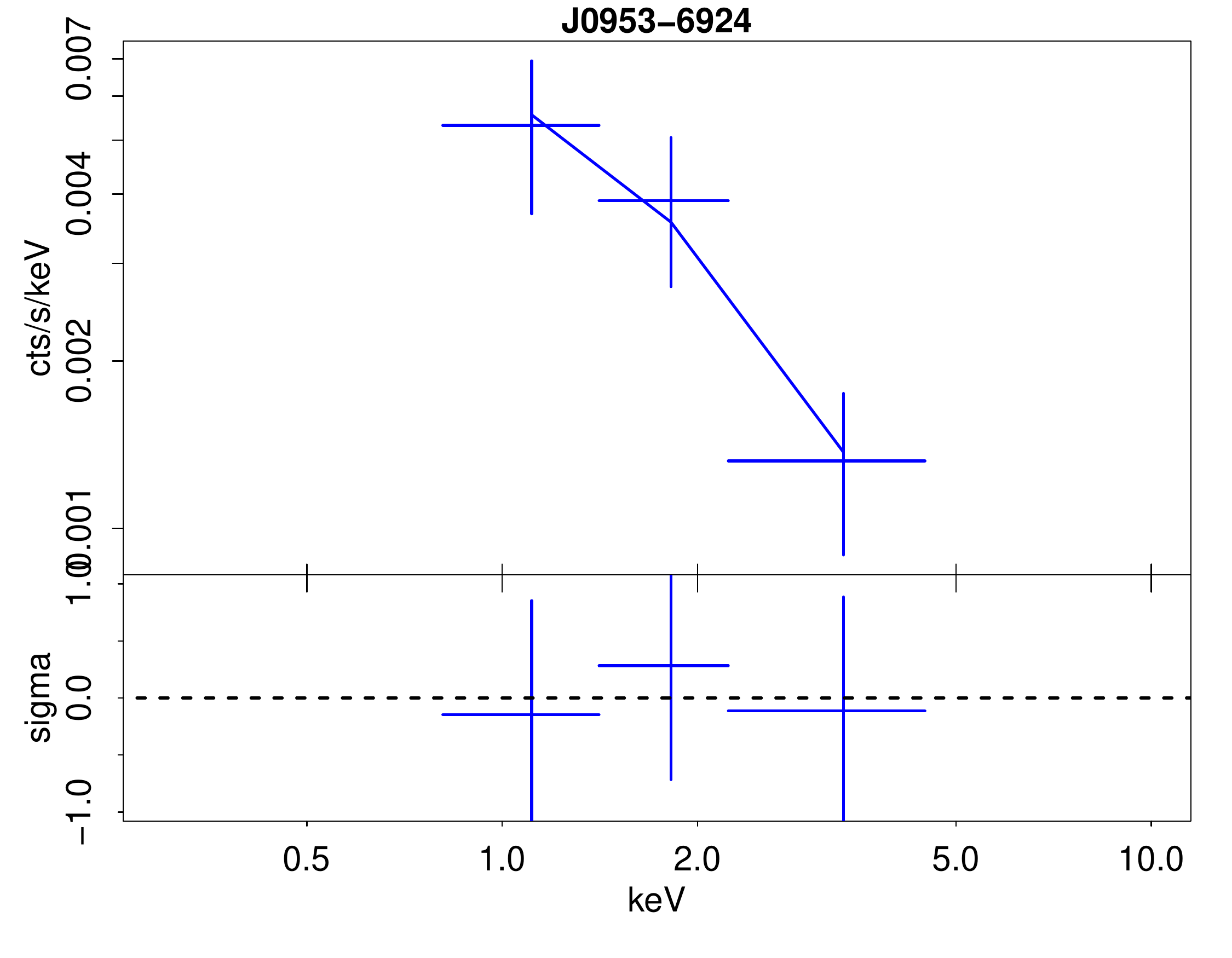}
\includegraphics[scale=0.19]{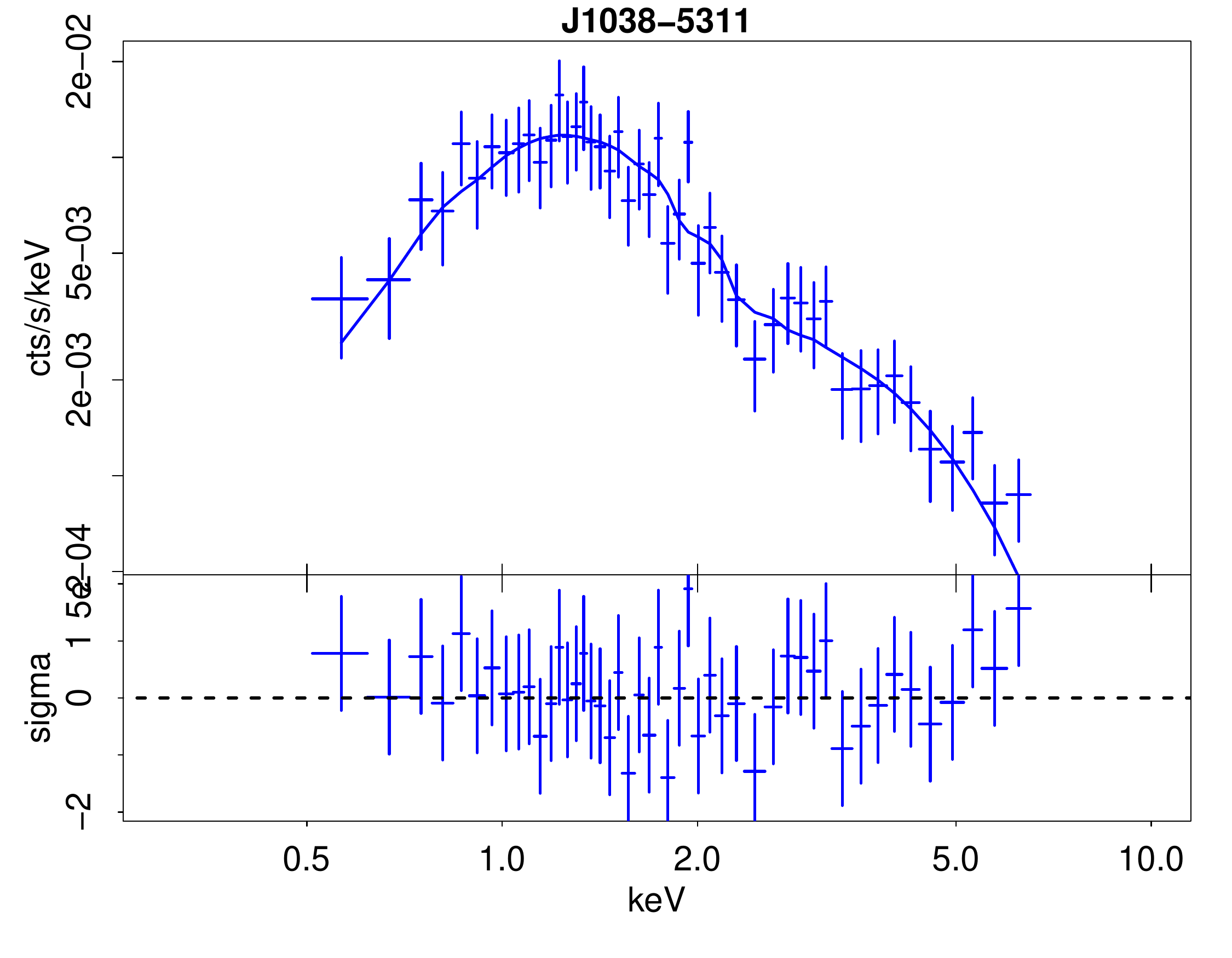}
\caption{\textit{Swift}-XRT spectra with their best-fit power law models (upper panels) and residuals (lower panels). Full set of figures available online.}\label{fig:xrt_spectra}
\end{figure*}

\begin{figure*}
   \centering
\includegraphics[scale=0.19]{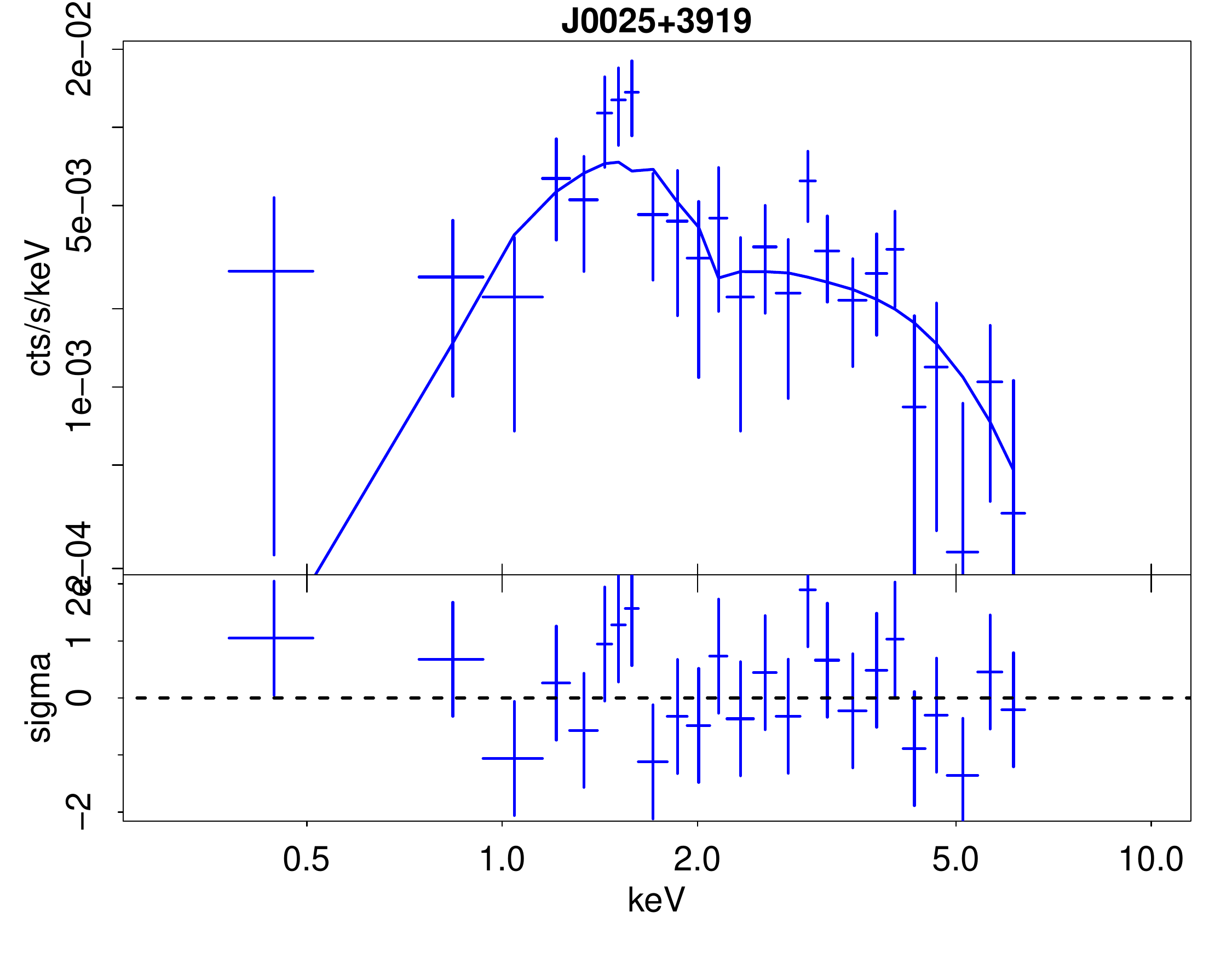}
\includegraphics[scale=0.19]{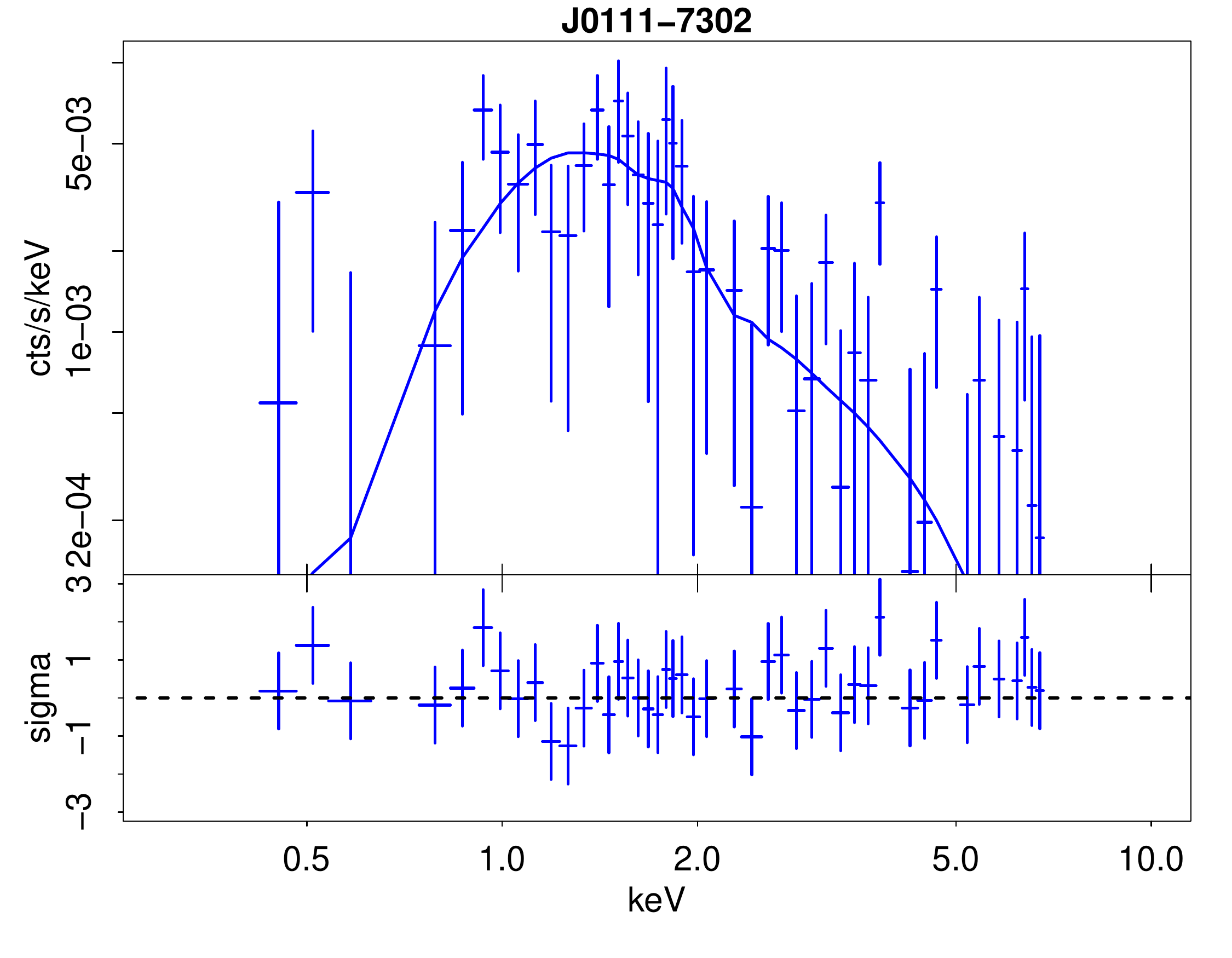}
\includegraphics[scale=0.19]{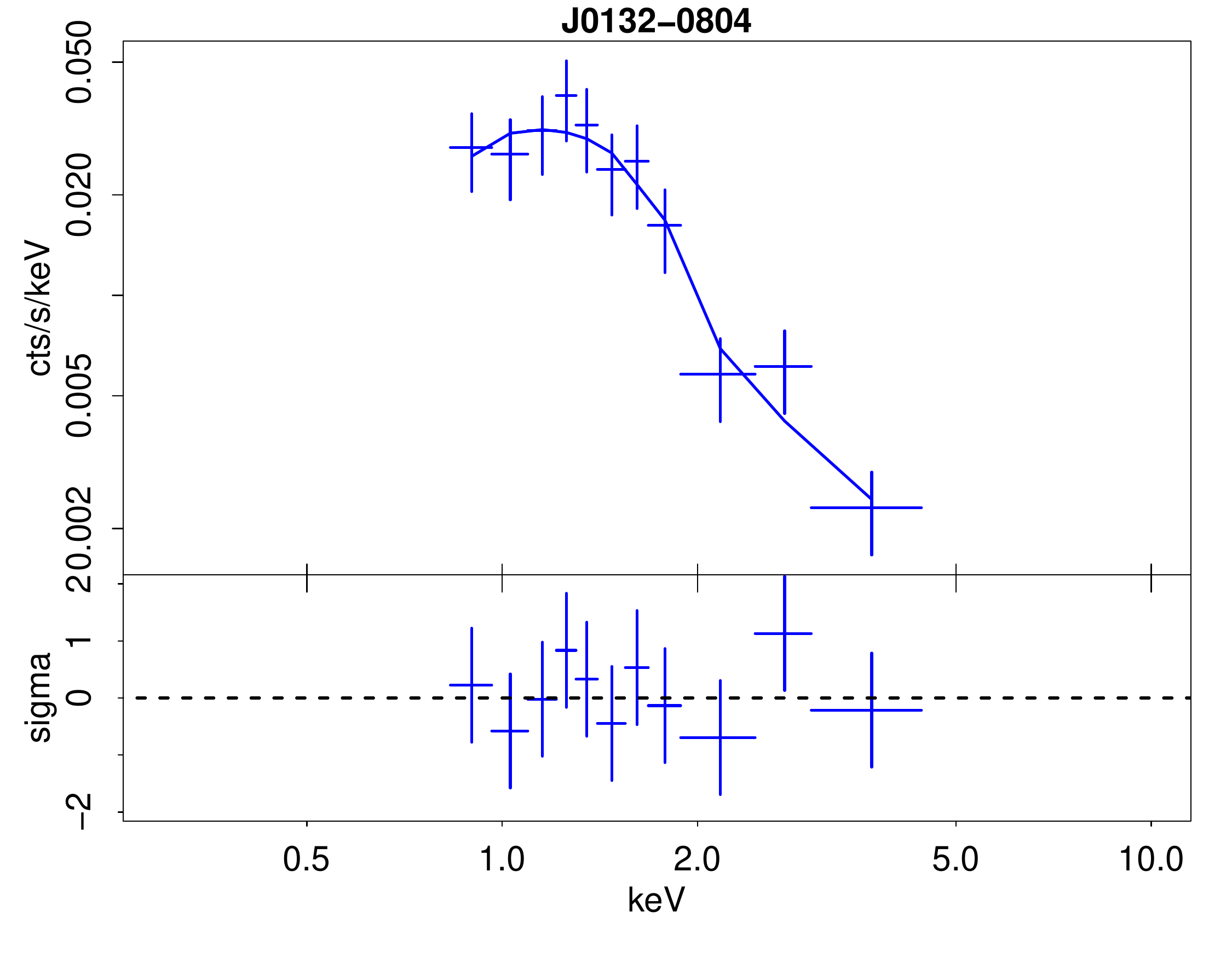}
\includegraphics[scale=0.19]{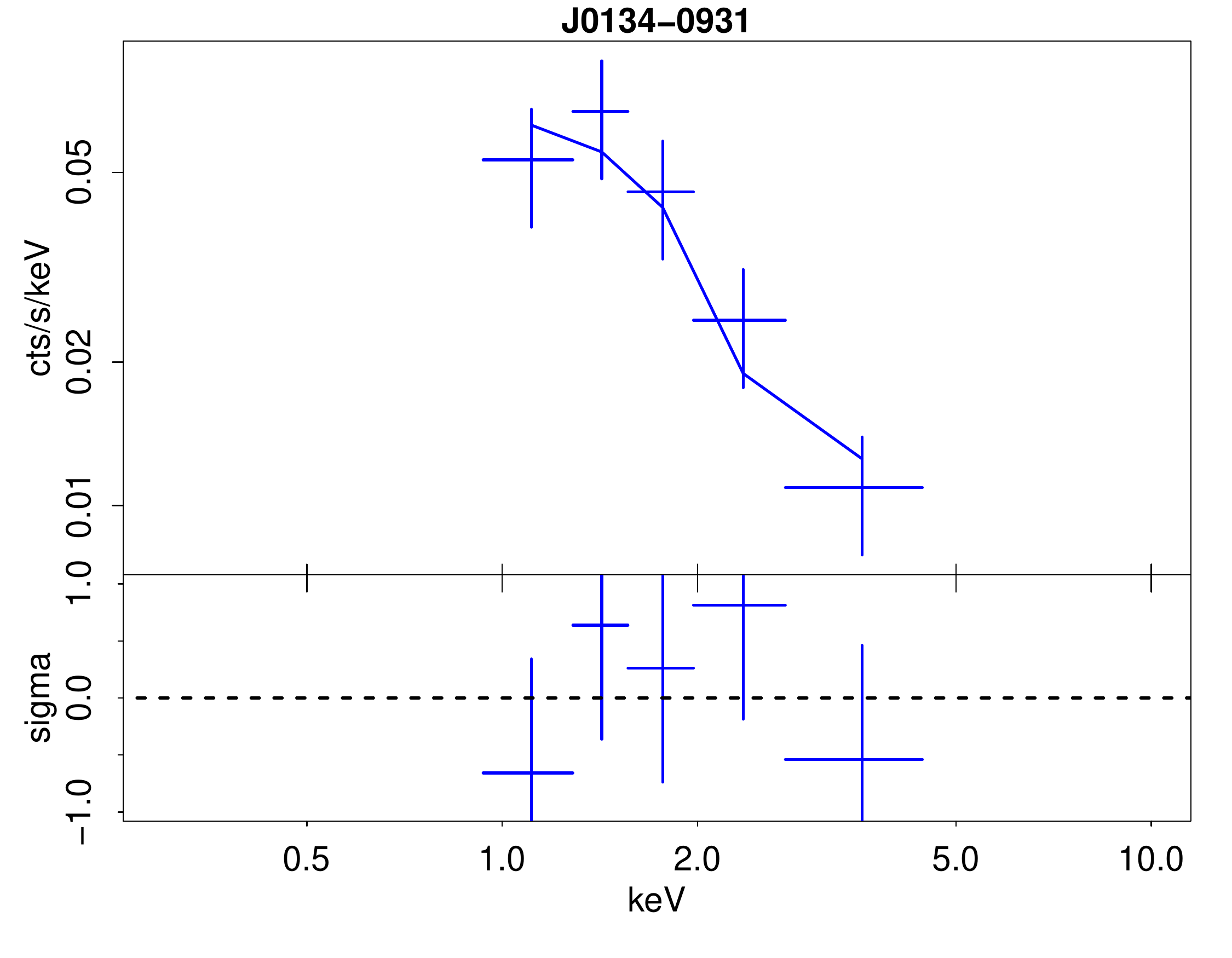}
\includegraphics[scale=0.19]{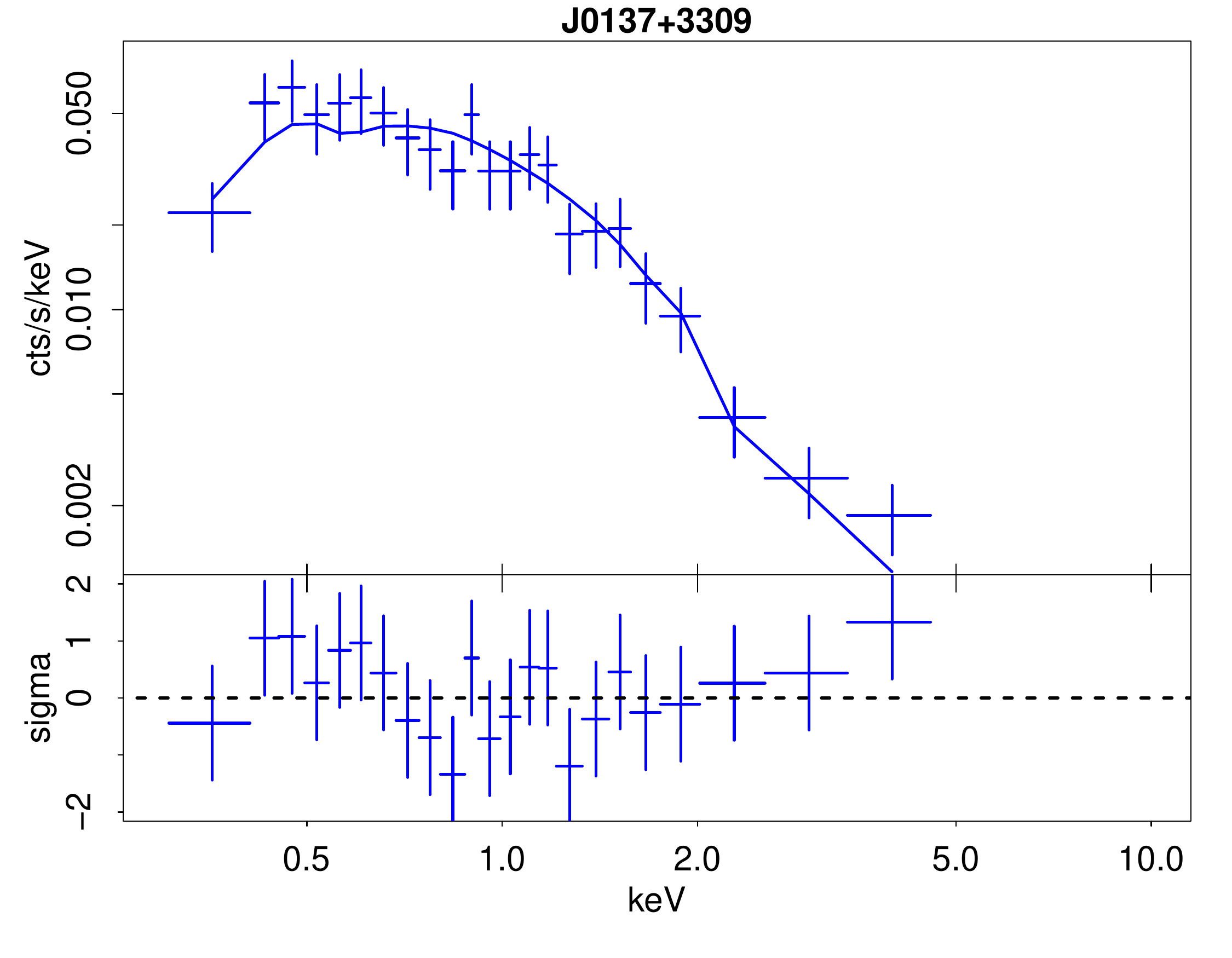}
\includegraphics[scale=0.19]{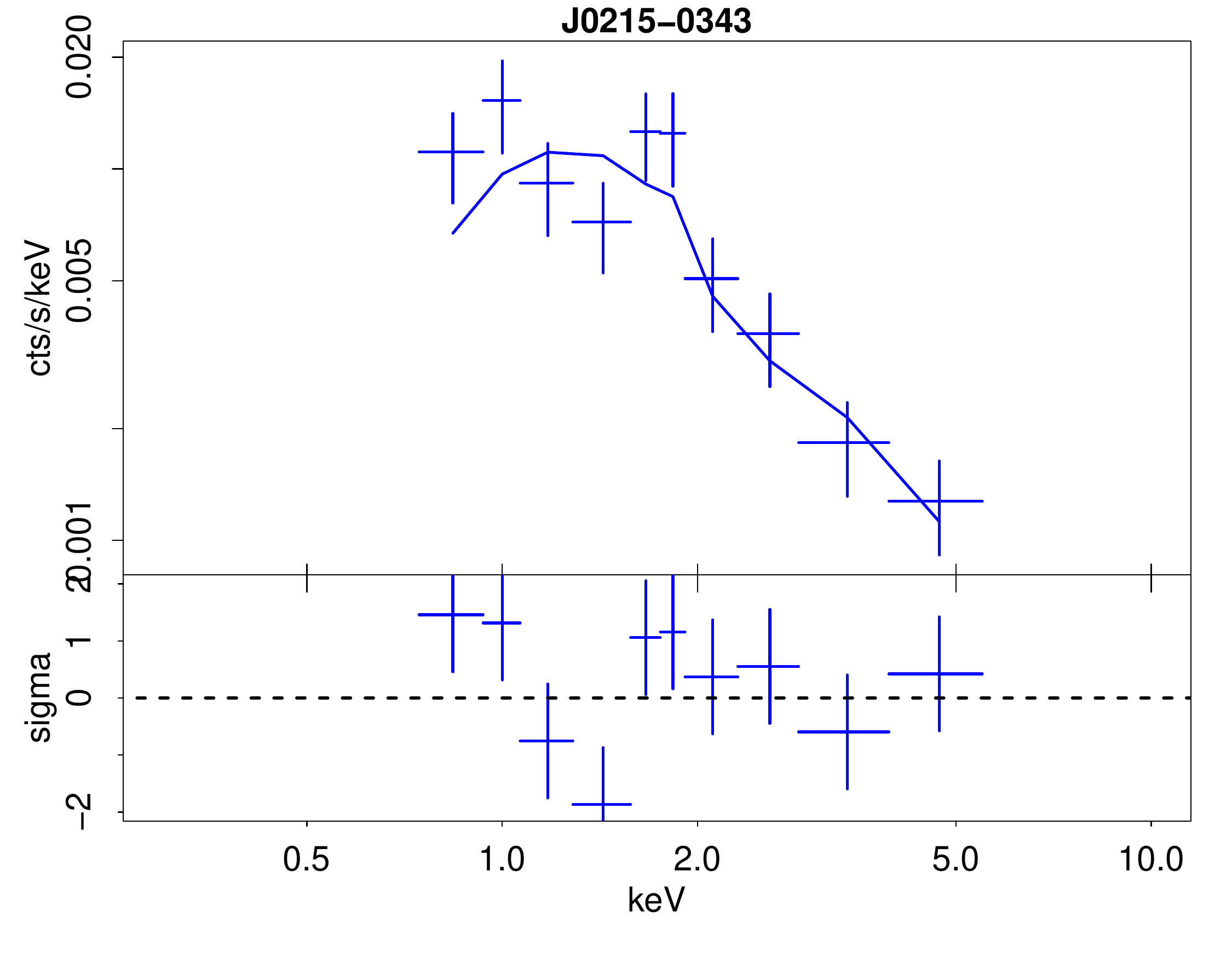}
\includegraphics[scale=0.19]{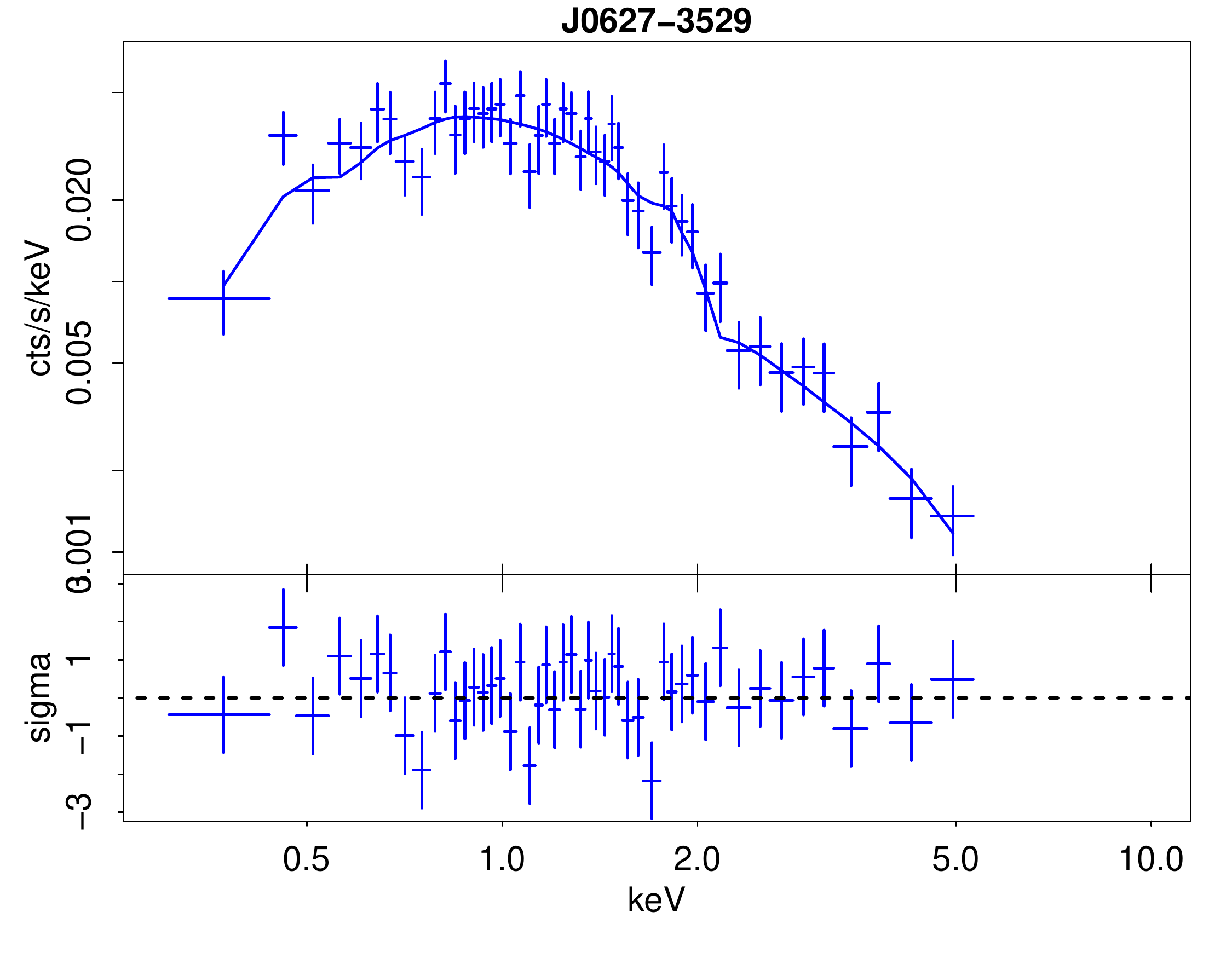}
\includegraphics[scale=0.19]{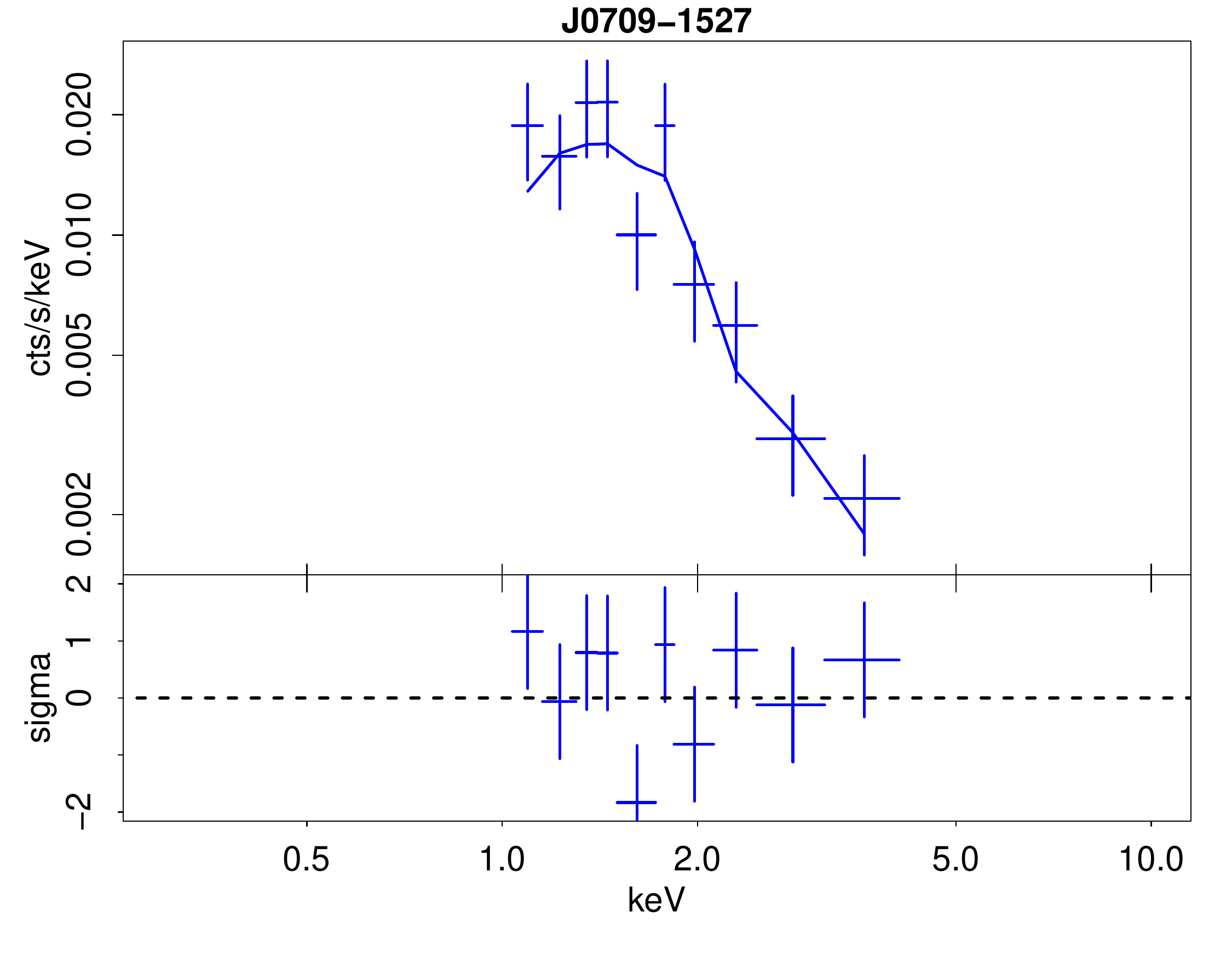}
\includegraphics[scale=0.19]{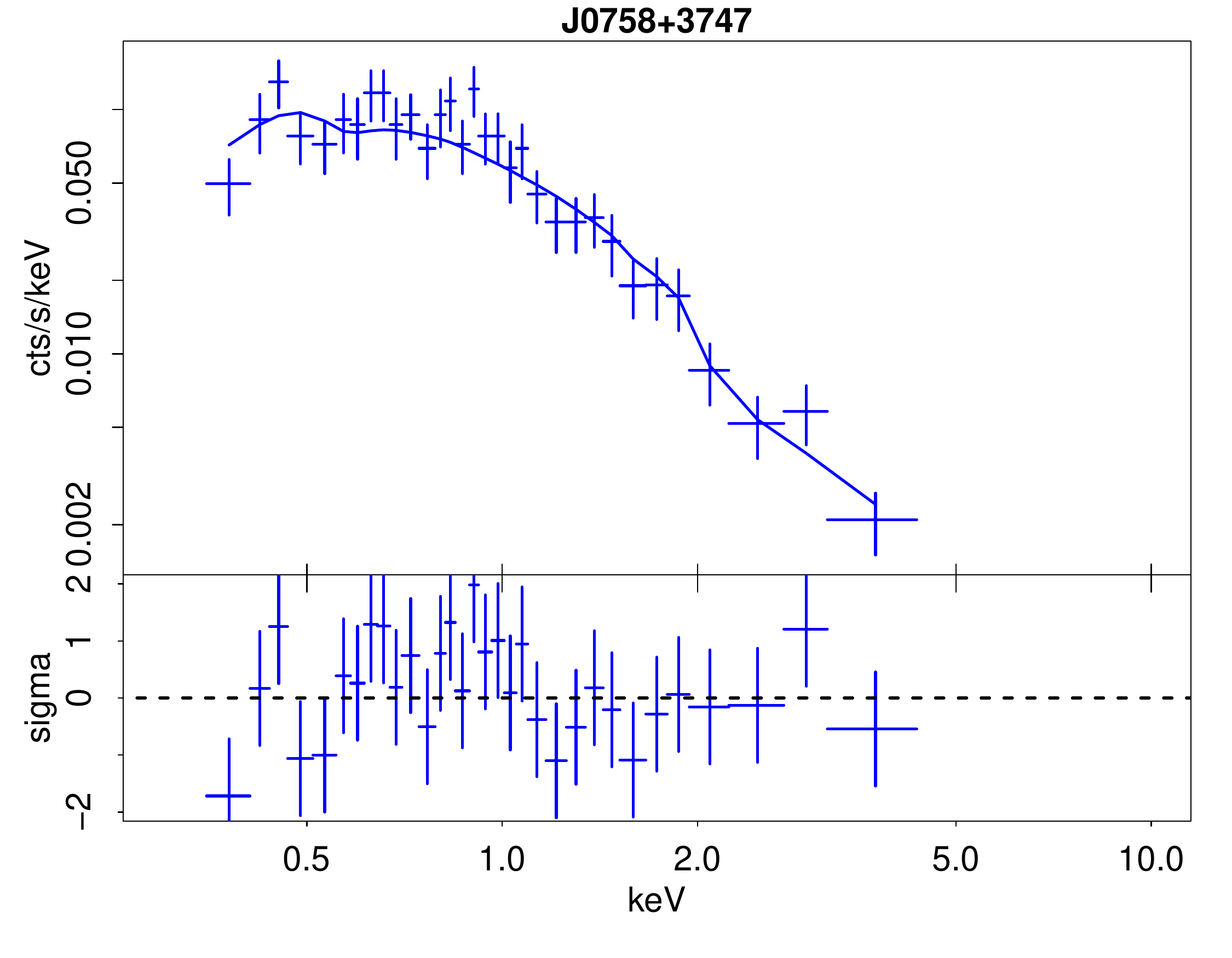}
\includegraphics[scale=0.19]{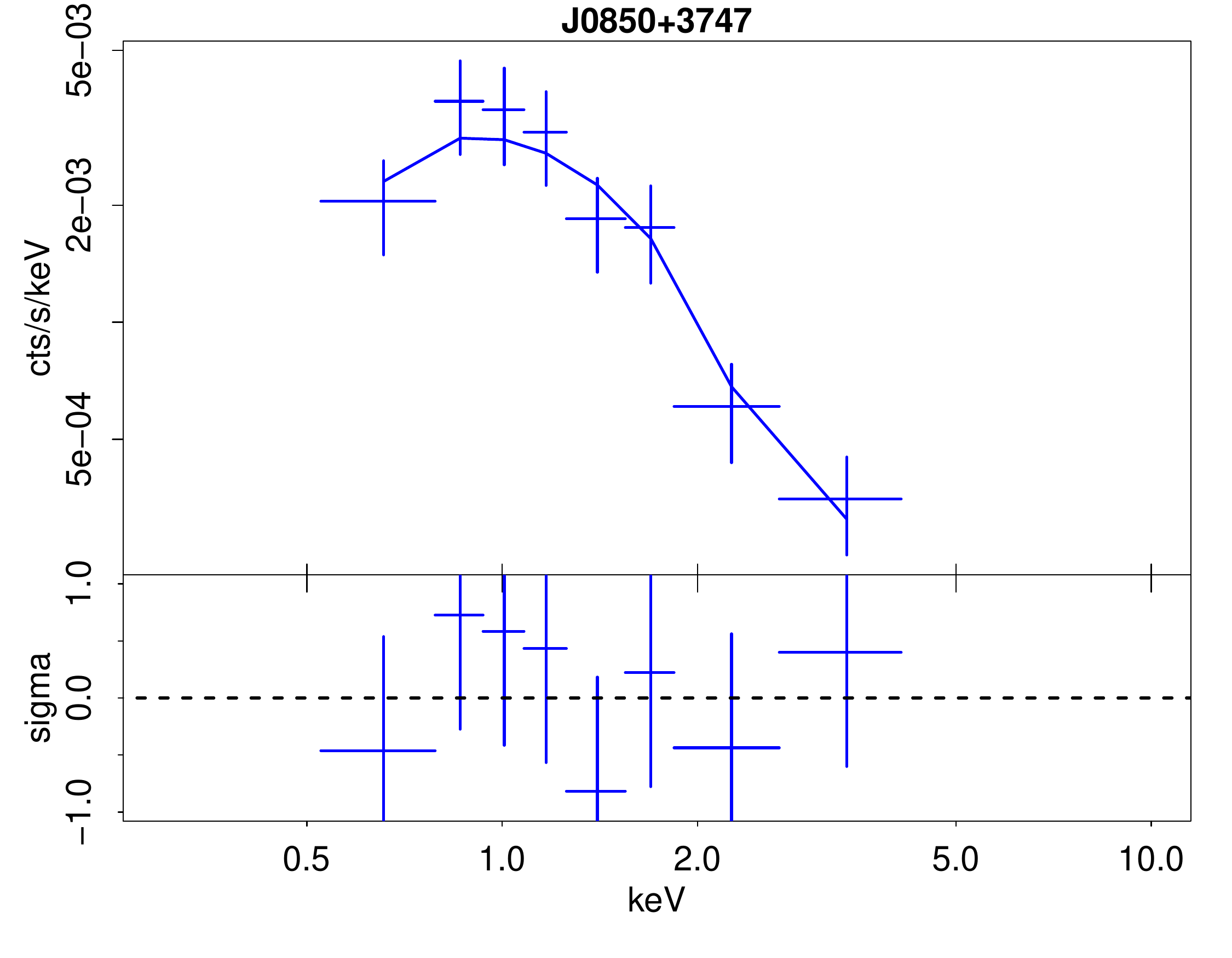}
\includegraphics[scale=0.19]{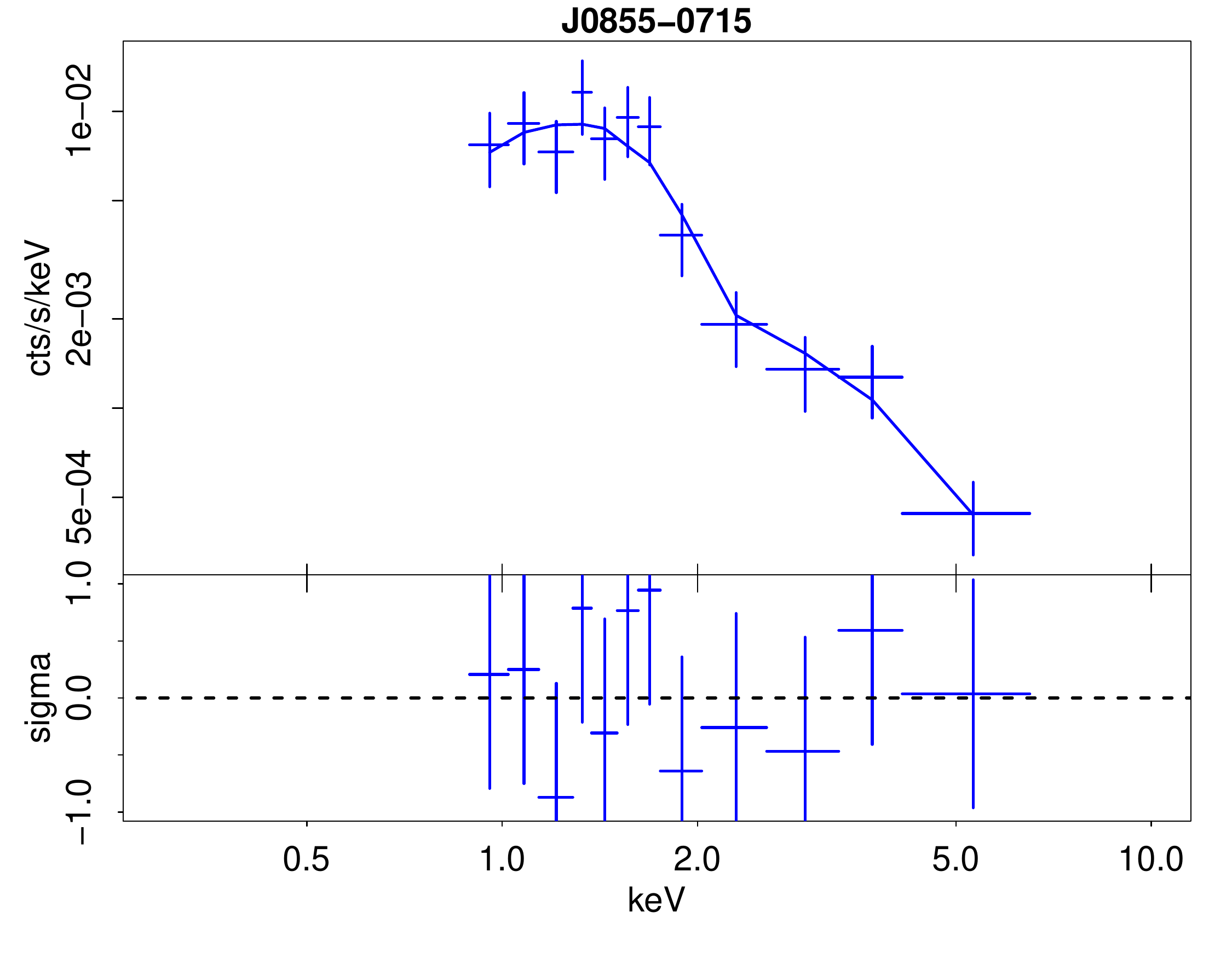}
\includegraphics[scale=0.19]{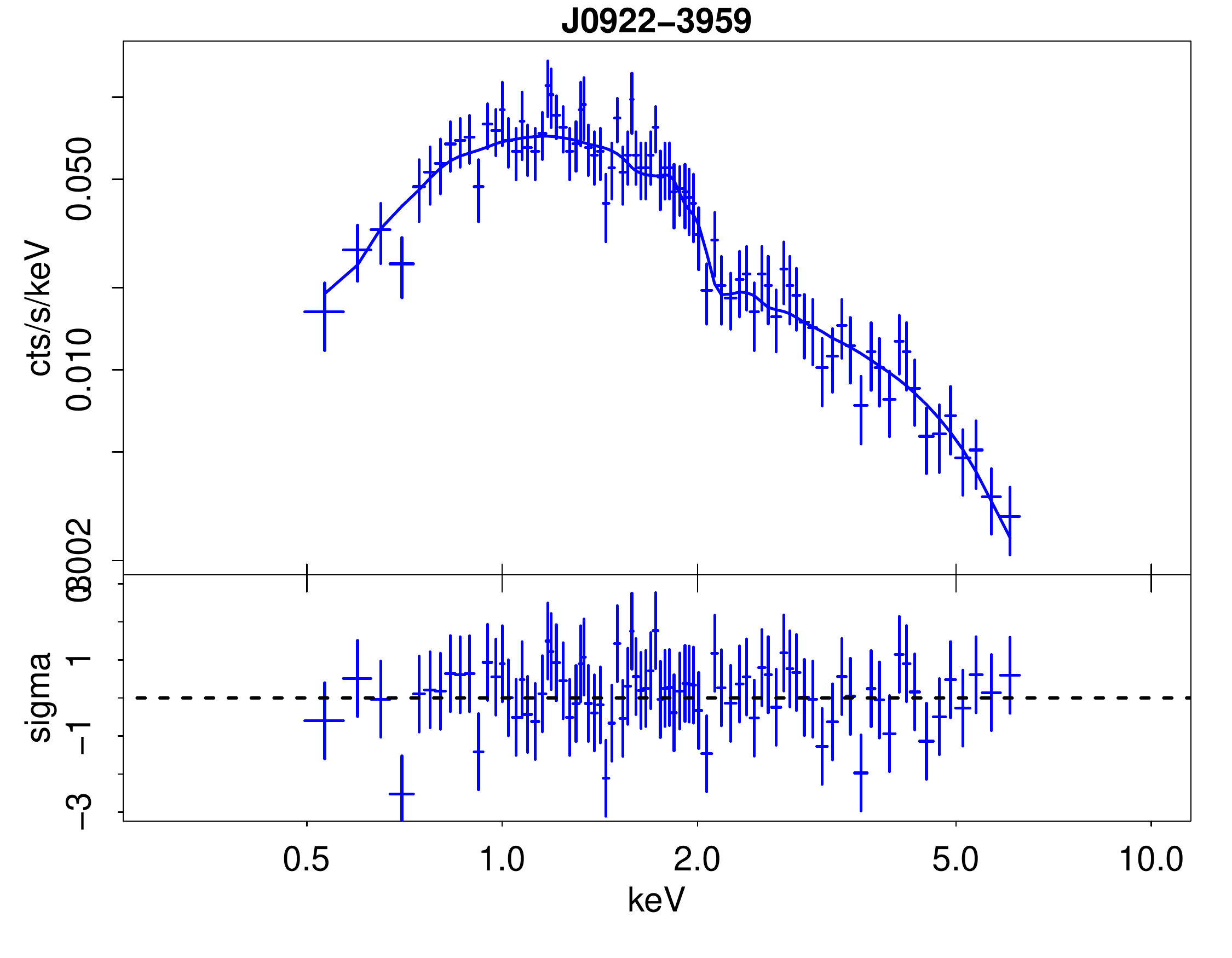}
\includegraphics[scale=0.19]{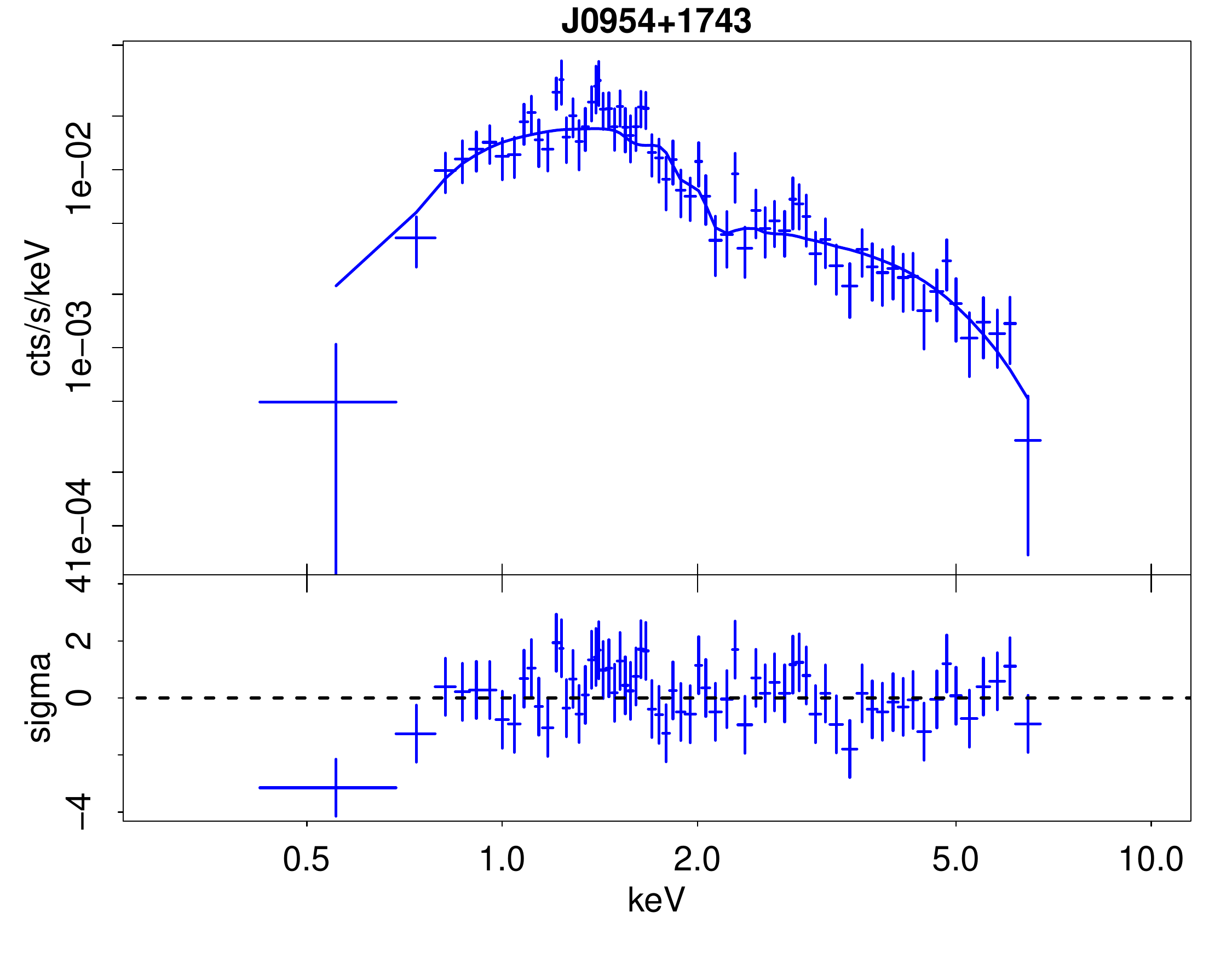}
\includegraphics[scale=0.19]{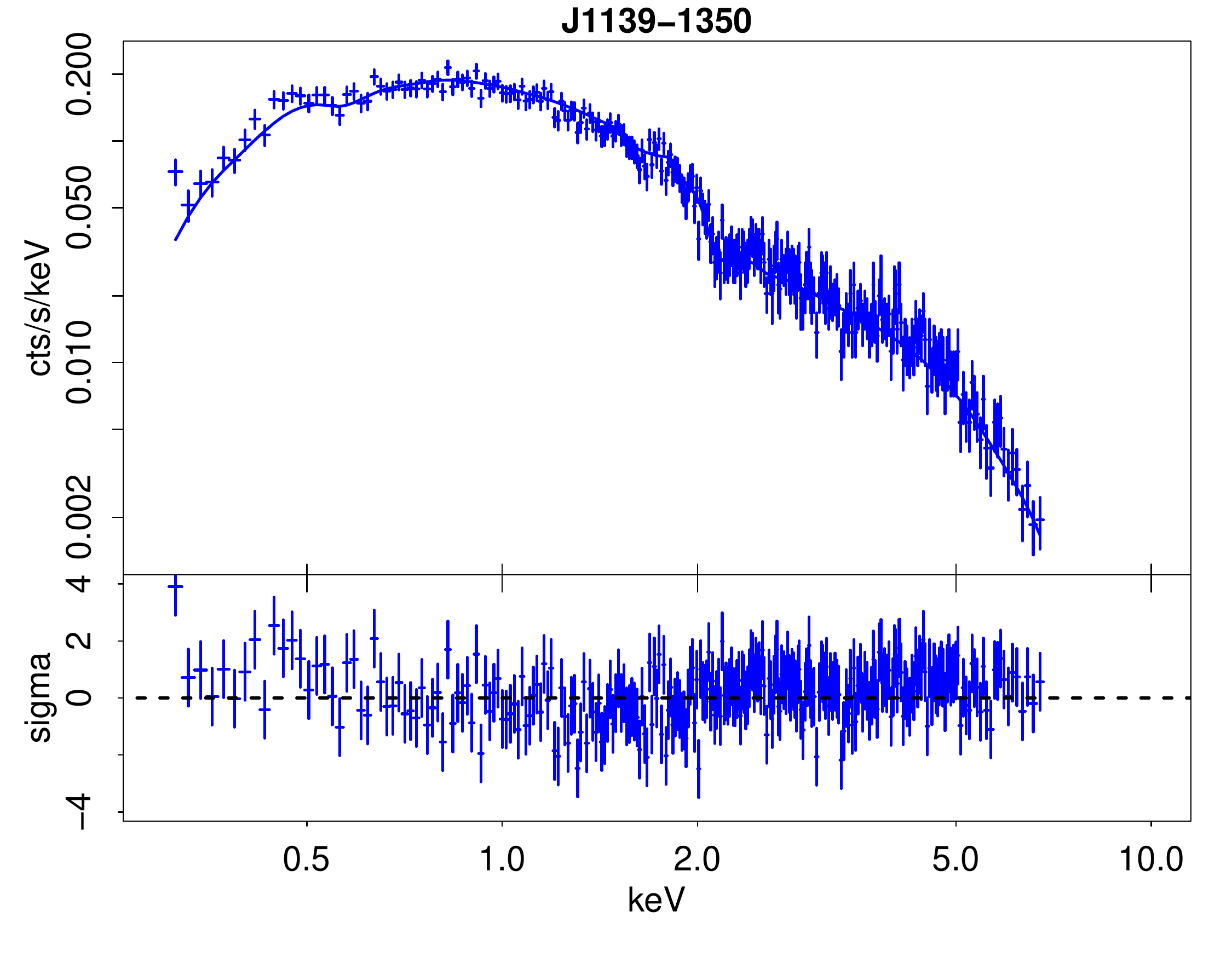}
\includegraphics[scale=0.19]{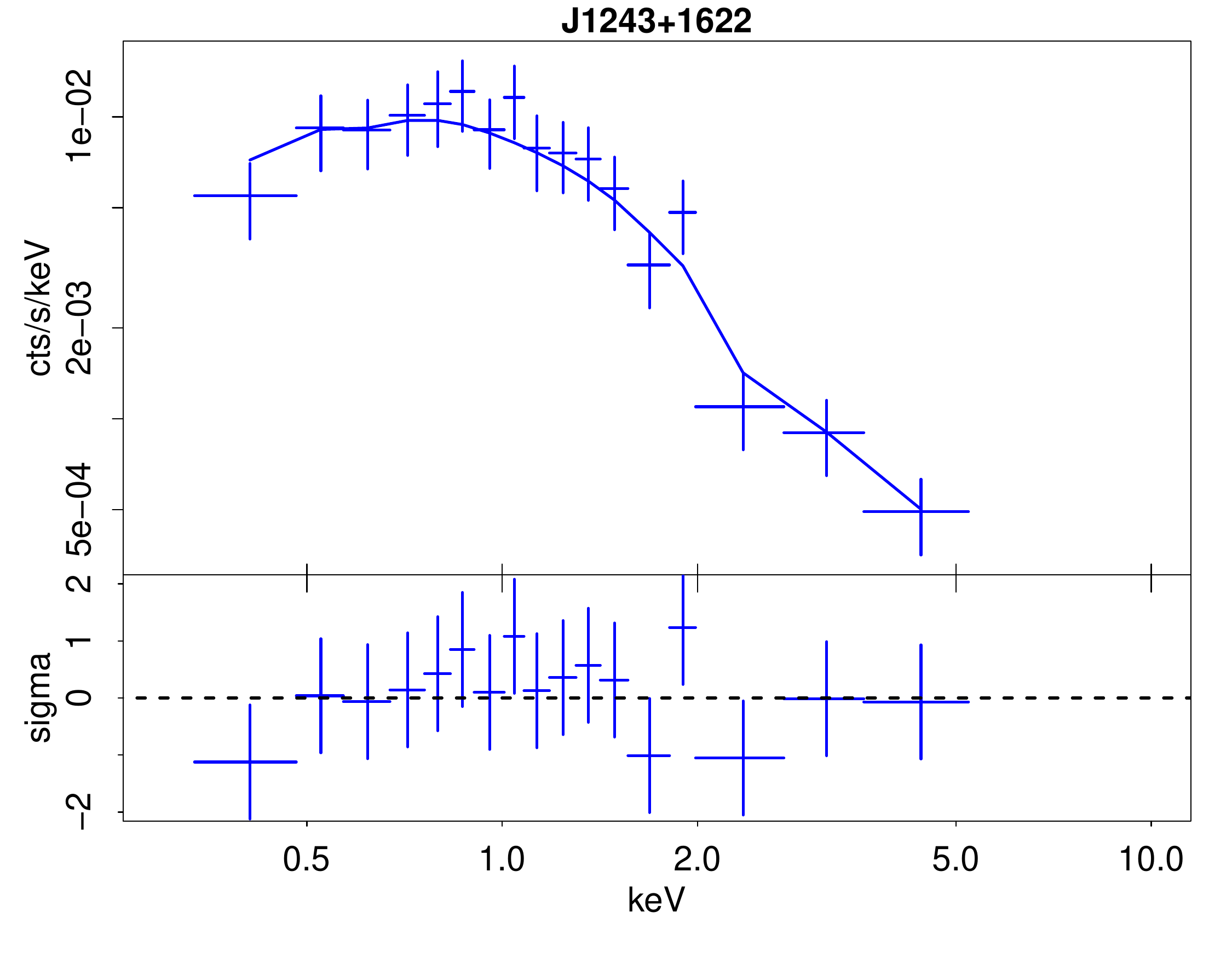}
\includegraphics[scale=0.19]{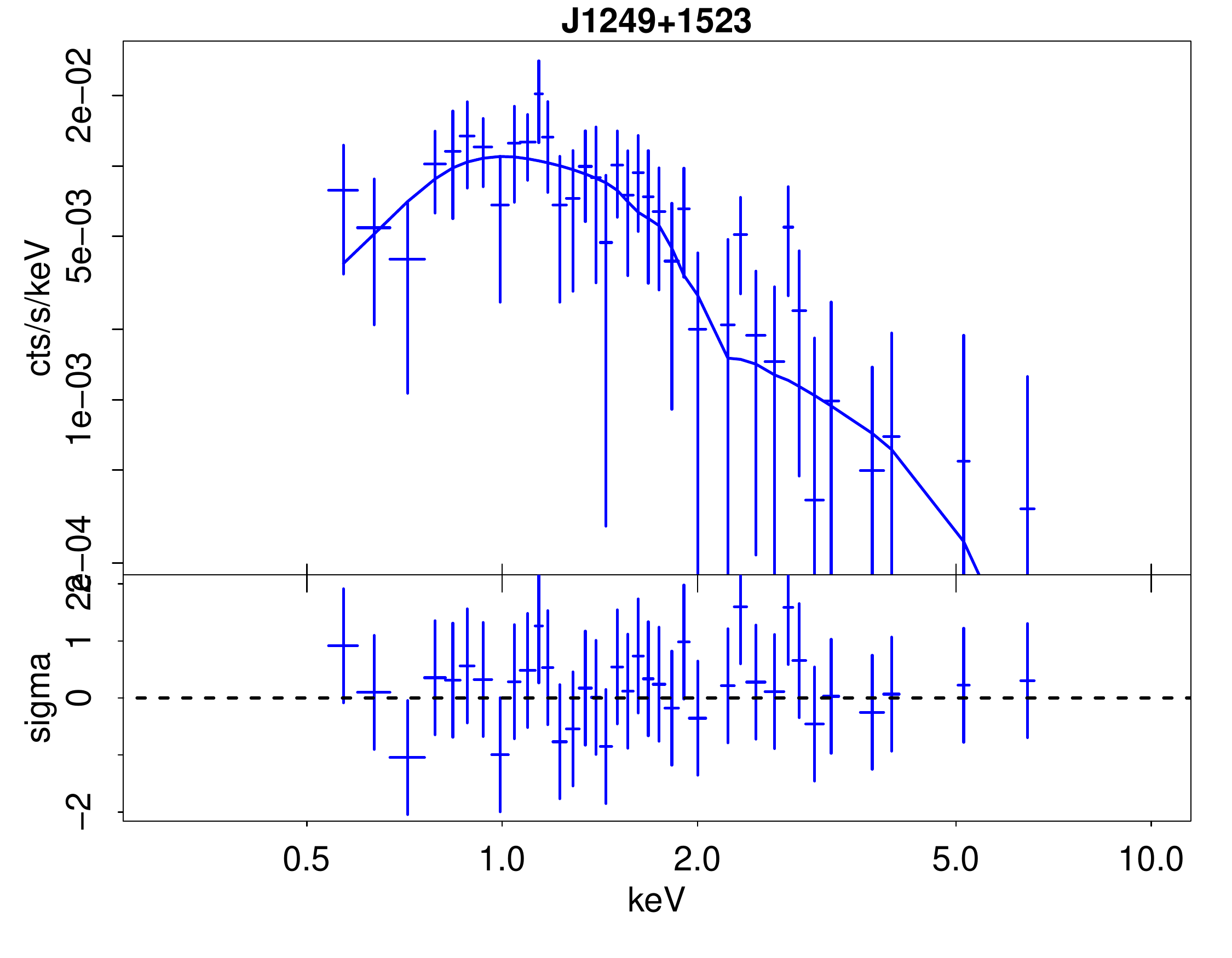}
\includegraphics[scale=0.19]{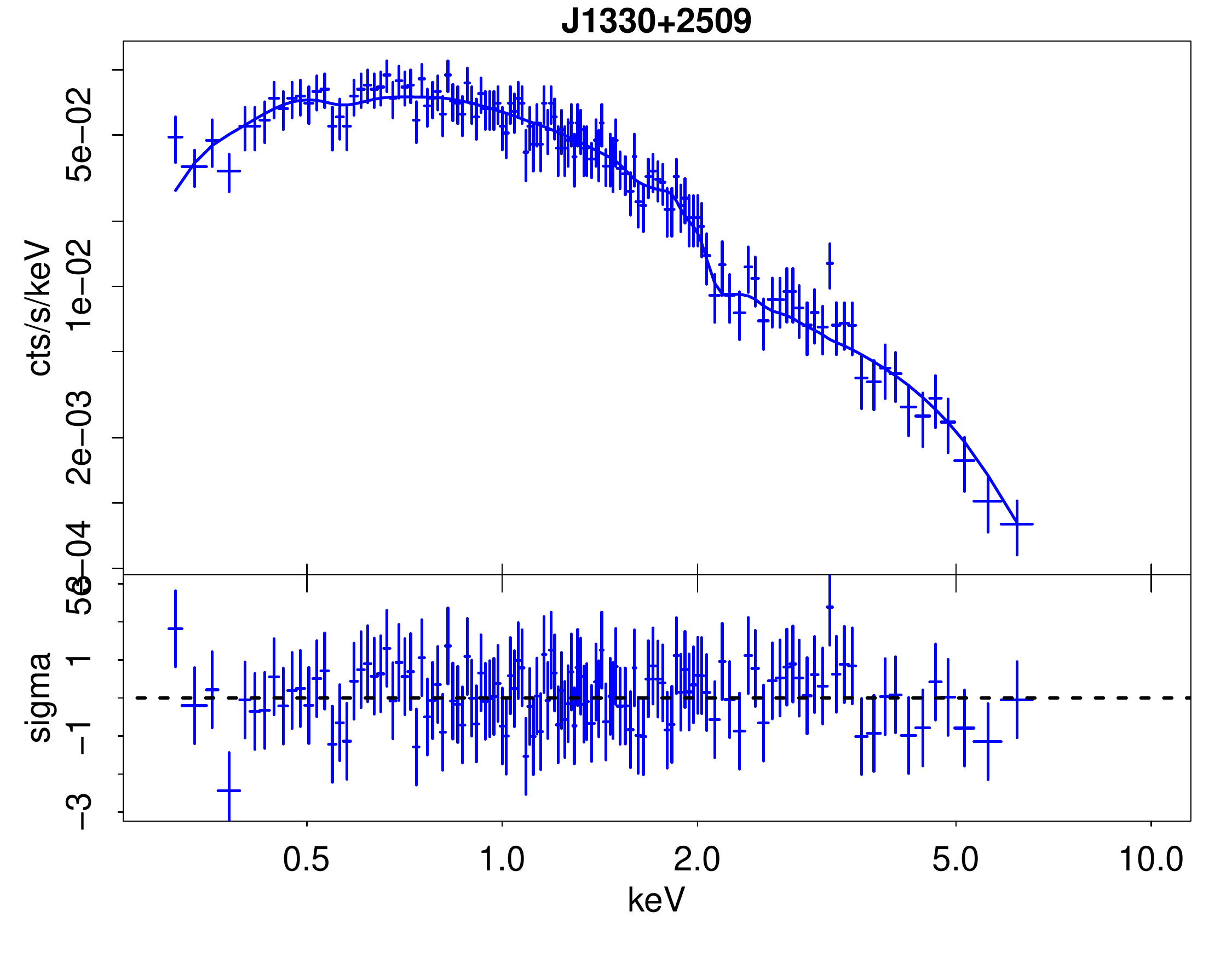}
\includegraphics[scale=0.19]{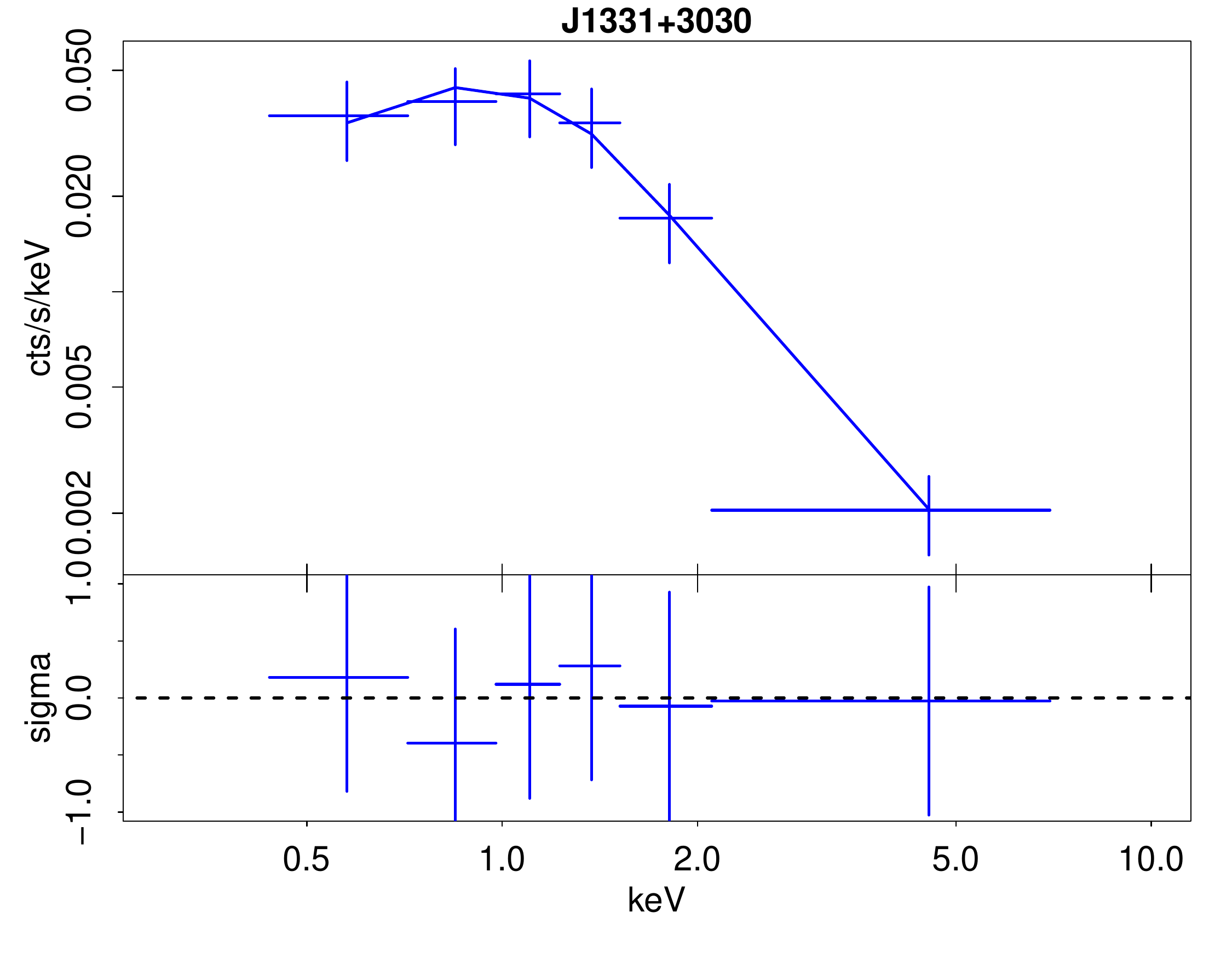}
\includegraphics[scale=0.19]{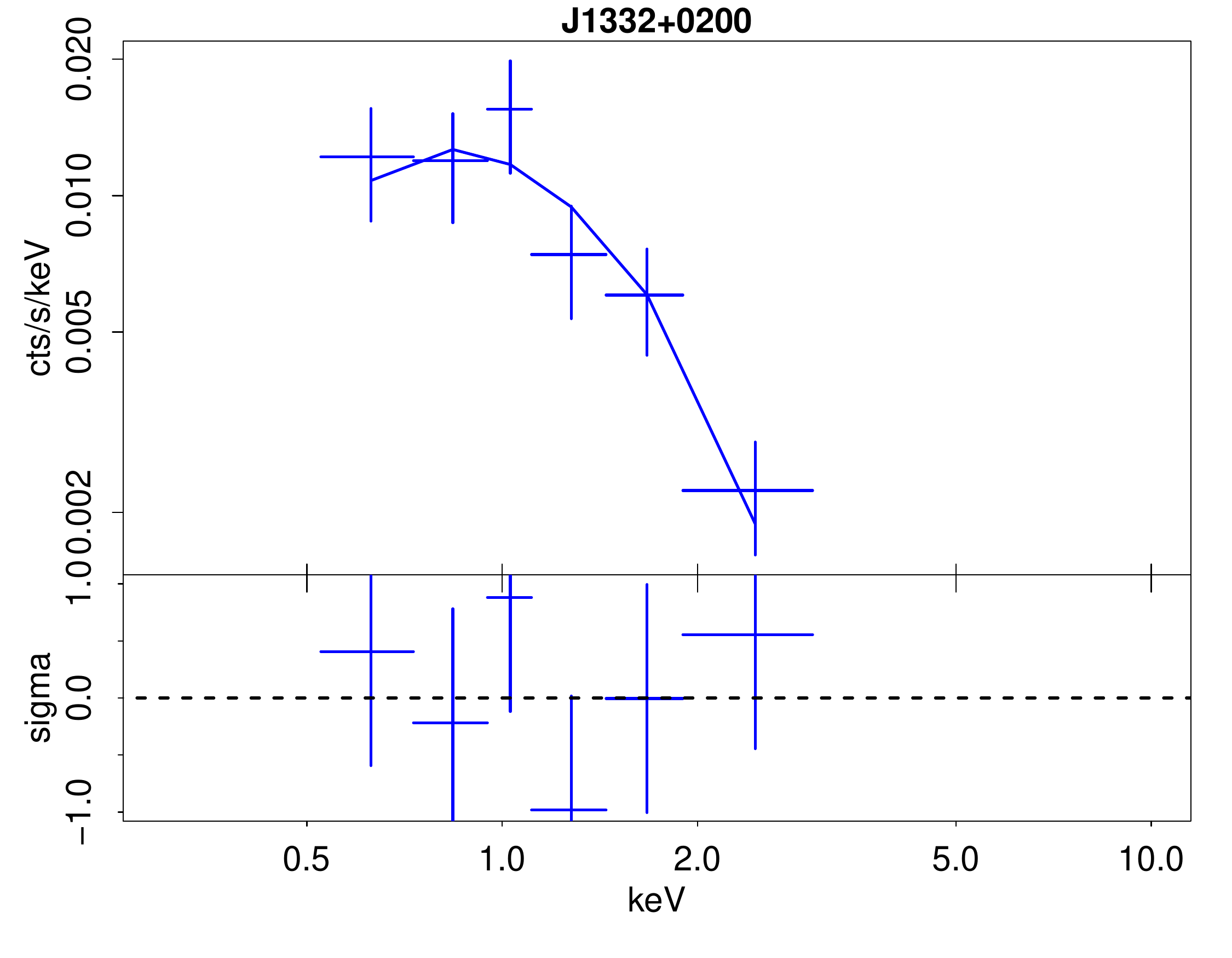}
\includegraphics[scale=0.19]{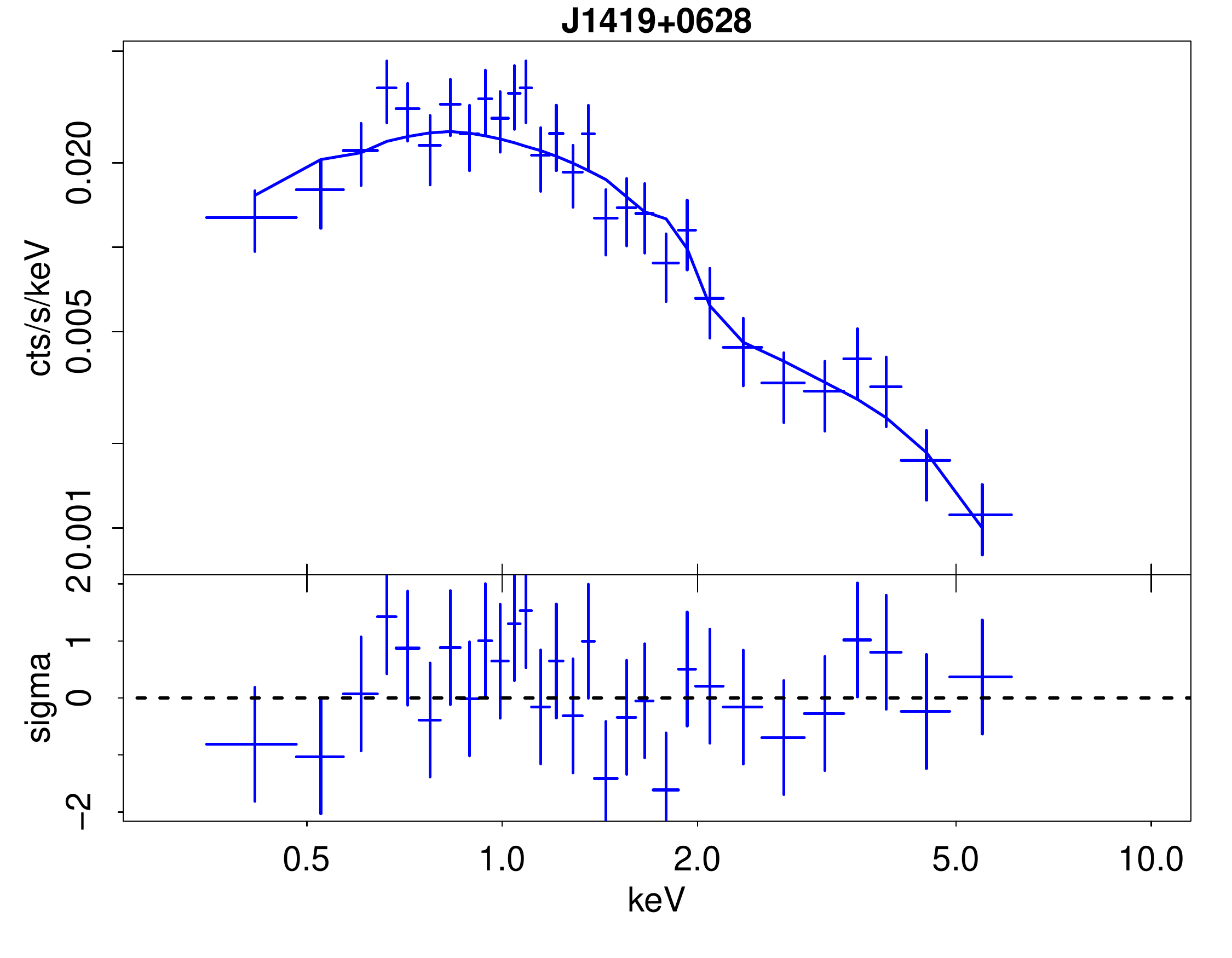}
\includegraphics[scale=0.19]{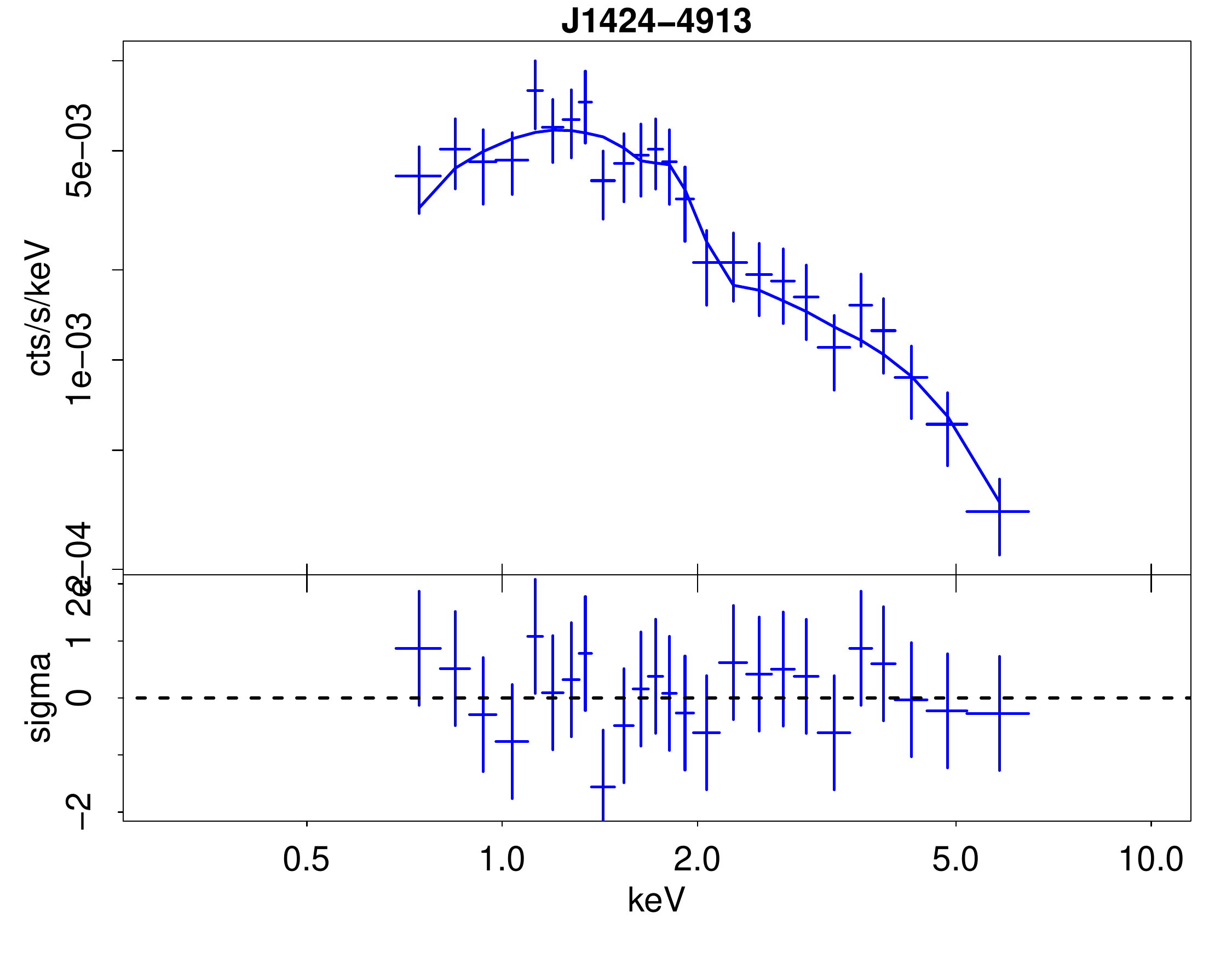}
\caption{\textit{Chandra}-ACIS spectra with their best-fit power law models (upper panels) and residuals (lower panels). Full set of figures available online.}\label{fig:acis_spectra}
\end{figure*}

\begin{figure*}
   \centering
\includegraphics[scale=0.19]{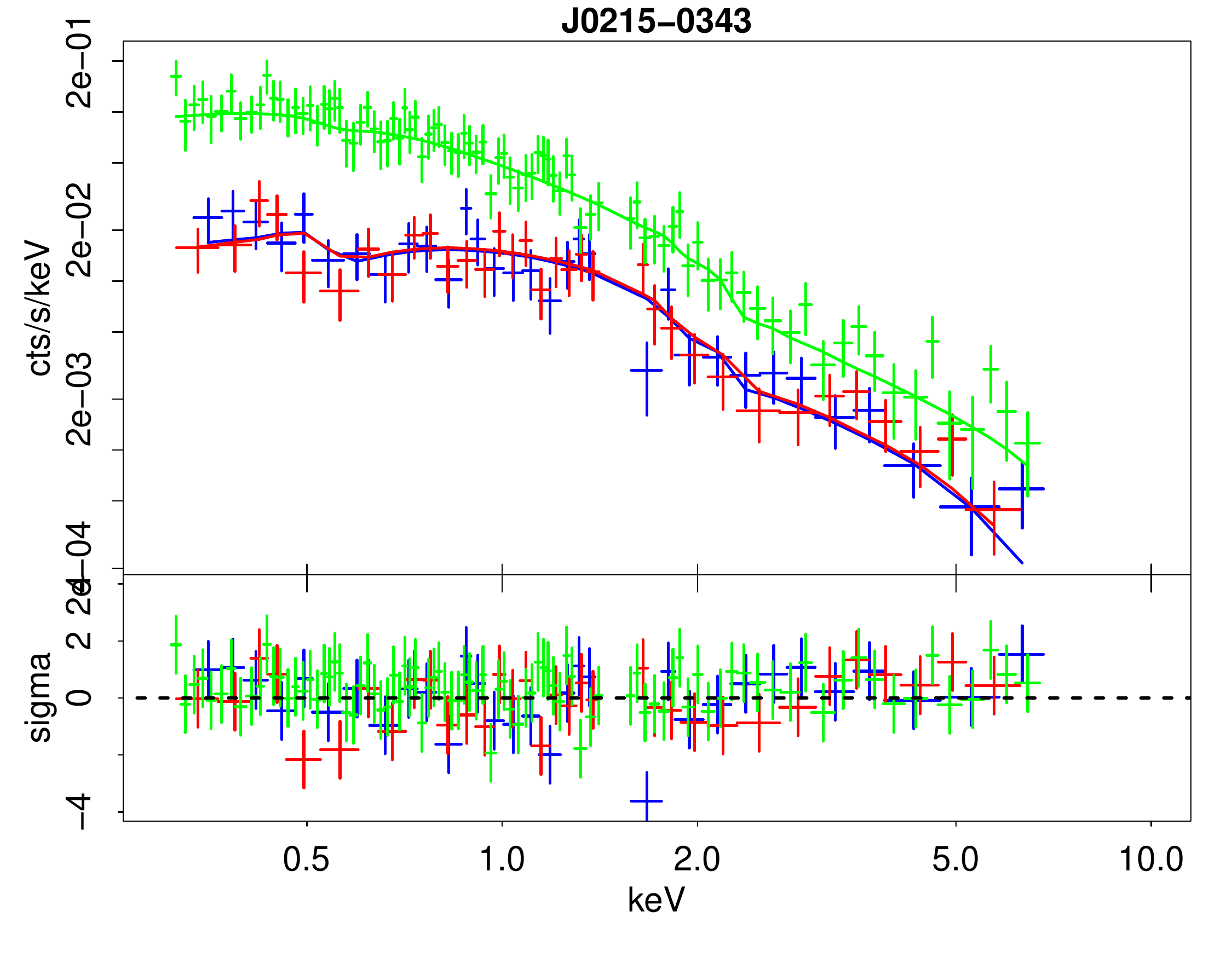}
\includegraphics[scale=0.19]{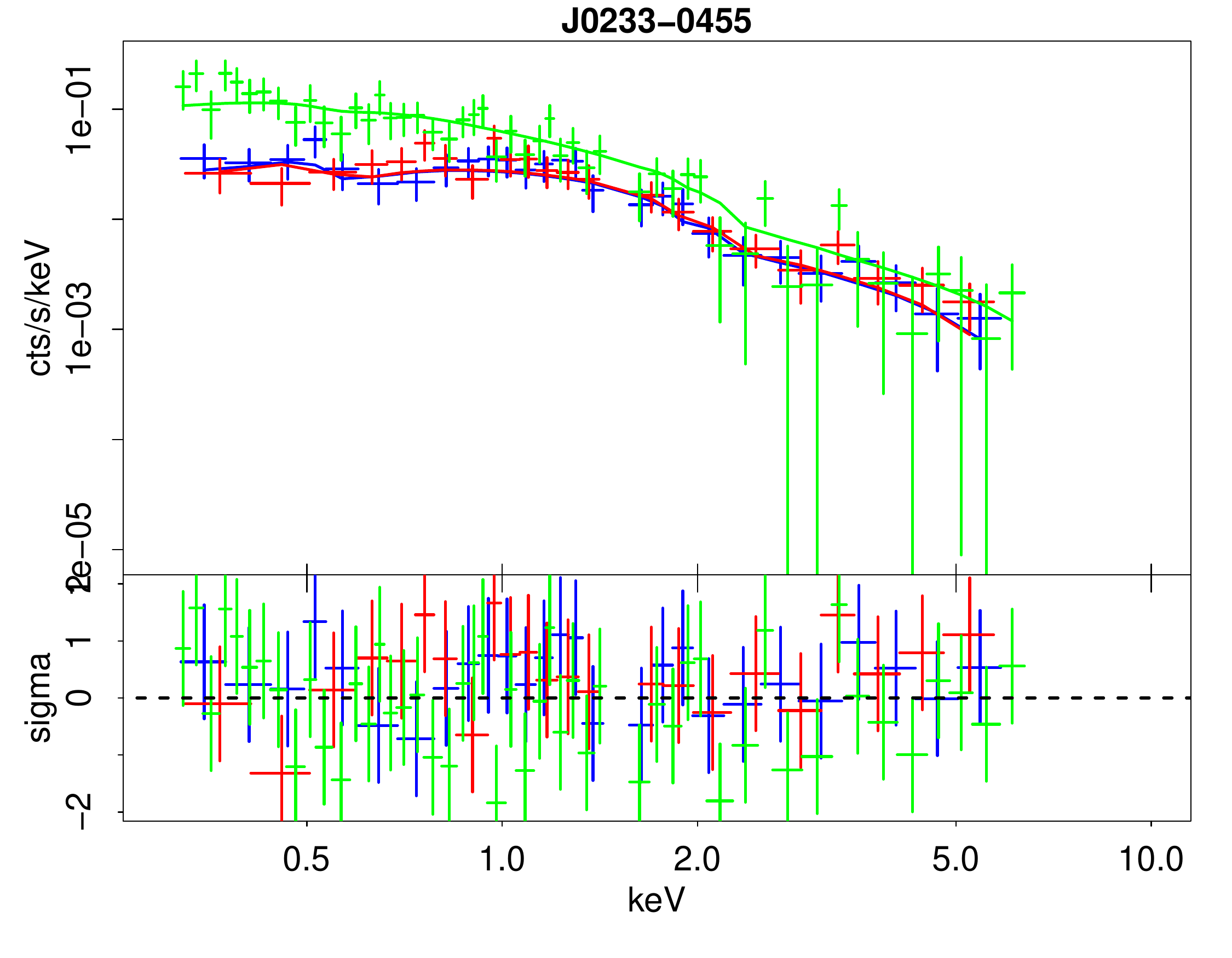}
\includegraphics[scale=0.19]{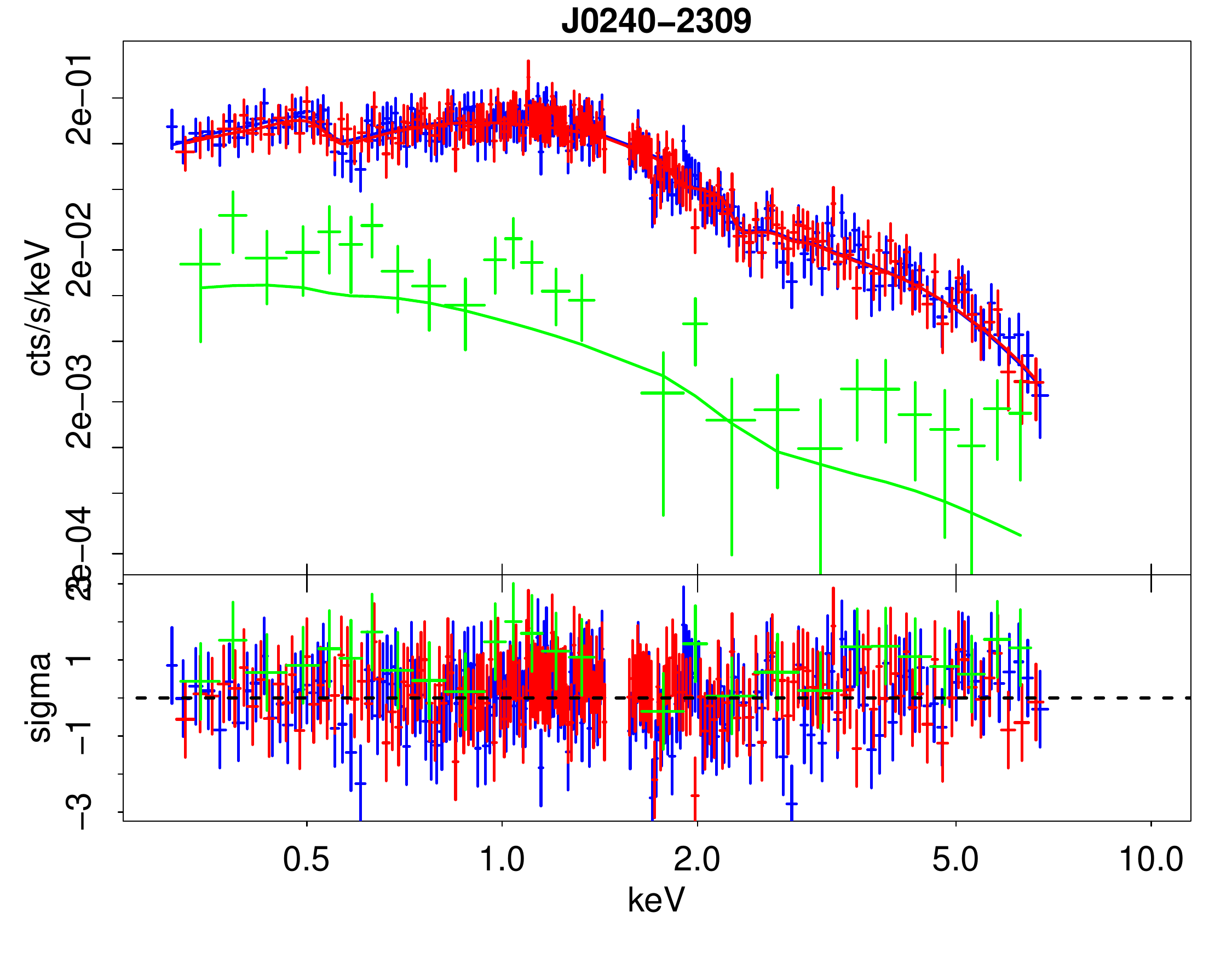}
\includegraphics[scale=0.19]{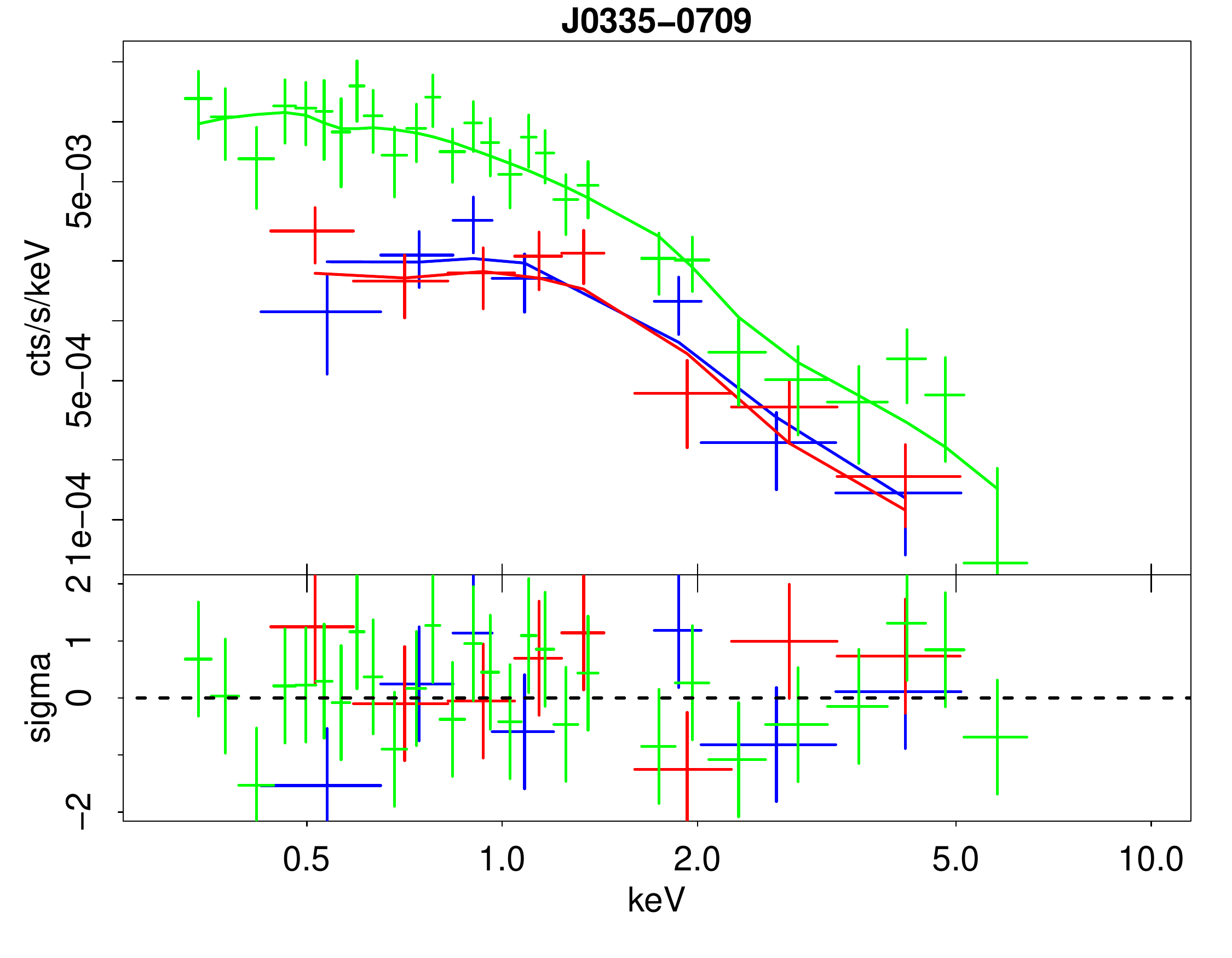}
\includegraphics[scale=0.19]{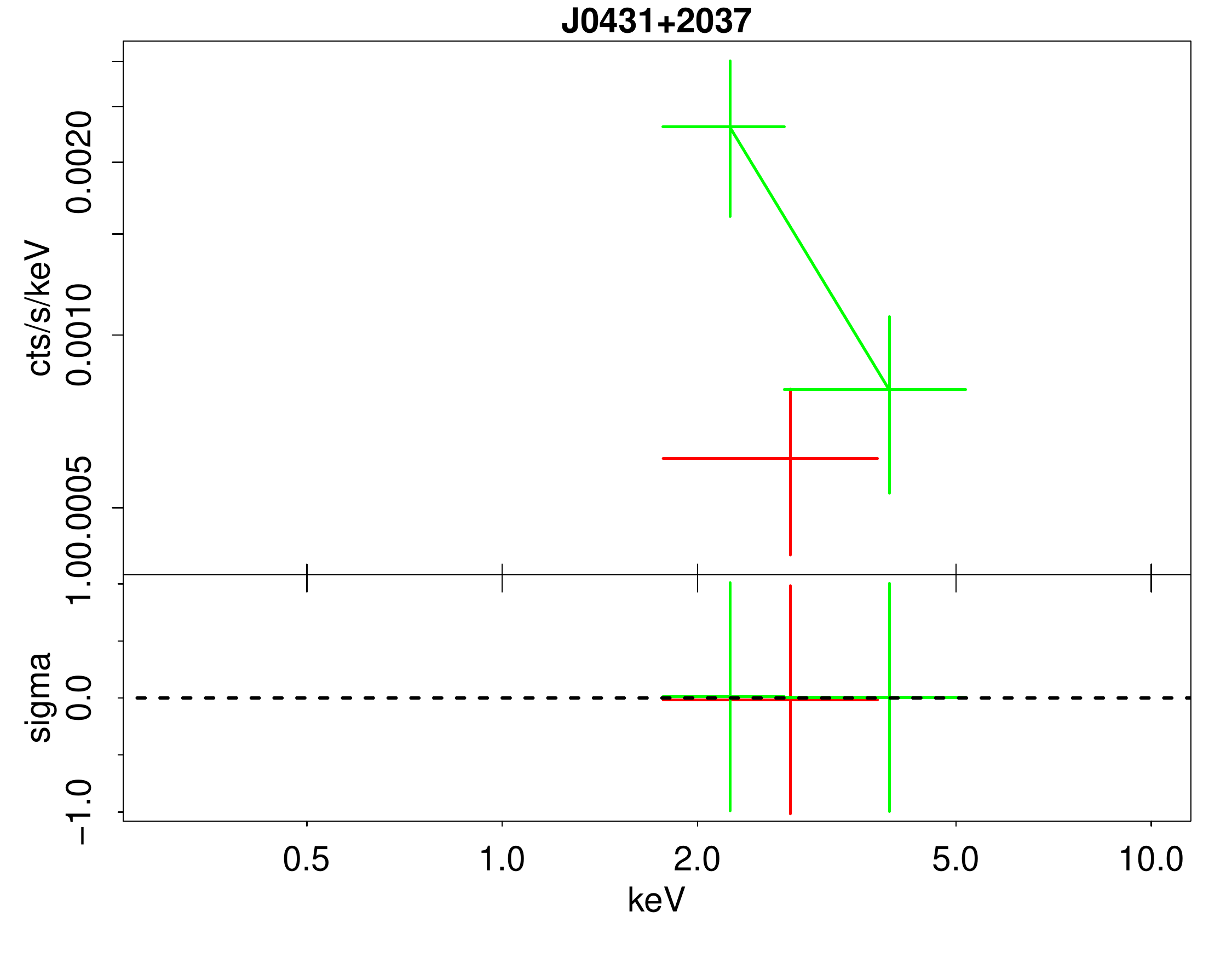}
\includegraphics[scale=0.19]{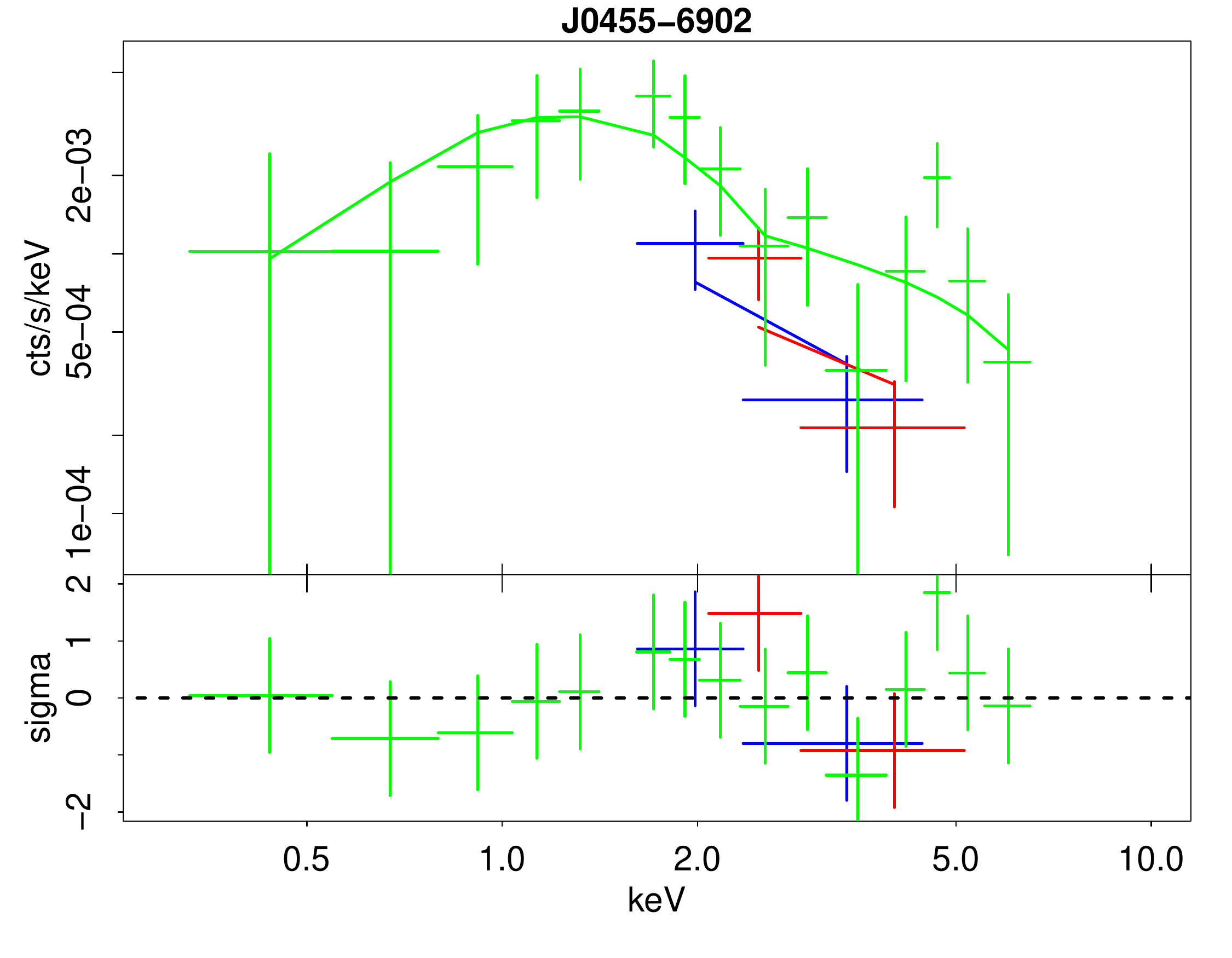}
\includegraphics[scale=0.19]{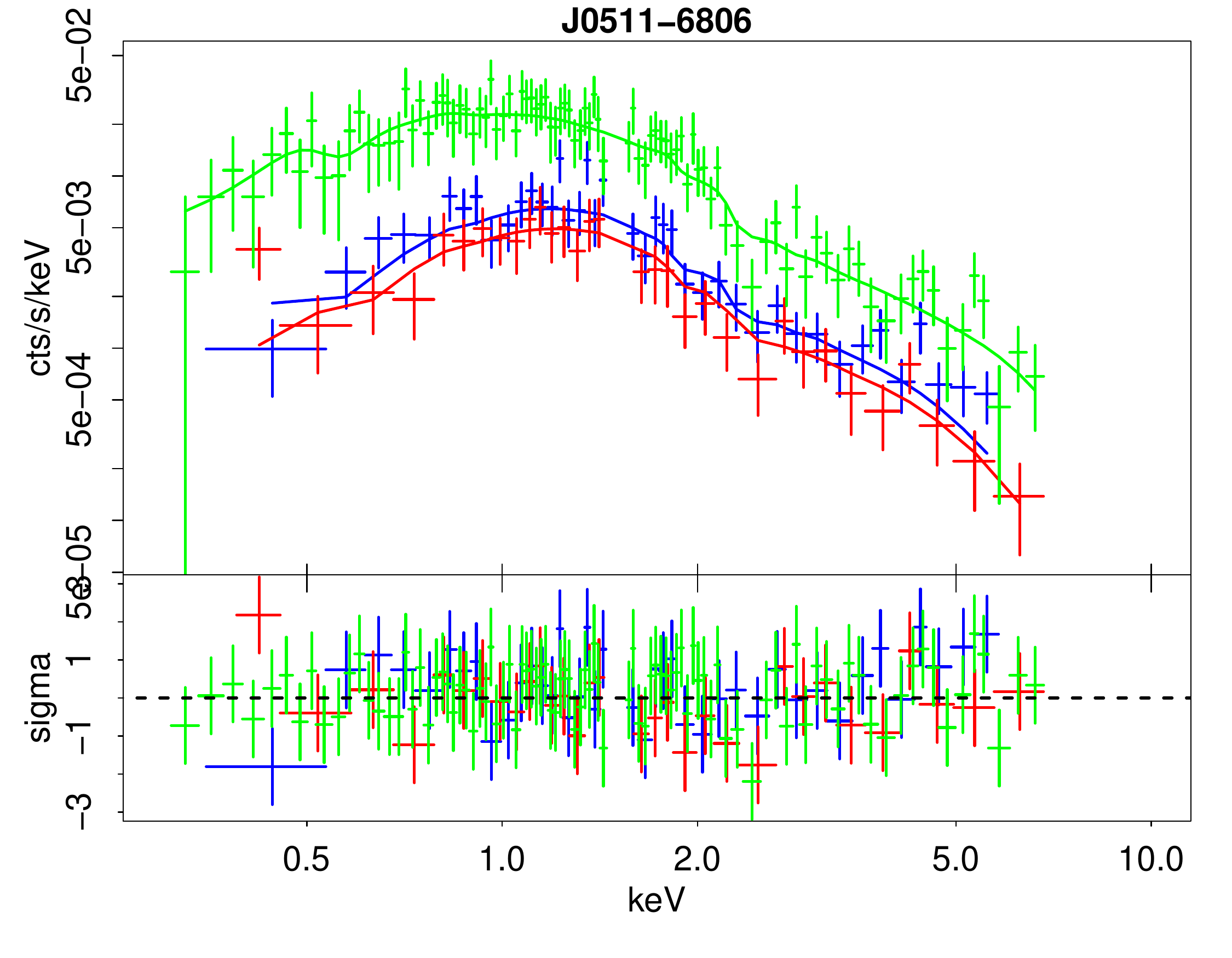}
\includegraphics[scale=0.19]{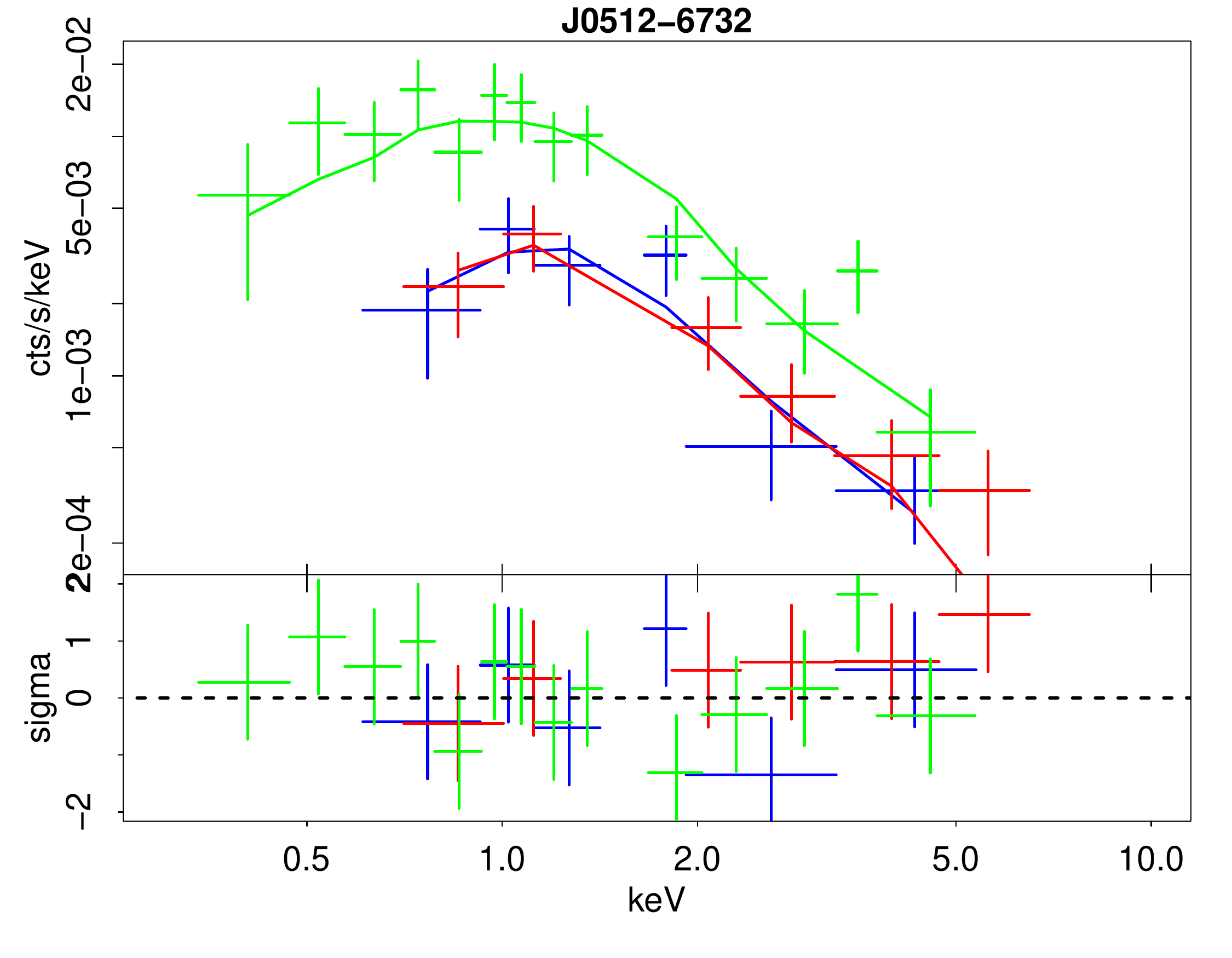}
\includegraphics[scale=0.19]{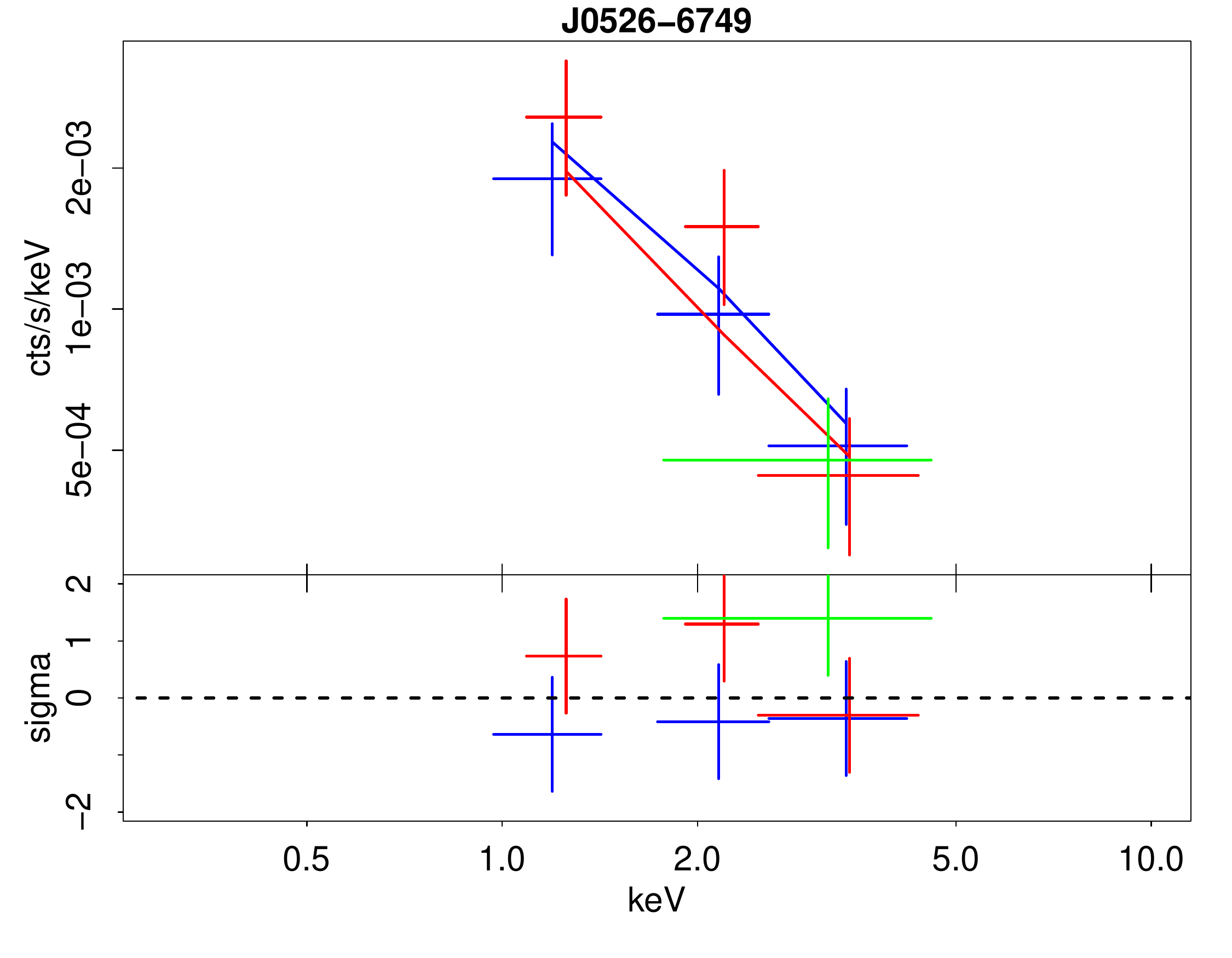}
\includegraphics[scale=0.19]{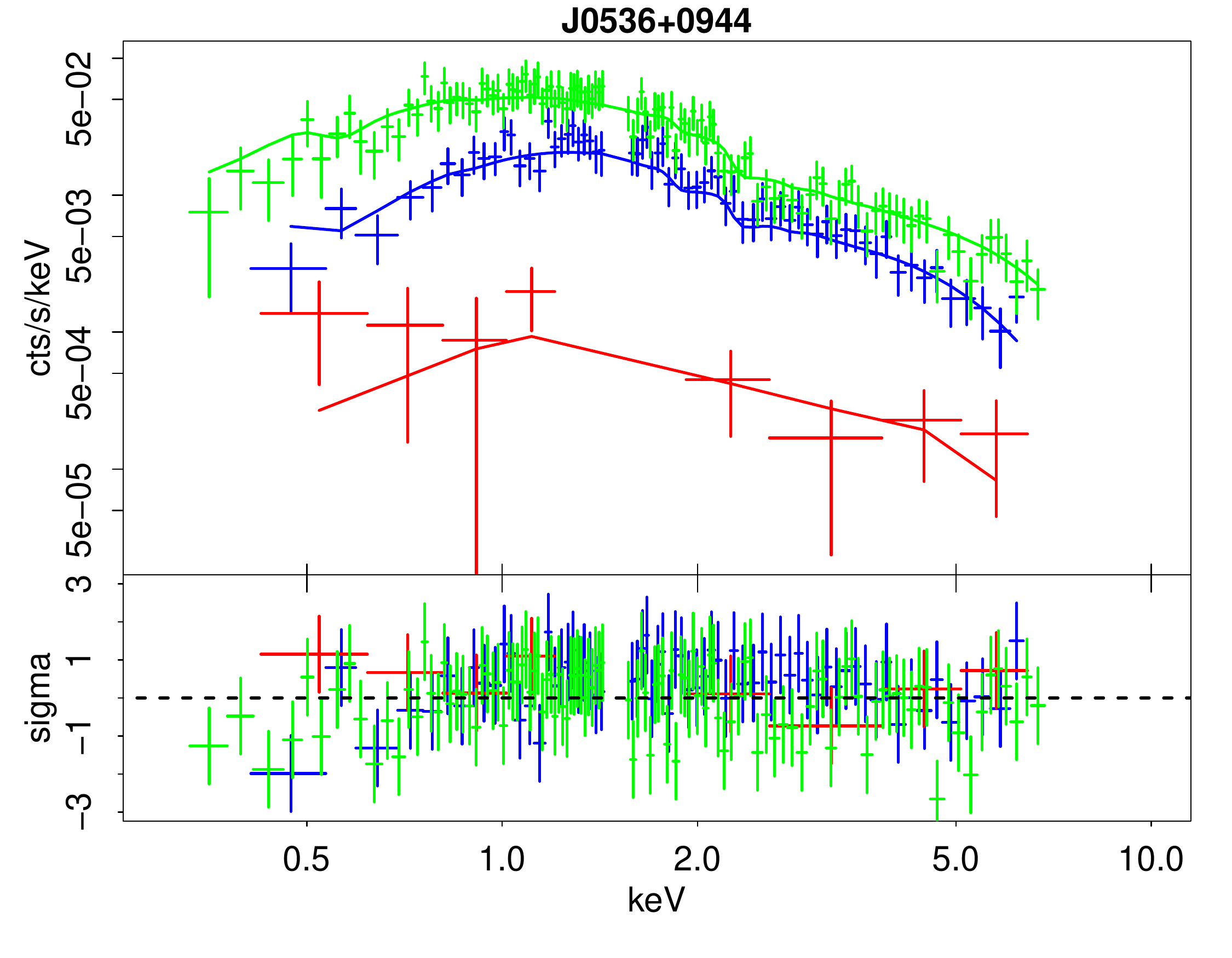}
\includegraphics[scale=0.19]{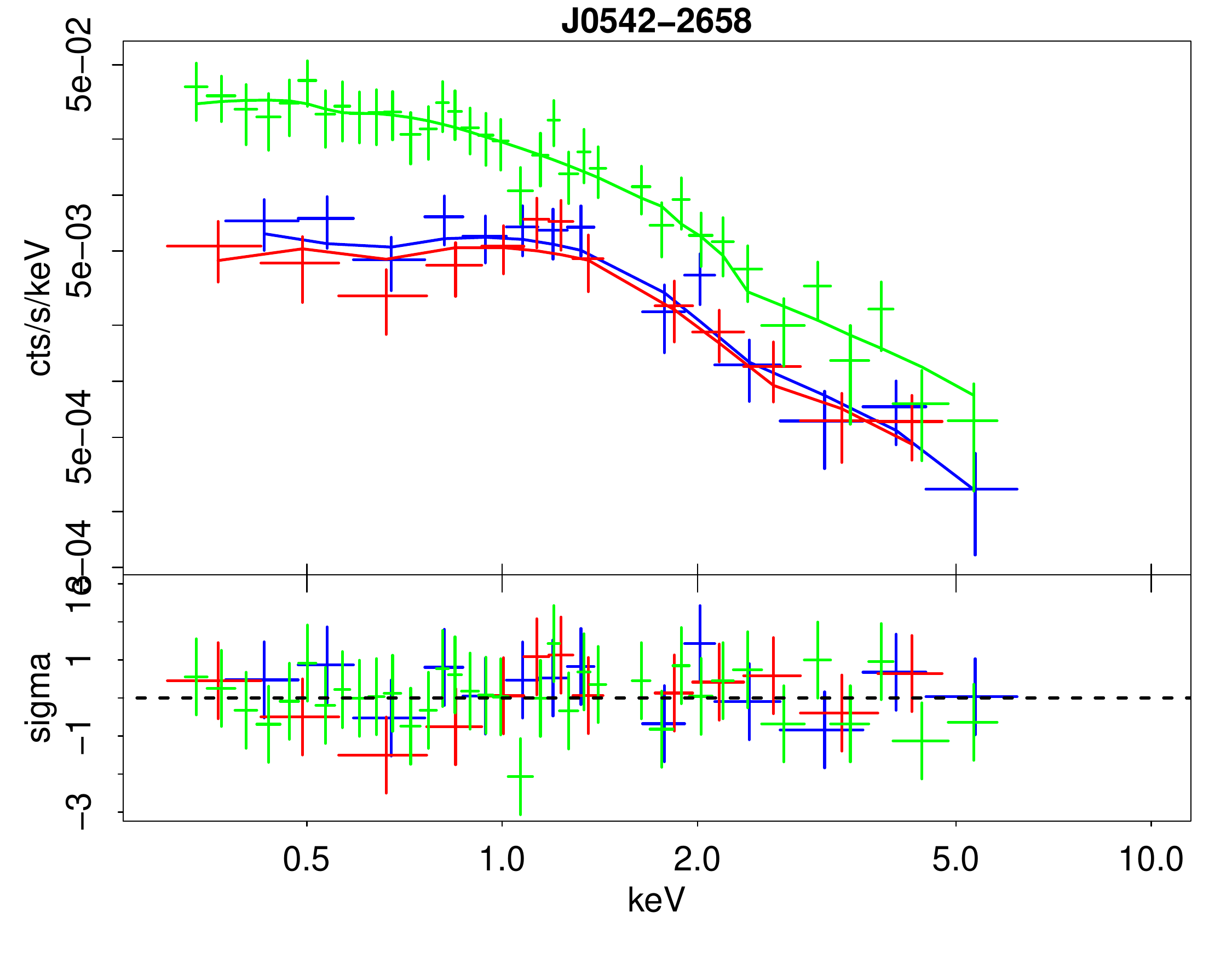}
\includegraphics[scale=0.19]{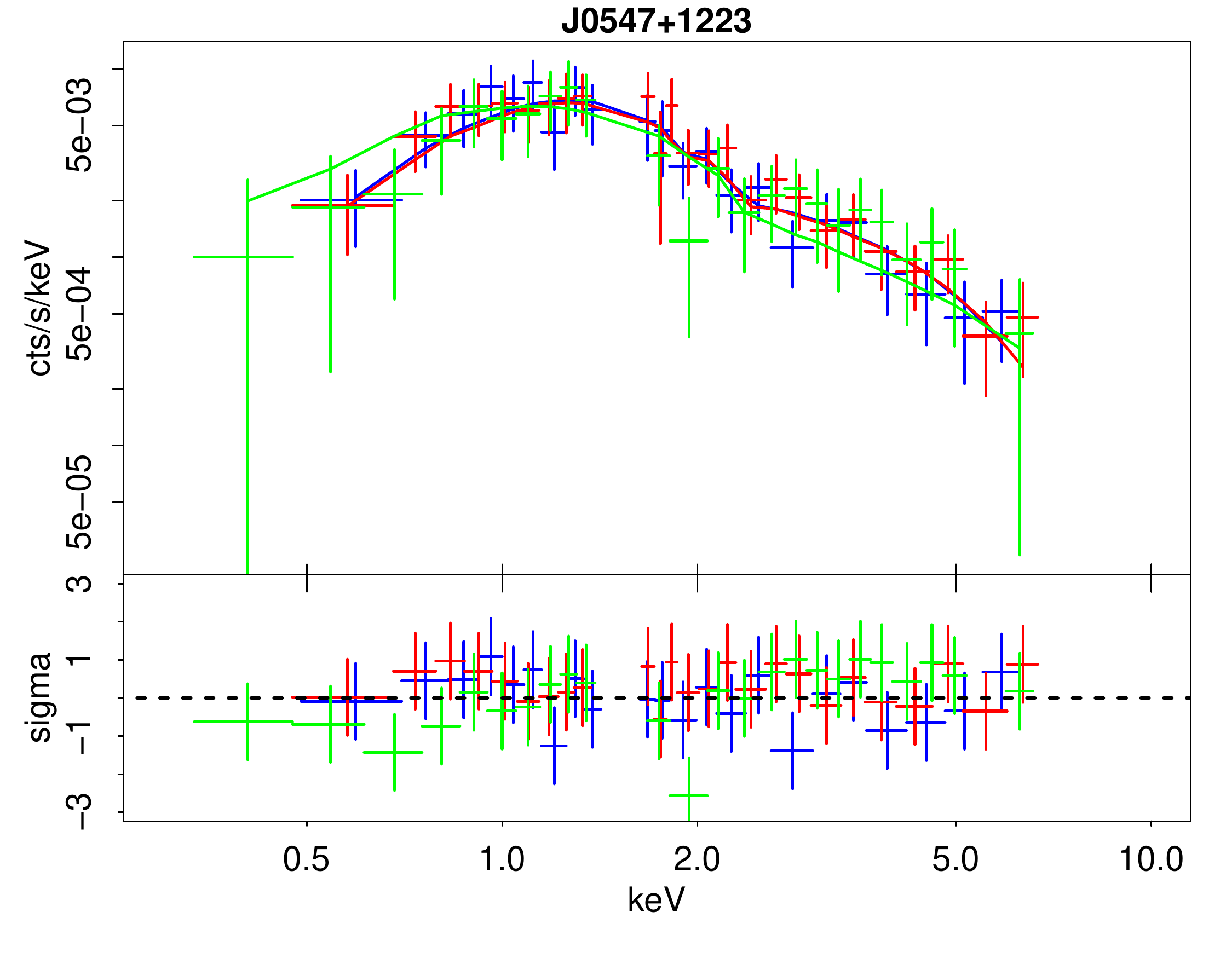}
\includegraphics[scale=0.19]{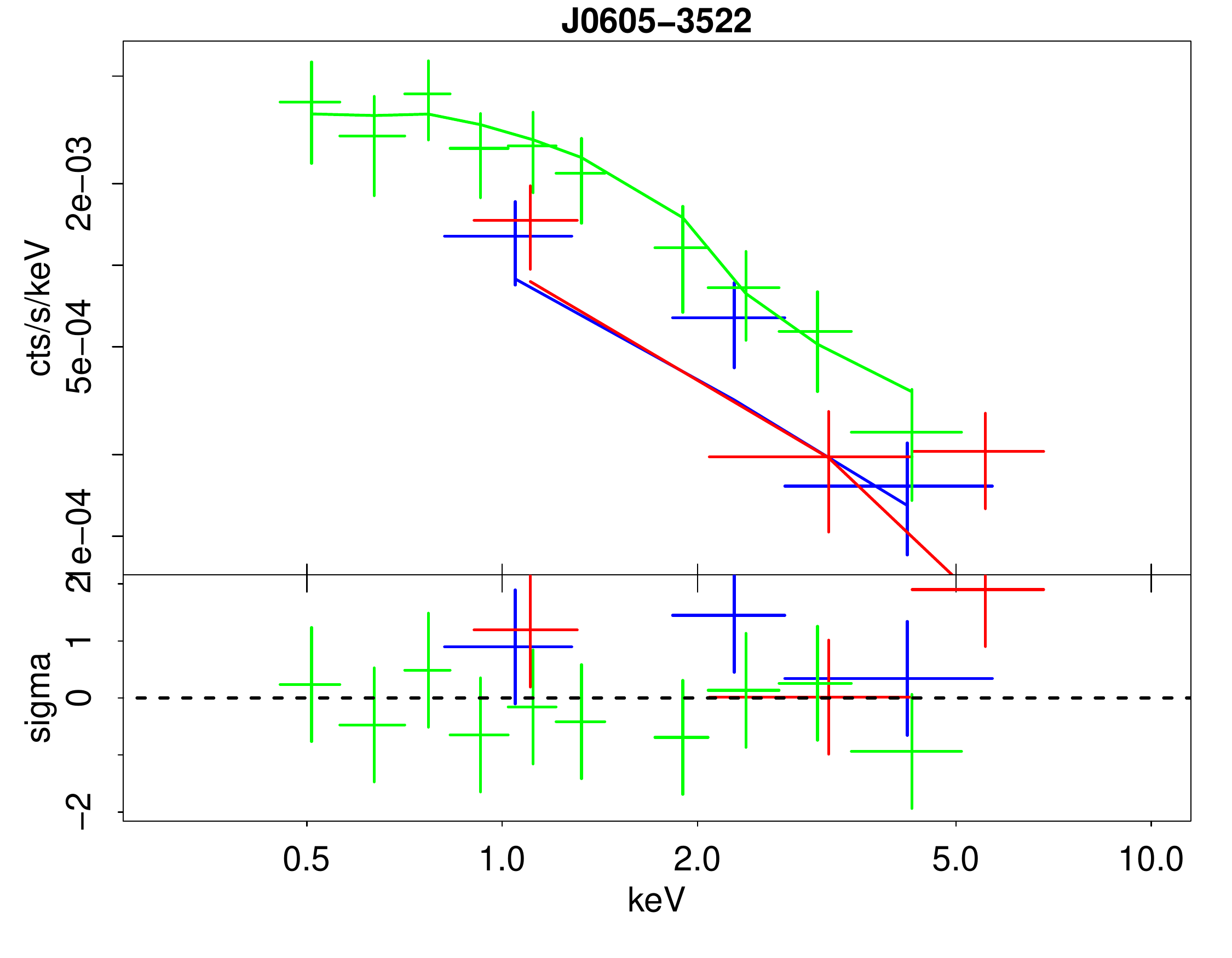}
\includegraphics[scale=0.19]{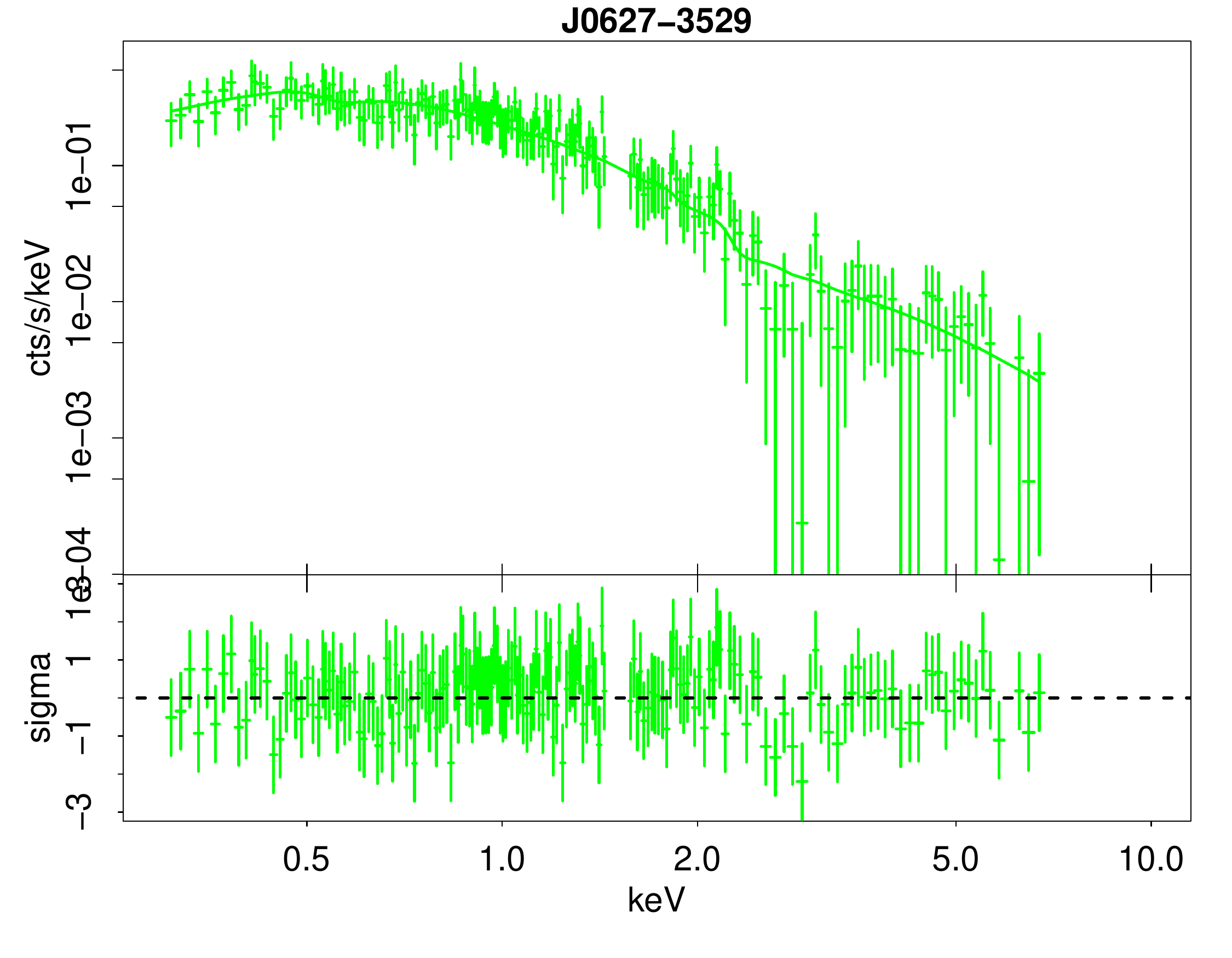}
\includegraphics[scale=0.19]{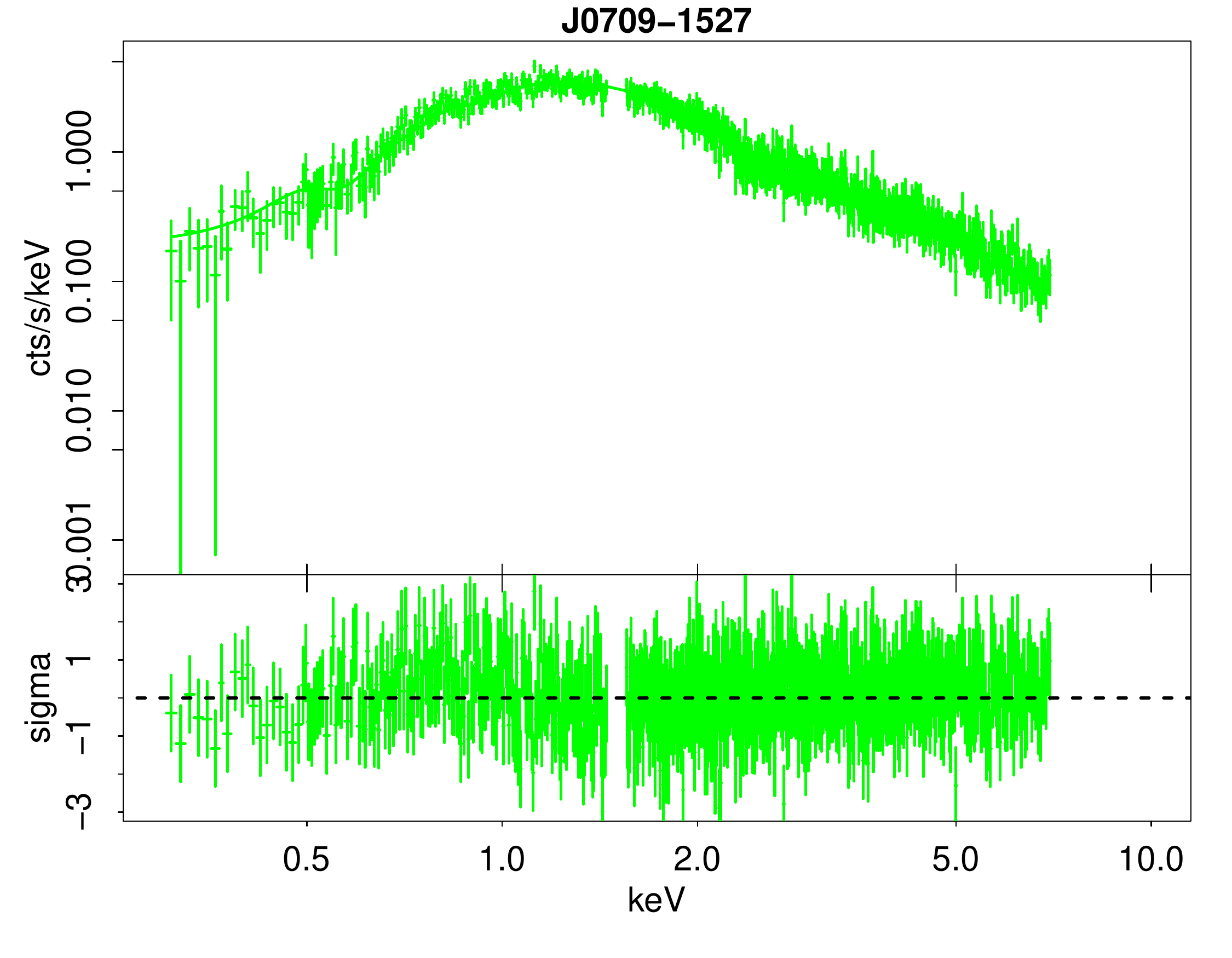}
\includegraphics[scale=0.19]{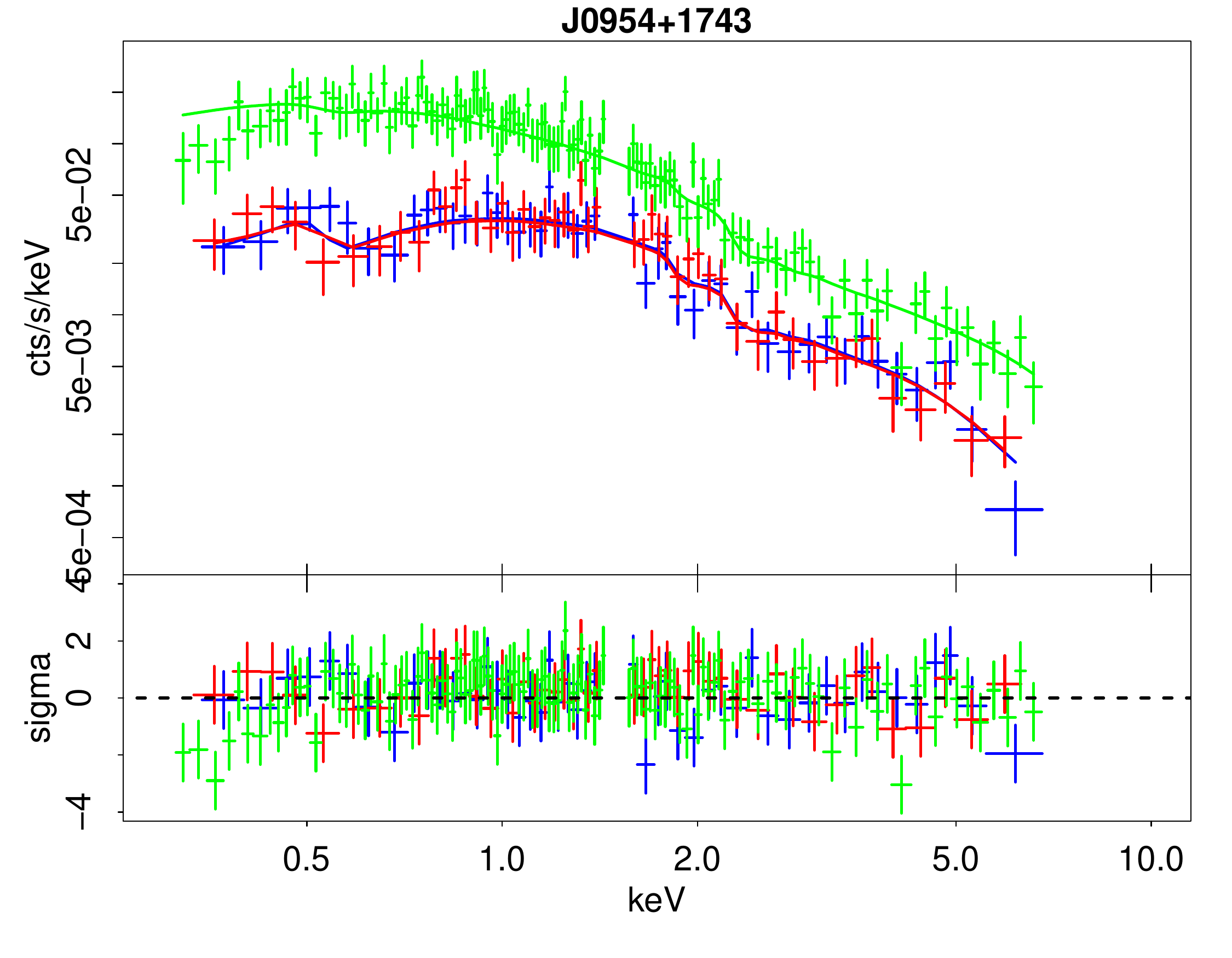}
\includegraphics[scale=0.19]{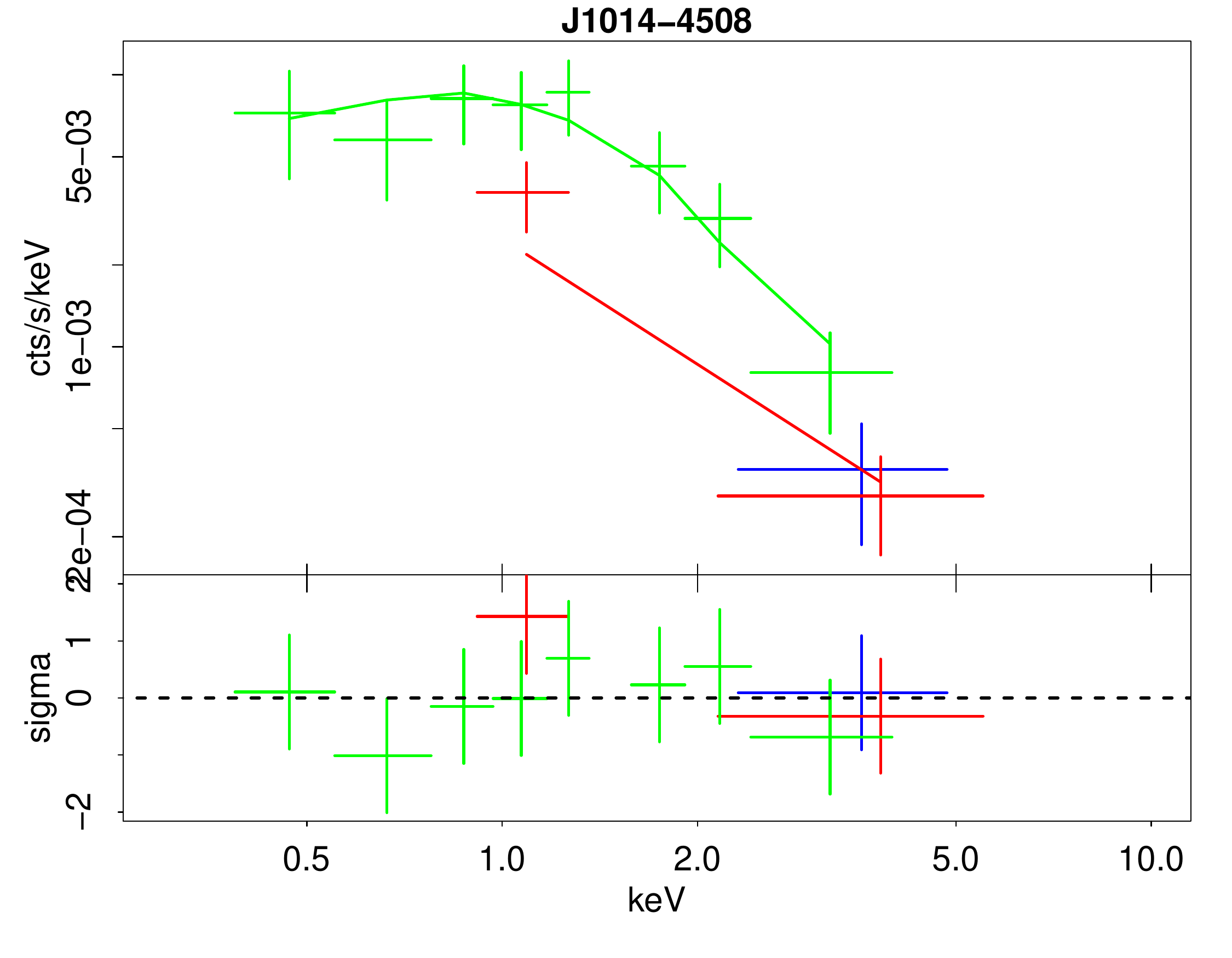}
\includegraphics[scale=0.19]{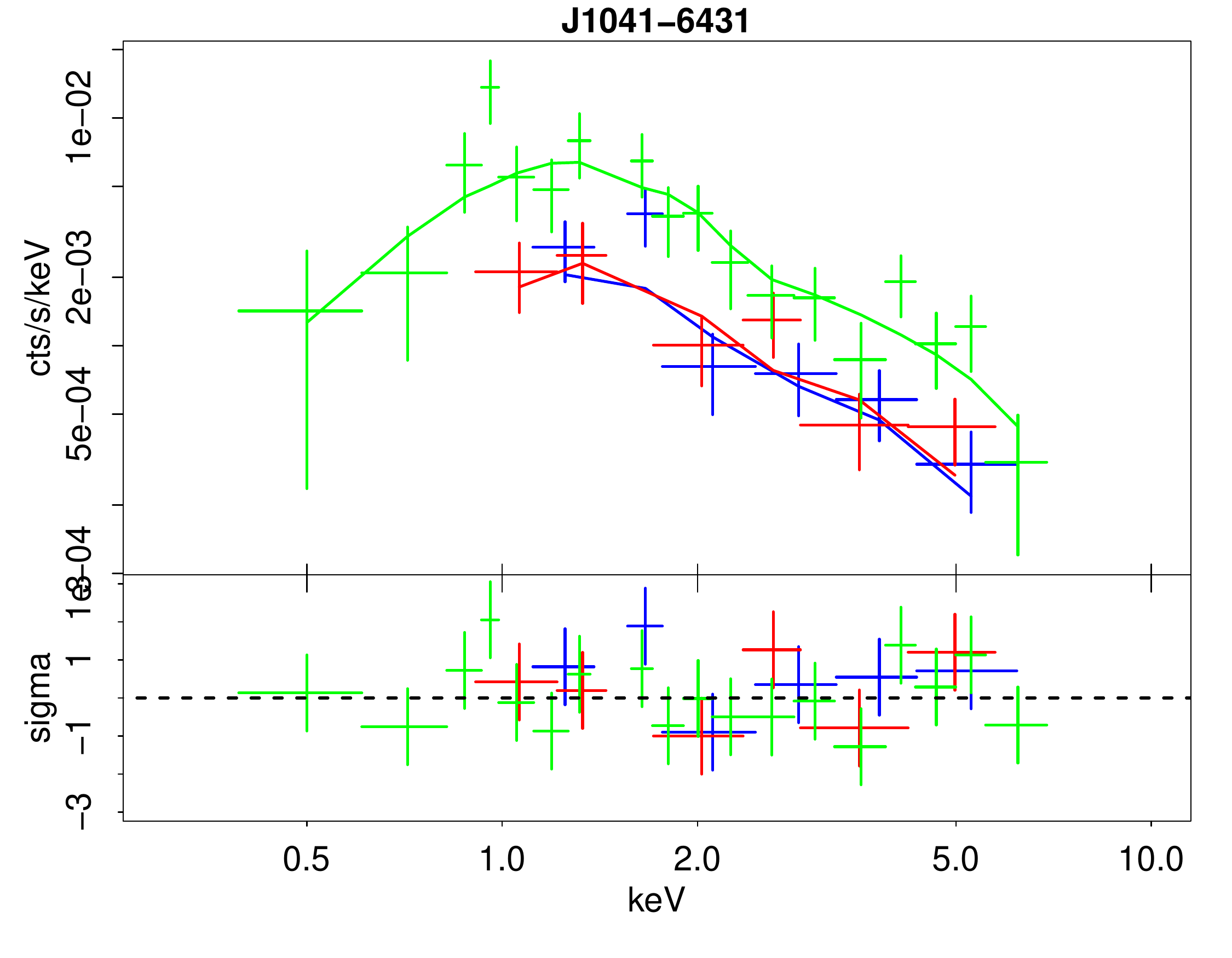}
\includegraphics[scale=0.19]{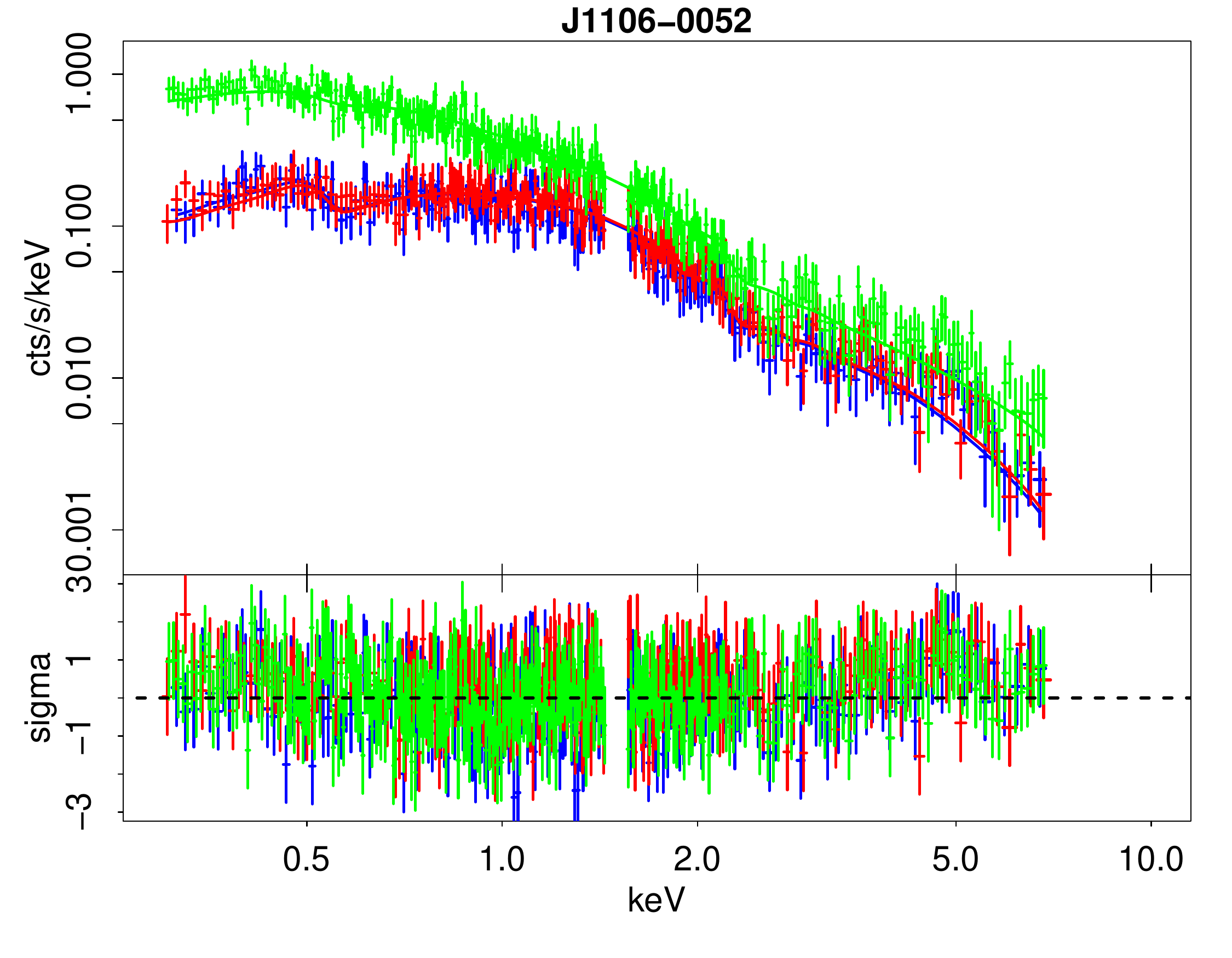}
\includegraphics[scale=0.19]{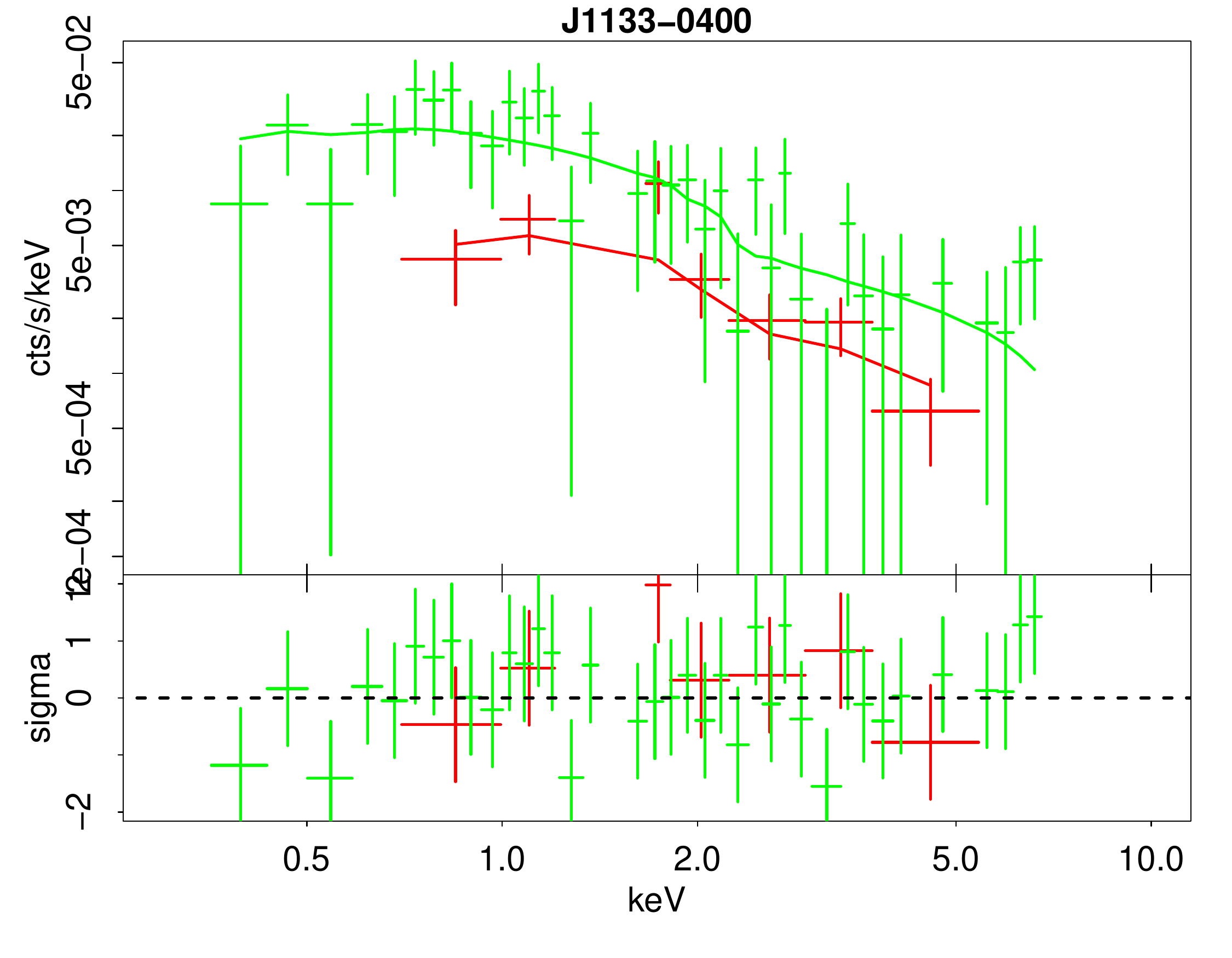}
\includegraphics[scale=0.19]{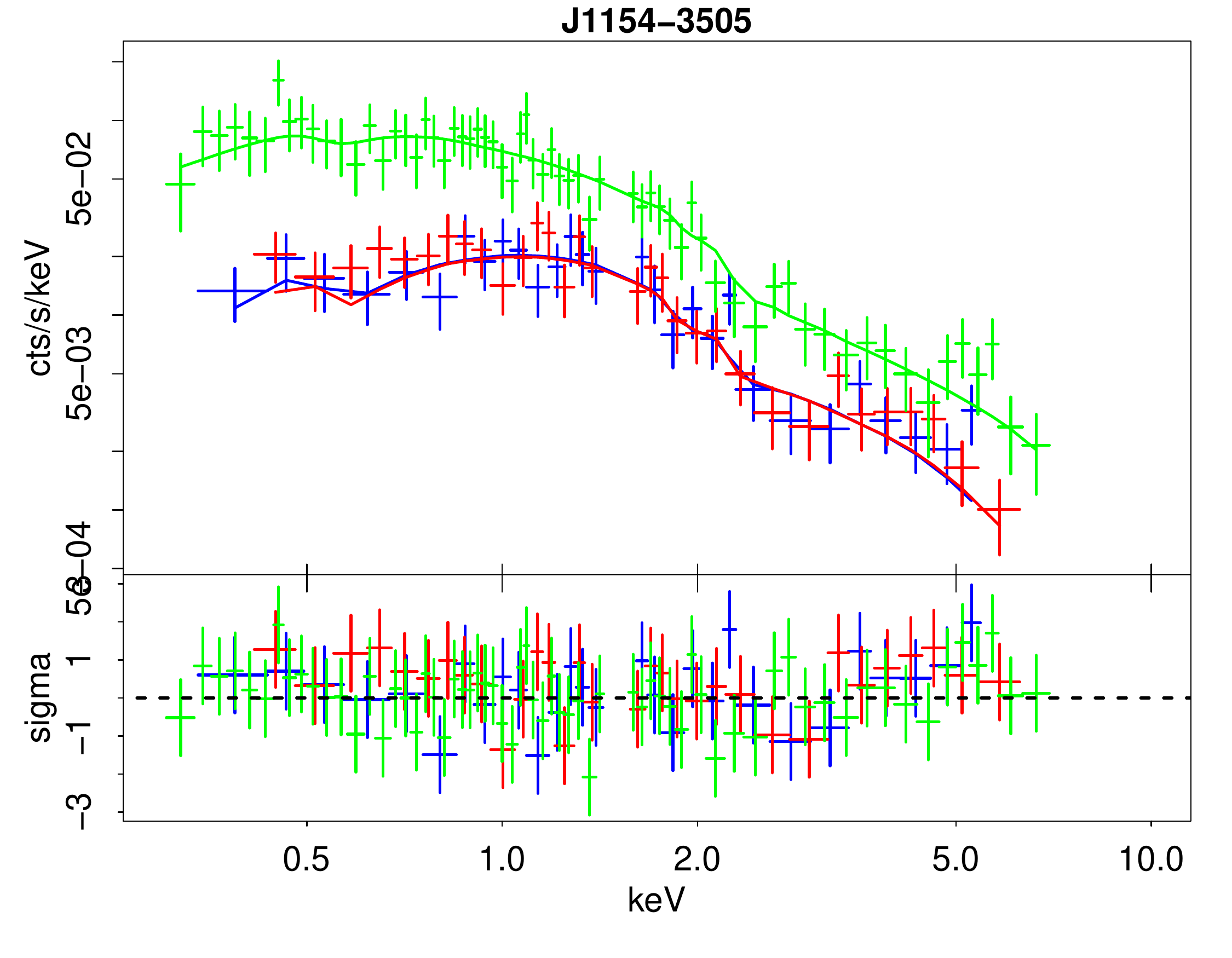}
\caption{\textit{XMM-Newton}-EPIC spectra with their best-fit power law models (upper panels) and residuals (lower panels). Lines and points in blue, red and green indicates models and data of MOS1, MOS2 and PN detectors, respectively. Full set of figures available online.}\label{fig:epic_spectra}
\end{figure*}

\begin{figure*}
   \centering
\includegraphics[scale=0.19]{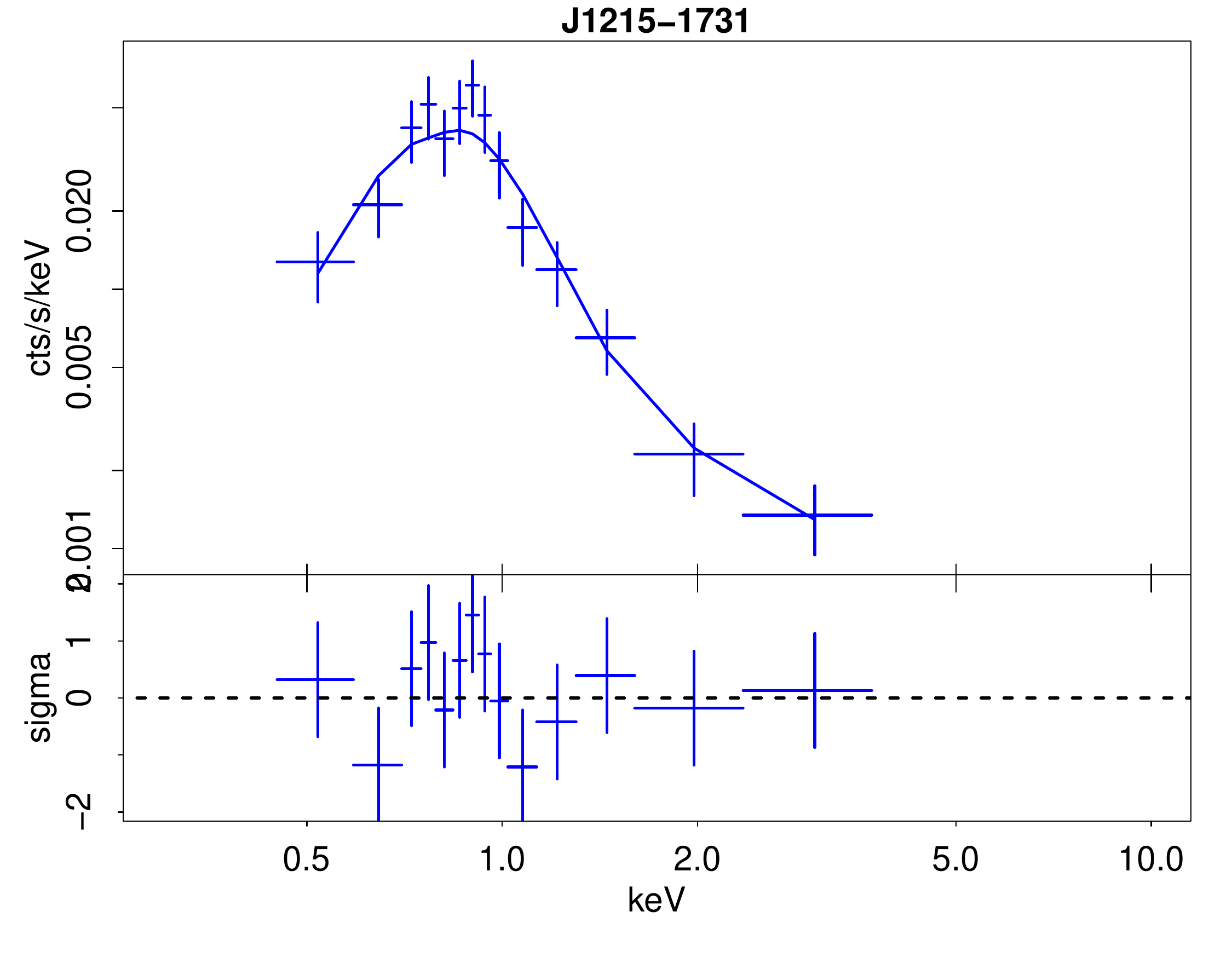}
\includegraphics[scale=0.19]{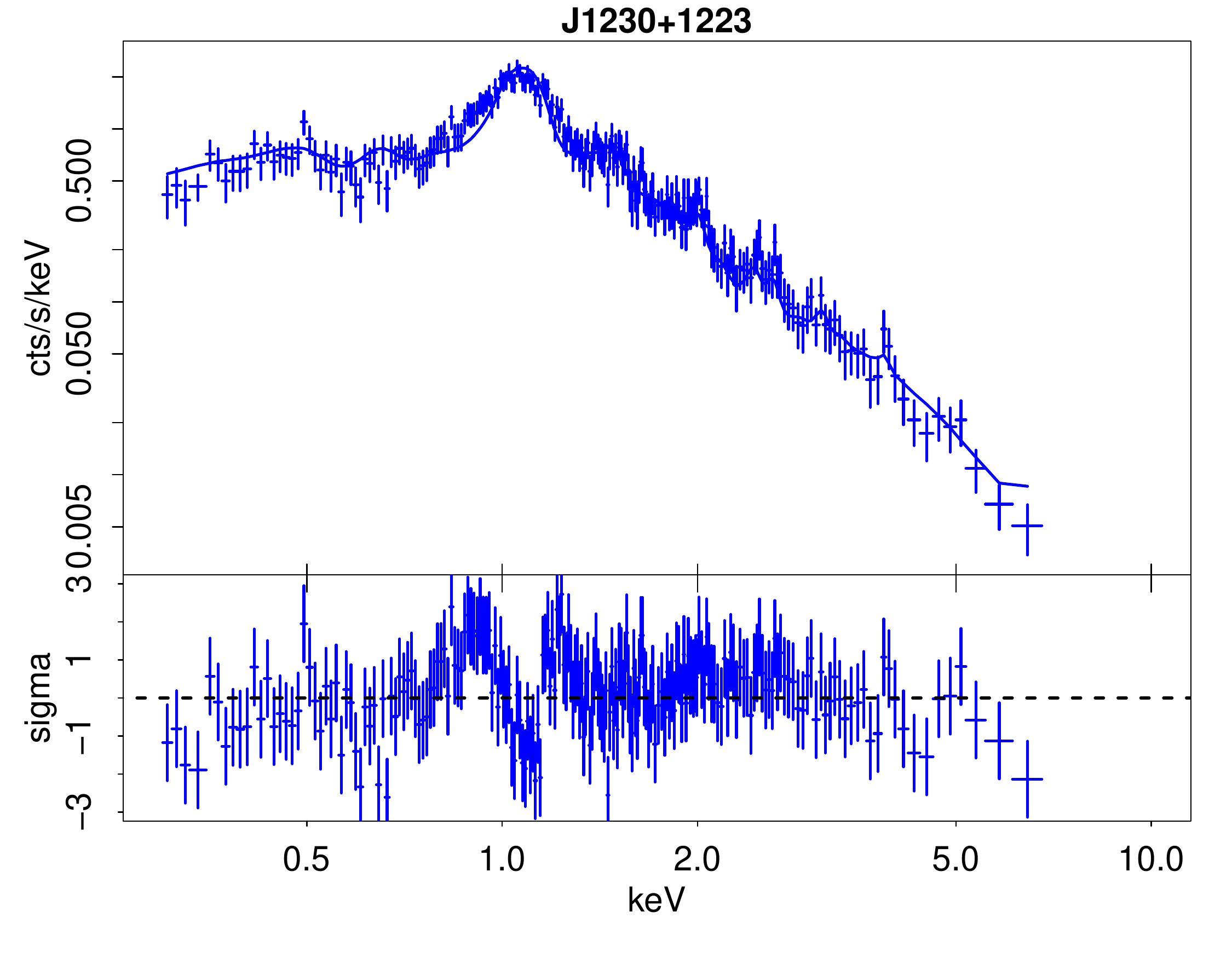}
\includegraphics[scale=0.19]{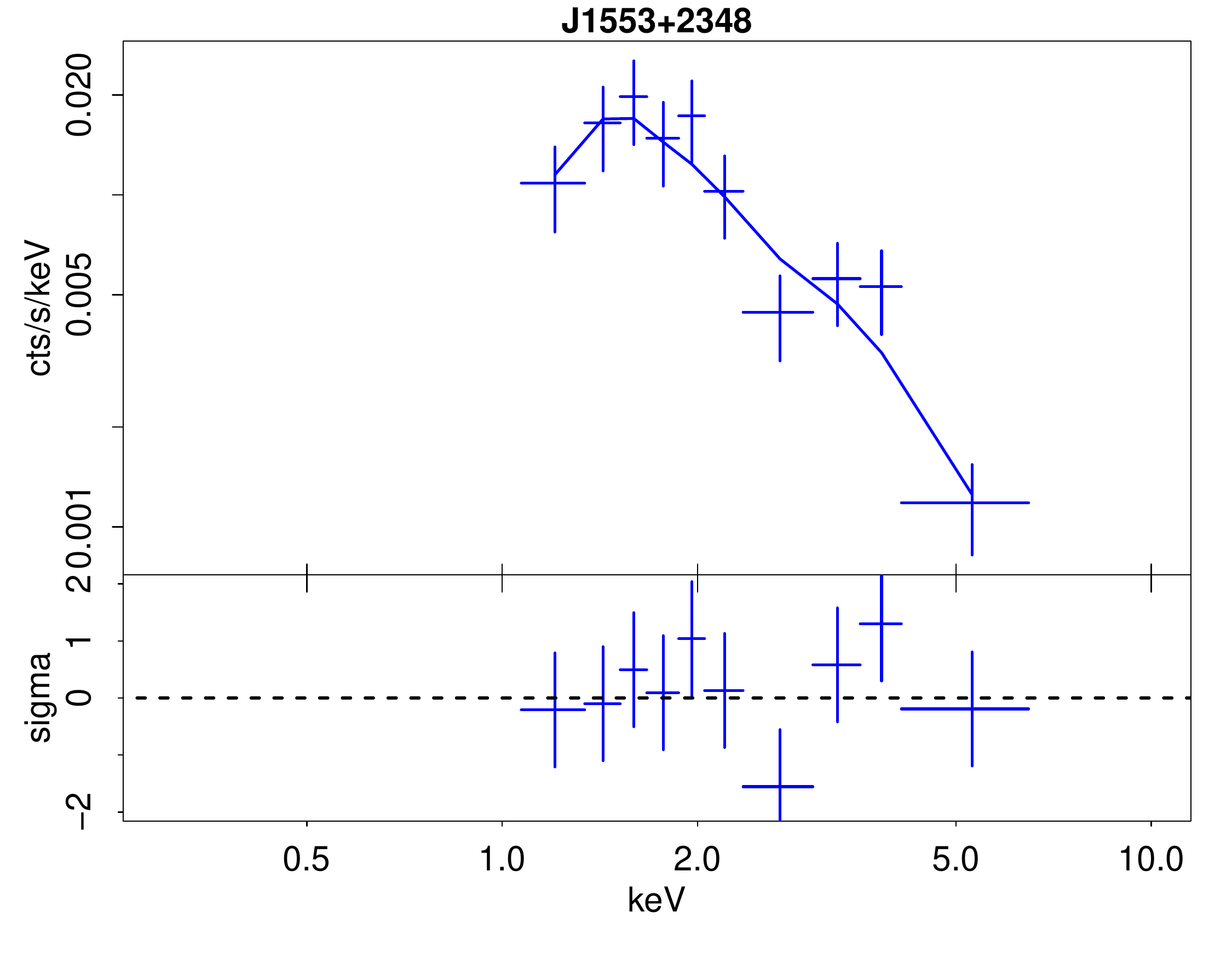}
\includegraphics[scale=0.19]{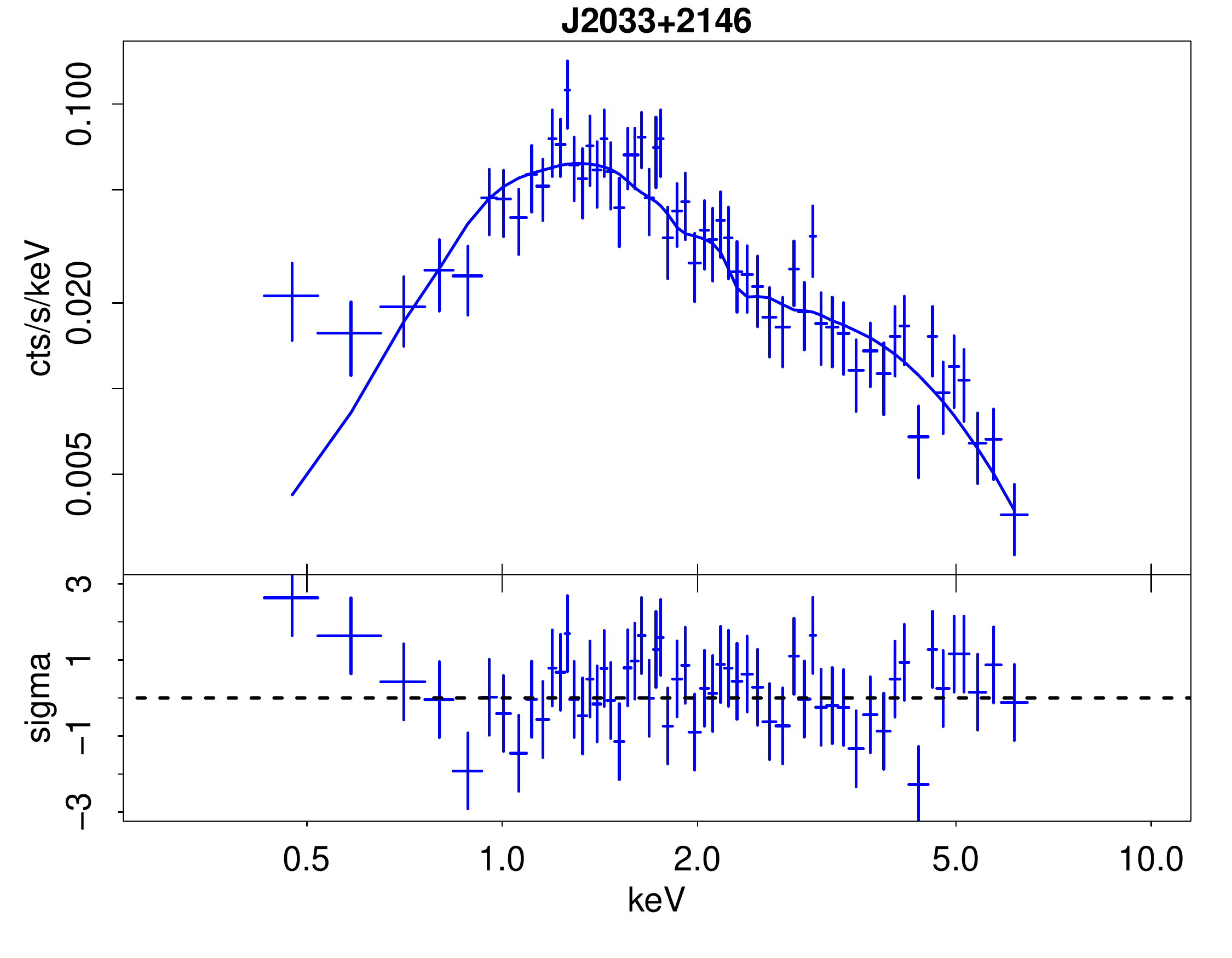}
\caption{\textit{Swift}-XRT spectra listed in Table \ref{tab:properties_peculiar} with their best-fit models (upper panels) and residuals (lower panels).}\label{fig:xrt_spectra_peculiar}
\end{figure*}

\begin{figure*}
   \centering
   \includegraphics[scale=0.19]{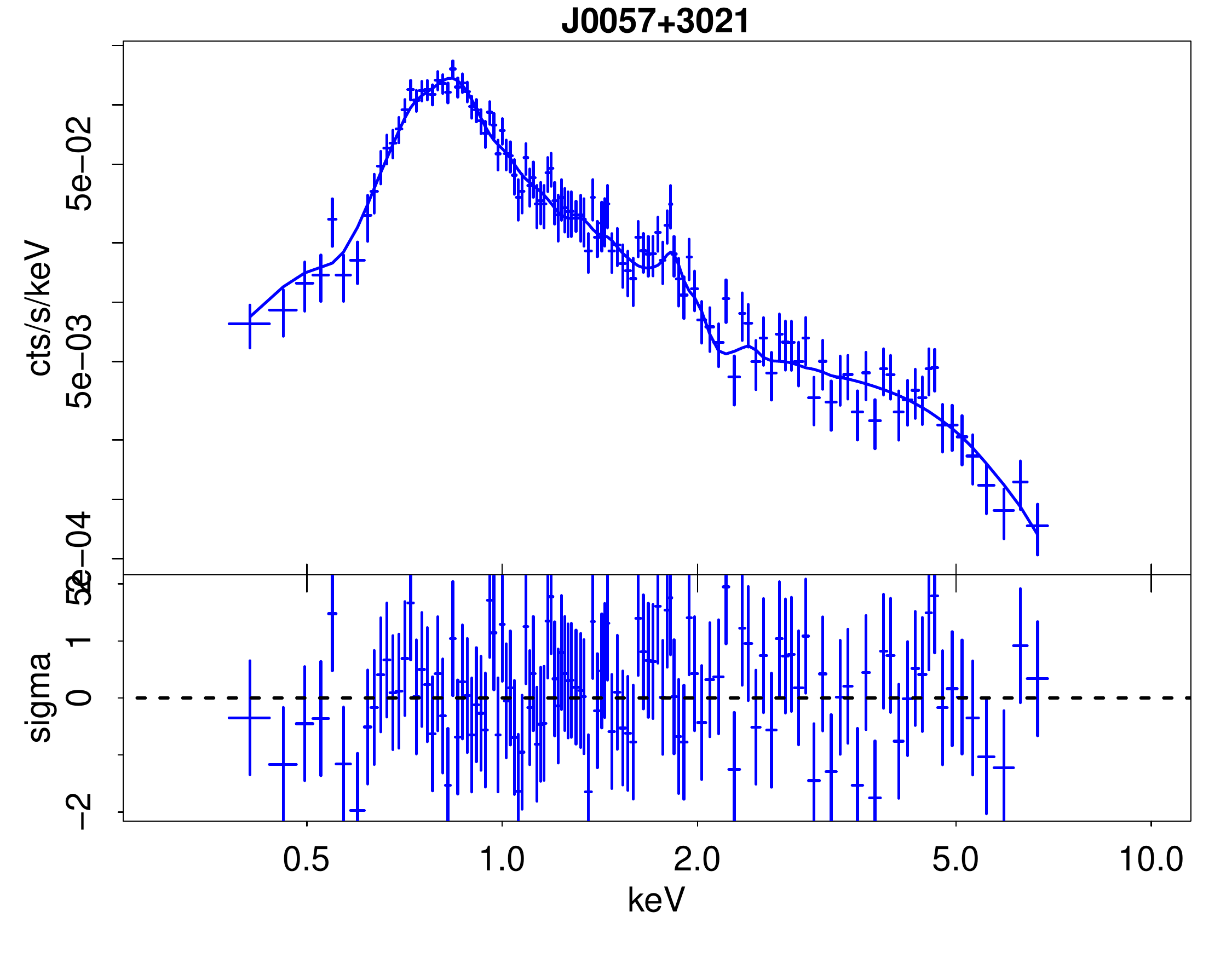}
   \includegraphics[scale=0.19]{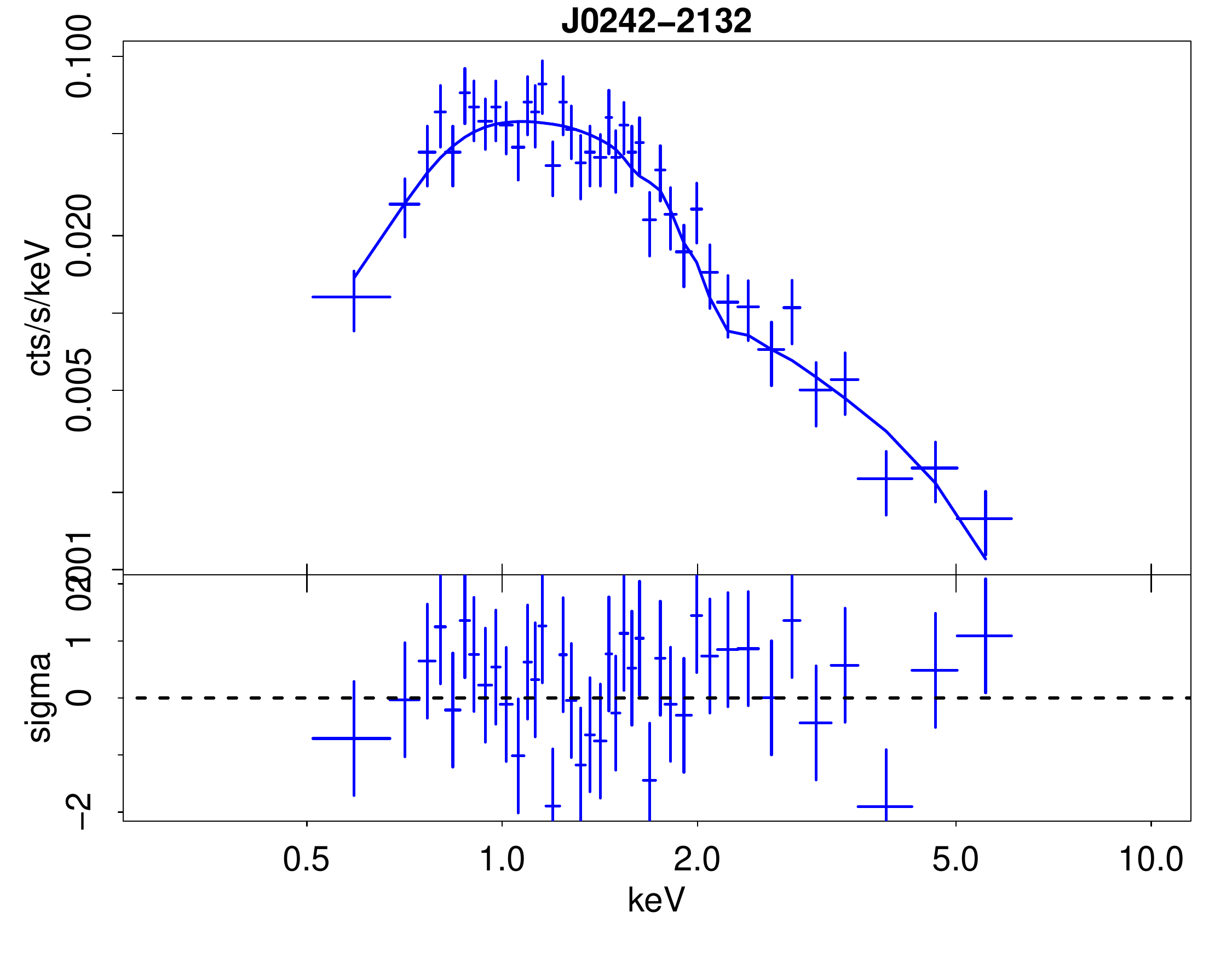}
   \includegraphics[scale=0.19]{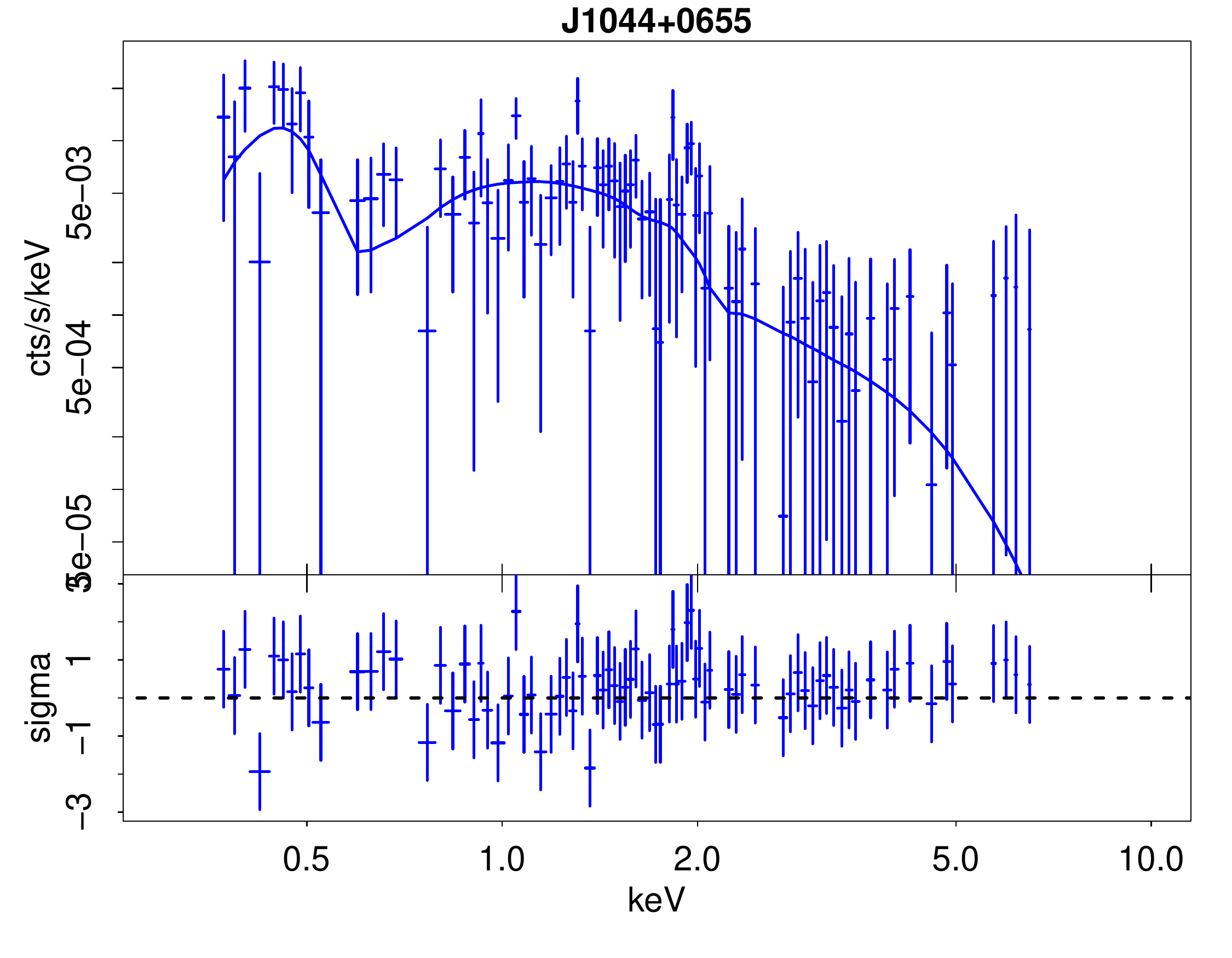}
   \includegraphics[scale=0.19]{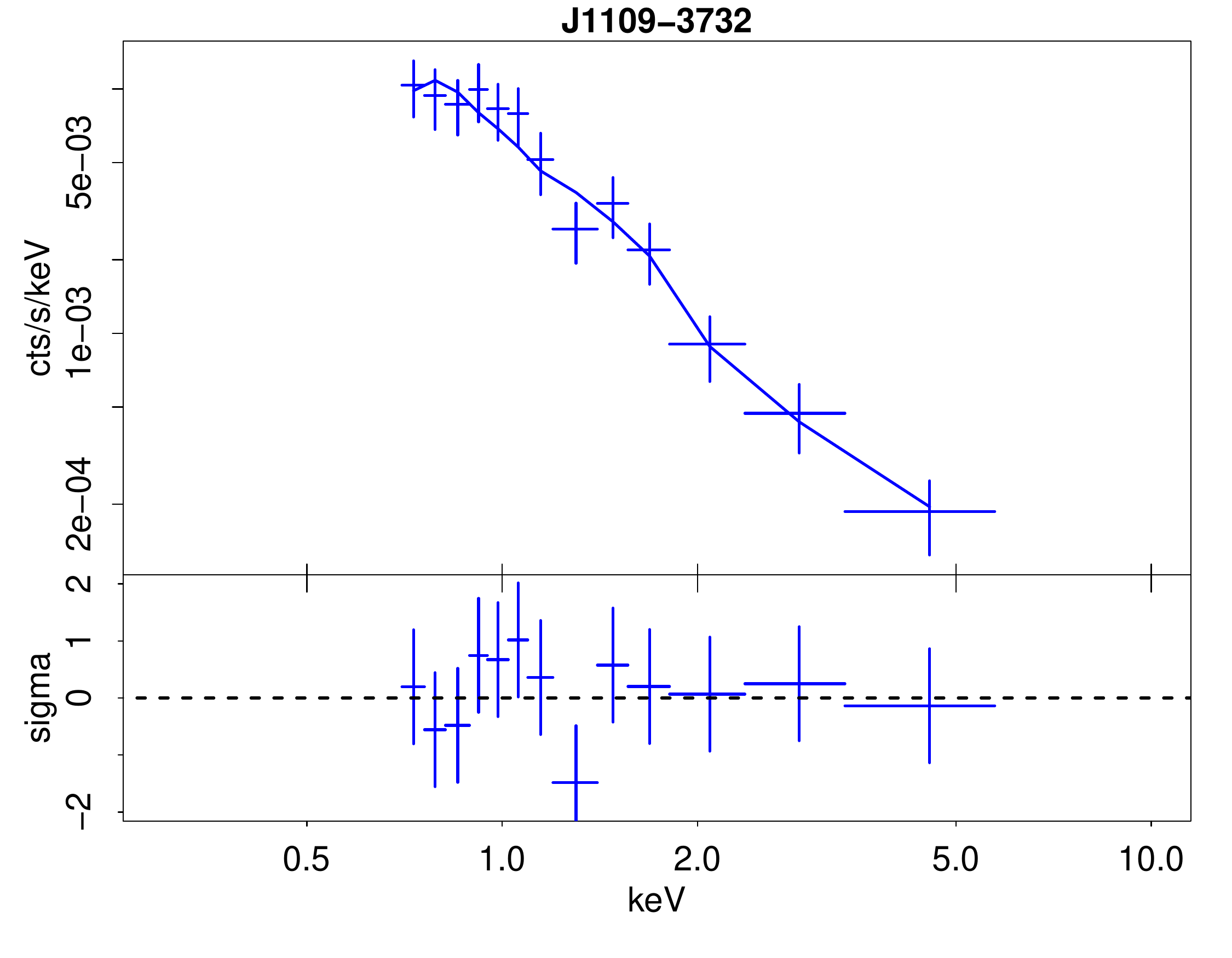}
   \includegraphics[scale=0.19]{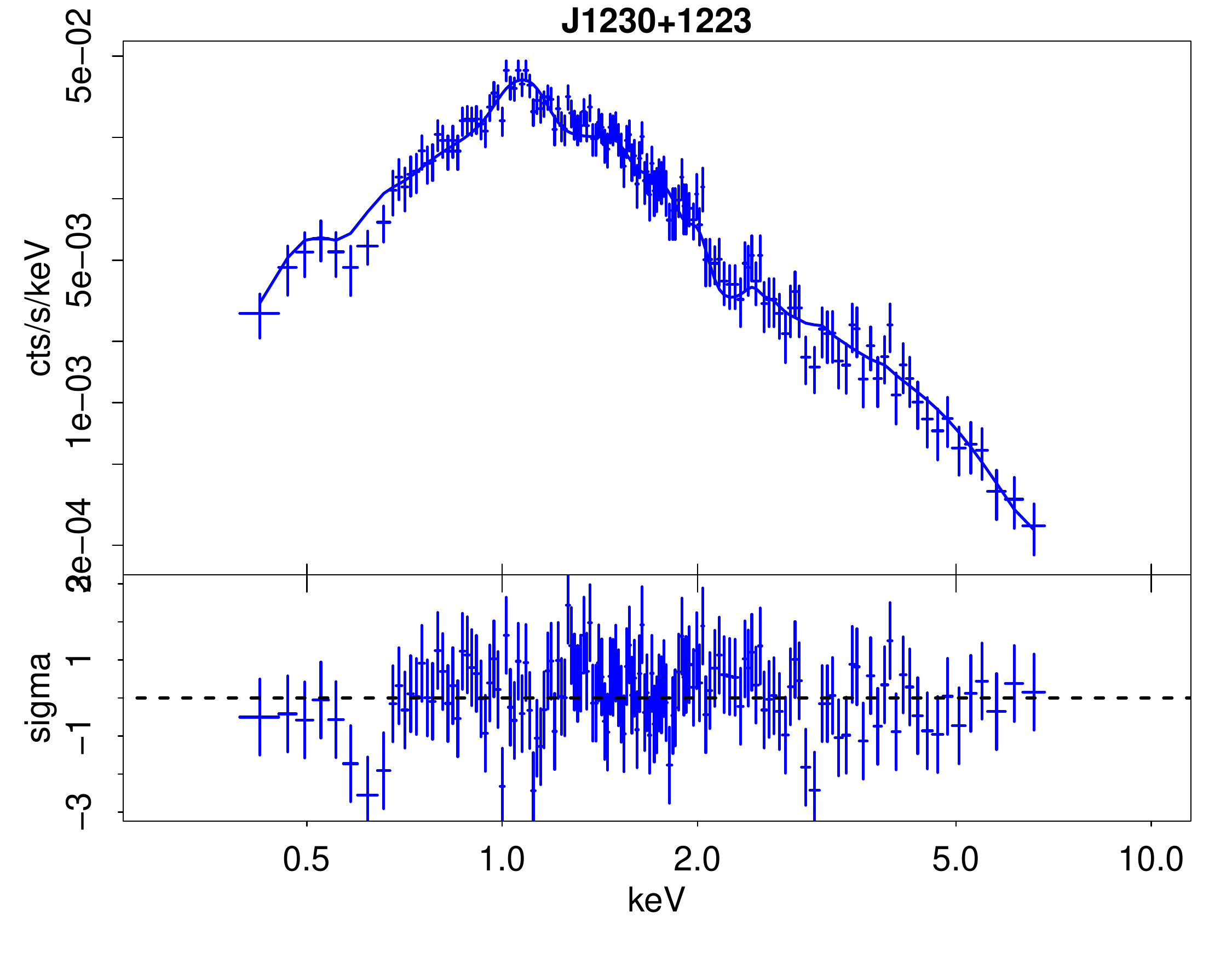}
   \includegraphics[scale=0.19]{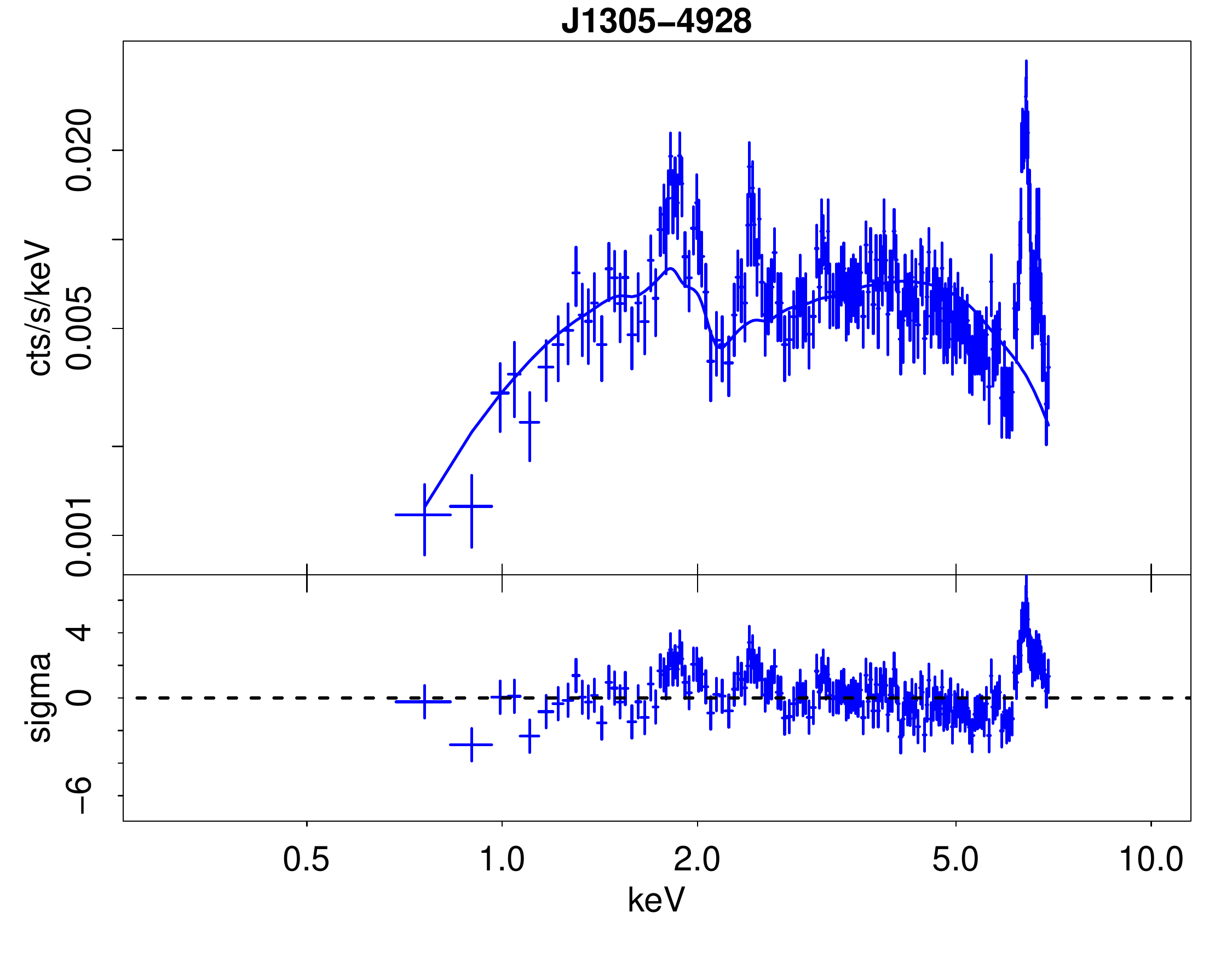}
   \includegraphics[scale=0.19]{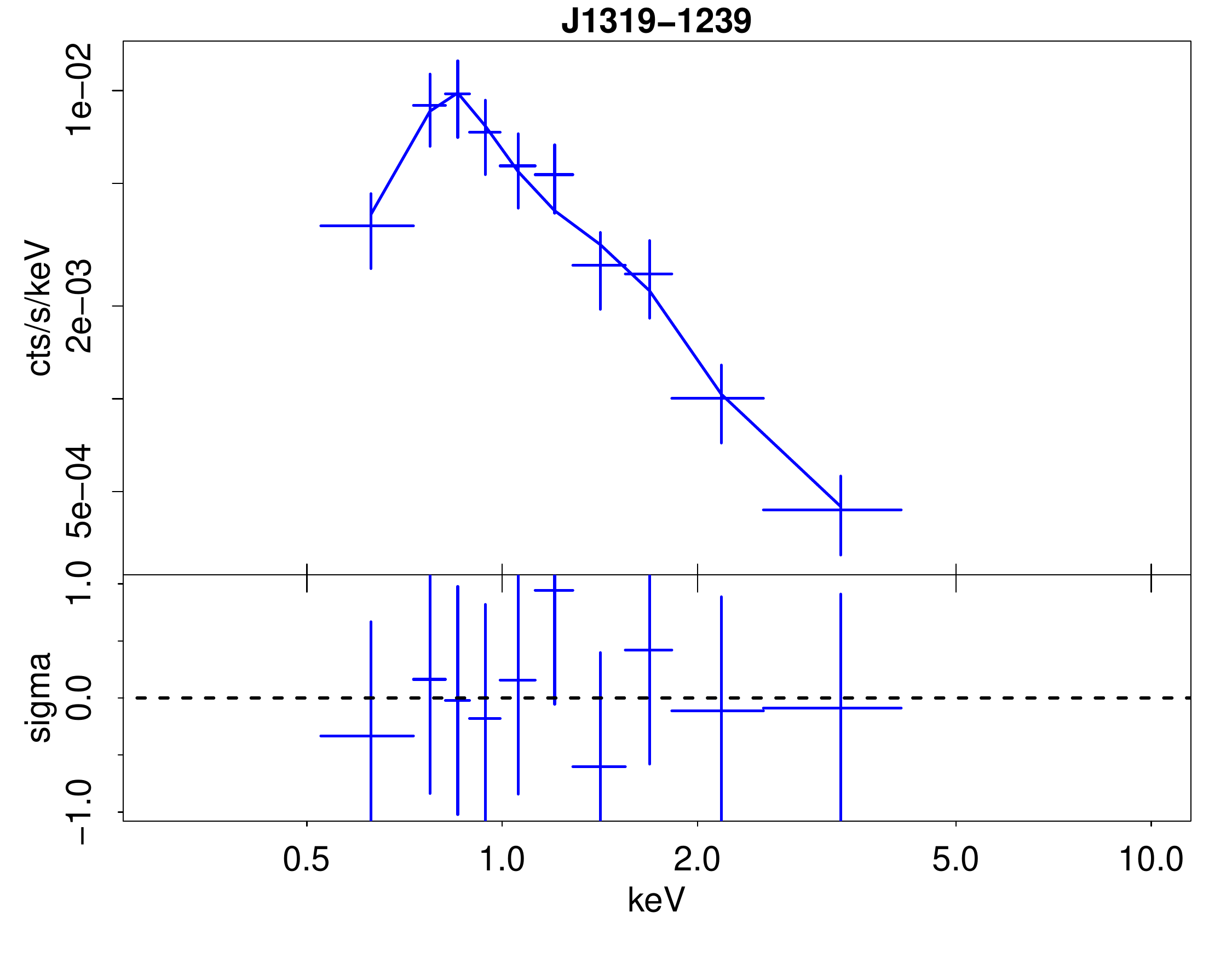}
   \includegraphics[scale=0.19]{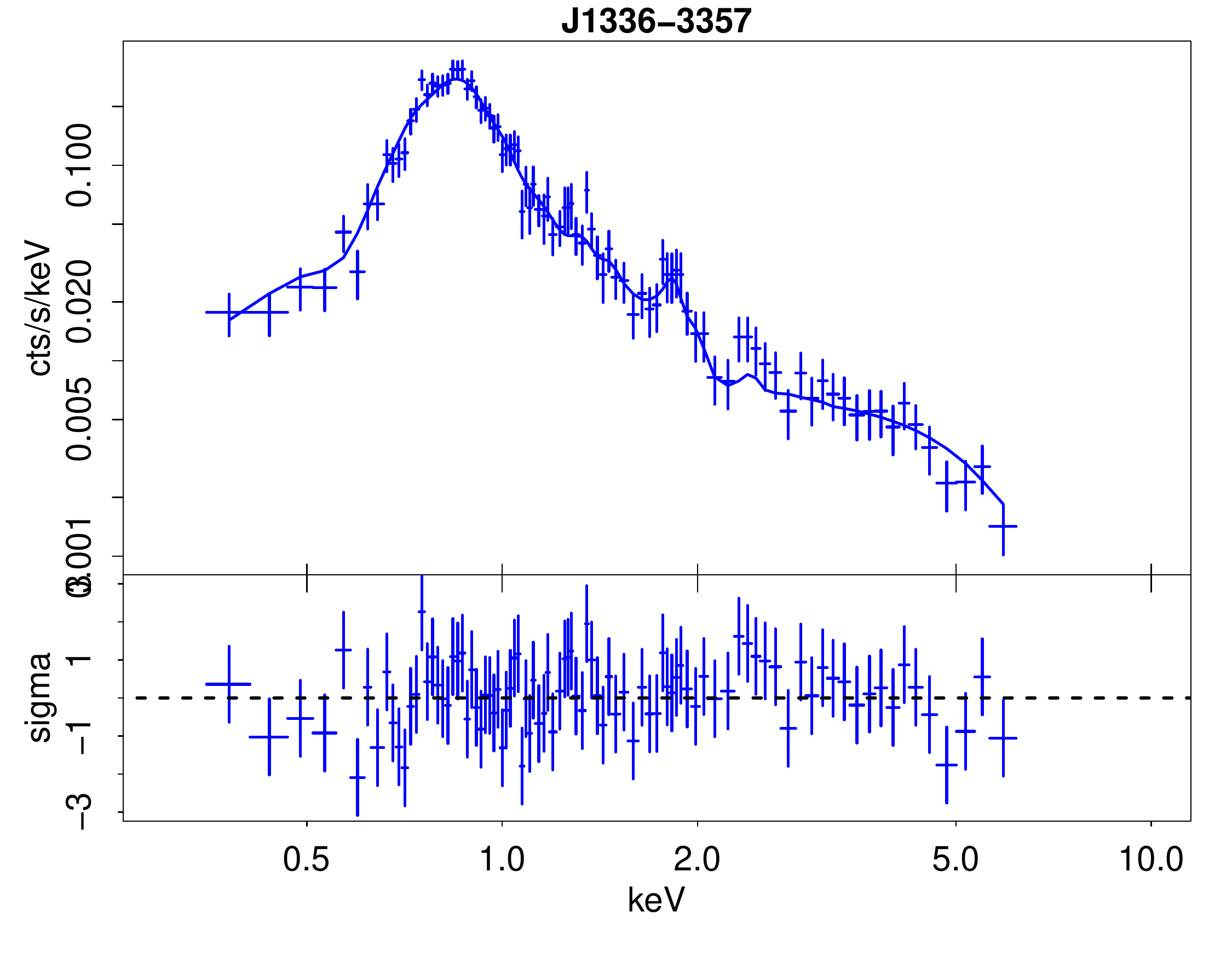}
   \includegraphics[scale=0.19]{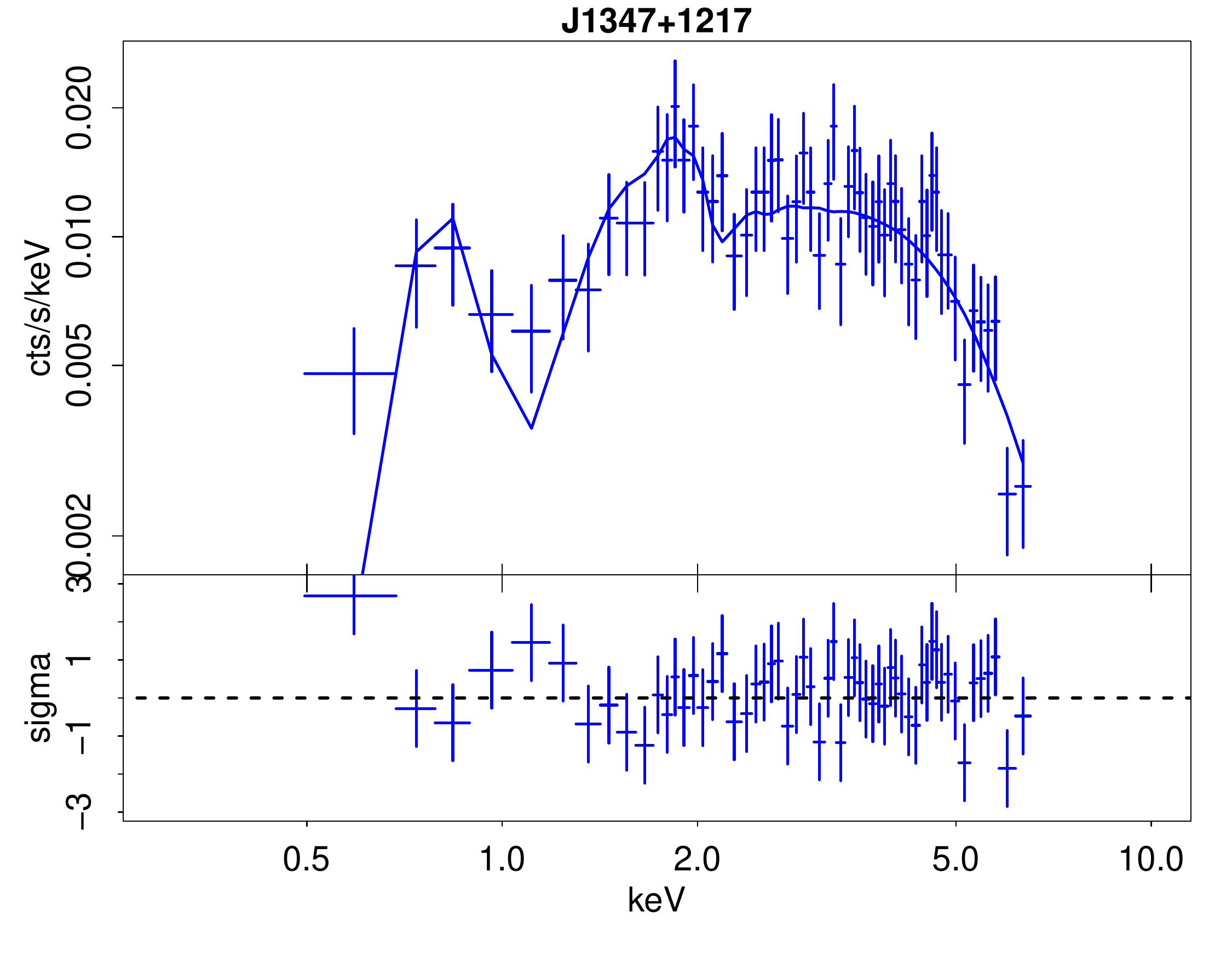}
   \includegraphics[scale=0.19]{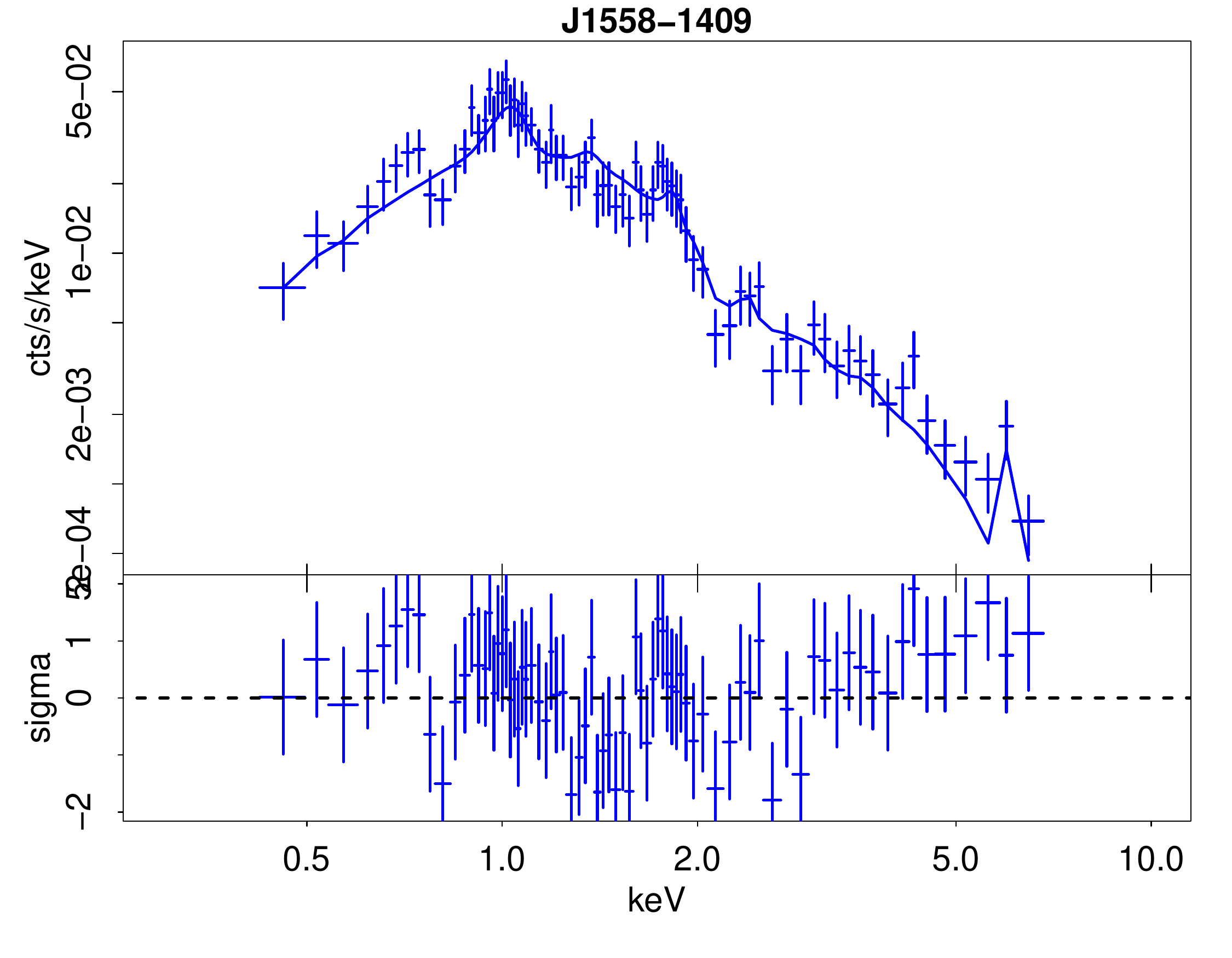}
   \includegraphics[scale=0.19]{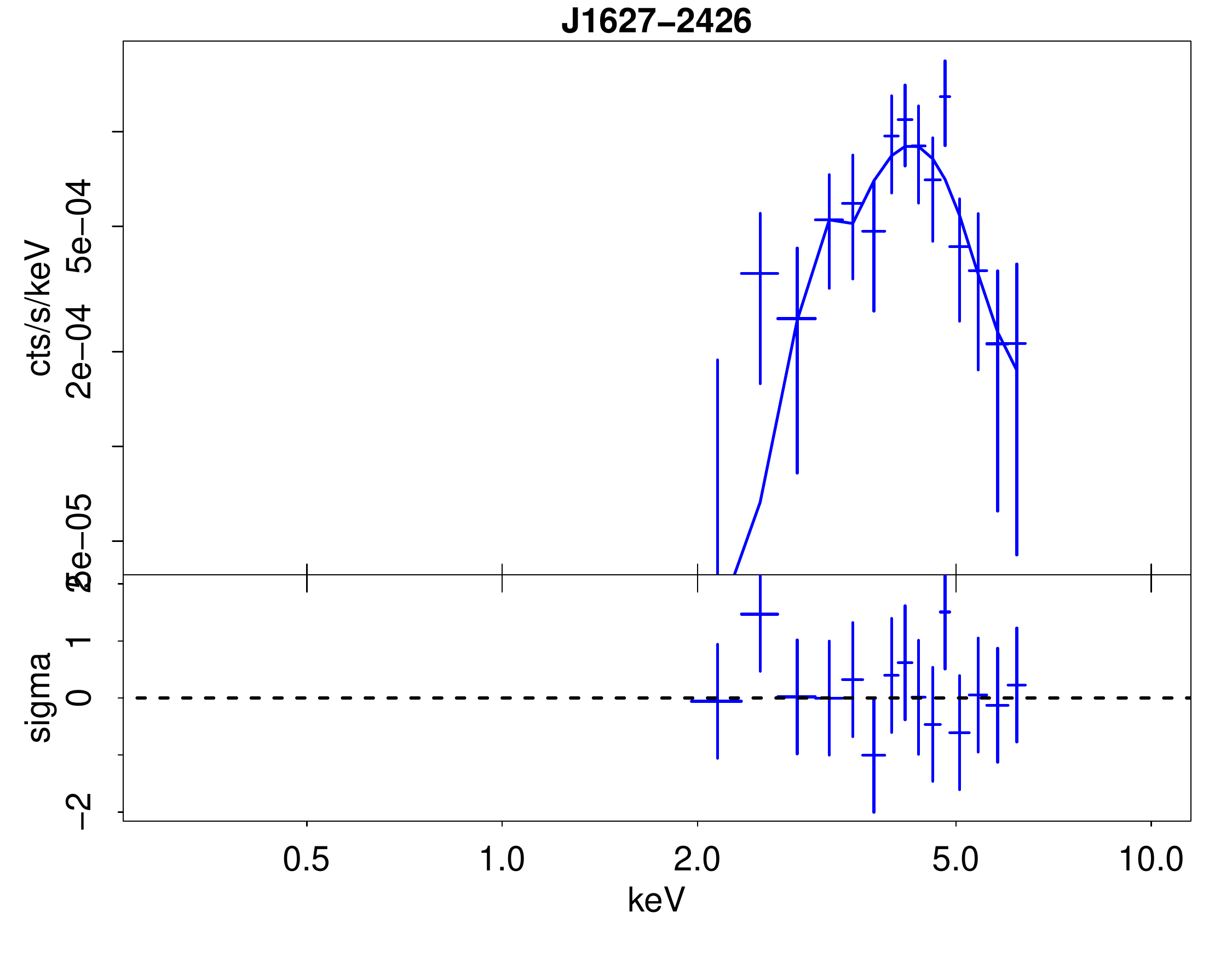}
   \includegraphics[scale=0.19]{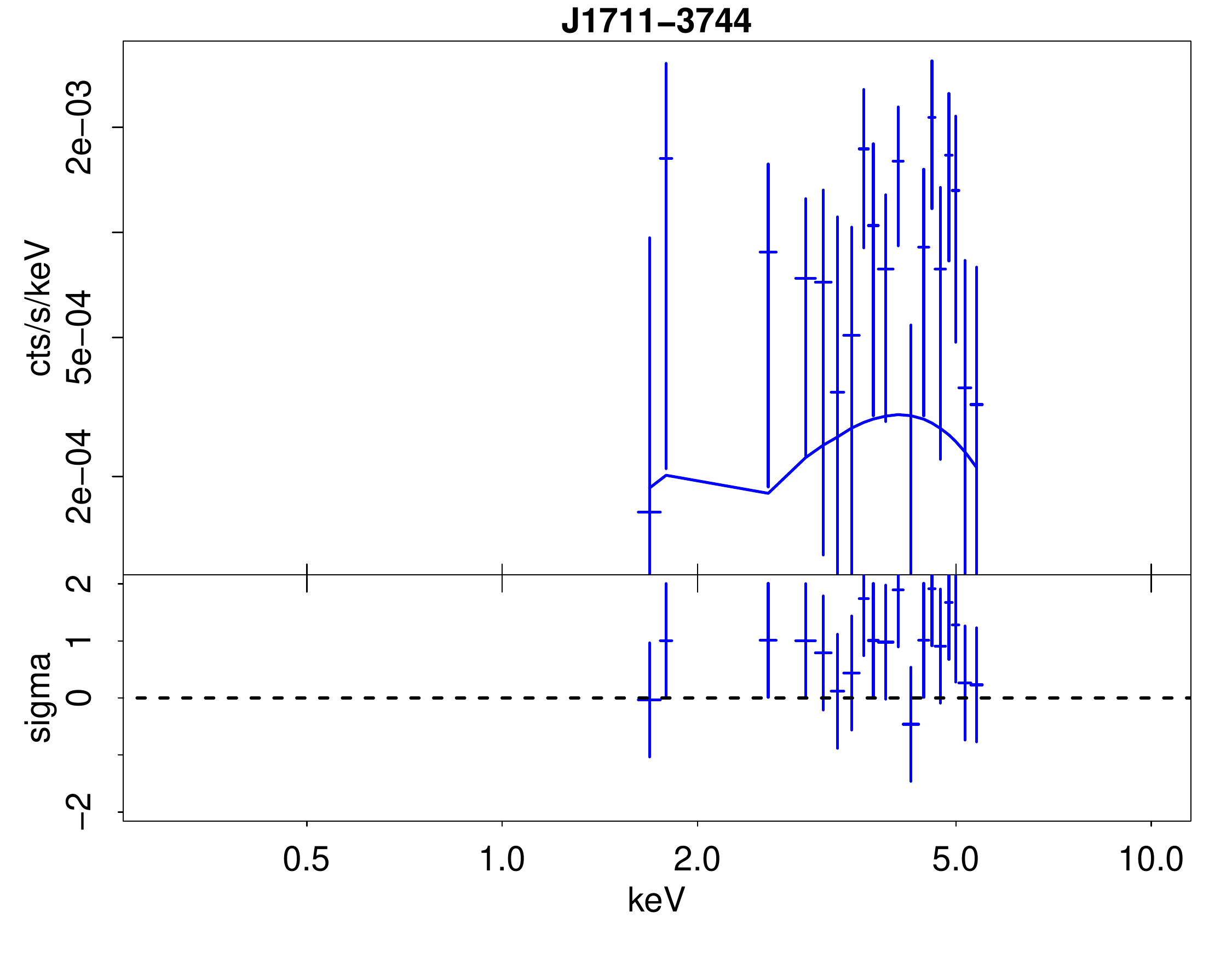}
   \includegraphics[scale=0.19]{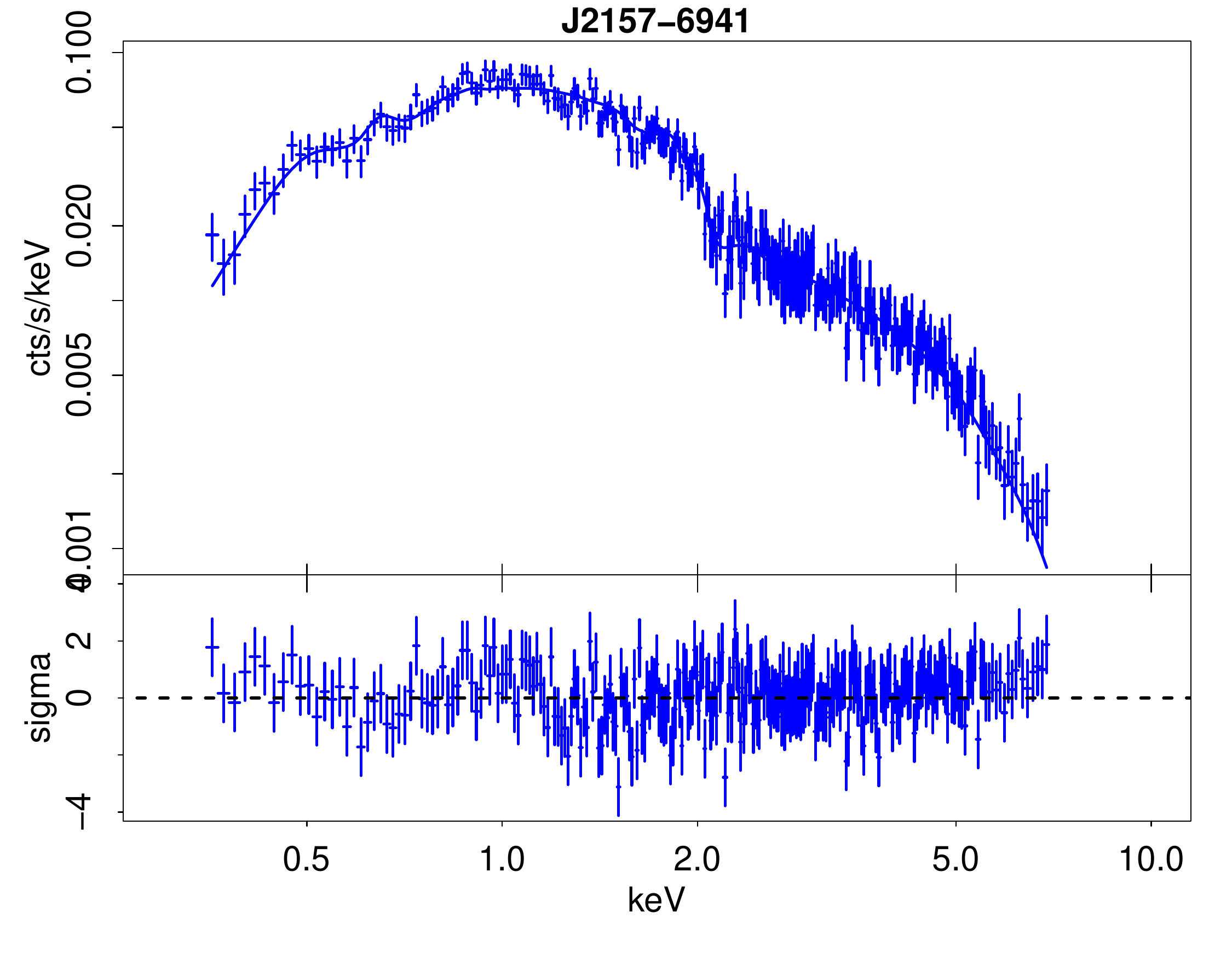}
   \includegraphics[scale=0.19]{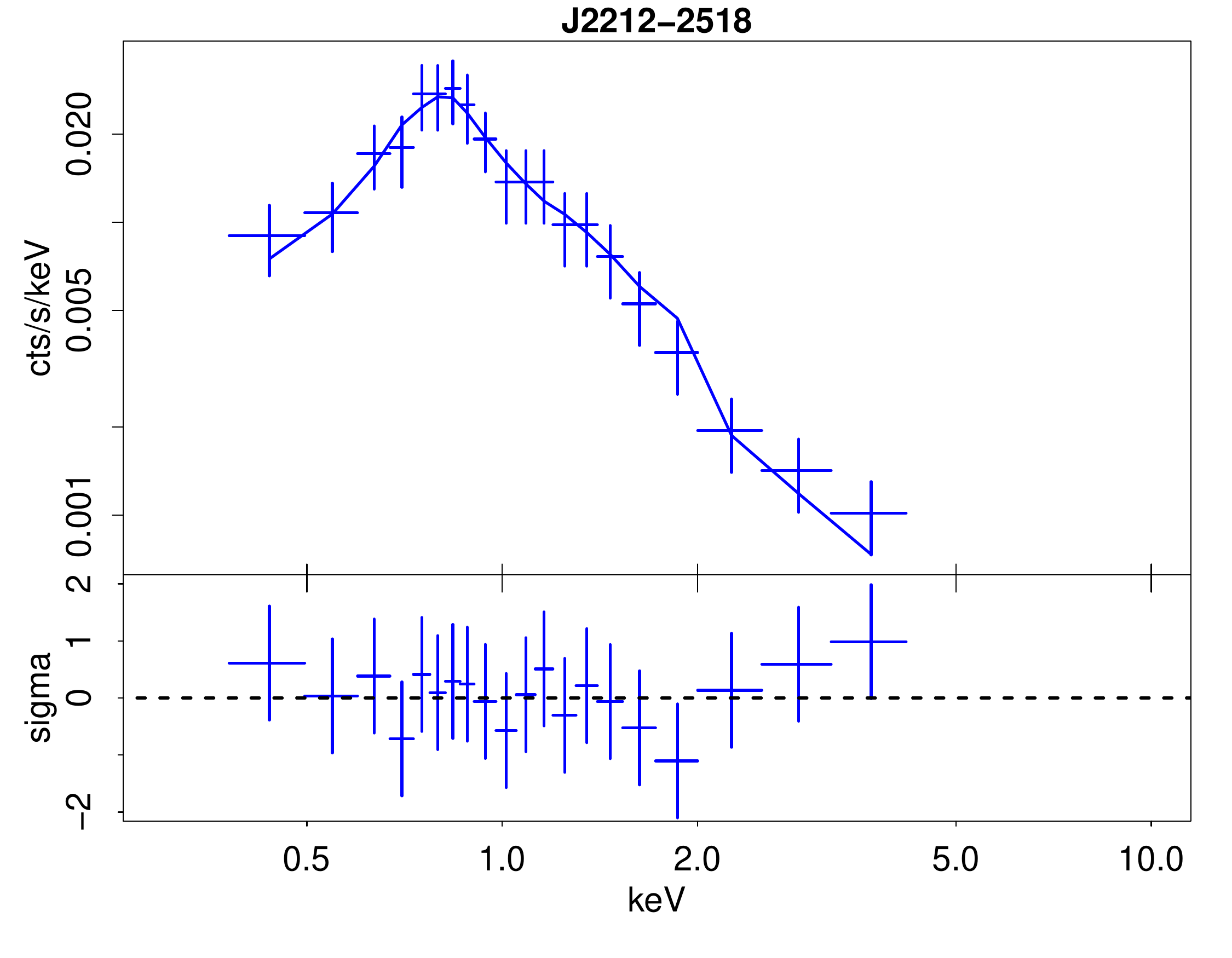}
\caption{\textit{Chandra}-ACIS spectra listed in Table \ref{tab:properties_peculiar} with their best-fit models (upper panels) and residuals (lower panels).}\label{fig:acis_spectra_peculiar}
\end{figure*}

\begin{figure*}
   \centering
   \includegraphics[scale=0.19]{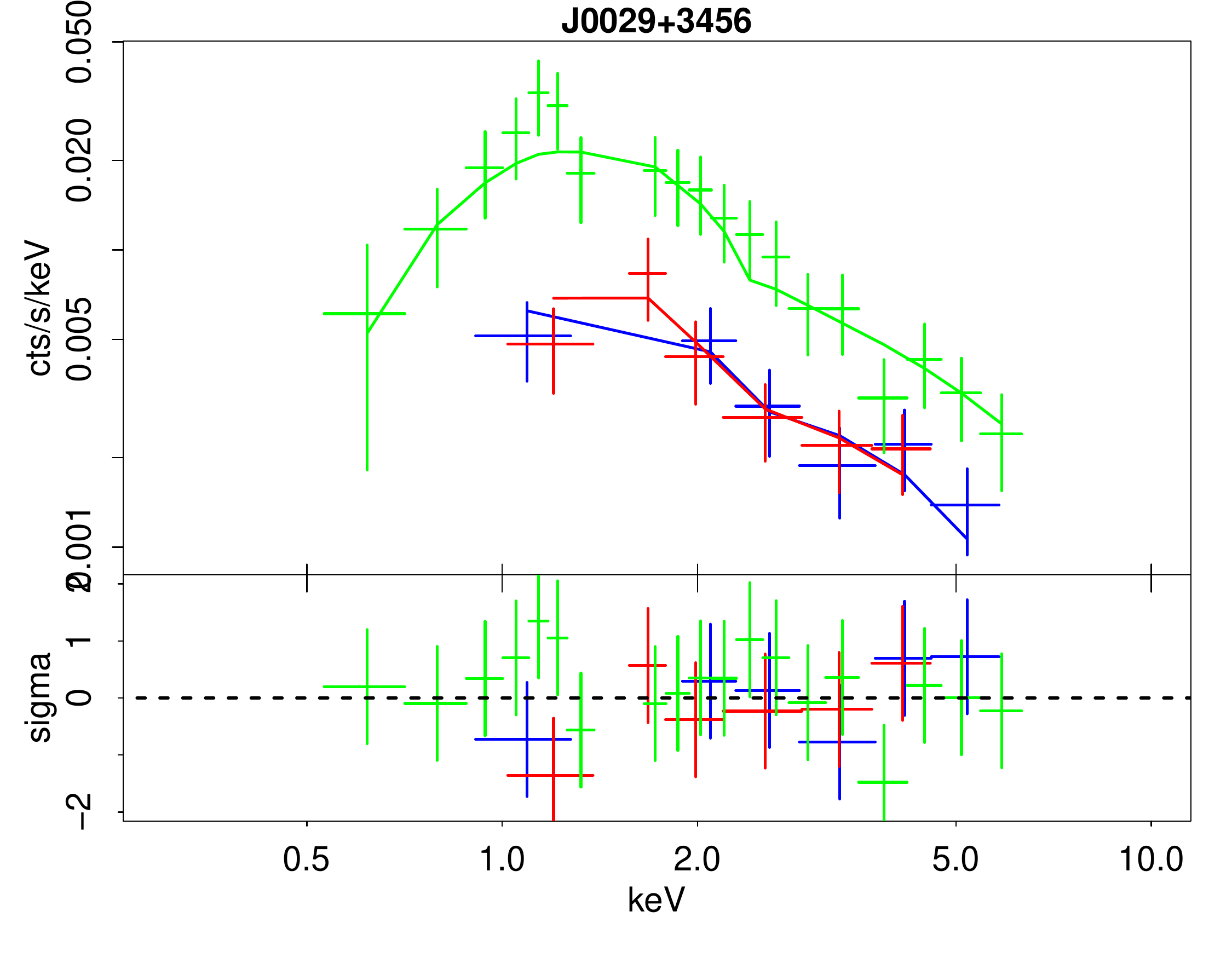}
   \includegraphics[scale=0.19]{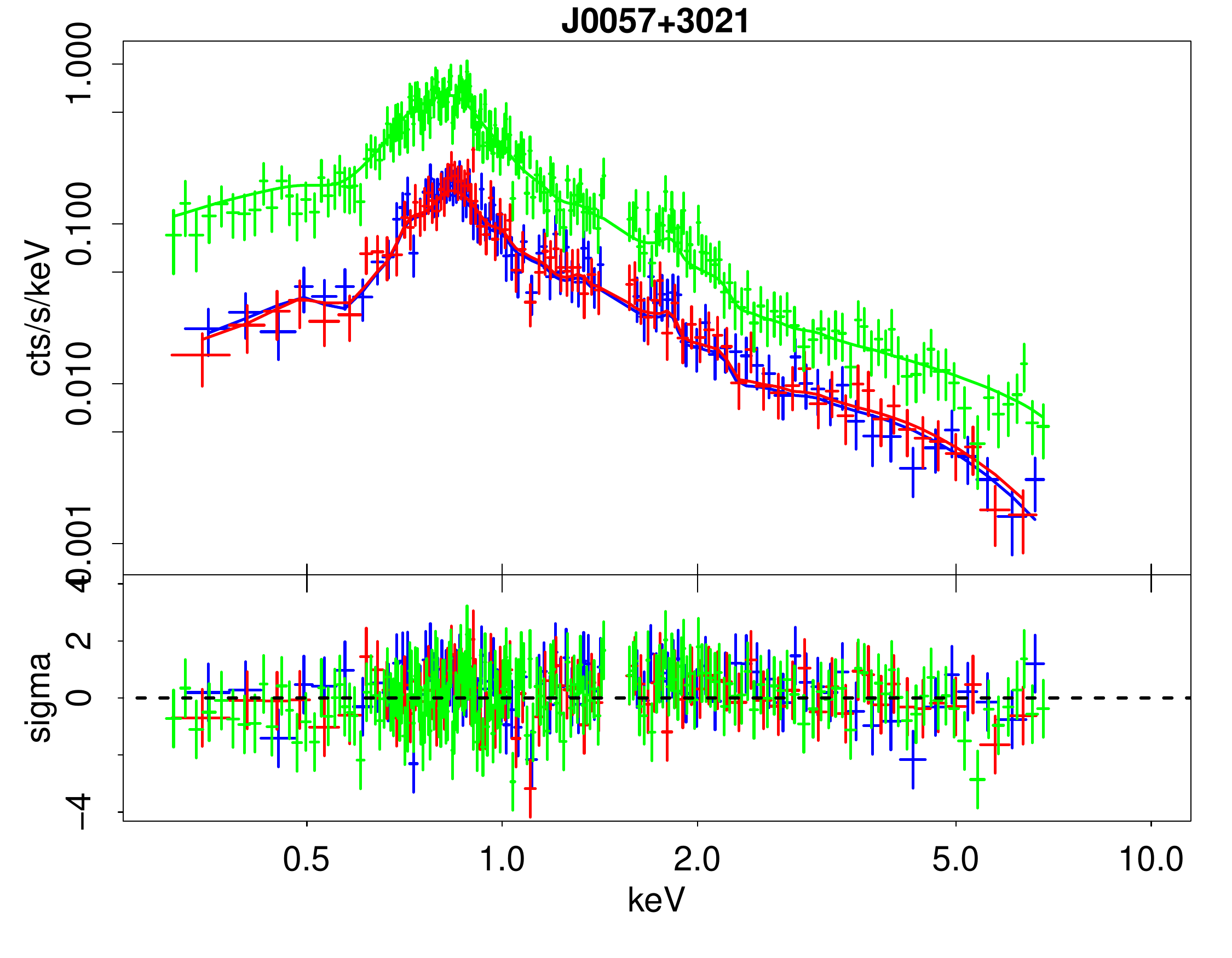}
   \includegraphics[scale=0.19]{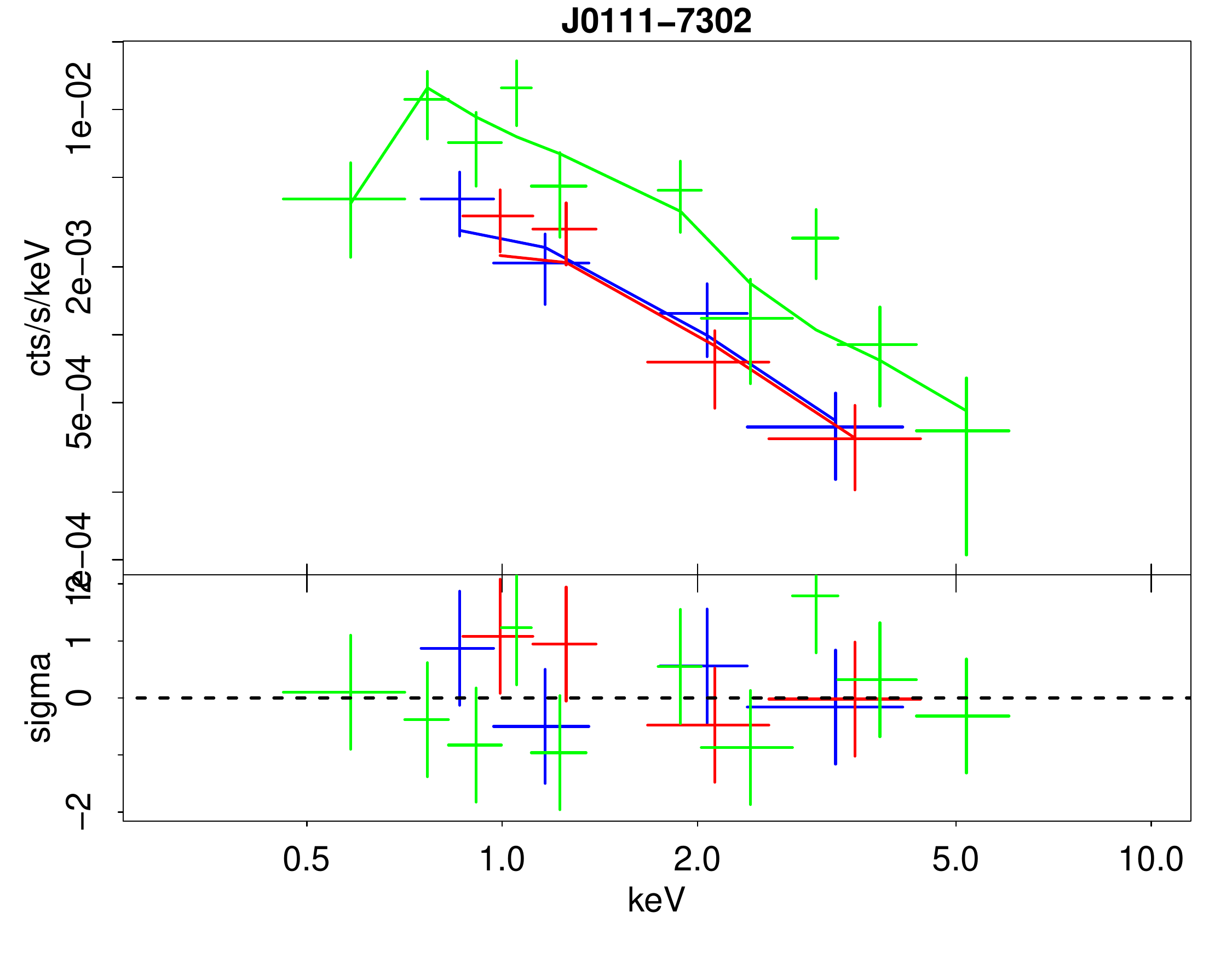}
   \includegraphics[scale=0.19]{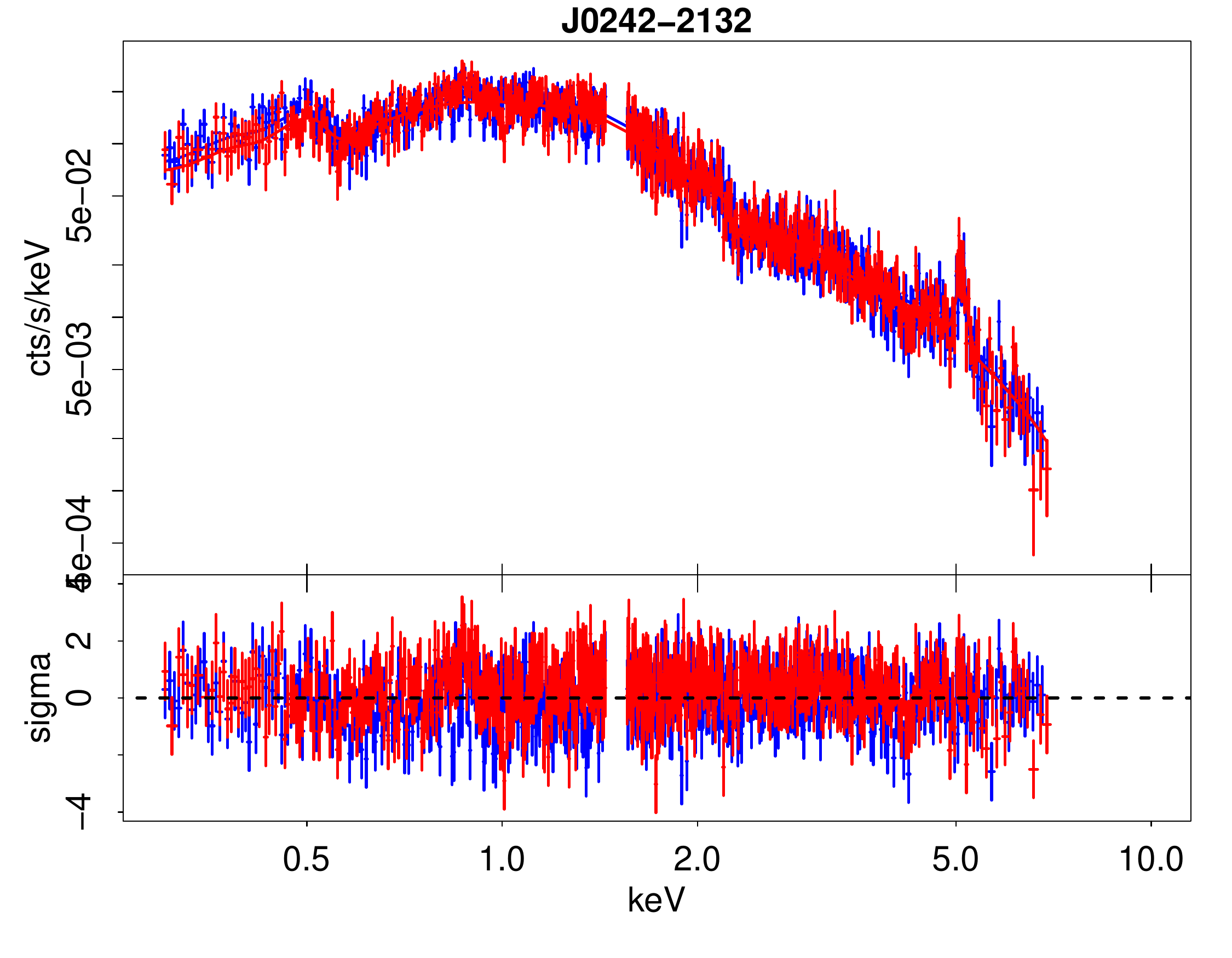}
   \includegraphics[scale=0.19]{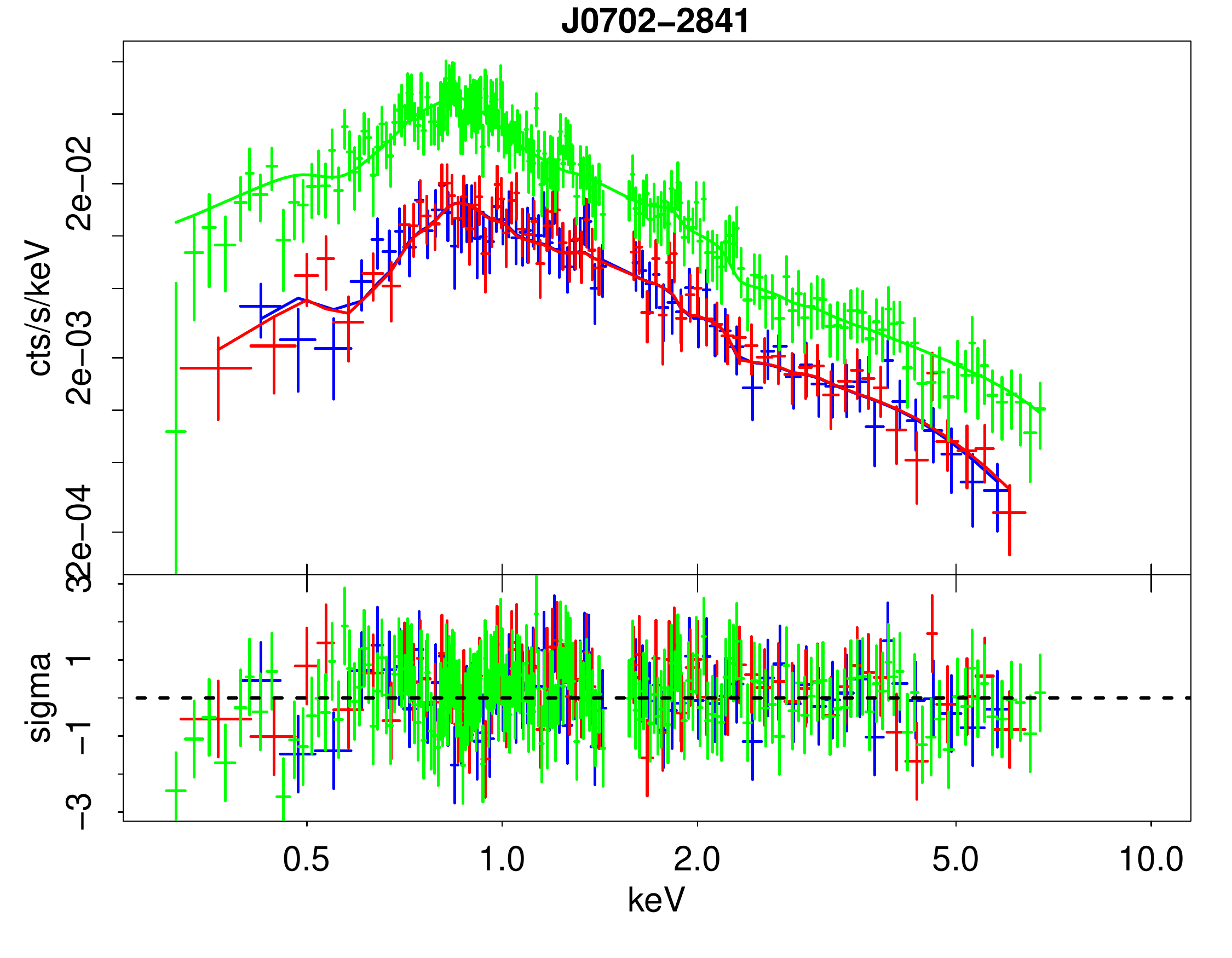}
   \includegraphics[scale=0.19]{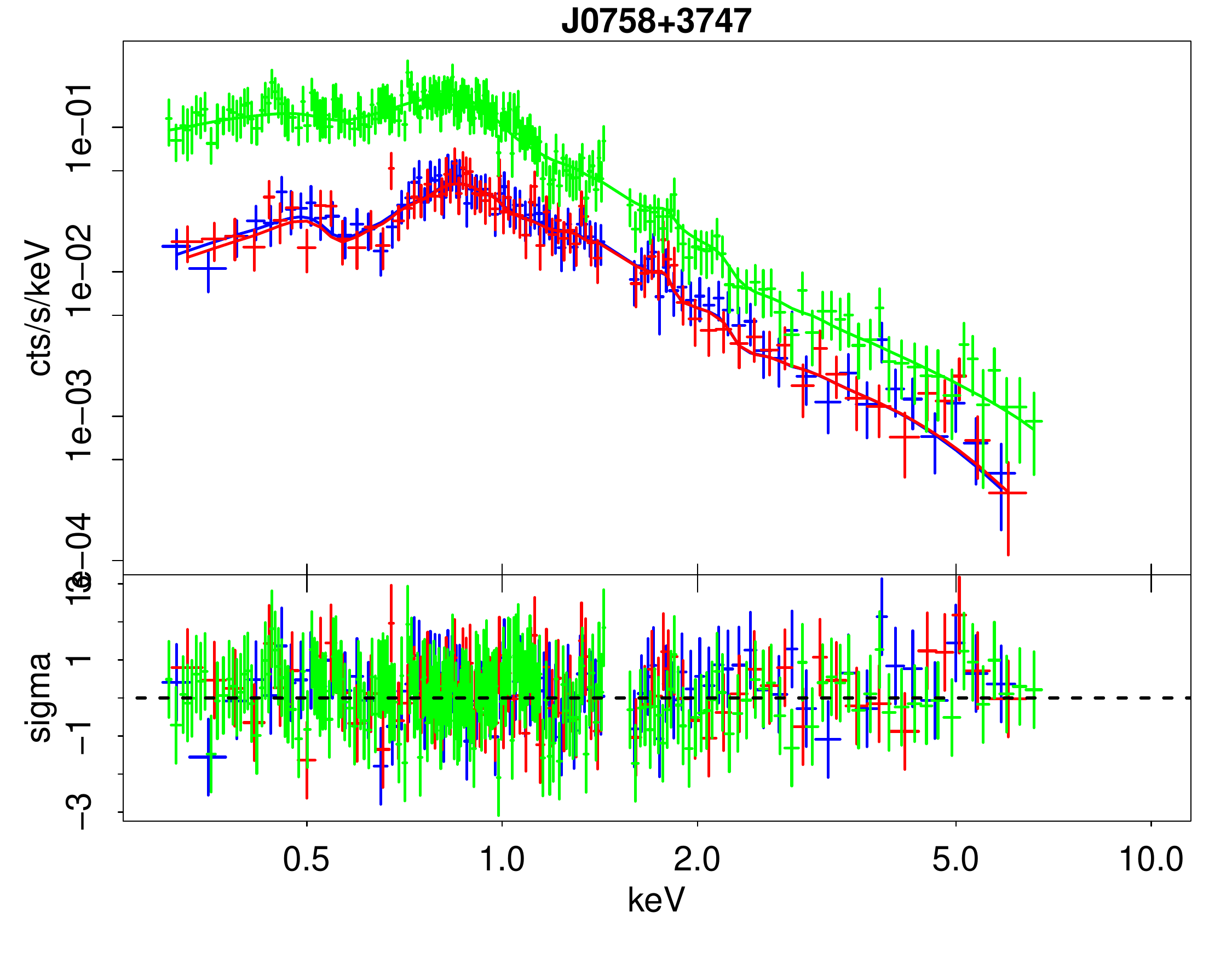}
   \includegraphics[scale=0.19]{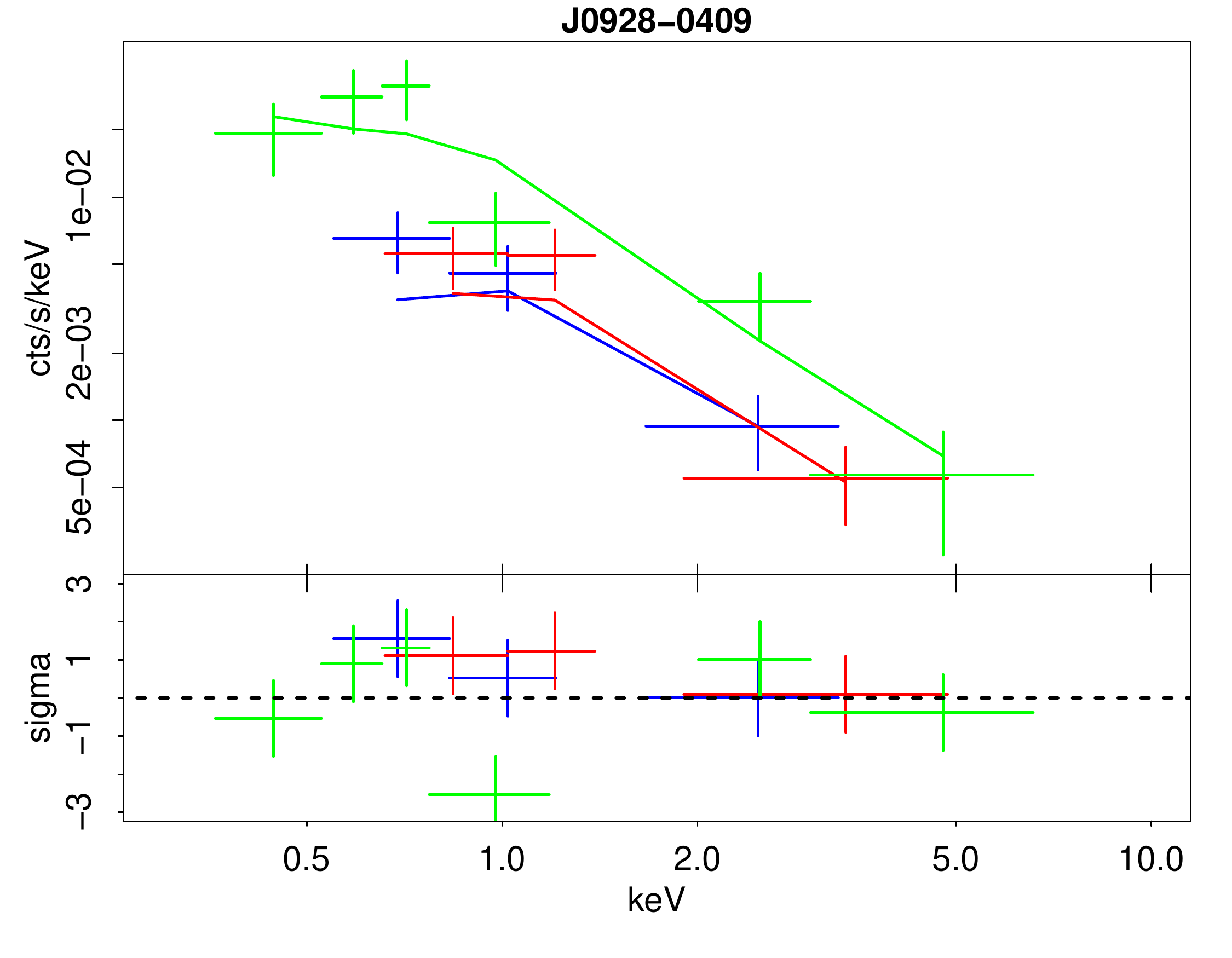}
   \includegraphics[scale=0.19]{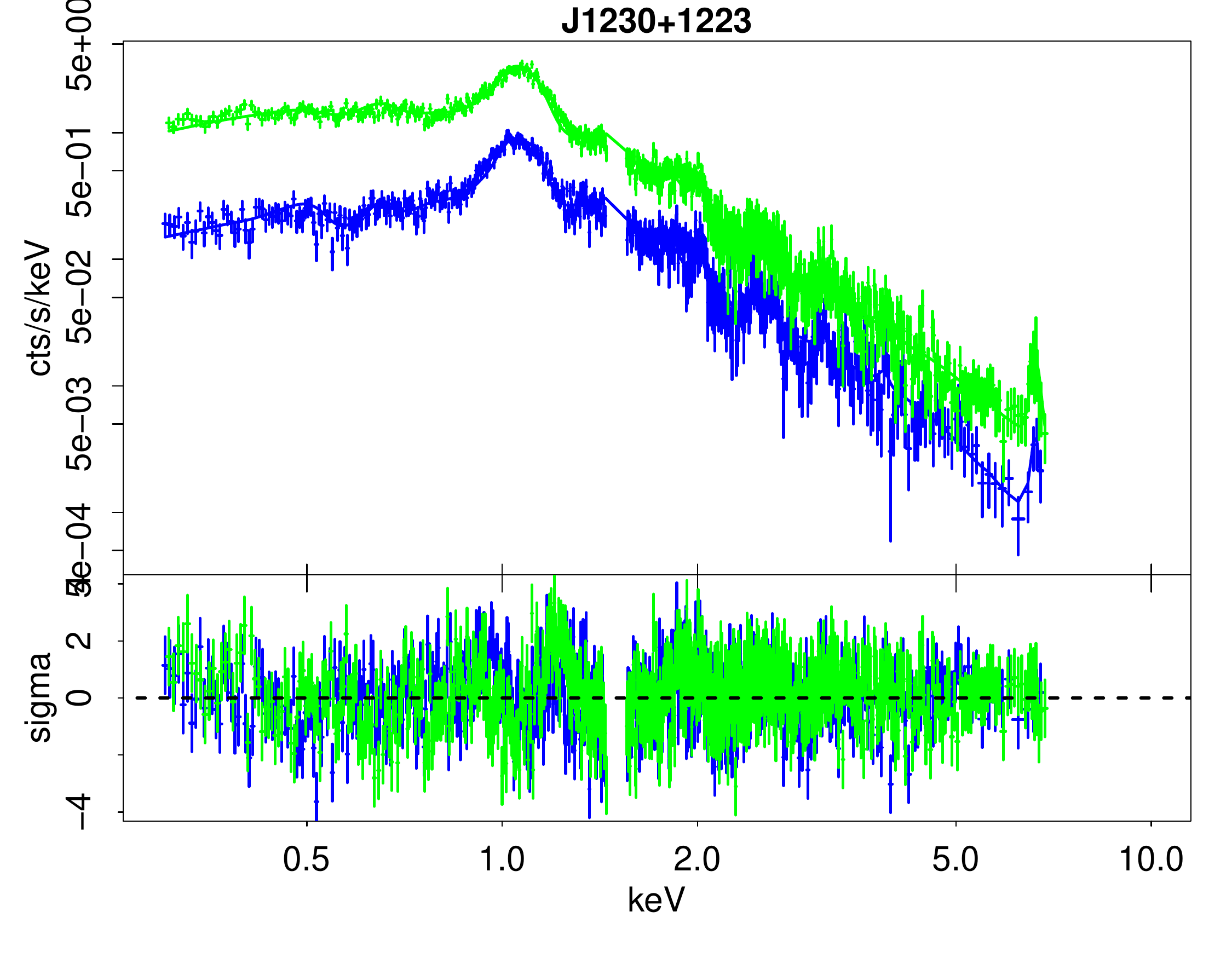}
   \includegraphics[scale=0.19]{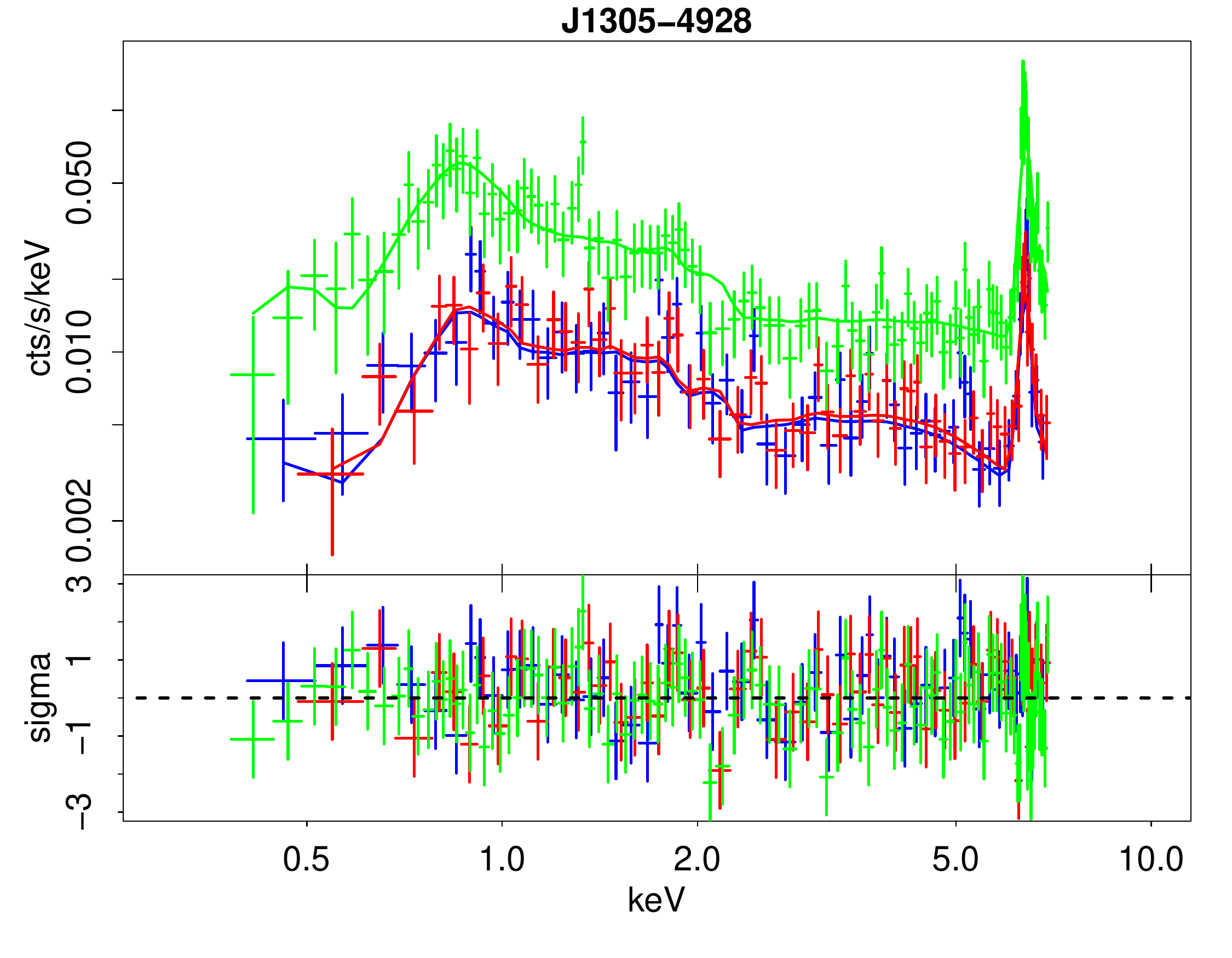}
   \includegraphics[scale=0.19]{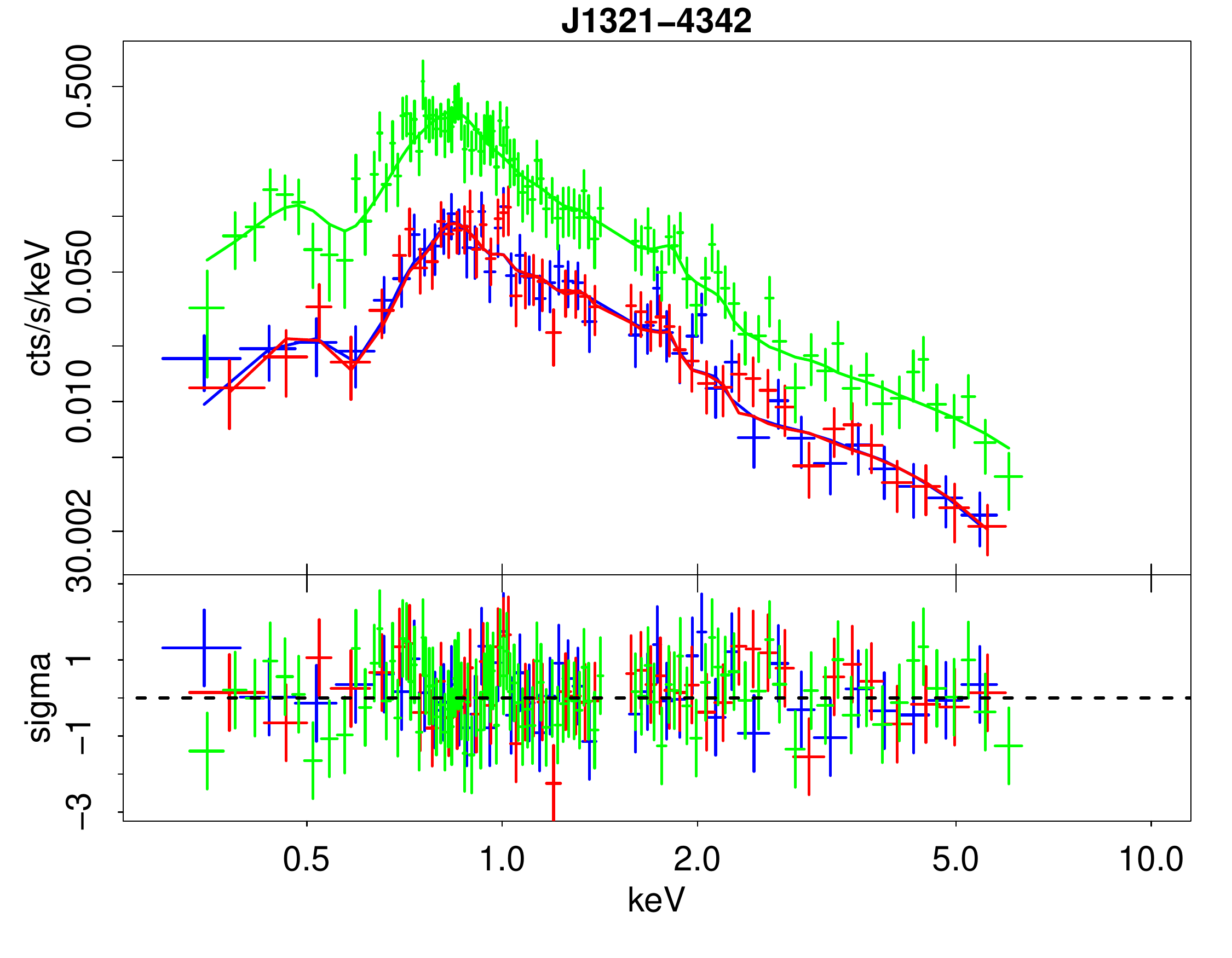}
   \includegraphics[scale=0.19]{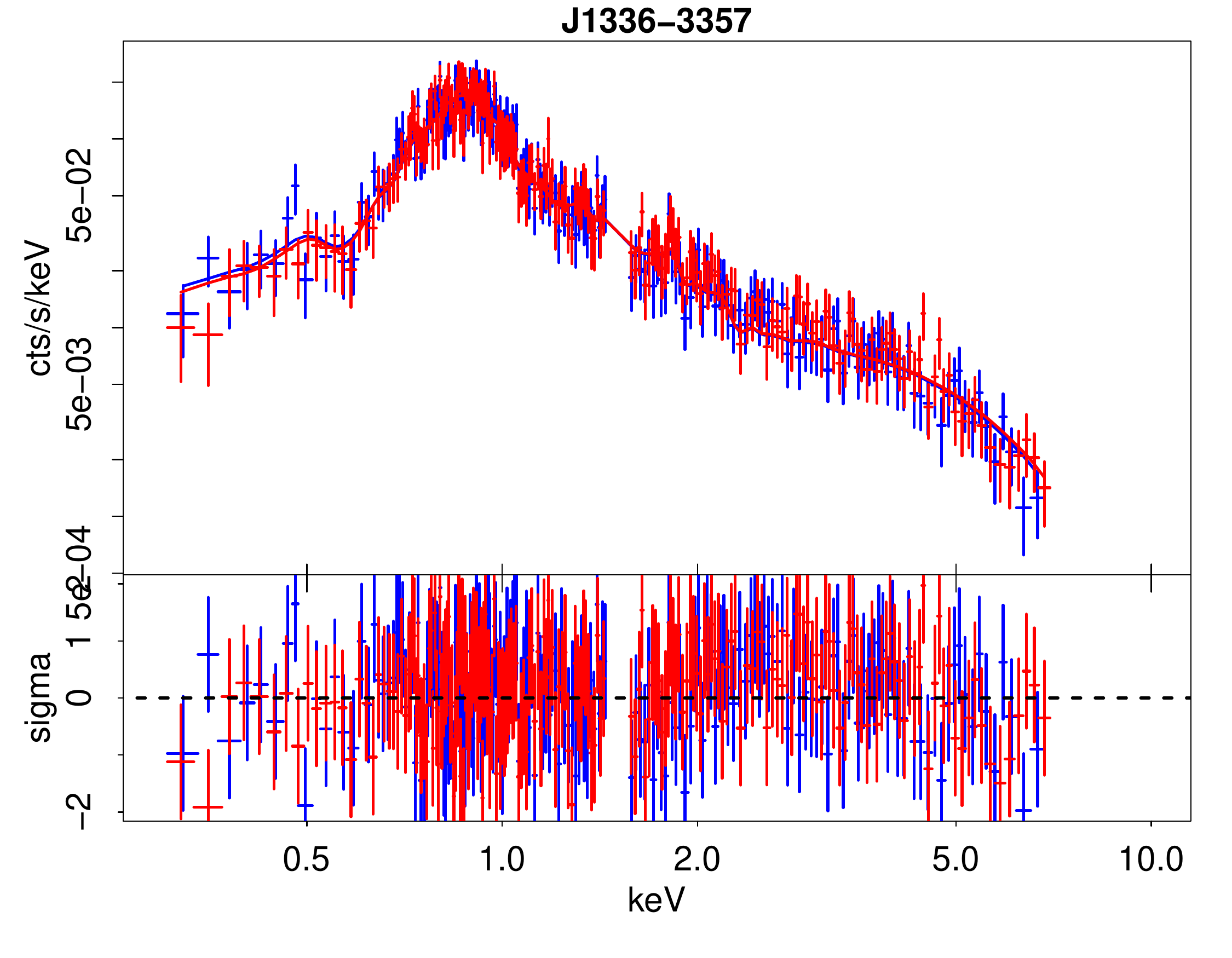}
   \includegraphics[scale=0.19]{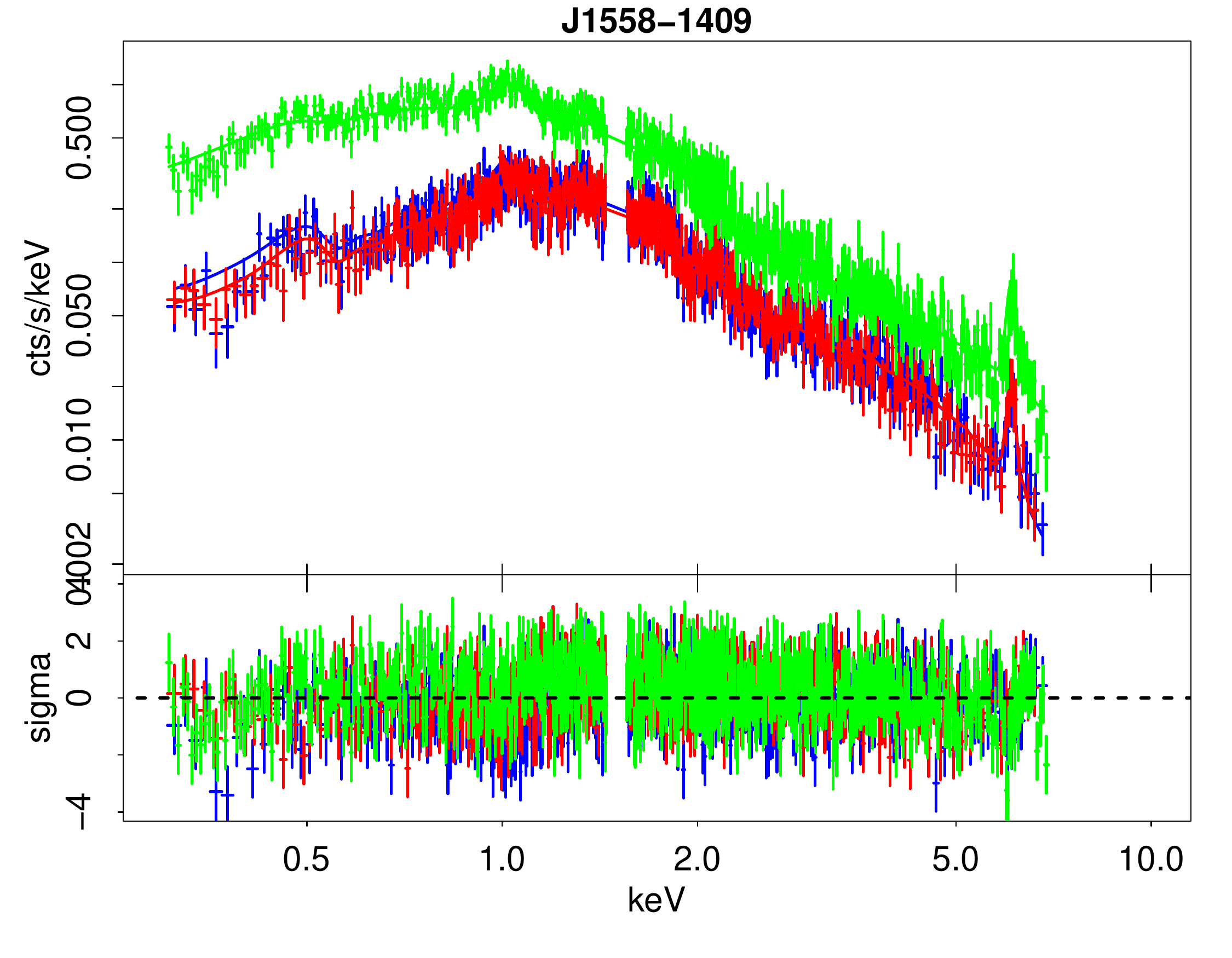}
   \includegraphics[scale=0.19]{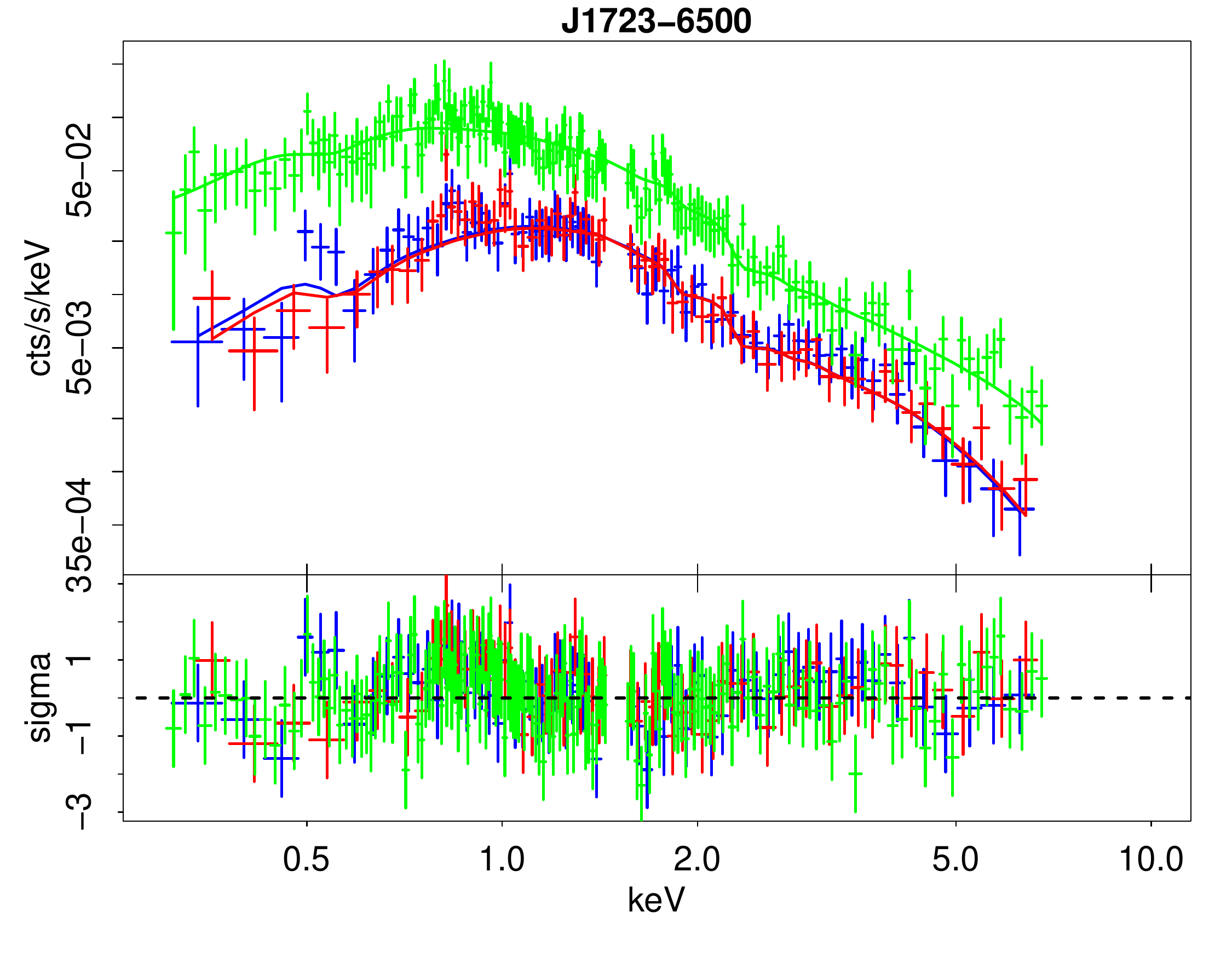}
\caption{\textit{XMM-Newton}-EPIC spectra listed in Table \ref{tab:properties_peculiar} with their best-fit models (upper panels) and residuals (lower panels). Lines and points in blue, red and green indicates models and data of MOS1, MOS2 and PN detectors, respectively.}\label{fig:epic_spectra_peculiar}
\end{figure*}

In this appendix we collect the results of the spectral fitting discussed in Sect. \ref{sec:xray-spectra}. In particular, in Figs. \ref{fig:xrt_spectra} and \ref{fig:acis_spectra}, \ref{fig:epic_spectra} we show the spectra fitted with a power-law model, while in Figs. \ref{fig:xrt_spectra_peculiar}, \ref{fig:acis_spectra_peculiar} and \ref{fig:epic_spectra_peculiar} we show spectra fitted with the models listed in Tab. \ref{tab:properties_peculiar}.

\section{Optical Spectra}
\begin{figure*}
   \centering
\stackunder[1pt]{J0024-0811}{\includegraphics[scale=0.15]{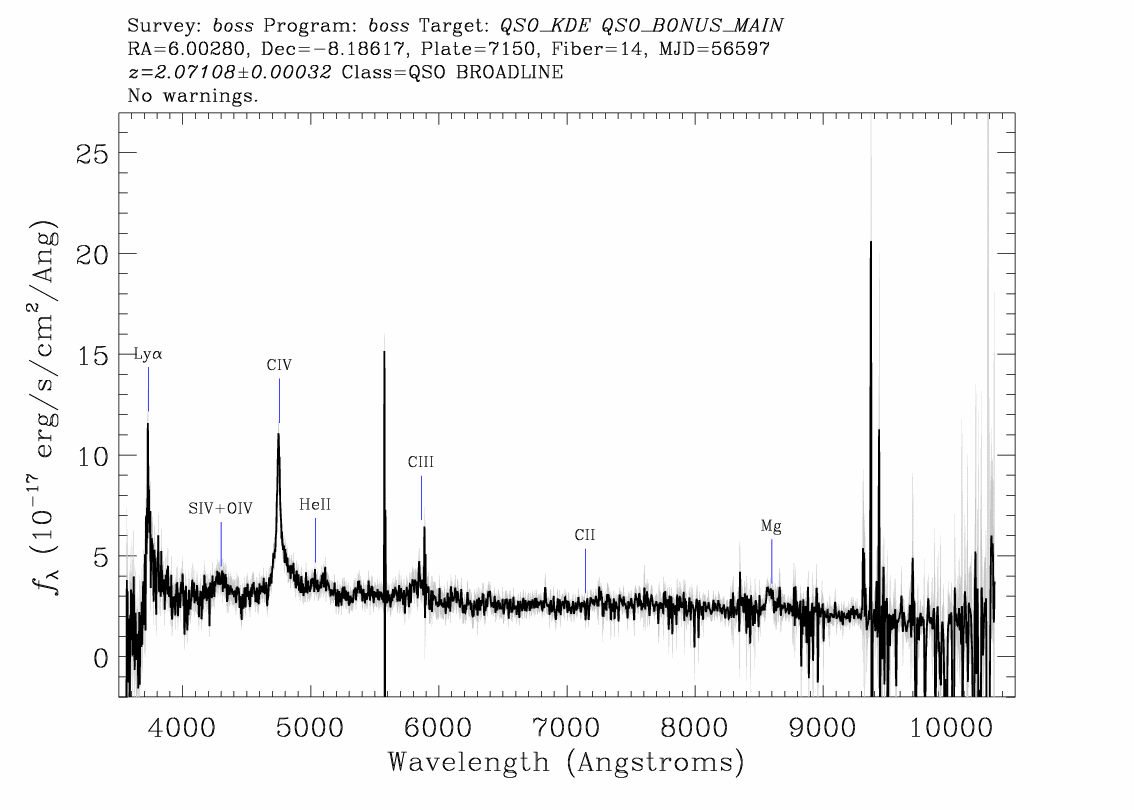}}
\stackunder[1pt]{J0028+2000}{\includegraphics[scale=0.15]{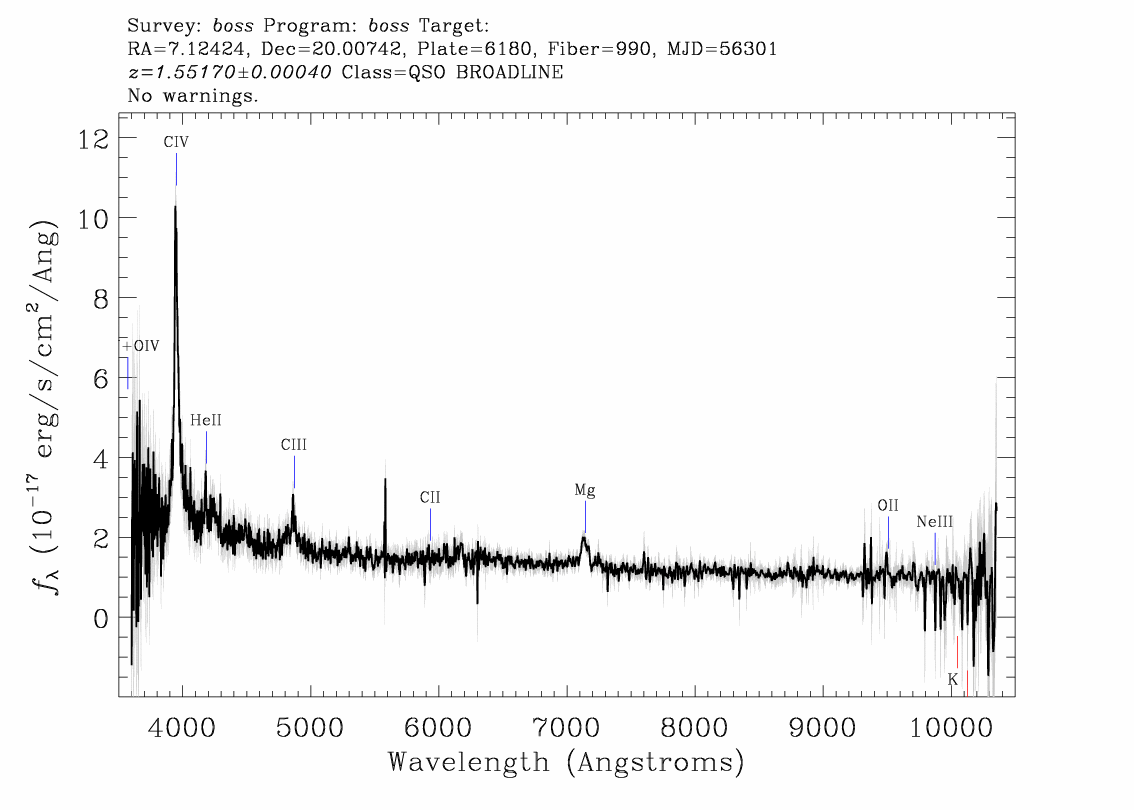}}
\stackunder[1pt]{J0029-0113}{\includegraphics[scale=0.15]{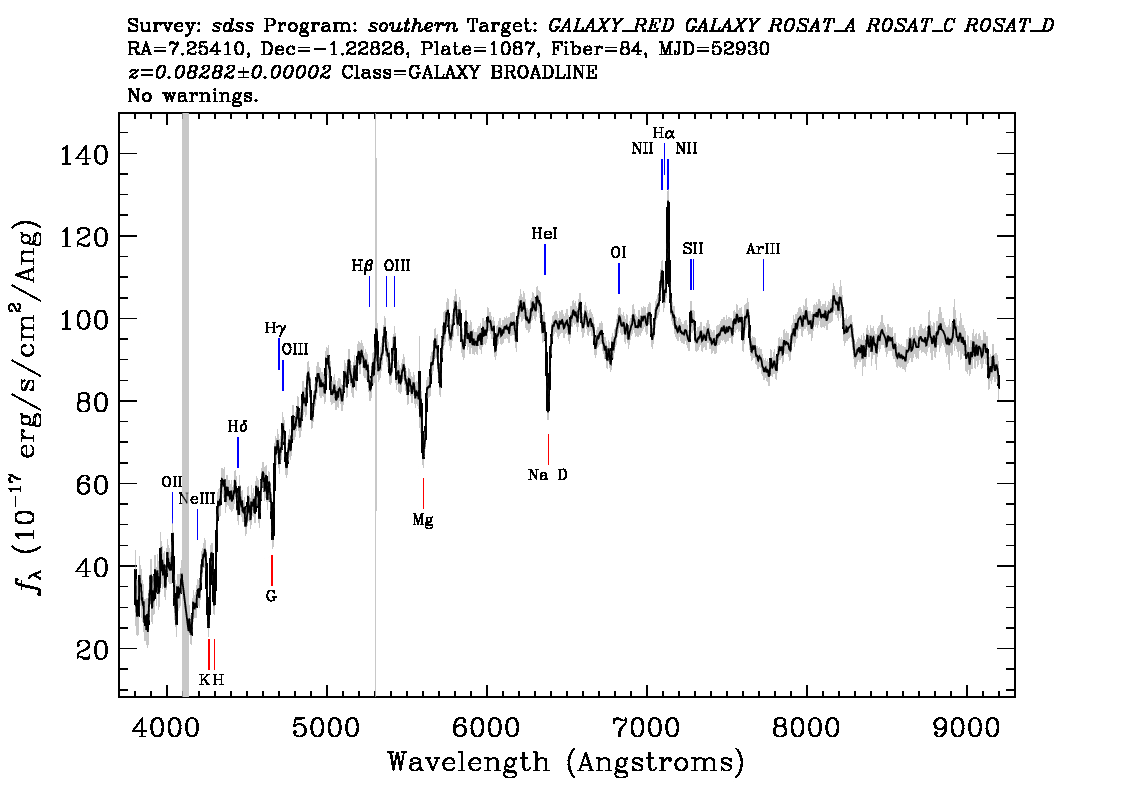}}
\stackunder[1pt]{J0029+3456}{\includegraphics[scale=0.15]{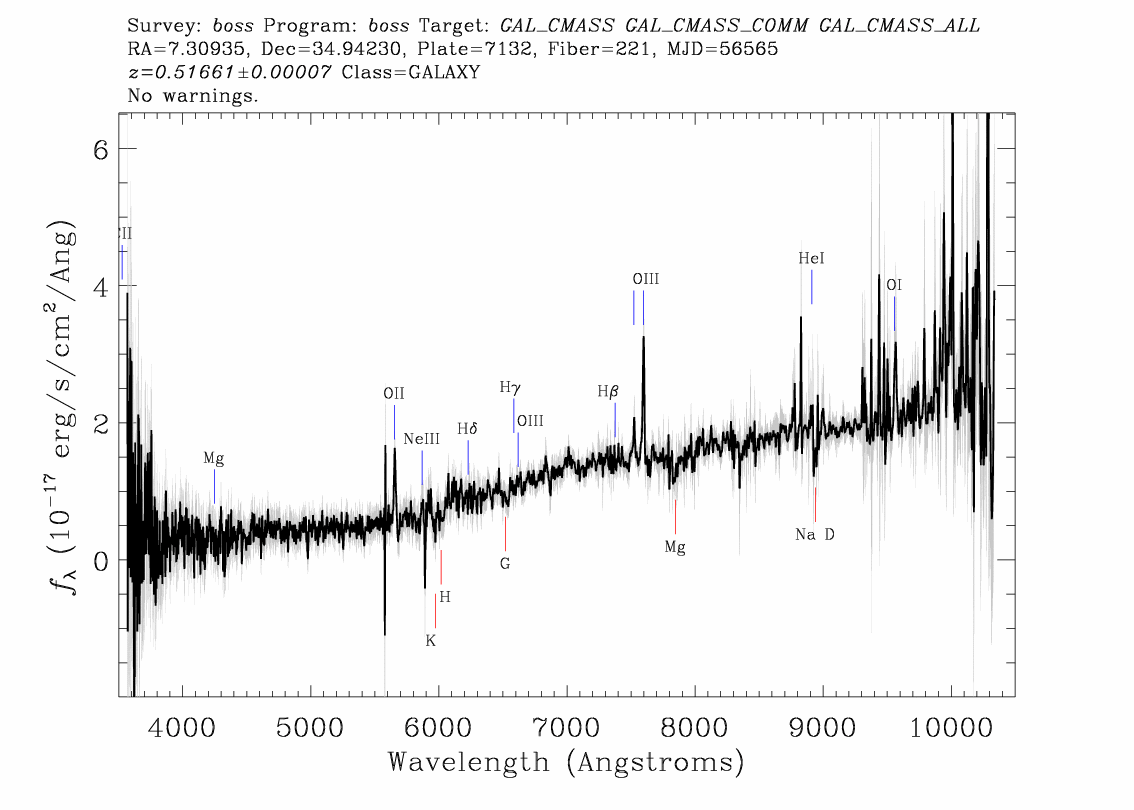}}
\stackunder[1pt]{J0034-0054}{\includegraphics[scale=0.15]{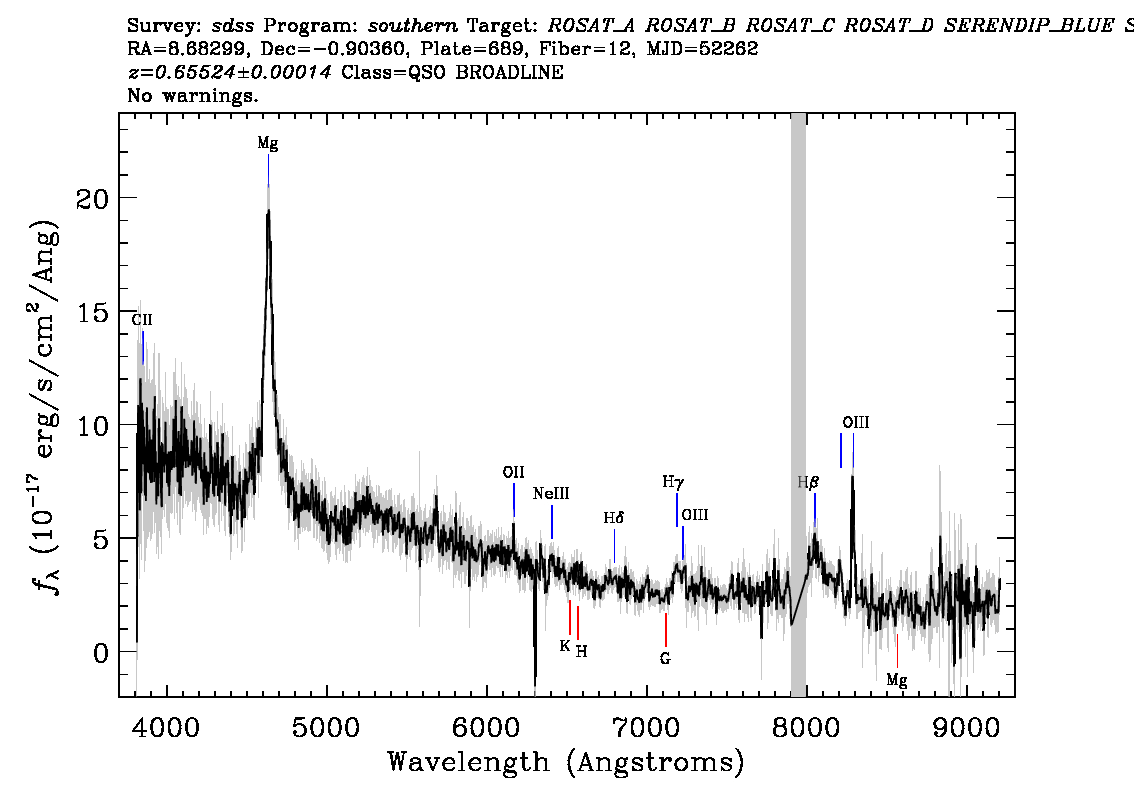}}
\stackunder[1pt]{J0107+1312}{\includegraphics[scale=0.15]{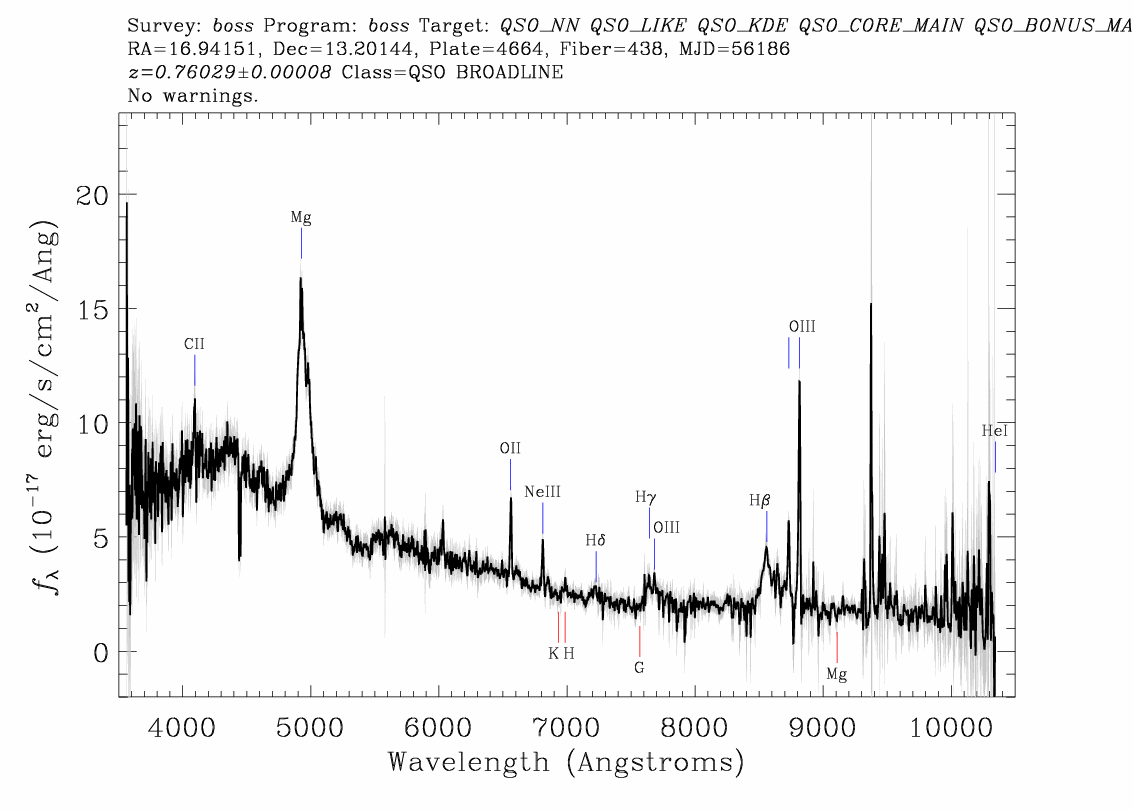}}
\stackunder[1pt]{J0112-0328}{\includegraphics[scale=0.15]{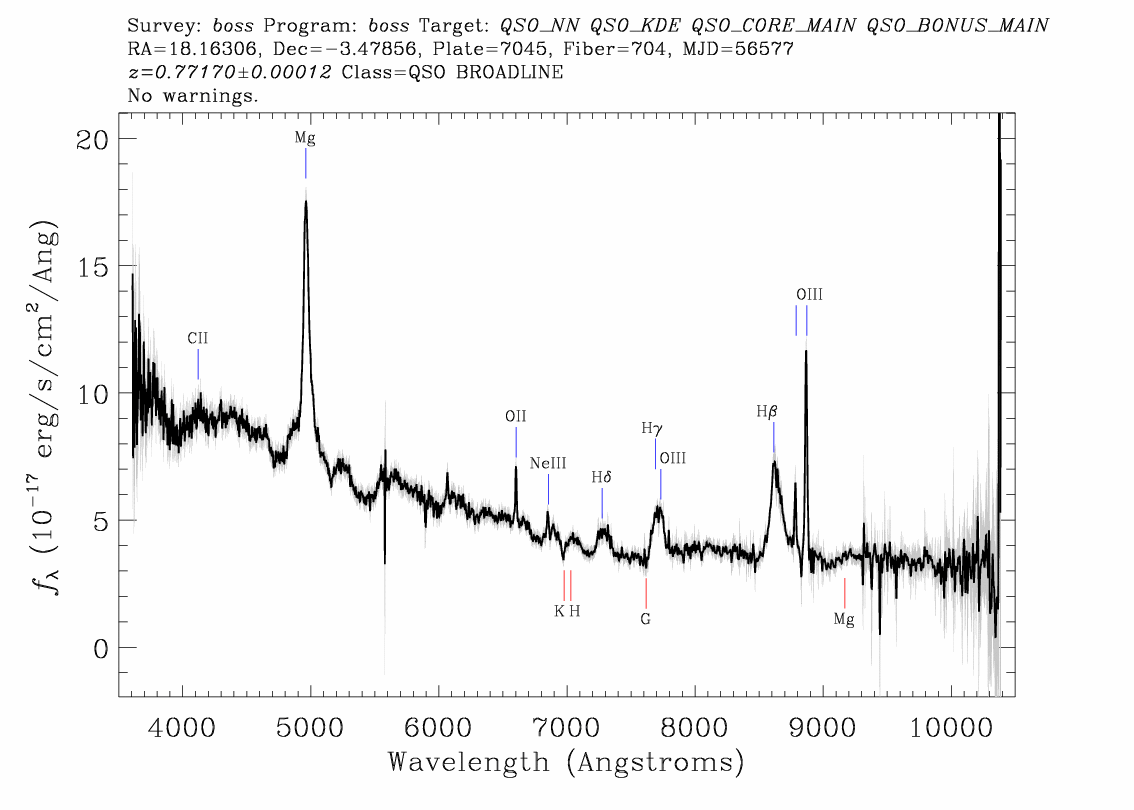}}
\stackunder[1pt]{J0123-0923}{\includegraphics[scale=0.15]{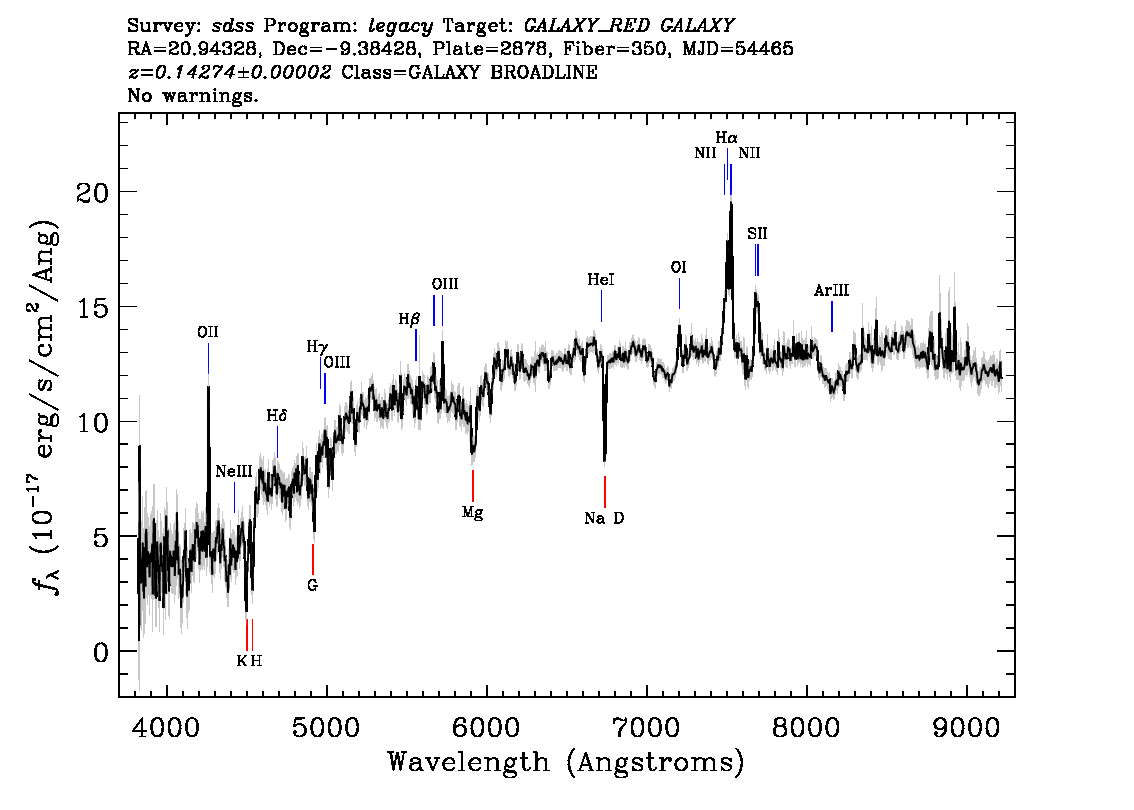}}
\stackunder[1pt]{J0134-0931}{\includegraphics[scale=0.15]{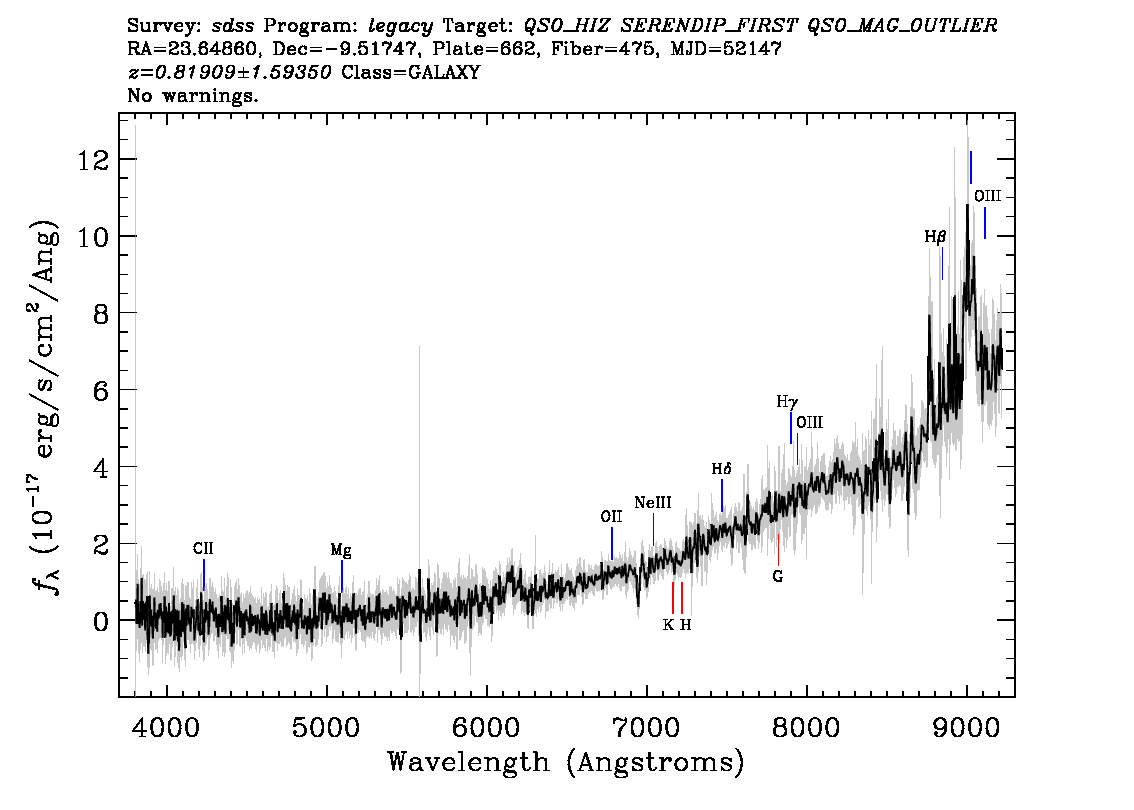}}
\stackunder[1pt]{J0139+1753}{\includegraphics[scale=0.15]{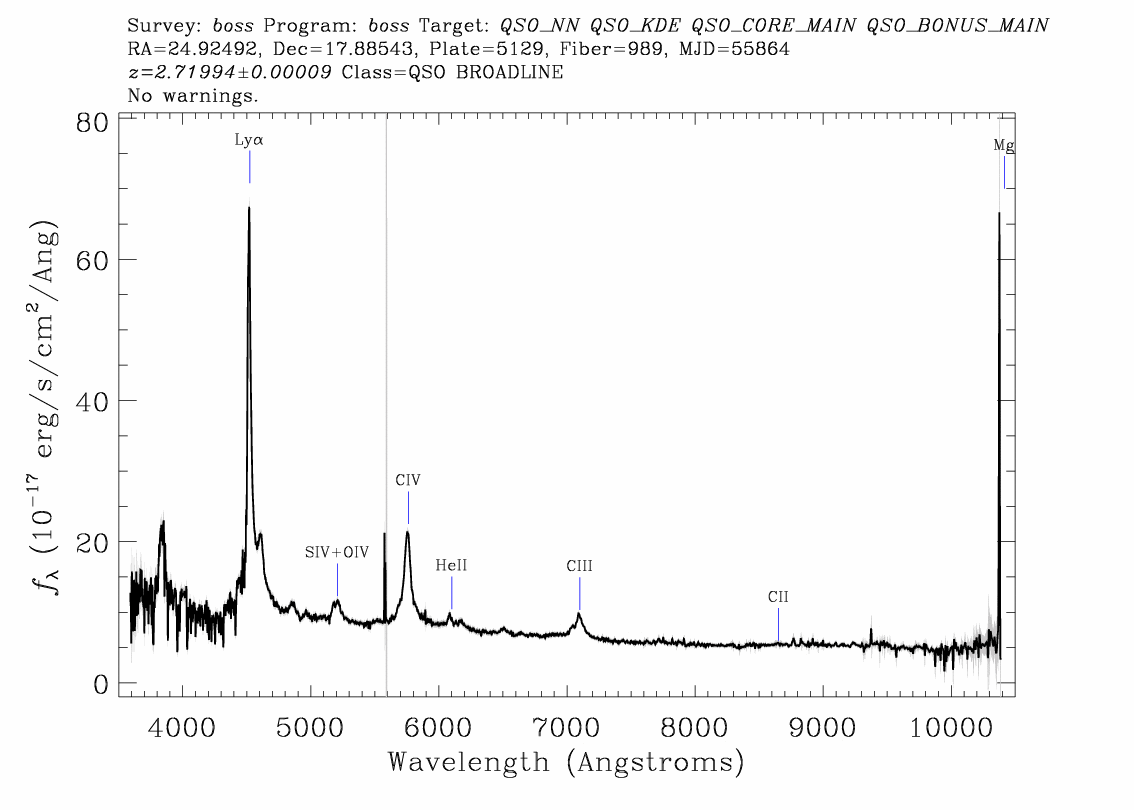}}
\caption{SDSS DR12 optical spectra for the 97 ABC sources discussed in Sect. \ref{sec:sdss}. Full set of figures available online.}\label{fig:sdss_spectra}%
\end{figure*}

\begin{figure*}
   \centering
\stackunder[1pt]{J0011+0823}{\includegraphics[scale=0.18]{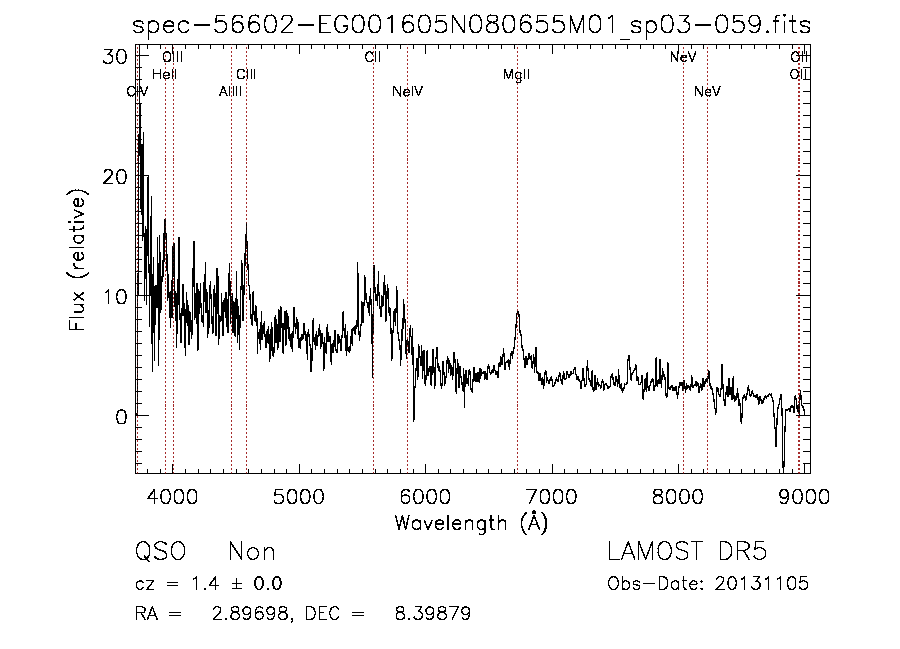}}
\stackunder[1pt]{J0119+3210}{\includegraphics[scale=0.18]{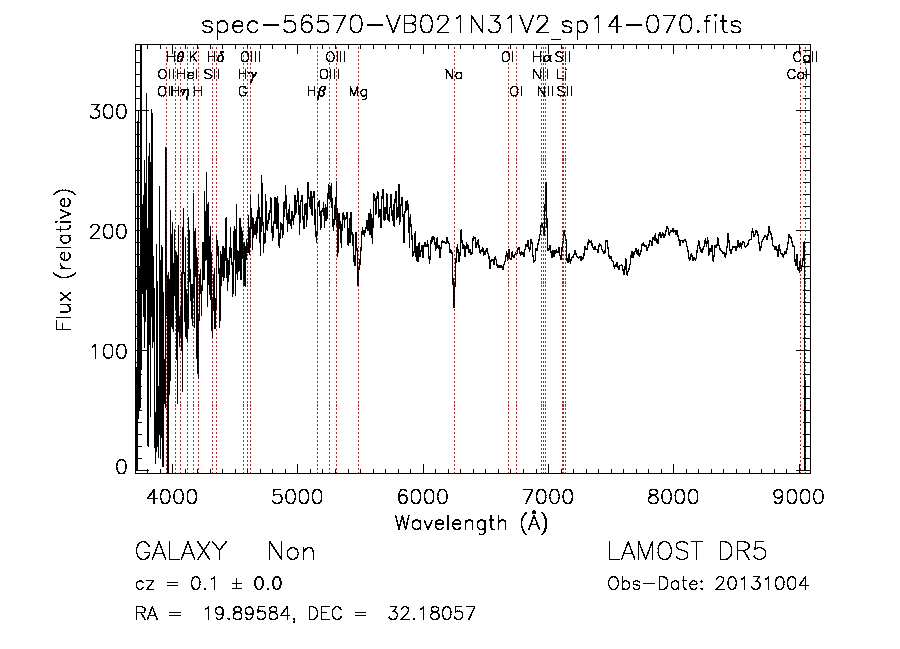}}
\stackunder[1pt]{J0137+3309}{\includegraphics[scale=0.18]{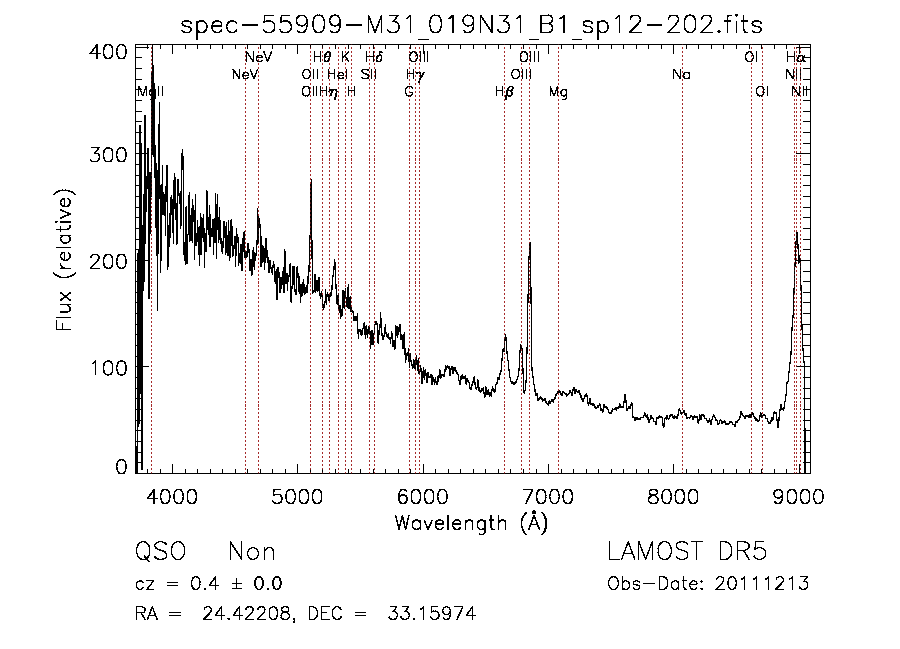}}
\stackunder[1pt]{J0152+3350}{\includegraphics[scale=0.18]{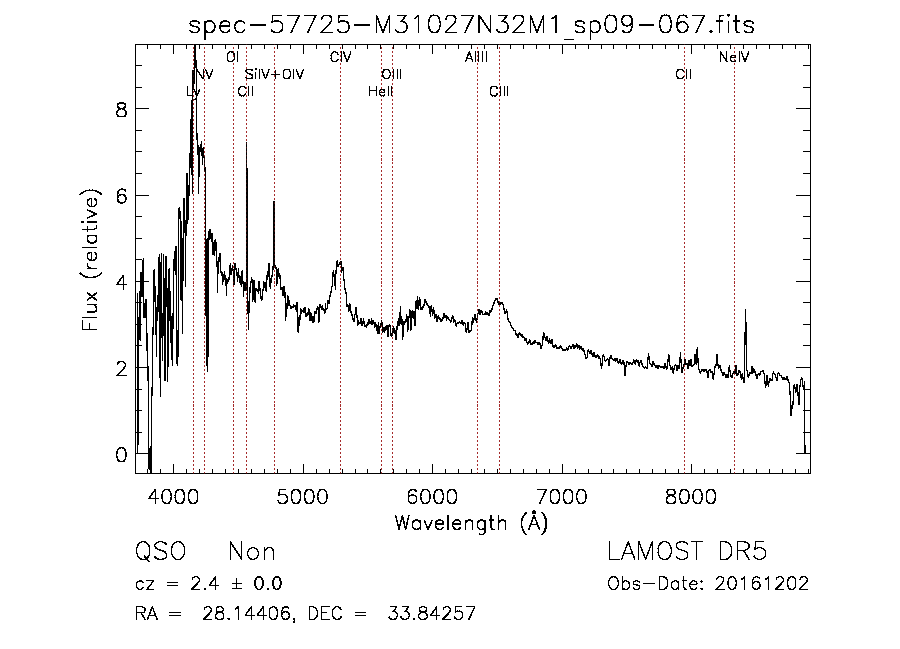}}
\stackunder[1pt]{J0156+3914}{\includegraphics[scale=0.18]{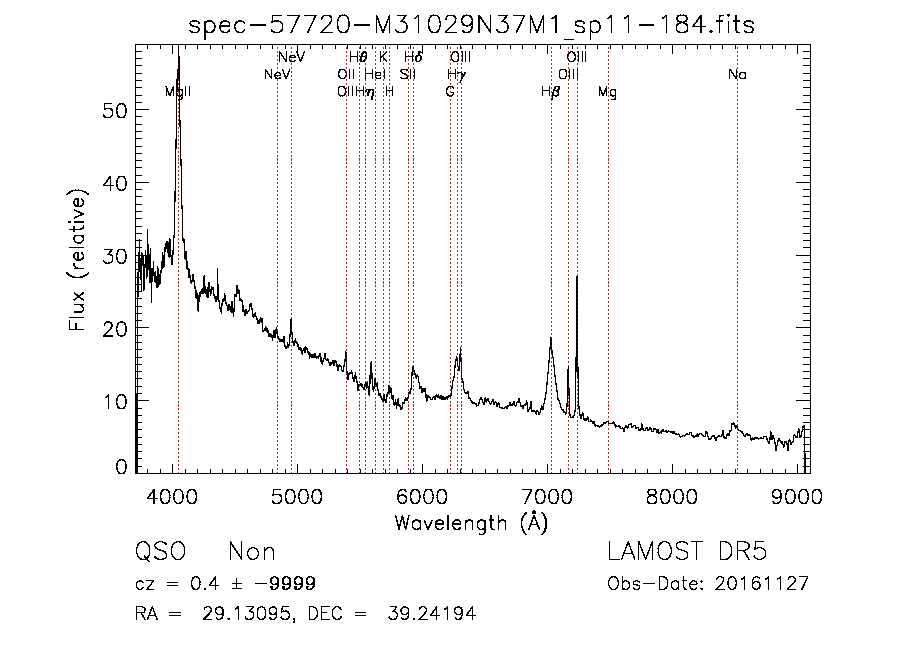}}
\stackunder[1pt]{J0204+3649}{\includegraphics[scale=0.18]{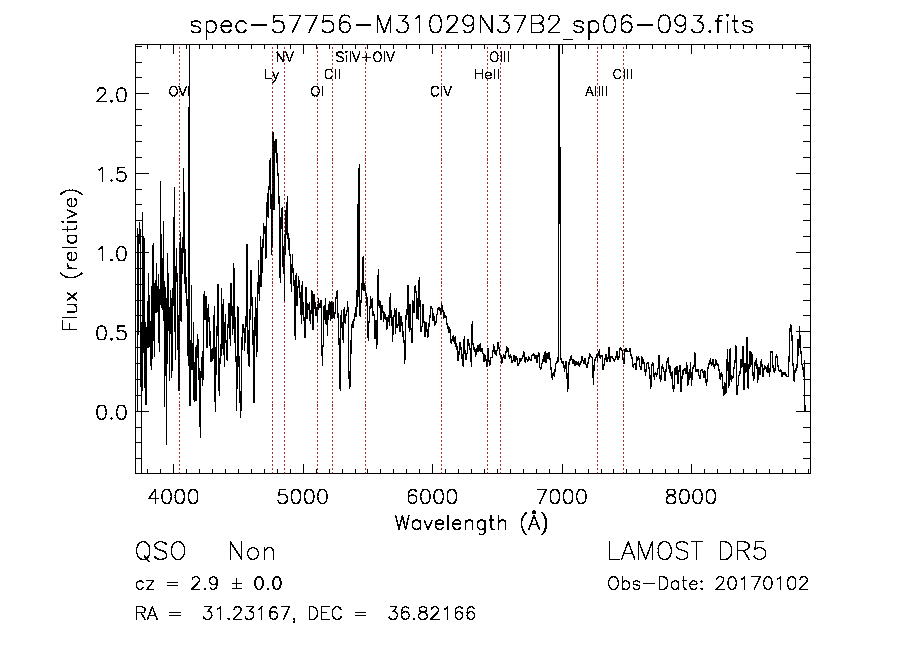}}
\stackunder[1pt]{J0954+1743}{\includegraphics[scale=0.18]{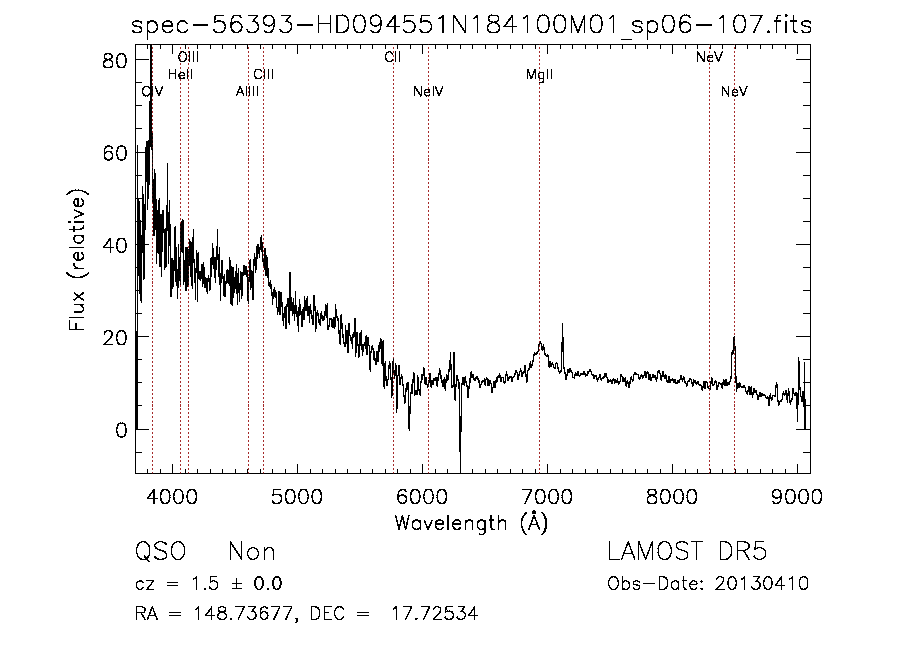}}
\stackunder[1pt]{J0945+3534}{\includegraphics[scale=0.18]{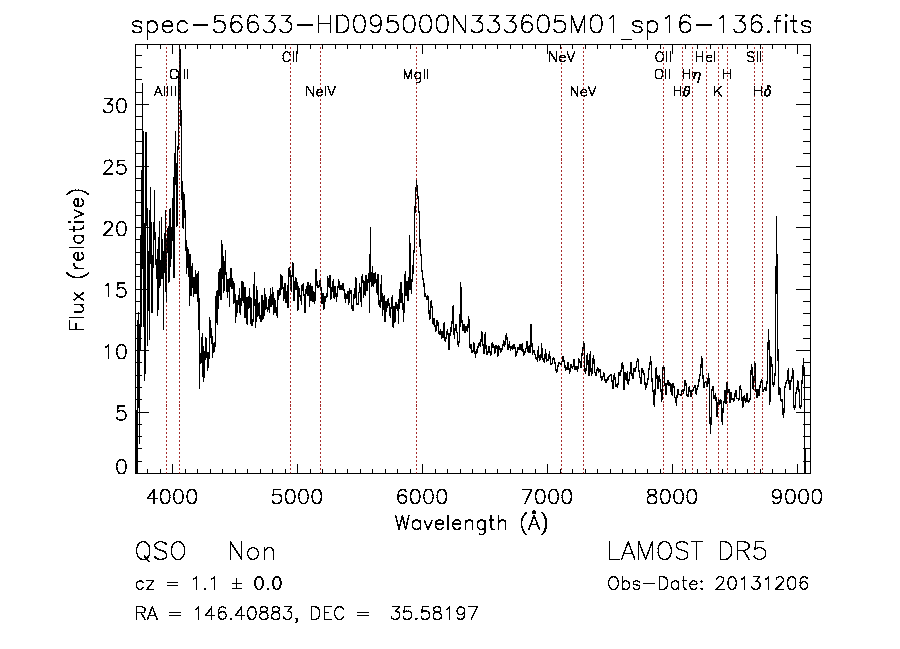}}
\stackunder[1pt]{J0958+3224}{\includegraphics[scale=0.18]{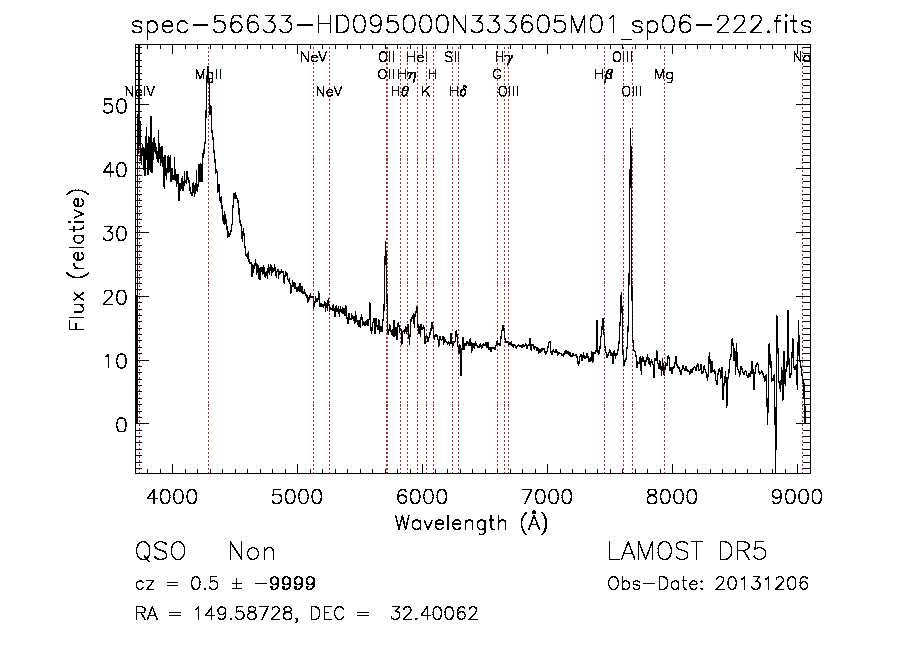}}
\stackunder[1pt]{J1000+0005}{\includegraphics[scale=0.18]{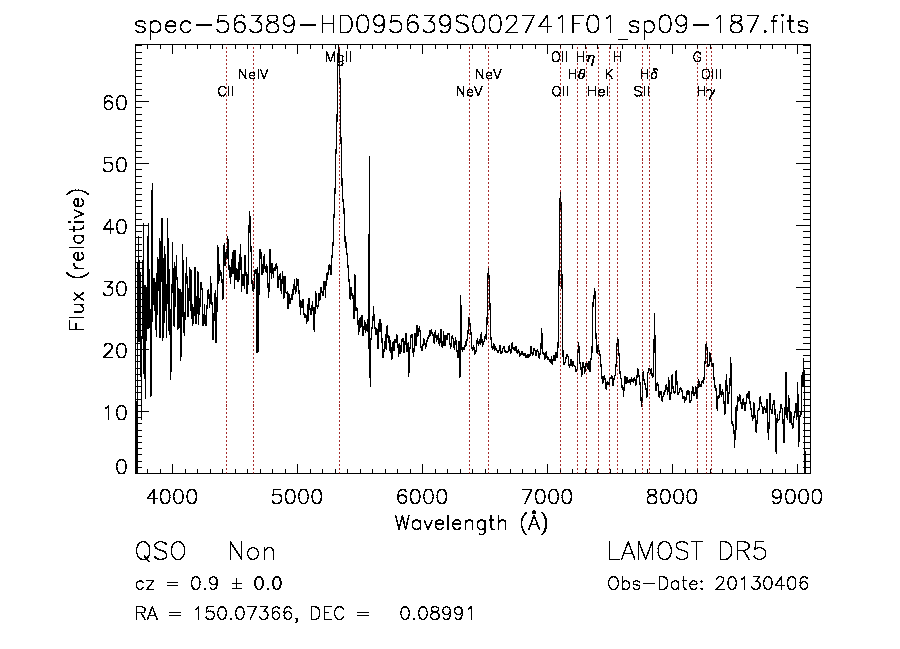}}
\caption{LAMOST DR5 optical spectra for the 27 ABC sources discussed in Sect. \ref{sec:sdss}. Full set of figures available online.}\label{fig:lamost_spectra}%
\end{figure*}

In this appendix we collect the optical spectra collected for ABC sources as discussed in Sect. \ref{sec:sdss}. In particular, in Fig. \ref{fig:sdss_spectra} we present the SDSS DR12 spectra, while in Fig. \ref{fig:lamost_spectra} we present the LAMOST DR5 spectra

\section{Catalog Comparison and Optical Properties}
In this appendix we present tables collecting properties of ABC sources. In particular in Table \ref{tab:comparison} we show the properties of the ABC sources found in other blazar candidate catalogs (4FGL, 3HSP, WIBRaLS2 and KDEBLLACS), while in Table \ref{tab:classification} we present the infrared and optical properties of ABC sources.

\begin{landscape}
\begin{table}[ht]
\caption{List of the ABC sources found in other blazar candidate catalogs. For each source we list the name (Name ALMA), the redshift (Redshift), the ABC source type (Object Type), the name (Name 4FGL) and source class (Type 4FGL) as listed in the 4FGL catalog, the name of the source as listed in the 3HSP catalog (Name 3HSP), the name (Name WIBRaLS2) and blazar candidate class (Type WIBRaLS2) as listed in the WIBRaLS2 catalog, and the name of the source as listed in the KDEBLLACS catalog (Name KDEBLLACS).}\label{tab:comparison}
\centering
\tiny{
\begin{tabular}{l|c|c|l|c|l|l|c|c|l}
\hline\hline
Name ALMA & Redshift & Object Type & Name 4FGL & Type 4FGL & Name 3HSP & Name WIBRaLS2 & Type WIBRaLS2 & Class WIBRaLS2 & Name KDEBLLACS \\
\hline
J0008-3945 & 1.9 & RS & 4FGL J0008.0-3937 & BCU & & J000809.17-394522.8 & BZQ & D & \\
J0011+0823 & 1.35 & QSO & & & & J001135.27+082355.5 & BZQ & D & \\
J0011-8443 & 0.6 & RS & & & & J001145.90-844320.0 & BZQ & C & \\
J0019-5641 & & RS & 4FGL J0019.2-5640 & BCU & & & & & \\
J0024-0811 & 2.067 & QSO & & & & J002400.67-081109.8 & BZQ & D & \\
J0024-4202 & 0.937 & QSO & & & & J002442.99-420203.9 & BZQ & D & \\
J0024-6820 & 0.354 & AGN & 4FGL J0023.7-6820 & BCU &  & J002406.72-682054.5 & BZQ & A & \\
J0025-4803 & & RS & 4FGL J0025.7-4801 & BCU & & J002545.81-480355.1 & BZQ & D & \\
J0028+2000 & 1.552 & QSO & 4FGL J0028.4+2001 & FSRQ & & J002829.81+200026.7 & BZQ & D & \\
J0030-0211 & 2.1 & QSO & 4FGL J0030.6-0212 & BCU & & & & & \\
J0034-4116 & 0.8 & RS & 4FGL J0034.0-4116 & BCU & & J003404.41-411619.4 & BZQ & C & \\
J0037+3659 & 0.366 & BL Lac & 4FGL J0037.6+3653 & FSRQ & & J003746.14+365910.9 & BZB & C & \\
J0039-2220 & 0.064 & BL Lac & 4FGL J0039.1-2219 & BCU & 3HSPJ003908.2-222001 & & & & \\
J0045-3705 & 1.0 & Blazar & 4FGL J0045.1-3706 & BCU & & & & & \\
J0046+2456 & 0.747 & QSO & & & & J004607.82+245632.5 & MIXED & C & \\
J0054-1953 & & RS & 4FGL J0054.8-1954 & BCU & & & & & \\
J0055-1217 &  & RS & 4FGL J0055.1-1219 & BCU &  &  &  &  & \\
J0056-4451 & 0.6 & Blazar & 4FGL J0056.6-4452 & BCU & & J005645.85-445102.0 & MIXED & C & \\
J0057+3021 & 0.017 & AGN & 4FGL J0057.7+3023 & RDG &  &  &  &  & 
\end{tabular}
}
\tablefoot{Full table available online.}
\end{table}
\end{landscape}

\begin{landscape}

\begin{table}[ht]
\caption{List of the infrared and optical properties of ABC blazar candidates. For each source we list the name (Name ALMA), the redshift (Redshift), the ABC source type (Object Type), the name of WISE counterpart (Name WISE), the blazar candidate class obtained from WISE colors (Type WISE), the name of the LAMOST DR5 counterpart (Name LAMOST), the redshift as evaluated from the LAMOST spectrum (redshift LAMOST) with the error in parenthesis, the name of the SDSS DR12 counterpart (Name SDSS), and the redshift as evaluated from the SDSS spectrum (redshift SDSS) with the error in parenthesis.}\label{tab:classification}
\centering
\tiny{
\begin{tabular}{l|c|c|l|c|l|c|c|l|c|c}
\hline\hline
Name ALMA & Redshift & Object Type & Name WISE & Type WISE & Name LAMOST & Type LAMOST & redshift LAMOST & Name SDSS & Type SDSS & redshift SDSS \\
\hline
J0002-2153 & & BL Lac & J000211.98-215310.0 & MIXED & & & & & & \\
J0006-2955 & 0.683 & QSO & J000601.12-295550.0 & BZU & & & & & & \\
J0008-3945 & 1.9 & RS & J000809.17-394522.8 & MIXED & & & & & & \\
J0011+0823 & 1.35 & QSO & J001135.27+082355.5 & MIXED & J001135.27+082355.6 & QSO & \(1.40175(0.0004989)\) & & & \\
J0011-4105 & & RS & J001152.39-410545.1 & MIXED & & & & & & \\
J0011-8443 & 0.6 & RS & J001145.90-844320.0 & MIXED & & & & & & \\
J0024-0811 & 2.067 & QSO & J002400.67-081109.8 & BZQ & & & & J002400.67-081110.2 & QSO & \(2.071(0.00032)\)\\
J0024-6820 & 0.354 & AGN & J002406.72-682054.5 & BZQ & & & & & & \\
J0024-4202 & 0.937 & QSO & J002442.99-420203.9 & BZQ & & & & & & \\
J0025+3919 & 1.946 & QSO & J002526.13+391935.6 & MIXED & & & & & & \\
J0025-4803 & & RS & J002545.81-480355.1 & MIXED & & & & & & \\
J0026-3512 & 1.299 & Blazar & J002616.39-351248.8 & MIXED & & & & & & \\
J0027+0929 & 1.4 & QSO & J002705.75+092957.4 & MIXED & & & & & & \\
J0028+2000 & 1.552 & QSO & J002829.81+200026.7 & MIXED & & & & J002829.81+200026.6 & QSO & \(1.552(0.0004)\)\\
J0029-0113 & 0.086 & QSO & J002900.97-011341.7 & BZG & & & & J002900.98-011341.7 & GALAXY & \(0.083(0.00002)\)\\
J0029+3456 & 0.517 & AGN & J002914.24+345632.2 & BZU & & & & J002914.24+345632.2 & GALAXY & \(0.517(0.00007)\) \\
J0034-0054 & 0.656 & QSO & & & & & & J003443.92-005412.9 & QSO & \(0.655(0.00014)\)\\
J0034-4116 & 0.8 & RS & J003404.41-411619.4 & MIXED & & & & & & \\
J0037+3659 & 0.366 & BL Lac & J003746.14+365910.9 & BZB & J003746.14+365910.8 & & & & & 
\end{tabular}
}
\tablefoot{Full table available online.}
\end{table}
\end{landscape}

\end{appendix}

\end{document}